\pdfoutput=1

\documentclass[twocolumn,notitlepage]{report}
\usepackage[left=2cm,right=1cm,top=2cm,bottom=2cm]{geometry}

\usepackage{authblk}

\usepackage{tikz}
\usetikzlibrary{decorations,decorations.pathmorphing,decorations.markings,calc,snakes,positioning,fixedpointarithmetic,shapes,shapes.geometric,patterns}

\usepackage{etoolbox}
\AtBeginEnvironment{tikzpicture}{\catcode`$=3 } 

\tikzset{alignmid/.style={baseline={([yshift=-.5ex]current bounding box.center)}}} 
\tikzset{every picture/.append style=alignmid}

\tikzset{
bottomzigzag/.style={postaction={draw,decorate, decoration={zigzag,amplitude=1pt,segment length=3pt,raise=1pt}}},
zigzag/.style={draw,decorate, decoration={zigzag,amplitude=1pt,segment length=3pt}},
rc/.style=rounded corners,
}

\tikzset{
    -|/.style={to path={-| (\tikztotarget)}},
    |-/.style={to path={|- (\tikztotarget)}},
}

\usepackage{ifthen}
\usepackage{xparse,pgffor}

\tikzset{enddots/.style= {postaction=decorate,decoration={name=markings,mark=at position 1 with{
    \ifdimless{\pgflinewidth}{1pt}{\newcommand\dotlw{1pt}}{\newcommand\dotlw{\pgflinewidth}}
    \ifdimless{\pgflinewidth}{1.5pt}{\newcommand\dotdist{1.5pt}}{\newcommand\dotdist{\pgflinewidth}}
    \fill (1*\dotdist,0) circle (0.5*\dotlw);
    \fill (2.5*\dotdist,0) circle (0.5*\dotlw);
    \fill (4*\dotdist,0) circle (0.5*\dotlw);
}}}}
\tikzset{startdots/.style= {postaction=decorate,decoration={name=markings,mark=at position 0 with{
   \ifdimless{\pgflinewidth}{1pt}{\newcommand\dotlw{1pt}}{\newcommand\dotlw{\pgflinewidth}}
   \ifdimless{\pgflinewidth}{1.5pt}{\newcommand\dotdist{1.5pt}}{\newcommand\dotdist{\pgflinewidth}}
   \fill (-1*\dotdist,0) circle (0.5*\dotlw);
   \fill (-2.5*\dotdist,0) circle (0.5*\dotlw);
   \fill (-4*\dotdist,0) circle (0.5*\dotlw);
}}}}

\tikzset{midlabeldistance/.store in=\midlabeldistance,midlabeldistance=0.2}
\tikzset{midlabelpos/.store in=\midlabelpos,midlabelpos=0.5}
\tikzset{midlabel/.style={decoration={markings, mark=at position \midlabelpos with {\node at (0,\midlabeldistance) {#1};}}, postaction={decorate}}}
\tikzset{midlabelr/.style={decoration={markings, mark=at position \midlabelpos with {\node at (0,-\midlabeldistance) {#1};}}, postaction={decorate}}}

\tikzset{cdirpos/.store in=\cdirpos,cdirpos=0.5}
\tikzset{cdir/.style={postaction=decorate,decoration={name=markings,mark=at position \cdirpos with{
    \coordinate (y0) at (#1 0.04,0.5*\pgflinewidth);
    \coordinate (y1) at (#1 0.04,-0.5*\pgflinewidth);
    \draw[solid,line width=0.7] (y0)--++(#1 -0.08,0.08) (y1)--++(#1 -0.08,-0.08);
  }}}}

\tikzset{ndirpos/.store in=\ndirpos,ndirpos=0.5}
\tikzset{ndir/.style={postaction=decorate,decoration={name=markings,mark=at position \ndirpos with{
    \draw[solid,semithick] (#1 -0.04,-0.08)arc(90-#1 180:90:0.08);
  }}}}

\tikzset{dualind/.style={postaction=decorate,decoration={name=markings,mark=at position \pgfdecoratedpathlength-0.08cm with{
    \draw[solid,line width=0.7] (-0.04,0)--(0.04,0.08) (-0.04,0)--(0.04,-0.08);
  }}}}

\tikzset{
}

\tikzset{
mark/.code={
\tikzset{postaction={/network/mark/.cd,#1,/tikz/.cd,decorate},decoration={name=markings,mark=at position \netmarkpos with{
\begin{scope}[netmarktrafo]
\netmarkcode
\end{scope}
}}}
\def\netmarkpos{0.5}
},
}

\def\netmarkpos{0.5}

\def\netmarkcode{}

\tikzset{
netmarktrafo/.style={},
netmarkstyle/.style={solid,semithick,sharp corners},
}

\pgfkeys{
/network/mark/.cd,
sty/.code={
\tikzset{netmarkstyle/.style={#1}}
},
asty/.code={
\tikzset{netmarkstyle/.append style={#1}}
},
f/.style={asty=fill},
p/.code={ 
\def\netmarkpos{#1}
},
poff/.code={ 
\def\netmarkposoff{#1cm}
},
e/.code={ 
\def\netmarkpos{\pgfdecoratedpathlength-0.005cm}

\tikzset{netmarktrafo/.append style={shift={(-\netmarkwidth,0)}}}
},
s/.code={ 
\def\netmarkpos{0.005cm}

\tikzset{netmarktrafo/.append style={shift={(\netmarkwidth,0)},xscale=-1,yscale=-1}}
},
a/.code={ 
\def\netmarkpos{\pgfdecoratedpathlength-0.005cm}

\tikzset{netmarktrafo/.append style={xscale=-1,shift={(-\netmarkwidth,0)}}}
},
b/.code={ 
\def\netmarkpos{0.005cm}

\tikzset{netmarktrafo/.append style={xscale=-1,shift={(\netmarkwidth,0),yscale=-1}}}
},
-/.code={ 
\tikzset{netmarktrafo/.append style={xscale=-1}}
},
r/.code={ 
\tikzset{netmarktrafo/.append style={yscale=-1}}
},
slab/.code={ 
\def\netmarkwidth{0}
\def\netmarkcode{
\node[inner sep=0.04cm,netmarkstyle,draw=none] (mylabelwidthtest) at (0,0){\phantom{#1}};
\path let \p1=(mylabelwidthtest.north east), \p2=(mylabelwidthtest.south east), \n1 = {max(abs(\y1),abs(\y2))} in node[inner sep=0.04cm,netmarkstyle] at (0,\n1) {#1};
}
},
lab/.code={ 
\def\netmarkwidth{0}
\def\netmarkcode{
\node[inner sep=0.04cm,anchor=\netmarkanchor] (mylabelwidthtest) at (0,0) {\phantom{#1}};
\draw[white] (mylabelwidthtest.\pgfdecoratedangle)--(mylabelwidthtest.\pgfdecoratedangle+180);
\node[inner sep=0.04cm,anchor=\netmarkanchor] at (0,0) {#1};
}
},
arr/.code={ 
\def\netmarkwidth{0.04}
\def\netmarkcode{\draw[netmarkstyle] (-0.04,0.08)--(0.04,0)--(-0.04,-0.08);}
},
rarr/.code={ 
\def\netmarkwidth{0.04}
\def\netmarkcode{\draw[netmarkstyle] (-0.04,-0.08)arc(90-180:90:0.08);}
},
dot/.code={ 
\def\netmarkwidth{0.08}
\def\netmarkcode{\draw[netmarkstyle] (0,0)circle(0.08);}
},
flag/.code={ 
\def\netmarkwidth{0.06}
\def\netmarkcode{\draw[netmarkstyle] (-0.06,0)--(0,0.09)--(0.06,0)--cycle;}
},
darr/.code={ 
\def\netmarkwidth{0.08}
\def\netmarkcode{\draw[netmarkstyle] (-0.04,0)--(0.04,0)--(-0.04,0.08)--cycle;}
},
hcirc/.code={ 
\def\netmarkwidth{0.1}
\def\netmarkcode{\draw[netmarkstyle] (-0.1,0) arc (180:0:0.1);}
},
bar/.code={ 
\def\netmarkwidth{0.05}
\def\netmarkcode{
\coordinate (a) at (0,-0.08cm-0.5*\pgflinewidth);
\coordinate (b) at (0,0.08cm+0.5*\pgflinewidth);
\draw[netmarkstyle] (a)--(b);
}
},
hbar/.code={ 
\def\netmarkwidth{0.05}
\def\netmarkcode{
\draw[netmarkstyle] (0, 0.5*\pgflinewidth)--++(0,0.12);
}
},
mill/.code={
\def\netmarkwidth{0.16}
\def\netmarkcode{
\draw[netmarkstyle] (0,-0.5*\pgflinewidth)--++(-0.08,-0.08)--++(0,0.08);
\draw[netmarkstyle] (0,0.5*\pgflinewidth)--++(0.08,0.08)--++(0,-0.08);
}
},
three dots/.code={
\def\netmarkwidth{0.2}
\def\netmarkcode{
\fill (-0.12,0) circle (0.5*0.05) (0,0) circle (0.5*0.05) (0.12,0) circle (0.5*0.05);
}
},
}

\tikzset{wid/.style={minimum width=#1cm}}
\tikzset{hei/.style={minimum height=#1cm}}

\tikzset{sx/.style={xshift=#1cm}}
\tikzset{sy/.style={yshift=#1cm}}

\tikzset{box/.style={draw,rectangle}}
\tikzset{fbox/.style={draw,rectangle, line width=1.1}}
\tikzset{roundbox/.style={draw,rectangle,rounded corners}}
\tikzset{froundbox/.style={draw,rectangle, rounded corners, line width=1.1}}
\tikzset{rounddiamond/.style={draw,diamond,rounded corners}}
\tikzset{dot/.style={draw, shape=circle, fill=black, scale=0.5}}

\tikzset{
netbox/.code={
\node[draw,netbdstyle] (\atomname) at (0,0) {#1};
\coordinate (\atomname-r) at (\atomname.east);
\coordinate (\atomname-l) at (\atomname.west);
\coordinate (\atomname-t) at (\atomname.north);
\coordinate (\atomname-b) at (\atomname.south);
\coordinate (\atomname-tr) at (\atomname.north east);
\coordinate (\atomname-br) at (\atomname.south east);
\coordinate (\atomname-tl) at (\atomname.north west);
\coordinate (\atomname-bl) at (\atomname.south west);
},
}

\tikzset{bdlw/.code={\tikzset{mybdstyle/.style={draw, line width=#1}}}}
\tikzset{bdcol/.code={\tikzset{mybdstyle/.append style={#1}}}}

\newcommand\setelements[1]{
\pgfkeys{/network/atom/.cd,#1}
}

\newcommand\setmarks[1]{
\pgfkeys{/network/mark/.cd,#1}
}

\newcommand\atoms[2]{
\foreach \name/\keys in {#2}{
\expandafter\atom\expandafter{\keys,#1}{\name}
}
}

\newcommand\atom[2]{
\def\atomname{#2}
\tikzset{
nettrafo/.style={},
netatompos/.style={},
netdeco/.style={},
netpostdeco/.style={},
}

\pgfkeys{/network/atom/.cd,#1}

\begin{scope}[netatompos] 
\begin{scope}[nettrafo] 
\netshapecoords 
\fill[netbackstyle] \netshapepath;
\clip \netshapepath;
\tikzset{netdeco}
\draw[netbdstyle] \netshapepath;
\end{scope}
\tikzset{netpostdeco} 
\end{scope}

}

\pgfkeys{
/network/atom/.cd,
square/.code={

\def\netshapepath{(-\tempsize,-\tempsize)rectangle (\tempsize,\tempsize)}
\def\netshapecoords{
\node[rectangle,wid=2*\tempsize,hei=2*\tempsize,inner sep=0,transform shape](\atomname)at(0,0){};
\coordinate(\atomname-c) at (0,0);
\coordinate(\atomname-r) at (\tempsize,0);
\coordinate(\atomname-l) at (-\tempsize,0);
\coordinate(\atomname-t) at (0,\tempsize);
\coordinate(\atomname-b) at (0,-\tempsize);
\coordinate(\atomname-br) at (\tempsize,-\tempsize);
\coordinate(\atomname-tr) at (\tempsize,\tempsize);
\coordinate(\atomname-bl) at (-\tempsize,-\tempsize);
\coordinate(\atomname-tl) at (-\tempsize,\tempsize);
}},
circ/.code={

\def\netshapepath{(0,0)circle(\tempsize)}
\def\netshapecoords{
\node[circle,wid=2*\tempsize,hei=2*\tempsize,inner sep=0,transform shape](\atomname)at(0,0){};
\coordinate(\atomname-c) at (0,0);
\coordinate(\atomname-r) at (\tempsize,0);
\coordinate(\atomname-l) at (-\tempsize,0);
\coordinate(\atomname-t) at (0,\tempsize);
\coordinate(\atomname-b) at (0,-\tempsize);
}},
triang/.code={

\def\netshapepath{(-30:\tempsize)--(90:\tempsize)--(-150:\tempsize)--cycle}
\def\netshapecoords{
\node[regular polygon,regular polygon sides=3,wid=2*\tempsize,inner sep=0,transform shape](\atomname)at(0,0){};
\coordinate(\atomname-c) at (0,0);
\coordinate(\atomname-cr) at (-30:\tempsize);
\coordinate(\atomname-cl) at (-150:\tempsize);
\coordinate(\atomname-ct) at (90:\tempsize);
\coordinate(\atomname-mb) at (-90:0.5*\tempsize);
\coordinate(\atomname-mr) at (30:0.5*\tempsize);
\coordinate(\atomname-ml) at (150:0.5*\tempsize);
}},
diamond/.code={

\def\netshapepath{(0,-\tempsize)--(\tempsize,0)--(0,\tempsize)--(-\tempsize,0)--cycle}
\def\netshapecoords{
\node[rotate=45,rectangle,wid=sqrt(2)*\tempsize,hei=sqrt(2)*\tempsize,inner sep=0,transform shape](\atomname)at(0,0){};
\coordinate(\atomname-c) at (0,0);
\coordinate(\atomname-r) at (\tempsize,0);
\coordinate(\atomname-l) at (-\tempsize,0);
\coordinate(\atomname-t) at (0,\tempsize);
\coordinate(\atomname-b) at (0,-\tempsize);
}},
pentagon/.code={

\def\netshapepath{(-126:\tempsize)--(-54:\tempsize)--(18:\tempsize)--(90:\tempsize)--(162:\tempsize)--cycle}
\def\netshapecoords{
\node[regular polygon,regular polygon sides=5,wid=2*\tempsize,inner sep=0,transform shape](\atomname)at(0,0){};
\coordinate(\atomname-c) at (0,0);
\coordinate (\atomname-mb)at(-90:{\tempsize*cos(36)});
\coordinate (\atomname-mbr)at(-18:{\tempsize*cos(36)});
\coordinate (\atomname-mtr)at(54:{\tempsize*cos(36)});
\coordinate (\atomname-mtl)at(126:{\tempsize*cos(36)});
\coordinate (\atomname-mbl)at(-162:{\tempsize*cos(36)});
\coordinate (\atomname-cbr)at(-54:\tempsize);
\coordinate (\atomname-cr)at(18:\tempsize);
\coordinate (\atomname-ct)at(90:\tempsize);
\coordinate (\atomname-cl)at(162:\tempsize);
\coordinate (\atomname-cbl)at(-126:\tempsize);
}},
void/.code={
\def\netshapepath{(0,0)}
\def\netshapecoords{
\coordinate(\atomname) at (0,0);
\coordinate(\atomname-c) at (0,0);
}},
labbox/.code={
\def\netshapepath{(0,0)}
\def\netshapecoords{}
\tikzset{netpostdeco/.append style={netbox=#1}}
},
}

\pgfkeys{
/network/atom/.cd,
p/.code={\tikzset{netatompos/.append style={shift={(#1)}}}},
xscale/.code={\tikzset{nettrafo/.append style={xscale=#1}}},
yscale/.code={\tikzset{nettrafo/.append style={yscale=#1}}},
scale/.code={\tikzset{nettrafo/.append style={scale=#1}}},
rot/.code={\tikzset{nettrafo/.append style={rotate=#1}}},
hflip/.style={xscale=-1},
vflip/.style={yscale=-1},
big/.style={scale=3/2},
small/.style={scale=2/3},
huge/.style={scale=9/4},
tiny/.style={scale=4/9},
flat/.style={yscale=2/3},
fflat/.style={yscale=4/9},
}

\tikzset{
netbdstyle/.style={line width=0.15em}, 
netdecstyle/.style={},
netpostdecstyle/.style={},
netbackstyle/.style={white},
}
\pgfkeys{
/network/atom/.cd,
bdstyle/.code={\tikzset{netbdstyle/.style={#1}}},
bdastyle/.code={\tikzset{netbdstyle/.append style={#1}}},
decstyle/.code={\tikzset{netdecstyle/.style={#1}}},
postdecstyle/.code={\tikzset{netpostdecstyle/.style={#1}}},
backstyle/.code={\tikzset{netbackstyle/.style={#1}}},
whiteblack/.style={decstyle=white,backstyle=black},
nobd/.code={bdstyle={line width=0}},
}

\tikzset{
netbscope/.code={\begin{scope}[#1]},
netescope/.code={\end{scope}},
}

\pgfkeys{
/network/atom/.cd,
bscope/.code={\tikzset{netdeco/.append style={netbscope={#1}}}},
escope/.code={\tikzset{netdeco/.append style={netescope}}},
dec/.style 2 args={bscope={#1},#2,escope}, 
vbar/.style={dec={rotate=90}{hbar}},
dcross/.style={dec={rotate=45}{hbar},dec={rotate=-45}{hbar}},
cross/.style={hbar,vbar},
sect6/.style={sect=60},
sect8/.style={sect=45},
sect10/.style={sect=36},
}

\def\regdec#1{\pgfkeys{/network/atom/.cd,#1/.code={\tikzset{netdeco/.append style={net#1}}}}}
\regdec{all}\regdec{rhalf}\regdec{rquart}\regdec{brquart}\regdec{dot}\regdec{spiral}\regdec{swirl}\regdec{hstripe}\regdec{hbar}\regdec{rrey}
\pgfkeys{/network/atom/.cd,sect/.code={\tikzset{netdeco/.append style={netsect=#1}}}}

\tikzset{
netall/.code={\fill[netdecstyle] (-0.3,-0.3)rectangle (0.3,0.3);}, 
netrhalf/.code={\fill[netdecstyle] (0,-0.3)rectangle (0.3,0.3);}, 
netrquart/.code={\fill[netdecstyle] (0.075,-0.3)rectangle (0.3,0.3);}, 
netbrquart/.code={\fill[netdecstyle] (0,0)rectangle (0.3,-0.3);}, 
netsect/.code={\fill[netdecstyle] (0,0)--(0,-0.3)arc(-90:-90+#1:0.3)--cycle;}, 
netdot/.code={\fill[netdecstyle] (0,0)circle(0.07);}, 
netspiral/.code={\draw[netdecstyle] plot [variable=\t,domain=0:4] ({0.075*\t*cos(pi*(\t-0.5) r)},{0.075*\t*sin(pi*(\t-0.5) r)});}, 
netswirl/.code={\fill[netdecstyle] plot [variable=\t,domain=0:2] ({0.15*\t*cos(pi*(\t-0.5) r)},{0.15*\t*sin(pi*(\t-0.5) r)}) arc(-90:-450:0.3)--cycle;}, 
nethstripe/.code={\fill[netdecstyle] (-0.3,-0.05)rectangle(0.3,0.05);}, 
nethbar/.code={\draw[netdecstyle] (-0.3,0)--(0.3,0);}, 
netrrey/.code={\draw[netdecstyle] (0,0)--(0.3,0);} 
}

\pgfkeys{
/network/atom/.cd,
postbscope/.code={\tikzset{netpostdeco/.append style={netbscope={#1}}}},
postescope/.code={\tikzset{netpostdeco/.append style={netescope}}},
postdec/.style 2 args={postbscope={#1},#2,postescope}, 
arc/.code={\tikzset{netpostdeco/.append style={netarc=#1}}},
lab/.code={\tikzset{netpostdeco/.append style={netlab={#1}}}},
shadecirc/.code={\tikzset{netpostdeco/.append style=netshadecirc}},
shaderect/.code={\tikzset{netpostdeco/.append style={netshaderect={#1}}}},
debug/.code={\tikzset{netpostdeco/.append style=netdebug}},
}

\tikzset{
netlab/.code={
\pgfkeys{/network/atom/lab/.cd,#1}
\node[netpostdecstyle] at (\ifdefined\netlabpos\netlabpos\else\netlabang:\netlabdist\fi) {\netlabwrap{\netlabtext}};
},
netarc/.code args={#1:#2:#3}{
\draw[netpostdecstyle] (#1:#3) arc (#1:#2:#3);
},
netshadecirc/.code= {
\fill[opacity=0.4,netpostdecstyle] (0,0)circle(0.4);
},
netshaderect/.code= {
\fill[rc,opacity=0.4,netpostdecstyle] ($-1*(#1)$) rectangle (#1);
},
netdebug/.code= {
\node[red] at (0,0){\atomname};
}
}

\def\netlabwrap#1{#1}
\pgfkeys{
/network/atom/lab/.cd,
t/.code={\def\netlabtext{#1}}, 
p/.code={\def\netlabpos{#1}}, 
ang/.code={\def\netlabang{#1}},
dist/.code={\def\netlabdist{#1}},
wrap/.code={\def\netlabwrap##1{#1}}, 
}

\usepackage{amsmath}
\usepackage{amssymb}
\usepackage{enumitem}
\usepackage{tabularx}
\usepackage{ltxtable}
\usepackage{longtable}
\usepackage{booktabs}
\usepackage{bbold}
\usepackage{braket}
\usepackage{mathtools}
\usepackage{cuted}
\usepackage{amsfonts}
\usepackage{blkarray}
\usepackage{mathabx}
\usepackage{babel}
\usepackage{xcolor}

\definecolor{darkblue}{RGB}{46,48,146}
\definecolor{quantumviolet}{HTML}{53257F} 
\definecolor{quantumgray}{HTML}{555555} 

\usepackage[
        colorlinks=true,
        linkcolor=quantumviolet,
        citecolor=quantumviolet,
        pdfborderstyle={0 0 0}
        ]{hyperref}

\usepackage{amsthm}
\theoremstyle{definition}
\newtheorem{myobs}{Observation}
\newtheorem{myrem}{Remark}
\newtheorem{mycom}{Comment}

\newtheorem{myexmp}{Example}
\newtheorem{mydef}{Definition}

\usepackage{enumitem}
\setlist[description]{topsep=5pt, itemsep=-3pt, font={\normalfont\itshape}}

\newcommand\id{\text{id}}
\newcommand\unitob{1}
\newcommand\catc{\mathcal{C}}
\newcommand\catk{\mathcal{K}}

\newcommand\homo{\text{hom}}
\newcommand\ob{\text{ob}}

\newcommand\cattwo{\textbf{two}}
\newcommand\catmat{\textbf{matrix}}
\newcommand\catfinset{\textbf{finset}}
\newcommand\catset{\textbf{set}}

\newcommand{\cop}{\operatorname{Copy}}

\newcommand\dat{\operatorname{Dat}}
\newcommand\idop{\operatorname{id}}
\newcommand\calg{\mathcal{G}}
\newcommand\calm{\mathcal{M}}
\newcommand\zz{\mathbb{Z}}

\newcommand{\mdef}[1]{{\bf #1}} 
\newcommand{\tdef}[2]{\hypertarget{#2}{{\bf #1}}} 

\newcommand\tpart[2]{#1_#2} 
\newcommand\schur{\operatorname{Schur}}
\newcommand\pfaff{\operatorname{Pf}}

\allowdisplaybreaks[1]

\tikzset{
ind/.style={mark={lab=$#1$,a}}, 
startind/.style={mark={lab=$#1$,b}}, 
mapping second 0dat/.style={red},
}

\setelements{
dlab/.style={lab/wrap=$##1$}, 
rotlab/.style={lab/wrap=\rotatebox{90}{$##1$}}, 
prod/.style={small,circ},
dual/.style={small,fflat,square},
copy/.style={small,circ,all},
data/.style={small,square,all},
mapping/.style={flat,backstyle={fill=black},decstyle={red},nobd,square,dec={rotate=-90}{rhalf}},
mapping second prod/.style={bdastyle=red,prod},
mapping second copy/.style={bdastyle=red,decstyle=red,copy},
mapping second data/.style={bdastyle=red,decstyle=red,data},
mapping second dual/.style={bdastyle=red,dual},
category data/.style={bdastyle=blue,decstyle=blue,data},
top half/.style={square,dec={rotate=90}{rhalf}},
}

\setmarks{
ar/.style={arr,asty=fill},
}

\setelements{
representation/.style={square,rhalf,dlab},
comcoalg/.style={small,circ,all,dlab},
comalg/.style={small,circ,dlab},
coalgebra/.style={whiteblack,circ,swirl,dlab},
antipode/.style={diamond,rhalf,dlab},
algebra/.style={circ,swirl,dlab},
tfrobenius/.style={circ,dlab},
hopfcounit/.style={small,circ,all,dlab},
hopfbraiding/.style={square,dcross,dec={rotate=45}{sect=90},dlab},
drinfeldelement/.style={dlab,lab={t=u},lab/dist=0.35,square},
idrinfeldelement/.style={dlab,lab={t=u^{-1}},lab/dist=0.5,square},
basis change/.style={square,rhalf,dlab},
}

\setelements{
1axiomlab/.style={draw,blue,inner sep=0,circle},
pairingdot/.style={tiny,circ,all,dlab,lab/p=-90:0.3},
}

\tikzset{irrep/.style={line width=1.5}} 
\tikzset{intern/.style={}} 
\tikzset{internal/.style={densely dotted}} 
\tikzset{fusion/.style={decorate, decoration={snake,amplitude=0.04cm, segment length=0.1cm}}} 
\tikzset{real/.style={}}
\tikzset{cstar/.style={}}
\tikzset{complex/.style={densely dotted}}
\tikzset{level/.style={densely dotted}} 
\tikzset{realified/.style={line width=1.5}}
\tikzset{multiplicity/.style={decorate, decoration={snake,amplitude=0.04cm, segment length=0.1cm}}}

\tikzset{mfat/.style={line width=1.3}} 
\tikzset{fat/.style={line width=2}} 
\tikzset{arrpos/.store in=\arrpos,arrpos=0.5}
\tikzset{arr/.style={postaction=decorate,decoration={name=markings,mark=at position \arrpos with{ 
    \draw[solid,line width=0.7] (#1 0.04,0)--(-#1 0.04,0.08) (#1 0.04,0)--(-#1 0.04,-0.08);
  }}}}

\tikzset{realife/.style={postaction=decorate,decoration={name=markings,mark=at position \pgfdecoratedpathlength -\realifeoff cm with{
    \draw[solid,semithick,sharp corners] (-#1 0.04,0.07)--(#1 0.04,0)--(-#1 0.04,-0.07);
  }}}}

\tikzset{staros/.style={postaction=decorate,decoration={name=markings,mark=at position 0.22cm with{
    \draw[fill,solid,semithick,sharp corners] (-0.06,0)--(0,#1 0.09)--(0.06,0)--cycle;
  }}}}
\tikzset{staroe/.style={postaction=decorate,decoration={name=markings,mark=at position \pgfdecoratedpathlength -0.22cm with{
    \draw[fill,solid,semithick,sharp corners] (-0.06,0)--(0,#1 0.09)--(0.06,0)--cycle;
  }}}}

\begin{document}
\title{Tensor types and their use in physics}
\author{Andreas Bauer\thanks{andibauer@zedat.fu-berlin.de} }
\author{Alexander Nietner}
\affil{Dahlem Center for Complex Quantum Systems, Freie Universit{\"a}t Berlin,
Arnimallee 14, 14195 Berlin}

\maketitle
\begin{strip}
\centering
\begin{minipage}{0.8\textwidth}
\begin{abstract}
The content of this paper can be roughly organized into a three-level hierarchy of generality. At the first, most general level, we introduce a new language which allows us to express various categorical structures in a systematic and explicit manner in terms of so-called \emph{2-schemes}. Although 2-schemes can formalize categorical structures such as symmetric monoidal categories, they are not limited to this, and can be used to define structures with no categorical analogue. Most categorical structures come with an effective graphical calculus such as string diagrams for symmetric monoidal categories, and the same is true more generally for interesting 2-schemes. In this work, we focus on one particular non-categorical 2-scheme, whose instances we refer to as \emph{tensor types}. At the second level of the hierarchy, we work out different flavors of this 2-scheme in detail. The effective graphical calculus of tensor types is that of tensor networks or Penrose diagrams, that is, string diagrams without a flow of time. As such, tensor types are similar to compact closed categories, though there are various small but potentially important differences. Also, the two definitions use completely different mechanisms despite both being examples of 2-schemes. At the third level of the hierarchy, we provide a long list of different families of concrete tensor types, in a way which makes them accessible to concrete computations, motivated by their potential use in physics. Different tensor types describe different types of physical models, such as classical or quantum physics, deterministic or statistical physics, many-body or single-body physics, or matter with or without symmetries or fermions.
\end{abstract}
\end{minipage}
\end{strip}

\tableofcontents
\chapter{Introduction}
Linear algebra is undoubtedly one of the most important mathematical tools applied in physics and other sciences, playing a major role in subjects as diverse as electrical engineering, classical field theory, statistical or quantum physics, data analysis, or machine learning. In many cases, it is necessary to use multi-index linear algebra, where the vectors under consideration live in vector spaces which are tensor products of smaller vector spaces, and are called \emph{tensors}. Linear expressions over such tensors are called \emph{tensor networks}, and can be denoted elegantly via a graphical notation often attributed to Penrose.

Large tensor networks and the corresponding Penrose notation have been used excessively for the analysis of quantum condensed matter systems in the last decades. Numerical algorithms like the \emph{density matrix renormalization group} (DMRG) rely on representing the ground state of a local Hamiltonian as a tensor networks known as \emph{MPS} or \emph{PEPS}, and have proven extremely powerful for the computation of ground state expectation values. Those highly successful methods originating from condensed matter physics have started to spread to many other fields. On the other hand, the general notion of tensor networks in particular in combination with Penrose notation is a very useful tool for getting analytical insights.

Diagrammatic calculi are also well-known in the seemingly uncorrelated field of \emph{category theory}. There, so-called \emph{string diagrams} represent sequences of compositions and tensor products of morphisms in a \emph{monoidal category}. Such string diagrams allow to define algebraic structures such as Frobenius algebras for arbitrary monoidal categories. Proofs which can be carried out on the diagrammatic level, such as the antipode of a Hopf algebra being unique, then automatically hold for any monoidal category.

The aim of this work is to combine the practicality of tensor networks with the abstractive power of category theory. Just like string diagrams can be interpreted in any monoidal category, the different ways to make sense of arbitrary tensor-network diagrams will be called \emph{tensor types}. In principle, category theory as such already does the job: Tensor networks are similar to string diagrams of \emph{symmetric monoidal categories}, and we could use the latter as a formalization of the notion of tensor types. However, those string diagrams still differ from tensor-network diagrams in that they need to obey a consistent ``flow of time''. While there are approaches to get rid of this flow of time via a ``closed compact'' structure, we choose a different path, and define tensor types in a way agnostic to any flow of time from the very beginning.

There is a natural way in which both category theory and the theory of tensor types can be used in physics: Different tensor types or (symmetric) monoidal categories formalize different types of theories, different types of physical descriptions, or different types of matter. Whether we have deterministic or statistical classical physics, quantum physics, whether we have many-body physics or single-particle physics, systems with bosonic or fermionic particles, systems with external symmetries, interacting of free models, any type of physical model can be formalized by a tensor type or monoidal category. Those models can be formalized as \emph{tensor-network models} consisting of a set of tensors. Different ways of probing the model are translated into different tensor networks, and the predictions of the model are obtained by simply evaluating the tensor network.

For physical systems with a flow of time and a notion of causality, symmetric monoidal categories with a terminal object are the correct kind of structure. The idea of formalizing physical models using category theory has been partly explored, especially in the context of \emph{categorical quantum mechanics}. However, when it comes to physical systems without time, such as models describing any sort of equilibrium like a classical or quantum thermal ensemble or ground state, tensor types are more natural. Any tensor type is also a symmetric monoidal category, but vice versa not all symmetric monoidal categories also yield tensor types.

This work is structured as follows. In this introduction, we review conventional tensor networks, develop some intuition for the general idea of tensor types, and provide an introduction to concepts from category theory which will play a role for the formal definition of tensor types. We also describe how tensor types can be used to formalize physical models.

In Chapter~\ref{sec:ttypes_definition}, we introduce the general mathematical framework of tensor types. To do so, we use a systematic graphical language which also allows us to express structures known from monoidal category theory. We also look at how the effective diagrammatic tensor-network calculus can be derived from the general theory of tensor types, introduce the notion of mappings between different tensor types, and compare our definition to structures familiar from category theory.

In Chapter~\ref{sec:enzyclopedia}, we give a large collection of concrete tensor types, for which we each give the definition, the mappings to other tensor types, and its significance in physics.

In Chapter~\ref{sec:outlook}, we mention a few more tensor types which we did not have time to work out in detail in the proceeding chapter.

\section{Introduction to array tensor networks}
\label{sec:intro_tensor_networks}
In this section we will recap the conventional notion of tensor networks and the associated graphical calculus, also known as \emph{Penrose notation}. To distinguish conventional tensors from the generalized notion of tensors of arbitrary \emph{tensor types}, we will sometimes explicitly refer to them as \emph{array tensors} here or in the following sections.

\subsection{Array tensors}
Simply put, an array tensor is a multi-dimensional array, that is, a list, matrix, or higher-dimensional block of real or complex numbers. More precisely, it is a collection consisting of a number $A_{i_0,i_1,i_2,\ldots}$ for each choice of $i_0\in \{0,\ldots,n_0\}$, $i_1\in \{0,\ldots,n_1\}$, $i_2\in \{0,\ldots,n_2\}$, $\ldots$. The subscript positions $i_0$, $i_1$, $i_2$, $\ldots$ are called \emph{indices} and the integers $n_0$, $n_1$, $n_2$, $\ldots$ are their \emph{bond dimensions}.


The simplest case is that of tensor with one index, which is also known as \emph{vector}. E.g., an example for a vector $v_i$ whose single index $i$ has a bond dimension $3$ is given by
\begin{equation}
v=(0.3,\ 0.25,\ 0.45)\;,
\end{equation}
meaning that $v_0=0.3$, $v_1=0.25$ and $v_2=0.45$. Physically, $v$ could for example represent a probability distribution over three elements. Abstractly, instead of writing $v_i$, we could also use a graphical notation,
\begin {equation}
\label{eq:vector_notation}
\begin{tikzpicture}
\node[draw,roundbox] (t) {$v$};
\draw (t.west)edge[ind=i]++(-0.3,0);
\end{tikzpicture}\;.
\end{equation}
It is common to express that, e.g., the $i=2$ entry of $v_i$ is $0.45$ as
\begin{equation}
\label{eq:vector_entry_notation}
\begin{tikzpicture}
\node[draw,roundbox] (t) {$v$};
\draw (t.west)edge[ind=2]++(-0.3,0);
\end{tikzpicture} = 0.45\;.
\end{equation}
We should bear in mind though that whereas Eq.~\eqref{eq:vector_notation} can be used for 1-index tensors of arbitrary tensor types, Eq.~\eqref{eq:vector_entry_notation} makes sense only for array tensors.

It doesn't matter how precisely we draw the diagram, only that there is a box for $v$ with a line sticking out. We also don't need the label $i$ at this point, so
\begin{equation}
\begin{tikzpicture}
\node[draw,roundbox] (t) {$v$};
\draw[rc, ind=i](t.west)--++(-0.4,0)--++(0.2,0.8)--++(0.4,-0.2);
\end{tikzpicture}
\qquad,\text{or}\qquad
\begin{tikzpicture}
\node[draw,roundbox, rotate=-45] (t) {$v$};
\draw[rc](t.north)--++(0.25, 0.25)--++(0.3,-0.2);
\end{tikzpicture}
\end{equation}
denote the same vector. Other vectors cannot only be distinguished by the name $v$, but also by the choice of shape, so
\begin{equation}
\begin{tikzpicture}
\node[draw,circle, fill=orange] (t) {$v$};
\draw[rc, ind=i](t.45)--++(0.4,0.4)--++(-0.2,0.2);
\end{tikzpicture}
\qquad,\text{or}\qquad
\begin{tikzpicture}
\node[draw,star, star points=7, rotate=-45,line width=2] (t) {$v$};
\draw[rounded corners, ind=i](t.283)--++(-0.25, -0.25)--++(0.3,-0.2);
\end{tikzpicture}
\qquad,\text{or}\qquad
\begin{tikzpicture}
\node[draw, kite, rc, fill=cyan!40!gray,line width=2] (t) {};
\draw[rc, ind=i](t.143)--++(-0.25, 0.25)--++(-0.3,-0.4);
\end{tikzpicture}
\end{equation}
denote three different vectors.

The next complicated example is a \emph{matrix}, such as
\begin{equation}
\label{eq:matrix_example}
\begin{gathered}
M = ((0, 1, 0), (1, 0, 0), (0, 0, 1))\quad ,\quad\text{or}\\
\begin{pmatrix}
0& 1& 0\\
1& 0& 0\\
0& 0& 1
\end{pmatrix}\;,
\end{gathered}
\end{equation}
such that $M_{01}=1$, $M_{11}=0$, etc. $M$ can also be seen as a linear map on $\mathbb{R}^3$, which happens to be a permutation of the three basis vectors. $M_{ij}$ is a tensor with two indices, each of bond dimension $3$.
In the graphical language, we represent $M$ by a shape with two lines sticking out. Again, the precise way in which we draw the diagram doesn't matter,
\begin{equation}
\begin{tikzpicture}
\node[draw,roundbox] (t) {$M$};
\draw[rounded corners, ind=i] (t.west)--++(-0.3,0);
\draw[rounded corners, ind=j] (t.east)--++(+0.2,0);
\end{tikzpicture}
\quad,\text{or}\quad
\begin{tikzpicture}
\node[draw,roundbox] (t) {$M$};
\draw[rounded corners, ind=i] (t.west)--++(-0.3,0)--++(0,0.3);
\draw[rounded corners, ind=j] (t.east)--++(+0.2,0)--++(0.1,-0.2);
\end{tikzpicture}\;.
\end{equation}
Note that the labels $i$ and $j$ are not strictly necessary. The two indices are not distinguished by those labels, but by the positions on the boundary of the shape where the lines are sticking out. This is not so different from the notation $M_{ij}$, where the two indices correspond to the first and second subscript position, whereas the labels $i$ and $j$ are arbitrary.
We could also write
\begin{equation}
\begin{tikzpicture}
\node[draw,roundbox] (t) {$M$};
\draw[rounded corners, ind=i] (t.west)--++(-0.3,0)--++(0,0.3);
\draw[rounded corners, ind=j] (t.east)--++(+0.2,0)--++(0.1,-0.2);
\end{tikzpicture}
=
\begin{cases}
1 & (i,j)\in\{(0,1), (1,0), (2,2)\}\\
0 & \text{else} 
\end{cases}
\end{equation}
for the concrete definition of $M$ in Eq.~\eqref{eq:matrix_example}. As this is a notation for the concrete contents of $M$, it is restricted to array tensors and not applicable to tensors of arbitrary tensor types.

Finally, let us consider a 3-dimensional array,
\begin{equation}
t=(((0,1),(2,3),(4,5)),((6,7),(8,9),(10,11)))\;,
\end{equation}
such that, e.g., $t_{020}=4$, $t_{001}=1$ or $t_{120}=10$. $t_{ijk}$ is a tensor with three indices of bond dimensions $2$, $3$, and $2$, respectively.
Graphically, we can write
\begin{equation}
\begin{tikzpicture}
\node[draw,roundbox] (t) {$t$};
\draw[red, rounded corners, ind=i] (t.west)--++(-0.3,0)--++(0,0.3);
\draw[rounded corners, ind=j] (t.north)--++(0,0.3)--++(.3,-0.1);
\draw[red, rounded corners, ind=k] (t.east)--++(+0.2,0)--++(0.1,-0.2);
\end{tikzpicture}
\quad,\text{or}\quad
\begin{tikzpicture}
\node[draw,roundbox] (t) {$t$};
\draw [red, ind=i, rounded corners](t.west)--++(-0.2,0)--++(-0.1, 0.4);
\draw [ind=j, rounded corners](t.north)--++(0,0.2)--++(-0.3,0)--++(-0.3,-0.6);
\draw [red, ind=k, rounded corners](t.east)--++(0.3,0)--++(0.2,0.2);
\end{tikzpicture}\;.
\end{equation}
The first, second and third indices are the ones sticking out of the left, top, and right of the box. We used a different color for the first and third index to indicate that they have a different bond dimension. We might also choose different line styles instead of colors.

In the conventional notation, we can change the order of indices, i.e., perform \emph{index permutations}, such as
\begin{equation}
\widetilde{t}_{ijk}=t_{jik}\;.
\end{equation}
In the graphical notation, there is no actual ordering of indices, however, we can change the positions where the indices stick out of the boundary of the shape, e.g.,
\begin{equation}
\begin{tikzpicture}
\atoms{labbox=$\widetilde{t}$, bdstyle=rc}{0/}
\draw[red] (0-l)edge[ind=k]++(180:0.4) (0-r)edge[ind=i]++(0:0.4);
\draw (0-b)edge[ind=j]++(-90:0.4);
\end{tikzpicture}
\coloneqq
\begin{tikzpicture}
\atoms{labbox=$t$, bdstyle=rc}{0/}
\draw[red] (0-l)edge[ind=i]++(180:0.4) (0-r)edge[ind=k]++(0:0.4);
\draw (0-t)edge[ind=j]++(90:0.4);
\end{tikzpicture}
\end{equation}

Another simple operation is given by grouping multiple indices of an array into a single one, which we'll refer to as \emph{blocking}. This is also known as \emph{flattening} or \emph{reshaping} an array. For example, we can block the indices labeled $j$ and $k$ of $t_{ijk}$ above, yielding a reshaped tensor
\begin{equation}
t'_{ix} = t_{i\Phi_0(x)\Phi_1(x)}\;.
\end{equation}
The bond dimension of the blocked index $x$ is the product of all the individual bond dimensions $j$ and $k$. Then, there is a canonical identification $\Phi$ which maps every configuration of $x$ to a configuration $\Phi_0(x)$ of $j$ and a configuration $\Phi_1(x)$ of $k$. In the above example, $j$ and $k$ have bond dimension $2$ and $3$, resulting in a bond dimension of $6$ for the blocked index. In the abstract graphical notation, we can denote the blocking by
\begin{equation}
\begin{tikzpicture}
\node[draw,roundbox] (t) {$t'$};
\draw[red, rounded corners, ind=a] (t.west)--++(-0.3,0)--++(0,0.3);
\draw[rounded corners, line width=2, ind=bc] (t.east)--++(0.3,0);
\end{tikzpicture}
\coloneqq
\begin{tikzpicture}
\node[draw,roundbox] (t) {$t$};
\draw[red, rounded corners, ind=a] (t.west)--++(-0.3,0)--++(0,0.3);
\draw[rounded corners, ind=b] (t.north)--++(0,0.3)--++(.5,0);
\draw[red, rounded corners, ind=c] (t.east)--++(+0.3,0);
\end{tikzpicture}\;.
\end{equation}
This new bond dimension of $6$ is indicated by a different, fat, line style.



\subsection{Array tensor networks}
Arrays can be seen as a data type, depending on the corresponding bond dimensions. Now we are going add operations for this data type, functions which map from one or multiple arrays to a new array. Specifically, we mean basic operations of linear algebra, such as matrix multiplication, applying a vector to a matrix, and so on. 
E.g., we can apply $M$ to $v$ to get a new vector,
\begin{equation}
\label{eq:matrix_vector_mult}
(Mv)_i=\sum_j M_{ij}v_j = (0.25,0.3,0.45)\;,
\end{equation}
which physically means that we're applying the permutation $M$ to the probability distribution $v$. Another example is taking the outer product of twice $v$,
\begin{equation}
\label{eq:vector_outer_product}
(v\otimes v)_{ij}=v_i v_j
=
\begin{pmatrix}
0.0625 & 0.075 & 0.1125\\
0.075 & 0.09 & 0.135\\
0.1125 & 0.135 & 0.2025
\end{pmatrix}
\;,
\end{equation}
which physically represents the joint probability distribution of two independent systems with probabilities $v$ each. Moreover, the inner product of $v$ with itself is
\begin{equation}
\label{eq:vector_inner_product}
\langle v,v\rangle=\sum_i v_iv_i= 0.355\;,
\end{equation}
and the trace of $M^2$ is
\begin{equation}
\label{eq:matrix_product_trace}
\operatorname{Tr}(M^2)=\sum_{i,j} M_{ij} M_{ji} = 3\;.
\end{equation}
All the above operations are of the following form: First, we take products of all entries of a collection of arrays, like $M$ and $v$, twice $v$, or twice $M$ in the examples above. For two arrays, such an operation is known as \emph{Kronecker product}. Second, we force some pairs of indices to have the same value, and then sum over all those values, known as \emph{Einstein summation}. We can also write down expressions involving tensors with more indices, such as
\begin{equation}
\label{eq:3index_trace}
m_i=\sum_at_{aia} =(7,11,15)\;,
\end{equation}
or
\begin{equation}
\label{eq:t_m_contraction}
\begin{multlined}
s_{abc}=\sum_iM_{ib}t_{aic}\\
=(((2,3),(0,1),(4,5)),((8,9),(6,7),(10,11)))\;.
\end{multlined}
\end{equation}
We should keep in mind that an Einstein summation over two indices only makes sense if they have the same dimensions. E.g., we can't Einstein sum over the first and second index of $t$, and also not over an index of $M$ and the first or third index of $t$. A more complicated example is given by
\begin{equation}
\label{eq:intro_tn_example}
r_{q\alpha} = \sum_{a,b,c,d,e,f,g,x,y}t_{xa\alpha} t_{xby} t_{ycq} M_{xbd} M_{fe} M_{fg} v_g \delta_{dace}\;,
\end{equation}
where
\begin{equation}
\delta_{dace}=
\begin{cases}
1 & d=a=c=e\\
0 & \text{otherwise} 
\end{cases}\;.
\end{equation}
All the above expressions can be very neatly expressed in a graphical notation. For each copy of a tensor in the expression, we draw one copy of the corresponding shape, and for each index pair that is summed over, we simply connect the two lines sticking out. The connecting lines will be called \emph{bonds}. Indices which are not summed over remain ``loose line ends'' and will be called \emph{open indices}. E.g., we can write the expression in Eq.~\eqref{eq:matrix_vector_mult} as
\begin{equation}
\begin{tikzpicture}
\atoms{labbox=$M$, bdstyle=rc}{M/}
\draw[rc, ind=i] (M-l)--++(-0.3,0)--++(0,0.3);
\atoms{labbox=$v$, bdstyle=rc}{v/p={1,0}}
\draw (v-l)--(M-r);
\end{tikzpicture}
\quad,\quad\text{or}\quad
\begin{tikzpicture}
\atoms{labbox=$M$, bdstyle=rc}{M/}
\draw[rc, ind=i] (M-l)--++(-0.3,0);
\atoms{labbox=$v$, bdstyle={rc,rotate=90}}{v/p={0.5,0.6}}
\draw[rc] (v-l)|-(M-r);
\end{tikzpicture}\;.
\end{equation}
Again, it doesn't matter how precisely we draw the diagram, analogously to how in the conventional notation it doesn't matter in which order we write down the individual tensors. The expression in Eq.~\eqref{eq:vector_outer_product} is denoted by
\begin{equation}
\begin{tikzpicture}
\atoms{labbox=$v$,bdstyle=rc}{0/, 1/p={0,0.6}}
\draw (0-l)edge[ind=i]++(180:0.4) (1-l)edge[ind=j]++(180:0.4);
\end{tikzpicture}\;,
\end{equation}
Eq.~\eqref{eq:vector_inner_product} by
\begin{equation}
\begin{tikzpicture}
\atoms{labbox=$v$,bdstyle=rc}{0/, 1/p={0,0.6}}
\draw[rc] (0-l)--++(180:0.4)|-(1-l);
\end{tikzpicture}\;,
\end{equation}
Eq.~\eqref{eq:matrix_product_trace} by
\begin{equation}
\begin{tikzpicture}
\atoms{labbox=$M$,bdstyle=rc}{0/, 1/p={1,0}}
\draw[rc] (0-r)--(1-l) (1-r)-|++(0.3,0.7)-|([sx=-0.3]0-l)--(0-l);
\end{tikzpicture}\;,
\end{equation}
Eq.~\eqref{eq:3index_trace} by
\begin{equation}\label{kindergarden:t_contracted}
\begin{tikzpicture}
\node[draw,roundbox] (m) {$m$};
\draw[rounded corners, ind=i](m.west)--++(-0.3,0);
\end{tikzpicture}
\coloneqq
\begin{tikzpicture}
\node[draw,roundbox] (t) {$t$};
\draw[rounded corners, ind=i] (t.north)--++(0,0.3)--++(-.3,-0.1);
\draw[red, rounded corners] (t.east)--++(+0.2,0)--++(0,-0.4)--++(-0.8,0)--++(0,0.4)--(t.west);
\end{tikzpicture}\;,
\end{equation}
and Eq.~\eqref{eq:t_m_contraction} by
\begin{equation}
\begin{tikzpicture}
\node[draw,roundbox, minimum width=1cm] (s) {$s$};
\draw [ind=b, rounded corners]([xshift=0.0cm] s.north)--++(0,0.33);
\draw [red, ind=c, rounded corners]([xshift=0.3cm] s.north)--++(0.0,0.3);
\draw [red, ind=a, rounded corners]([xshift=-0.3cm] s.north)--++(0.0,0.3);
\end{tikzpicture}
\coloneqq
\begin{tikzpicture}
\node[draw,roundbox] (t) {$t$};
\node[draw, roundbox, above right=0.3 and 0.3 of t] (M) {$M$};
\draw[red, rounded corners, ind=a] (t.west)--++(-0.3,0)--++(0,0.3);
\draw[rounded corners] (t.north)--++(0,0.545)--(M.west);
\draw[red, rounded corners, ind=c] (t.east)--++(+0.2,0)--++(0.1,-0.2);
\draw[ind=b] (M.east)--++(0.3,0);
\end{tikzpicture}\;.
\end{equation}
This example shows that in an equation, the index labels tell us how to match up the open indices on both sides. The more complicated example in Eq.~\eqref{eq:intro_tn_example} can be drawn as
\begin{equation}
\begin{tikzpicture}
\atoms{labbox=$r$,bdstyle=rc}{0/}
\draw[red] (0-l)edge[ind=\alpha]++(180:0.4) (0-r)edge[ind=q]++(0:0.4);
\end{tikzpicture}
\coloneqq
\begin{tikzpicture}
\node[draw, roundbox] (t1) at (0,0) {$t$};
\node[draw, roundbox] (t2) at (1,2) {$t$};
\node[draw, roundbox, rotate=180] (t3) at (-.7,0) {$t$};
\node[draw, roundbox, rotate=45] (M1) at (1.2,0.3) {$M$};
\node[draw, roundbox, rotate=90] (M2) at (0,0.8) {$M$};
\node[draw, roundbox, rotate=45] (M3) at (-1,1.1) {$M$};
\node[draw, roundbox, rotate=45] (v) at (1.7,0.8) {$v$};
\atoms{circ,small,all,lab={t=$\delta$,p=135:0.3}}{delta/p={0,1.8}}

\draw[red] (t1.west)--(t3.west);
\draw (t1.north)--(M2.west);
\draw (M2.east)--(delta);
\draw[rounded corners] (M3.east)--++(0.2,0.2)--(delta);
\draw[rounded corners] (delta)--++(0,1.0)--++(1,0)--(t2.north);
\draw[red, rounded corners] (t1.east)--++(0.5,0)--++(0,2)--(t2.west);
\draw[rounded corners] (M1.west)--++(-0.2,-0.2)--++(0,-0.8)--++(-2.2,0)--++(0,1.6)--(M3.west);
\draw (v.west)--(M1.east);
\draw[red, rounded corners, ind=\alpha] (t3.east)--++(-0.25,0)--++(0,-0.2);
\draw[red, ind=q] (t2.east)--++(0.3,0);
\draw[rounded corners] (t3.north)--++(0,-0.2)--++(1.2,0)--++(0,2)--(delta);
\end{tikzpicture}\;.
\end{equation}
This example shows that two lines crossing doesn't mean anything. All that matters is how many copies of which tensors are involved, and which indices of which tensors are connected. The example also demonstrates how the use of different colors or line styles for different bond dimensions helps us to prevent accidentally writing Einstein summations between indices with unequal bond dimensions.

\subsection{Consistency of the diagrammatic calculus}
In this section we will talk about the reason why we can use the diagrammatic calculus of tensor networks. In general, the computation described by a tensor network consists of two elementary operations, together with blocking and index permutations, namely an individual Kronecker product, and an individual Einstein summation. A tensor network can be translated into a sequence of elementary operations in different ways. To ensure that the diagrammatic notation makes sense, all the different sequences have to yield the same result.

Consider the graphical notation for the tensor product of, e.g., a vector $v$ and a matrix $M$,
\begin{equation}
\begin{tikzpicture}
\node[draw,roundbox] (M) {$M$};
\draw[rounded corners, ind=i] (M.west)--++(-0.3,0)--++(0,0.3);
\draw[rounded corners, ind=j] (M.east)--++(+0.2,0)--++(0.1,-0.2);

\node[draw,roundbox, right=0.7 of M] (v) {$v$};
\draw[rounded corners, ind=k](v.west)--++(-0.4,0)--++(0.2,0.8)--++(0.4,-0.2);
\end{tikzpicture}
\quad,\quad\text{or}\quad
\begin{tikzpicture}
\node[draw,roundbox, rotate=45] (M) {$M$};
\draw[rounded corners, ind=i] (M.west)--++(-0.25,-0.25)--++(0.2,-0.2);
\draw[rounded corners, ind=j] (M.east)--++(+0.25,0.25)--++(0.1,-0.2);

\node[draw,roundbox, above left=0.2 and 0.1 of M] (v) {$v$};
\draw[rounded corners, ind=k](v.west)--++(-0.4,0);
\end{tikzpicture}\;.
\end{equation}
From the diagrams it is impossible to determine any order of $v$ and $M$, which is fine as we have
\begin{equation}
M_{ij}v_k=v_kM_{ij}
\end{equation}
in conventional notation, due to the commutativity of the multiplication of real or complex numbers.

As another example, consider the tensor product of three vectors $v$, $w$, $x$,
\begin{equation}
\begin{tikzpicture}
\atoms{labbox=$v$,bdstyle=rc}{v/}
\atoms{labbox=$w$,bdstyle=rc}{w/p={0.8,0}}
\atoms{labbox=$x$,bdstyle=rc}{x/p={1.6,0}}
\draw (v-t)edge[ind=a]++(90:0.4) (w-t)edge[ind=b]++(90:0.4) (x-t)edge[ind=c]++(90:0.4);
\end{tikzpicture}\;.
\end{equation}
Even if we fix the ordering of $v$, $w$, $x$, in terms of elementary operations, this could either be decomposed as
\begin{equation}
v_a (w_b x_c)\;,
\end{equation}
or as
\begin{equation}
(v_a w_b) x_c\;,
\end{equation}
which is fine as the two expressions are the same.


As a next example, let us consider a new $4$-index tensor $T_{abcd}$,
\begin{equation}
\begin{tikzpicture}
\node[draw,roundbox] (T) {$T$};
\draw[rounded corners, ind=a] (T.west)--++(-0.3,0);
\draw[rounded corners, ind=b] (T.north)--++(0,0.3);
\draw[rounded corners, ind=c] (T.east)--++(0.3,0);
\draw[rounded corners, ind=d] (T.south)--++(0,-0.3);
\end{tikzpicture}\;,
\end{equation}
and a tensor network with Einstein summations over both the index pair $a$ and $c$, as well as $b$ and $d$,
\begin{equation}
\begin{tikzpicture}
\node[draw,roundbox] (T) {$T$};
\draw[rounded corners] (T.south)--++(0,-0.25)--++(0.35,0)--++(0, 1.)--++(-0.35,0)--(T.north);
\draw[rounded corners] (T.west)--++(-0.25,0)--++(0,0.35)--++(1.,0)--++(0,-0.35)--(T.east);
\end{tikzpicture}\;.
\end{equation}
The result of the tensor network might be different depending on which of the two Einstein summations we perform first. Luckily, this is not the case, since we have
\begin{equation}
\sum_i\left(\sum_jT_{jiji}\right)=\sum_{i,j} T_{jiji} = \sum_j\left(\sum_iT_{jiji}\right)\;.
\end{equation}

As a final example let us consider a tensor network which includes both a Kronecker product and an Einstein summation,
\begin{equation}\label{kindergarden:tensor_product_contraction}
\begin{tikzpicture}
\node[draw,roundbox] (t) {$t$};
\draw[red, rounded corners] (t.east)--++(0.25,0)--++(0,-0.5)--++(-.9,0)--++(0,0.5)--(t.west);
\draw[ind=i] (t.north)--++(0,0.3);

\node[draw,roundbox, above right=0.2 and 0.4 of t] (M) {$M$};
\draw[rounded corners, ind=k] (M.east)--++(0.3,0)--++(0.1,-0.2);
\draw[rounded corners, ind=j] (M.west)--++(-0.25,0)--++(0,0.2);
\end{tikzpicture}\;.
\end{equation}
Again, the result of the tensor network could depend on whether we first perform the Einstein summation or the tensor product, but it doesn't, as
\begin{equation}
\sum_l \left(t_{lil}M_{jk}\right) = \left(\sum_lt_{lil}\right)M_{jk}\;.
\end{equation}


\section{Introduction to tensor types}
The idea rough behind tensor types is a very simple one. Ordinary array tensor networks as revised in Section~\ref{sec:intro_tensor_networks} describe calculations on arrays involving Einstein summations and Kronecker products. The notation is consistent because those two operations obey certain properties, such that it doesn't matter in which order we apply the different operations. That way, we will get the same result when evaluating a tensor network no matter how precisely we perform the computation.

We can now choose different data structures than arrays and different operations than Kronecker products and Einstein summations. The generalized arrays will be called \emph{tensors}, the generalized Kronecker product \emph{tensor product}, and the generalized Einstein summation \emph{contraction}. Assume that the tensor product and contraction have the same properties as described for the Kronecker product and Einstein summation in the section before. Then can use the same tensor-network diagrams to also describe computations of those different data structures and operations. Such a structure, consisting of a generalized definition of tensor, tensor product, and contraction will be called a \emph{tensor type}.

Let us consider a very simple example for this, where the data structures themselves can be represented by diagrams. We will call those tensors \emph{pairing tensors}. First, for arrays each index carries a bond dimensions, whereas for pairing tensors each index carries a \emph{number of dots}. The data of a pairing tensor is then a way to pair up all the dots. E.g., let us consider a 5-index tensor
\begin{equation}
\begin{tikzpicture}
\atoms{labbox=$T$, bdastyle={circle}}{t/}
\draw (t)edge[ind=a]++(90:0.7) (t)edge[ind=b]++(162:0.7) (t)edge[ind=c]++(-126:0.7) (t)edge[ind=d]++(-54:0.7) (t)edge[ind=e]++(18:0.7);
\end{tikzpicture}
\end{equation}
where $a,b,c,d,e$, have numbers of dots $2,3,1,5,3$. Such a tensor is a way to pair up the overall set of $2+3+1+5+3$ dots, which can be represented graphically, e.g.,
\begin{equation}
\begin{tikzpicture}
\foreach \i/\x/\b in {a/0/1,b/1/2,c/2.3/0,d/3/4,e/4.9/2}{
\foreach \e in {0,...,\b}{\atoms{lab={t=\e},pairingdot}{\i\e/p={\x+0.3*\e,0}}};
\draw[decorate,decoration={brace,amplitude=3pt,mirror,raise=4pt},yshift=0.05cm](\x-0.2,-0.5)--++(\b*0.3+0.4,0) node [midway,yshift=-0.5cm] {$\i$};
};
\draw[out=90, in=90](a0)to(c0);
\draw[out=90, in=90](a1)to(d3);
\draw[out=90, in=90](b0)to(d1);
\draw[out=90, in=90](b1)to(e1);
\draw[out=90, in=90](b2)to(d0);
\draw[out=90, in=90](d2)to(d4);
\draw[out=90, in=90](e0)to(e2);
\end{tikzpicture}\;.
\end{equation}
For each index we draw the corresponding number of dots, and connect paired dots by lines. Be aware that this diagram is not a tensor-network diagram, but a concrete representation of the data of a pairing tensor, analogous to the list of all the numbers in a concrete array. Note that in contrast to a continuous set of array tensors for fixed bond dimensions, there is only a finite set of pairing tensors for a fixed numbers of dots. Concretely, for an even total number of dots $2n$, there are $(2n)!/(n!2^n)$ pairing tensors, and for odd total numbers there are no pairing tensors at all.

Even though the bond dimension for array tensors and the number of dots of pairing tensors are analogues, there is one important difference: When we block two array tensor indices into one, the bond dimension of the combined index is the product of the individual bond dimensions. However, when we block two pairing tensor indices, the number of dots of the combined index is the \emph{sum} of the two individual numbers. E.g., the $5$-dot index $d$ and $3$-dot index $e$ together yield a $8$-dot index. Similarly, the total number of dots of a pairing tensor is the sum of the numbers of the individual indices, whereas the total bond dimension of an array tensor is the product of all the bond dimensions.

We still have to specify what the tensor product and contraction for pairing tensors are. The tensor product simply consists in taking the disjoint union of the two pairing diagrams. The contraction of two indices $x$ and $y$ of a tensor consists in connecting all the dots of $x$ with the corresponding dots of $y$. Erasing the dots from the diagram yields a new pairing diagram for the remaining indices, after possibly removing disconnected loops. E.g., consider contracting the indices $b$ and $e$ of the pairing tensor $T$ above. First, we connect the dots of $b$ and $e$ with dashed lines,
\begin{equation}
\begin{tikzpicture}
\foreach \i/\x/\b in {a/0/1,c/2.3/0,d/3/4}{
\foreach \e in {0,...,\b}{\atoms{lab={t=\e},pairingdot}{\i\e/p={\x+0.3*\e,0}}};
};
\foreach \i/\x/\b in {b/1/2,e/4.9/2}{
\foreach \e in {0,...,\b}{\atoms{pairingdot}{\i\e/p={\x+0.3*\e,0}}};
};
\draw[out=90, in=90](a0)to(c0);
\draw[out=90, in=90](a1)to(d3);
\draw[out=90, in=90](b0)to(d1);
\draw[out=90, in=90](b1)to(e1);
\draw[out=90, in=90](b2)to(d0);
\draw[out=90, in=90](d2)to(d4);
\draw[out=90, in=90](e0)to(e2);
\draw[dashed,out=-90, in=-90](b0)to(e0);
\draw[dashed,out=-90, in=-90](b1)to(e1);
\draw[dashed,out=-90, in=-90](b2)to(e2);
\end{tikzpicture}\;.
\end{equation}
Removing the dots and continuing he solid lines via the dashed lines we get a new pairing,
\begin{equation}
\begin{tikzpicture}
\foreach \i/\x/\b in {a/0/1,c/1/0,d/1.7/4}{
\foreach \e in {0,...,\b}{\atoms{lab={t=\e},pairingdot}{\i\e/p={\x+0.3*\e,0}}};
\draw[decorate,decoration={brace,amplitude=3pt,mirror,raise=4pt},yshift=0.05cm](\x-0.2,-0.5)--++(\b*0.3+0.4,0) node [midway,yshift=-0.5cm] {$\i$};
};
\draw[out=90, in=90](a0)to(c0);
\draw[out=90, in=90](a1)to(d3);
\draw[out=90, in=90](d0)to(d1);
\draw[out=90, in=90](d2)to(d4);
\end{tikzpicture}\;.
\end{equation}

In fact, the line of $T$ connecting dot number $1$ of $b$ with dot number $1$ of $e$ yields a loop together with the dashed line, which we simply neglect.

As a last step we need to argue why pairing tensors with those operations fulfill the same diagrammatic calculus as array tensors. As for array tensors themselves, this is very easy to see. If we take the disjoint union of a bunch of pairing diagrams, the result will always look the same, independent on the way in which we build the overall disjoint union from single pair-wise disjoint unions. Also, connecting dots of a pairing diagram with dashed lines is obviously independent from taking the disjoint union with another pairing diagram. Alternatively, there is a description of the evaluation a tensor network of pairing tensors which does not require representing it as a sequence of individual contractions and tensor products. First, we replace every tensor by the corresponding pairing diagram. Then, we replace every bond in the tensor network by dashed lines connecting the corresponding pairing dots. In the end, we join up lines such that we end up with a pairing diagram for only the dots at the open indices.

Let us consider \emph{prefactor} pairing tensors, a variation of pairing tensors which will be useful for an example of a tensor mapping in Section~\ref{sec:tensor_mappings}. The data of a normalized pairing tensor does not only consist of a pairing diagram, but also of a \emph{prefactor}, which is a single real number. Now, taking the tensor product of two prefactor pairing tensors consists in taking the disjoint union of the two pairings, as well as the product of the two prefactors. When performing a contraction, we have to multiply the prefactor by a factor of $2$ for every loop that we neglect in the process. This again defines a tensor type due to the commutativity and associativity of the product of real numbers, and the fact that the number of neglected loops in the evaluation of a tensor network does not depend on how we represent the network as a sequence of individual contractions and tensor products.

\section{Introduction to monoidal categories}
In the previous section we described a diagrammatic calculus which has different interpretations in terms of different tensor types. An area of mathematics where such abstract diagrammatic calculi are already established and widely in use is the field of \emph{(monoidal) category theory}. Let us therefore give us here a quick introduction to category theory, formulated in a way that already makes it more compatible to the language we use to define tensor types later in Chapter~\ref{sec:ttypes_definition}.

\subsection{Categories}

\begin{mydef}
\label{def:category}
A \emph{category} $\catc$ is defined by the following.
\begin{itemize}
\item A collection \footnote{Formally, this ``collection'' is a \emph{class}, such that we can make sense of categories like the category of all sets. However, if we want to use the category for concrete computations as is the focus of this paper, we should think of the objects as forming some data type, which would be a set.} of \emph{objects} $\ob$.
\item For every object $o\in O$, a collection (or, class) of \emph{morphisms} $\hom(o)$.
\item For every triple of objects $s, m, t\in \ob$, a function $\circ$ called \emph{composition},
\begin{equation}
\begin{aligned}
\circ_{s,m,t}: \hom(s,m)\times\hom(m,t)&\rightarrow\hom(s,t)\;,\\
(g,f)&\mapsto g\circ f\;.
\end{aligned}
\end{equation}
\item For every object $o\in \ob$, a morphism
\begin{equation}
\idop_o\in\hom(o,o)
\end{equation}
called the \emph{identity morphism}.
\item For all $s,m,n,t\in\ob$ and $h\in\hom(s,m)$, $g\in\hom(m,n)$ and  $f\in\hom(n,t)$, the composition must be \emph{associative},
\begin{equation}
(h\circ g)\circ f=h\circ(g\circ f)\;.
\end{equation}
\item For every $s, t\in \ob$ and every $f\in\hom_\catc(s,t)$, the identity and composition must satisfy
\begin{equation}
\label{eq:identity_property}
f=\idop_s\circ f = f\circ \idop_t\;.
\end{equation}
\end{itemize}
\end{mydef}

We notice that the definition of a category involves entities on two levels: The ``first'' level of objects, and the ``second'' level of morphisms. We will therefore say that categories are an example of a \emph{2-scheme}. In general, we will call any sort of data on the first level of a 2-scheme (for a category, the objects) \emph{0-data}, and any sort of data on the second level (for a category, the morphisms) \emph{1-data}. In general, the 1-data depends on a choice of 0-data. For a category, the morphisms depend on a choice of two objects. On each level of a 2-scheme, we also have functions acting on 0-data, or 1-data, called \emph{1-functions} and \emph{2-functions}, respectively. For a category, there are no 1-functions \footnote{We can always copy and discard objects, which might be viewed as two 1-functions. However, those are not functions which can be freely chosen when defining a category, but they are fixed as soon as we write down the collection of objects.} acting on objects, but two 2-functions acting on (tuples of) morphisms, namely the composition and the identity \footnote{The identity is not a function on the morphism but a special choice of morphism. However, it can be formally interpreted as a function from an empty tuple of morphisms, which is an 1-element set, to a single morphism, such that the domain of this function is exactly the identity morphism.}. Note that the 2-functions are not individual functions, but depend on a choice of 0-data. E.g., the composition in a category depends on a choice of three objects $s, m, t$. At last, we can also have axioms at each level, called \emph{1-axioms}, and \emph{2-axioms}, respectively. For categories, there are no 1-axioms, but two 2-axioms, namely associativity of the composition, and the defining property of the identity morphism in Eq.~\eqref{eq:identity_property}. Again, the 2-axioms are equations that hold for all different choices of 0-data and 1-data.

Let us give some examples for categories.




\begin{myexmp}
A very simple example is the finite category $\cattwo$ given by the following.
\begin{itemize}
\item The objects/0-data form a finite set,
\begin{equation}
\ob=\{\bullet, \star\}\;.
\end{equation}
\item Also the morphisms/1-data form finite sets,
\begin{equation}
\begin{aligned}
\homo(\star,\bullet)=\{\bigtriangleup \}\;,\\
\homo(\bullet,\star)=\{\}\;,\\
\homo(\star,\star)=\{\idop_\star\}\;,\\
\homo(\bullet,\bullet)=\{\idop_\bullet\}\;.
\end{aligned}
\end{equation}
\item The identity morphisms were indicated in the previous point.
\item There is only one possibility for the composition as for any two objects, the set of morphisms contains at most one element.
\item It is easy to see that the morphisms are chosen in precisely that way that the $2$-axioms are satisfied.
\end{itemize}
\end{myexmp}


\begin{myexmp}
An important example is the category $\catset$ of sets.
\begin{itemize}
\item The objects/0-data are sets, so the collection of objects is the class of all sets.
\item The morphisms/$1$-data are given by functions. That is, for two sets $x,y\in\ob_\catset$, a morphism $F\in \homo_\catset(x,y)$ is a function
\begin{equation}
F:x\rightarrow y\;,
\end{equation}
so $\homo_\catset(x,y)$ is a proper set.
\item The composition of morphisms is their composition as functions.
\item The identity morphism on a set $x$ is the identity function $\idop_x$.
\end{itemize}
The category of sets can be viewed as the generic example of a category. Many categories arise from this category by restricting to functions which preserve certain structures defined on the sets, such as homeomorphisms preserving topologies, or group homomorphisms preserving a group multiplication.
\end{myexmp}


\begin{myexmp}
Another important example which formalizes finite-dimensional linear algebra (over some field $K$) is the category $\catmat$ of matrices.
\begin{itemize}
\item The objects/0-data are natural numbers,
\begin{equation}
\ob_\catmat=\mathbb{N}_0\;.
\end{equation}
\item The morphisms/$1$-data are given by matrices: For two numbers $m,n\in \ob_{\catmat}$ a morphism $M\in \homo_\catmat(m,n)$ is an $m\times n$ matrix,
\begin{equation}
(M_{ij}\in K)_{i\in\{0,\ldots,m-1\}, j\in\{0,\ldots,n-1\}}\;.
\end{equation}
\item The composition of two morphisms is their multiplication as matrices,
\begin{equation}
(M\circ N)_{ij}=\sum_k M_{ik}N_{kj}\;.
\end{equation}
\item For a number $n$, the identity morphism is the $n\times n$ identity matrix,
\begin{equation}
(\idop_n)_{ij}=\delta_{ij}\;.
\end{equation}
\end{itemize}
$\catmat$ can be seen as the practical version of the category whose objects finite-dimensional vector spaces, and whose morphisms are linear maps. Formally, the former is the \emph{skeleton} of the latter, which means that the objects of the former (numbers) are the equivalence classes of the objects of the latter (vector spaces) under isomorphism.
\end{myexmp}

For any 2-scheme, sequences of 2-functions can be considered computations in that 2-scheme. For a category, those computations are thus sequences of compositions and identity morphisms, as well as copy operations. E.g.,
\begin{equation}
(f\circ(g\circ\idop))\circ f
\end{equation}
is a computation taking two morphisms $f$ and $g$ as an input. Due to the associativity 2-axiom, we can arbitrarily change the brackets, and due to the 2-axiom for the identity, we can drop any identity. So any computation is just a sequence of morphisms, e.g., for the above computation we could simply write
\begin{equation}
f\circ g\circ f\;.
\end{equation}
Likewise, we can represent this in a more ``diagrammatic'' way by replacing the morphisms by shapes and the $\circ$ symbols by lines connecting the shapes, e.g.,
\begin{equation}
\begin{tikzpicture}
\node[roundbox](f){$f$};
\node[roundbox, right=0.5 of f](g){$g$};
\node[roundbox, right=0.5 of g](h){$f$};

\draw[<-](f.west)--++(-0.3,0);
\draw[->](f)--(g);
\draw[->](g)--(h);
\draw[->](h.east)--++(0.3,0);
\end{tikzpicture}\;.
\end{equation}
At this stage the new notation might seem a little arbitrary. It will make more sense later when we introduce a monoidal structure yielding 2-dimensional diagrams.

A computation can also consist of a single identity, which then cannot be dropped. This single identity can be elegantly denoted by just a loose line,
\begin{equation}
\begin{tikzpicture}
\node[roundbox](id){$\idop_a$};

\draw[<-](id.west)--++(-0.3,0);
\draw[->](id.east)--++(0.3,0);
\end{tikzpicture}
=
\begin{tikzpicture}
\draw[->](0,0)--++(0.5,0);
\end{tikzpicture}\;.
\end{equation}
This is consistent with being able to drop the identity when composed with another morphism.

We can now use equations between diagrams to express statements which have different interpretations for different categories. E.g., we can write
\begin{equation}
\label{eq:right_inverse_liquid}
\begin{tikzpicture}
\node[roundbox] (a){$A$};
\node[roundbox, right=0.4 of a] (b){$B$};

\draw[<-](a.west)--++(-0.3,0);
\draw[->](a)--(b);
\draw[->](b.east)--++(0.3,0);
\end{tikzpicture}
=
\begin{tikzpicture}
\draw[->](0,0)--++(0.5,0);
\end{tikzpicture}
\end{equation}
in the diagrammatic calculus to state that a morphism $B$ is a left inverse of a morphism $A$. E.g., in the category $\catset$, this equation holds for $A$ being the exponential function $\exp\in\hom_\catset(\mathbb{R},\mathbb{R}^+)$ and $B$ being the logarithm function $\log\in\hom_\catset(\mathbb{R}^+,\mathbb{R})$, and expresses that they are inverses as functions. In the category $\catmat$, this would hold for the both $A$ and $B$ being the Pauli matrix $\sigma_x\in \hom_\catmat(2,2)$, and expresses that $\sigma_x$ is its own inverse as a matrix.

The matrix $\sigma_x$ can be viewed as a permutation of basis vectors.  So the fact that $\sigma_x^2=\idop$ can be traced back to the fact that $\tau^2=\idop$ where $\tau$ is the function which permutes the elements of the set $\{0,1\}$. In general, functions between finite sets can be made into matrices with only $0$ and $1$ entries and permutation functions can be thus represented as permutation matrices. If some diagrammatic equation like Eq.~\eqref{eq:right_inverse_liquid} holds for the composition of functions, then the same equation will hold for the multiplication of the corresponding matrices. So we have found a way to embed the category $\catset$ (restricted to functions on finite sets) into the category $\catmat$. The following definition formalizes this notion of embedding.

\begin{mydef}
\label{def:functor}
Just like categories, \emph{functors} form a 2-scheme. A functor $F:\catc\rightarrow \catk$ from a category $\catc$ to a category $\catk$ is defined by the following.
\begin{itemize}
\item A map on objects/0-data
\begin{equation}
F:\ob_\catc\rightarrow\ob_\catk\;.
\end{equation}
This is the first example of a 1-function.
\item For every $s,t\in \ob_{\mathcal{C}}$, a map on morphisms/1-data
\begin{equation}
F:\homo_\catc(s,t)\rightarrow\homo_\catk(F(s),F(t))\;.
\end{equation}
This function is another example of a 2-function. We hope no confusion arises using the same letter for the 2-function and the 1-function above.
\item The 2-function $F$ needs to be compatible with the composition. That is, for any objects $s,m,t\in \ob_{\mathcal{C}}$ and morphisms $f\in \hom_\catc(s,m)$ and $g\in \hom_\catc(m,t)$, we have
\begin{equation}
\label{eq:2axiom_functor_composition}
F(f\circ g)=F(f)\circ F(g)\;.
\end{equation}
This axiom is an axiom for the 2-functions, thus it is a 2-axiom.
\item Another 2-axiom is given by
\begin{equation}
F(\idop_a)=\idop_{F(a)}
\end{equation}
for all $a\in \ob_\catc$.
\end{itemize} 
\end{mydef}

The example above defines a functor in the following way:
\begin{myexmp}
\label{exmp:linearization_functor}
Consider the category $\catfinset$ being the subcategory of $\catset$ restricted to only sets of the form $\{0,\ldots,n-1\}=[n]$ for natural numbers $n$. Then, there is the following \emph{linearization} functor from the category $\catfinset$ to the category $\catmat$:
\begin{itemize}
\item The 1-function $F$ acting on objects simply maps the set $[n]$ to the number $n$.
\item The 2-function $F$ maps a function $A\in \hom_\catset([n],[m])$ to a matrix
\begin{equation}
(F(A))_{ij} = 
\begin{cases}
1 &\text{if}\quad j=A(i)\\
0 &\text{otherwise}
\end{cases}\;.
\end{equation}
\end{itemize}
It can be easily seen that the composition of such matrices with $0$ and $1$ entries is equivalent to the composition of the corresponding functions. So the 2-axioms in Definition~\ref{def:functor} hold.
\end{myexmp}

\subsection{Monoidal structure}
\label{kindergarden_categories:Monoidal_structure}
In order to generalize the one dimensional graphical calculus just discussed to a graphical calculus similar to that of tensor networks we need to first make sense of morphisms with multiple outputs or inputs, such as
\begin{equation}
\begin{tikzpicture}
\node[roundbox, minimum height=1.15cm] (u){$U$};

\draw[<-,ind=a] ([yshift=0.3cm]u.west)--++(-0.3,0);
\draw[<-,ind=b] ([yshift=-0.3cm]u.west)--++(-0.3,0);
\draw[->,ind=c] ([yshift=0.3cm]u.east)--++(0.3,0);
\draw[->,ind=d] ([yshift=-0.3cm]u.east)--++(0.3,0);
\end{tikzpicture}\;.
\end{equation}
We do this by combining the objects at the different inputs/outputs into a single one, such as $a$ with $b$, and $c$ with $d$ in the example above. Second, we need to be able to compose morphisms not only in sequence but also ``in parallel''. This will yield diagrams like
\begin{equation}\label{kindergarden_categories:monoidal_exp1}
\begin{tikzpicture}
\node[roundbox] (h) at(0,0){$h$};
\node[roundbox] (g) at(.8,0){$g$};
\node[roundbox] (f) at(.3,-.8){$f$};

\draw[<-, ind=s_h] (h.west)--++(-0.3,0);
\draw[->](h.east)--(g.west);
\draw[->, ind=t_g](g.east)--++(0.3,0);
\draw[<-, ind=s_f](f.west)--++(-0.3,0);
\draw[->,ind=t_f](f.east)--++(0.3,0);
\end{tikzpicture}\;.
\end{equation}
Combining both abilities, we can interpret proper 2-dimensional diagrams, such as
\begin{equation}
\label{eq:2d_circuit_diagram}
\begin{tikzpicture}
\def\X{1}

\def\Y{2}
\foreach \x in {0,...,\X}{
\foreach \y in {0,...,\Y}{
\node[roundbox, minimum height=1.15cm] (a\x\y) at (2*\x,1.5* \y) {$U$};
}}

\foreach \x in {0,...,\X}{
\foreach \y in {0,...,\Y}{
\node[roundbox, minimum height=1.15cm] (b\x\y) at ($(2*\x ,1.5* \y)+(1,0.75)$) {$U$};
}}

\foreach \x in {0,...,\X}{
\foreach \w in {0,...,\Y}{
\draw[->] ([yshift=0.3cm]a\x\w.east)--([yshift=-0.3cm]b\x\w.west);
}}

\foreach \x in {0,...,\X}{
\foreach \w in {1,...,\Y}{
\pgfmathsetmacro\z{int(\w-1)}
\draw[->] ([yshift=-0.3cm]a\x\w.east)--([yshift=0.3cm]b\x\z.west);
}}

\pgfmathsetmacro\Z{int(\X-1)}
\foreach \x in {0,...,\Z}{
\pgfmathsetmacro\w{int(\x+1)}
\foreach \y in {0,...,\Y}{
\draw[->] ([yshift=-0.3cm]b\x\y.east)--([yshift=0.3cm]a\w\y.west);
}}

\pgfmathsetmacro\Z{int(\X-1)}
\foreach \x in {0,...,\Z}{
\pgfmathsetmacro\W{int(\Y-1)}
\pgfmathsetmacro\w{int(\x+1)}
\foreach \y in {0,...,\W}{
\pgfmathsetmacro\z{int(\y+1)}
\draw[->] ([yshift=0.3cm]b\x\y.east)--([yshift=-0.3cm]a\w\z.west);
}}

\pgfmathsetmacro\Z{int(\X-1)}
\foreach \x in {0,...,\Z}{
\pgfmathsetmacro\w{int(\x+1)}
\draw [->] ([yshift=-0.3cm]a\x0.east)--([yshift=-0.3cm]a\w0.west);
\draw [->] ([yshift=0.3cm]b\x\Y.east)--([yshift=0.3cm]b\w\Y.west);
}

\foreach \y in {0,...,\Y}{
\draw [<-] ([yshift=0.3cm]a0\y.west)--++(-0.3,0);
\draw [<-] ([yshift=-0.3cm]a0\y.west)--++(-0.3,0);
}

\foreach \y in {0,...,\Y}{
\draw [->] ([yshift=0.3cm]b\X\y.east)--++(0.3,0);
\draw [->] ([yshift=-0.3cm]b\X\y.east)--++(0.3,0);
}

\draw [<-] ([yshift=0.3cm]b0\Y.west)--++(-1.3,0);

\draw [->] ([yshift=-0.3cm]a\X0.east)--++(1.3,0);
\end{tikzpicture}\;.
\end{equation}

To achieve this, we have to add a \emph{monoidal structure} which consists of a 1-function and a 2-function both called the \emph{tensor product}, corresponding to the combination of objects, and morphisms each. There are different ways to define a monoidal structure, and in the following we will introduce the simplest possible version needed to justify the diagrammatic calculus.

\begin{mydef}
\label{def:strict_monoidal_structure}
A \emph{strict monoidal category} is another example of a 2-scheme, which extends the 2-scheme of a category by the following:
\begin{itemize}
\item There is a function
\begin{equation}
\otimes:\ob_\catc\times\ob_\catc\rightarrow\ob_\catc
\end{equation}
called the \emph{tensor product}. This is an example for a 1-function.
\item There is a distinct object/$0$-data $\unitob\in\ob_\catc$ called the \emph{monoidal unit}. This can be formally interpreted as a 1-function whose domain is the empty tuple of objects.
\item For all objects $a,b,c,d\in \ob_\catc$, there is a function
\begin{equation}
\otimes:\homo_\catc(a,b)\times\homo_\catc(c,d)\rightarrow\homo_\catc(a\otimes c,b\otimes d)
\end{equation}
called the \emph{tensor product}. This is a 2-function, in contrast to the 1-function with equal name and symbol above.
\item We require that the tensor product 1-function is associative,
\begin{equation}
a\otimes(b\otimes c)=(a\otimes b)\otimes c
\end{equation}
for every $a$, $b$ and $c\in\ob_\catc$. This is an axiom for the 1-functions, so it is an example of a 1-axiom.
\item For every $o\in\ob_\catc$, we have another 1-axiom,
\begin{equation}
o\otimes \unitob=\unitob\otimes o=o\;.
\end{equation}
\item There is a $2$-axiom requiring that the tensor product 2-function is associative,
\begin{equation}
f\otimes(g\otimes h)=(f\otimes g)\otimes h
\end{equation}
for all $f\in \hom(s_f,t_f)$, $g\in \hom(s_g, t_g)$ and $h\in \hom(s_h, t_h)$. Note that the target and source object on the left and right hand side are equal due to the associativity of the tensor product 1-function.
\item Another 2-axiom states that the identity $\id_\unitob$ for the monoidal unit object acts as a unit for the tensor product 2-function,
\begin{equation}
f\otimes\id_\unitob=\id_\unitob\otimes f = f
\end{equation}
for all morphisms $f$.
\item Moreover, there is a 2-axiom which ensures that the tensor product 2-function is compatible with the composition,
\begin{equation}
(f_1\circ g_1)\otimes(f_2\circ g_2)=(f_1\otimes f_2)\circ(g_1\otimes g_2)
\end{equation}
for all morphisms $f_1$, $f_2$, $g_1$, and $g_2$ with suitable source and target objects.
\item Also the identity must be compatible with $\otimes$, which is another 2-function,
\begin{equation}
\id_a\otimes\id_b=\id_{a\otimes b}
\end{equation}
for all $a$ and $b\in\ob_\catc$.
\end{itemize}
\end{mydef}



Before we go to the more general notion of a (non-strict) \emph{monoidal category}, let us make some motivating examples.

\begin{myexmp}
The category $\cattwo$ from the previous section can be extended into a monoidal category in a unique way:
\begin{itemize}
\item The unit object is given by $1=\bullet$.
\item The tensor product 1-function is
\begin{equation}
\begin{gathered}
\bullet\otimes\bullet=\bullet\;,\\
\bullet\otimes\star=\star\otimes\bullet=\star\otimes\star=\star\;.
\end{gathered}
\end{equation}
\item The tensor product 2-function is
\begin{equation}
\begin{gathered}
\idop_\bullet\otimes\idop_\bullet=\idop_\bullet\;,\\
\idop_\bullet\otimes\bigtriangleup=\bigtriangleup\otimes\idop_\bullet=\bigtriangleup\;,
f\otimes\idop_\star=\idop_\star\otimes f=\idop_\star\;,
\end{gathered}
\end{equation}
for any morphism $f\in\homo_\cattwo$.
\end{itemize}
\end{myexmp}

\begin{myrem}
The two-dimensional diagrammatic calculus is clearly a bit of an overkill for the category $\cattwo$, as any diagram evaluates to one of three morphisms between one of two objects each. E.g., if we interpret the diagram in Eq.~\eqref{eq:2d_circuit_diagram} with all objects being $\star$, then the tensor product of all input objects is $\star$ again, so the evaluation must be the only morphism between $\star$ and $\star$, $\idop_\star$. In more non-trivial examples, tensoring objects will make them ``larger''. This can be seen in the following example.
\end{myrem}

\begin{myexmp}
\label{expl:catmattimes}
We can make $\catmat$ into a monoidal category $\catmat_\times$ as follows:
\begin{itemize}
\item The monoidal unit object is given by the number $1$.
\item The tensor product 1-function on two objects/$0$-data $m,n\in\ob_\catmat$ is given by the product of numbers,
\begin{equation}
m\otimes n=mn\;.
\end{equation}
\item The tensor product 2-function on two matrices $A\in\homo(m,n)$ and $B\in\homo(p,q)$ is given by the Kronecker product
\begin{equation}
A\otimes B=
\begin{pmatrix}
A_{00}B & \cdots & A_{m0}B\\
\vdots & \ddots & \vdots \\
A_{0n}B & \cdots & A_{mn}B
\end{pmatrix}\;.
\end{equation}
\end{itemize}
Clearly, the product of natural numbers is associative with $1$ as unit, and thus fulfills the $1$-axioms. Moreover, the Kronecker product is associative, and compatible with composition.
\end{myexmp}

\begin{myrem}
We can now interpret the diagram in Eq.~\eqref{eq:2d_circuit_diagram} in the monoidal category $\catmat_\times$ with all objects being the number $2$, and
\begin{equation}\label{cat_mono_permutation_U}
\begin{tikzpicture}
\node[roundbox, minimum height=1.15cm] (u){$U$};

\draw[<-] ([yshift=0.3cm]u.west)--++(-0.3,0);
\draw[<-] ([yshift=-0.3cm]u.west)--++(-0.3,0);
\draw[->] ([yshift=0.3cm]u.east)--++(0.3,0);
\draw[->] ([yshift=-0.3cm]u.east)--++(0.3,0);
\end{tikzpicture}
=
\begin{pmatrix}
1 & 0 & 0 & 0\\
0 & 0 & 1 & 0\\
0 & 1 & 0 & 0\\
0 & 0 & 0 & 1
\end{pmatrix}
\end{equation}
Because $2\otimes2=4$, we find that $U$ corresponds to a 4-dimensional matrix. Physically, $U$ can be seen as a 2-qubit gate, and the diagram can be seen as a quantum circuit consisting of those gates. In contrast to the example for  $\cattwo$, however, the combination of all source or target objects is a pretty large number, namely $2^7$. So, the evaluation is far from trivial and yields a large matrix.
\end{myrem}

While the monoidal category above is very important in physics and can be considered the categorical analogue of array tensors, this is not the only monoidal structure that we can define on $\catmat$.

\begin{myexmp}
The strict monoidal category $\catmat_+$ is based on the category $\catmat$ and is defined as follows:
\begin{itemize}
\item The unit object is the number $0$.
\item The tensor product 1-function on two objects $m,n\in\ob_\catmat$ is the sum of natural numbers,
\begin{equation}
m\otimes n=m+n\;.
\end{equation}
\item The tensor product 2-function on two matrices $A\in\homo(m_1,n_1)$ and $B\in\homo(m_2,n_2)$ is given by their direct sum,
\begin{equation}
A\otimes B=A\oplus B \coloneqq
\begin{pmatrix}
A&0\\0&B
\end{pmatrix}\;.
\end{equation}
\end{itemize}
\end{myexmp}

\begin{myrem}
Again, we can interpret Eq.~\eqref{eq:2d_circuit_diagram} for the monoidal category $\catmat_+$. The evaluation is still a large matrix, but whereas the combined input objects (and therefore the size of the evaluation of the diagram as a single matrix) grow exponentially in the number of component objects for $\catmat_\times$, it only grows linearly in $\catmat_+$.
\end{myrem}


We did not yet give a monoidal structure for $\catset$. This has a reason: A natural choice of the tensor product 1-function would be the cartesian product of sets. However, for three sets $a, b, c$ we have in general
\begin{equation}
\label{eq:cartesian_product_associative}
a\times (b\times c)\neq (a\times b)\times c\;.
\end{equation}
So, the associativity 1-axiom would not be satisfied. We can still make $\catset$ into a monoidal category with the cartesian product if we take a non-strict version of the monoidal structure. In particular, we have to \emph{weaken} the associativity 1-axioms. That is, instead of demanding that the objects in Eq.~\eqref{eq:cartesian_product_associative} are equal, we only demand that there exists an invertible morphism between the objects up to which they are equivalent.

\begin{mydef}
\label{inf_def:monoid}
A (non-strict) \emph{monoidal category} $\catc$ is like a strict monoidal category, with the following differences:
\begin{itemize}
\item We do not demand associativity of $\otimes$ as a 1-axiom. Instead, for all objects $a,b,c\in \ob_\catc$, we have a morphism
\begin{equation}
\alpha_{a,b,c}\in \hom_\catc(a\otimes (b\otimes c), (a\otimes b)\otimes c)\;.
\end{equation}
This collection of morphisms is conventionally called the \emph{associator} and is an example of a \emph{natural isomorphism}. Formally, $\alpha$ can be interpreted as a $2$-function with trivial domain.
\item Similarly, instead of the 1-axiom $a\otimes 1=a$ we have a collection of morphisms
\begin{equation}
\lambda_a\in \hom_\catc(1\otimes a, a)
\end{equation}
called the \emph{left unitor}, and analogously, there is a \emph{right unitor} $\rho$. Both of these collections $\lambda$ and $\rho$ are formally $2$-functions.
\item Additionally we have corresponding collections of morphisms $\alpha^{-1}$, $\lambda^{-1}$, and $\rho^{-1}$ going in the opposite directions, together with three obvious 2-axioms such as
\begin{equation}
\alpha_{a,b,c}^{-1}\circ \alpha_{a,b,c} = \idop_{a\otimes(b\otimes c)}\;.
\end{equation}
\item There is a $2$-axiom called the \emph{pentagon equation} ensuring the equivalence of the different sequences for applying the monoidal product $\otimes$. It is most instructive to state this as a commuting diagram,
\begin{equation}
\begin{tikzpicture}
\node[] (a_b_cd__)at(0,0){$a\otimes(b\otimes(c\otimes c))$};
\node[above right=1.6cm and 0.2cm of a_b_cd__](_ab__cd_){$(a\otimes b)\otimes(c\otimes d)$};
\node[below right= 1.6cm and 0.2cm of _ab__cd_](__ab_c_d){$((a\otimes b)\otimes c)\otimes d$};
\node[below right=2.0cm and -1.8cm of a_b_cd__](a__bc_d_){$a\otimes((b\otimes c)\otimes d)$};
\node[right= 1.8cm of a__bc_d_](_a_bc__d){$(a\otimes(b\otimes c))\otimes d$};

\draw (a_b_cd__)edge[->,mark={slab=$\alpha_{a,b,c\otimes d}$}](_ab__cd_);
\draw (_ab__cd_)edge[->,mark={slab=$\alpha_{a\otimes b,c,d}$}](__ab_c_d);
\draw (a_b_cd__)edge[->,mark={slab=$\idop_a\otimes\alpha_{b,c,d}$}](a__bc_d_);
\draw (a__bc_d_)edge[->,mark={slab=$\alpha_{a,b\otimes c, d}$}](_a_bc__d);
\draw (_a_bc__d)edge[->,mark={slab=$\alpha_{a,b,c}\otimes\idop_d$}](__ab_c_d);
\end{tikzpicture}
\end{equation}
for all $a,b,c,d\in \ob_\catc$. The diagram denotes an equation between two different sequences of the 2-functions composition, identity, tensor product, and most notably the associator $\alpha$, appearing three times on one side and two times on the other side.
\item There is a $2$-axiom called the \emph{triangle equation} ensuring the triviality of introducing the monoidal unit $\unitob$ in a tensor product,
\begin{equation}
\begin{tikzpicture}
\node[] (a_ib_){$a\otimes(1\otimes b)$};
\node(_ai_b) at (4,0) {$(a\otimes1)\otimes b$};
\node(ab) at (2,-2) {$a\otimes b$};

\draw(a_ib_)edge[->,mark={slab=$\alpha_{a,1,b}$}](_ai_b);
\draw(a_ib_)edge[->,mark={r,slab=$\idop_a\otimes\lambda_b$}](ab);
\draw(_ai_b)edge[->,mark={slab=$\rho_a\otimes\idop_b$}](ab);
\end{tikzpicture}
\end{equation}
for all $a,b\in \ob_\catc$.
\end{itemize}

\end{mydef}

\begin{myrem}
In the conventional language, its compatibility with the composition and identity morphism make the tensor product 1-function and 2-function a \emph{bifunctor}. $\alpha$ is then a \emph{natural isomorphism} between the functors $\cdot \otimes (\cdot \otimes \cdot)$ and $(\cdot\otimes \cdot)\otimes\cdot$, and similarly for $\lambda$ and $\rho$. In general, in the conventional categorical language, structures and axioms are often expressed in a very implicit manner. This makes it a rather efficient language for those already fluent in it, but is very confusing for newcomers and makes a systematic understanding more difficult. In our language we try to use as minimal vocabulary as necessary, and we insist in spelling out all structures and axioms explicitly. For the conventional definitions of a monoidal structure via bifunctors and a natural isomorphisms we refer the reader to, e.g., \cite{coecke2009, heunen2019}.
\end{myrem}

\begin{myrem}
Note that while general categories make sense also without identity morphisms, the latter are needed in the formulation of the pentagon and triangle equations of the monoidal structure. Tensor types as defined in Section~\ref{sec:ttypes_definition} are a 2-scheme just like monoidal categories. However, the definition of this 2-scheme is completely different, which allows us to define tensor types without an analogue of an identity morphism.
\end{myrem}

\begin{mycom}
Note that a strict monoidal category automatically yields a monoidal category, by choosing $\alpha$, $\lambda$ and $\rho$ to be identity morphisms.
\end{mycom}


Now equipped with a more general notion of a monoidal structure we can introduce a monoidal structures for the category $\catset$.

\begin{myexmp}
\label{kindergarden_categories:exmp_monoidal_matrix_times}
$\catset_\times$ is the following monoidal category based on the category $\catset$.
\begin{itemize}
\item The monoidal unit object is an arbitrary one element set,
\begin{equation}
1=\{\bullet\}\;.
\end{equation}
\item The tensor product 1-function of two sets $a,b\in \ob_\catset$ is the cartesian product of sets,
\begin{equation}
a\otimes b=a\times b
\end{equation}
\item The tensor product 2-function is given by the cartesian product of functions. This is, for $f\in\homo_\catset(s_f, t_f)$ and  $g\in\homo_\catset(s_g, t_g)$ we have
\begin{equation}
(f\otimes g)(x,y) = (f(x), g(y))
\end{equation}
for $x\in s_f$ and $y\in s_g$.
\item The associator $\alpha_{a,b,c}$ is the canonical bijection
\begin{equation}
\begin{gathered}
\alpha_{a,b,c}: a\times(b\times c)\rightarrow (a\times b)\times c\;,\\
\alpha_{a,b,c}((i,(j,k))) = ((i,j),k)\;.
\end{gathered}
\end{equation}
\item Similarly, the left unitor $\lambda$ is the canonical bijection between $\{\bullet\}\times a$ and $a$, and analogously for the right unitor.
\end{itemize}
Morally speaking, $a\times (b\times c)$ and $(a\times b)\times c$ are the same, both sets consists of triples of elements of the sets $a$, $b$, and $c$. However, the two expressions are formally different as sets. The ``moral equality'' can be mathematically formulated as isomorphism, and this isomorphism is exactly given by $\alpha$. Note however that not all non-strict monoidal categories are of this kind. There are practically relevant categories where $\alpha$ is a very non-trivial operation instead of a very simple reinterpretation.
\end{myexmp}

\begin{myrem}
It is easy to see that the monoidal structure of $\catset_\times$ is similar to the monoidal structure of $\catmat_\times$. Formally, those monoidal structures are consistent with the linearization functor given in Example~\ref{exmp:linearization_functor}. E.g., on the level of objects, we find
\begin{equation}
|[n]\times[m]| = |[n]| |[m]|=mn\;.
\end{equation}
In order to make this fully rigorous, we would have to introduce the notion of a \emph{monoidal functor}, which is another 2-scheme.
\end{myrem}

Similar to how we defined two different monoidal structures on $\catmat$, we can define a second monoidal structure in $\catset$.

\begin{myexmp}
\label{kindergarden_categories:exmp_monoidal_matrix_plus}
$\catset_\sqcup$ is the following monoidal category based on the category $\catset$.
\begin{itemize}
\item The monoidal unit object is given by the empty set,
\begin{equation}
1=\{\}\;.
\end{equation}
\item The tensor product 1-function applied to two sets $a,b\in \ob_\catset$ is the disjoint union of sets,
\begin{equation}
a\otimes b = a\sqcup b\;.
\end{equation}
For the definition of the disjoint union and the notation we use, see Eq.~\eqref{eq:disjoint_union_definition}.
\item The tensor product $2$-function maps to $f\in\homo_\catset(s_f, t_f)$ and $g\in\homo_\catset(s_g, t_g)$ to the concatenation
\begin{equation}
f\otimes g = f\hat\sqcup g\;.
\end{equation}
as defined in Eq.~\eqref{eq:function_concatenation_diffout}.
\item The associator $\alpha$ is given by the canonical bijection $\Phi_\sqcup^\alpha$ defined in Eq.~\eqref{eq:disjoint_union_alpha_bijection}.
\item The right unitor $\rho$ is given by $\Phi_\sqcup^U$ from Eq.~\eqref{eq:disjoint_union_u_bijection}, and analogously for the left unitor $\lambda$.
\end{itemize}




\end{myexmp}

\begin{myrem}
Just as there is a monoidal functor from $\catset_\times$ to $\catmat_\times$ (restricted to sets $\{0,\ldots,n-1\}$), there is a monoidal functor from $\catset_\sqcup$ to $\catmat_+$.
\end{myrem}

\begin{myrem}
Let us interpret the diagram in Eq.~\eqref{eq:2d_circuit_diagram} in the category $\catset_\sqcup$, with all objects being the set $\{x,y\}$ and $U$ being the function
\begin{equation}
\begin{gathered}
U((0,x)) = (0,x),\quad U((1,x)) = (0,y)\;,\\
U((0,y)) = (1,x),\quad U((1,y)) = (1,y)
\end{gathered}
\end{equation}
from $\{x,y\}\sqcup \{x,y\}$ to that same set again. The evaluation of the diagram yields a function from a set with $2\cdot 7=14$ elements to that same set. We see that the two different monoidal structures on $\catset$ are very different in the same way as the two different monoidal structures on $\catmat$. For the monoidal structure induced by $\times$ the objects ``grow'' exponentially when taking tensor products, whereas the objects ``grow'' only linearly for the monoidal structure $+$.

\end{myrem}

\subsection{Braiding}
In the previous section we have seen how diagrams that look like 2-dimensional planar graphs can be interpreted as computations in any monoidal category. If we want to describe physical models beyond two space(-time) dimensions, we need to be able to make sense also of non-planar diagrams. We can go there from the planar diagrams by adding an interpretation to crossing lines, such as
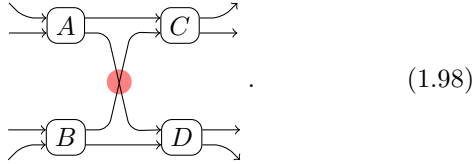
\begin{equation}
\begin{tikzpicture}
\node[roundbox] (a){$A$};
\node[roundbox, below=of a](b){$B$};
\node[roundbox, right=of a](c){$C$};
\node[roundbox,right=of b](d){$D$};

\node[circle,fill=red!50!white]at(.71,-.75){};

\draw[<-, rounded corners]([yshift=.1cm]a.west)--++(-.3,0)--++(-.2,.2);
\draw[<-]([yshift=-.1cm]a.west)--++(-.5,0);
\draw[<-]([yshift=.1cm]b.west)--++(-.5,0);
\draw[<-, rounded corners]([yshift=-.1cm]b.west)--++(-.3,0)--++(-.2,-.2);
\draw[->, rounded corners]([yshift=-.1cm]a.east)--++(.3,0)--++(.3,-1.3)--([yshift=.1cm]d.west);
\draw[->]([yshift=.1cm]a.east)--([yshift=.1cm]c.west);
\draw[->]([yshift=-.1cm]b.east)--([yshift=-.1cm]d.west);
\draw[->, rounded corners]([yshift=.1cm]b.east)--++(.3,0)--++(.3,1.3)--([yshift=-.1cm]c.west);
\draw[->, rounded corners]([yshift=.1cm]c.east)--++(.3,0)--++(.2,.2);
\draw[->]([yshift=-.1cm]c.east)--++(.5,0);
\draw[->]([yshift=.1cm]d.east)--++(.5,0);
\draw[->, rounded corners]([yshift=-.1cm]d.east)--++(.3,0)--++(.2,-.2);
\end{tikzpicture}\;.
\end{equation}
This can be done by introducing a \emph{braiding} morphism for every pair of objects, which exchanges those two objects.

\begin{mydef}
\emph{Symmetric monoidal categories} are another 2-scheme. A braided monoidal category is a monoidal category with the following additional structure.
\begin{itemize}
\item For every $a,b\in \ob_\catc$, there is a morphism
\begin{equation}
\gamma_{a,b}\in \hom_\catc(a\otimes b,b\otimes a)\;.
\end{equation}
The collection is called the \emph{braiding} and it is a 2-function with trivial domain.
\item In the conventional language, $\gamma$ is a \emph{natural isomorphism} between two \emph{bi-functors}, which defines a 2-axiom. For all $a, b, c, d\in \ob$, and $F\in \hom(a,b)$, $G\in \hom(c,d)$, we have
\begin{equation}
\gamma_{c,a}\circ (F\otimes G) = (G\otimes F)\circ \gamma_{d,b}\;.
\end{equation}
\item A $2$-axiom called the \emph{(first) hexagon equation},
\begin{equation}
\begin{tikzpicture}
\node[] (_ab_c) at(0,0) {$(a\otimes b)\otimes c$};
\node (a_bc_) at (0.5,1.5) {$a\otimes(b\otimes c)$};
\node(_bc_a) at (4.5,1.5) {$(b\otimes c)\otimes a$};
\node (b_ca_) at (5,0) {$b\otimes(c\otimes a)$};
\node (_ba_c) at (0.5,-1.5) {$(b\otimes a)\otimes c$};
\node (b_ac_) at (4.5,-1.5){$b\otimes(a\otimes c)$};

\draw(_ab_c)edge[->, mark={r,slab=$\alpha$}](a_bc_);
\draw(a_bc_)edge[->,mark={slab=$\gamma$}](_bc_a);
\draw(_bc_a)edge[->,mark={r,slab=$\alpha$}](b_ca_);
\draw(_ab_c)edge[->,mark={slab=$\gamma\otimes\idop$}](_ba_c);
\draw(_ba_c)edge[->,mark={slab=$\alpha$}](b_ac_);
\draw(b_ac_)edge[->,mark={slab=$\idop\otimes\gamma$}](b_ca_);
\end{tikzpicture}\;.
\end{equation}
\item For all $a,b\in \ob$, we have
\begin{equation}
\gamma_{a,b}\circ \gamma_{b,c}=\idop_{a\otimes b}\;,
\end{equation}
which defines another 2-axiom.
\end{itemize}
\end{mydef}

\begin{myrem}
If we do \emph{not} require the following last 2-axiom in the definition of a symmetric monoidal category above, and instead add an explicit inverse of $\gamma$ as well as a second hexagon equation for $\alpha^{-1}$ instead of $\alpha$, we obtain what is called a \emph{braided monoidal category}.
\end{myrem}

The interpretation of the hexagon equation for the diagrammatic calculus is the following. Consider one line crossing over two other lines,
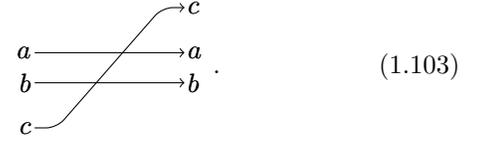
\begin{equation}
\begin{tikzpicture}
\draw[rc] (0,0)edge[->,startind=a, ind=a]++(2,0) (0,-0.4)edge[->,startind=b, ind=b]++(2,0);
\draw[->,rc,startind=c, ind=c] (0,-1)--++(0.3,0)--(1.7,0.6)--++(0.3,0);
\end{tikzpicture}
\;.
\end{equation}
This can be either interpreted exchanging first $c$ with $b$ using $\gamma_{c,b}$ and then $c$ with $a$ using $\gamma_{c,a}$. Or, it can be viewed as simultaneously exchanging $c$ with $a$ and $b$ together using $\gamma_{c,a\otimes b}$.

The ``symmetric'' 2-axiom $\gamma^2=\idop$ also has an important consequence for the diagrammatic calculus. It implies that crossing over two lines twice is the same as doing nothing,
\begin{equation}\label{kindergarden_categories:symmetric_braiding}
\begin{tikzpicture}
\draw[->, rounded corners, startind=a, ind=a] (0,0)--++(0.65,0)--++(0.8,-0.75)--++(0.8,0)--++(0.8,0.75)--++(0.65,0);
\draw[->, rounded corners,startind=b, ind=b] (0,-0.75)--++(0.65,0)--++(0.8,0.75)--++(0.8,0)--++(0.8,-0.75)--++(0.65,0);
\end{tikzpicture}
=
\begin{tikzpicture}
\draw[->, rounded corners, startind=a, ind=a] (0,0)--++(2,0);
\draw[->, rounded corners, startind=b, ind=b] (0,-0.75)--++(2,0);
\end{tikzpicture}\;.
\end{equation}
If we would not impose this axiom such as for a braided monoidal category, we would end up with a different diagrammatic calculus where for each crossing we have to indicate whether a line crosses \emph{over} or \emph{under} another line. Whereas the diagrams of monoidal categories have a 2-dimensional character, those of a braided monoidal categories have a 3-dimensional character, and those of a symmetric monoidal categories can live in spaces of arbitrary dimensions.

\begin{myexmp}
The category $\catmat_+$ from Example~\eqref{kindergarden_categories:exmp_monoidal_matrix_plus} can be made into a symmetric monoidal category. $\gamma_{m,n}$ is an element of
\begin{equation}
\homo(m\otimes n, n\otimes m) = \homo(m+n,m+n)
\end{equation}
and given by
\begin{equation}
\gamma_{m,n}=
\begin{pmatrix}
0 & \idop_n \\
\idop_m & 0
\end{pmatrix}\;.
\end{equation}
\end{myexmp}

\begin{myexmp}
Similarly, we can define a braiding on $\catset_\sqcup$ with $\gamma_{A,B}\in\homo(A\sqcup B, B\sqcup A)$ defined as
\begin{equation}
\gamma_{A,B}((i,x))=(\neg i,x)
\end{equation}
with $\neg 0=1$ and $\neg 1=0$.
\end{myexmp}



\begin{myexmp}
The monoidal category $\catset_\times$ can be made symmetric monoidal with $\gamma_{A,B}\in\homo(A\times B,B\times A)$ given by
\begin{equation}
\gamma_{A,B}((a,b)) = (b,a)\;.
\end{equation}
\end{myexmp}


\subsection{Compact structure}
The diagrams of symmetric monoidal categories already look a lot like the Penrose diagrams for tensor networks. However, there is one important difference: Diagrams of symmetric monoidal categories have to respect a global flow of time, as indicated by the arrows. That is, all lines in the diagrams have to be directed from a target object to a source object, and there must not be any cyclic loops of lines. E.g., the diagram
\begin{equation}
\begin{tikzpicture}
\node[roundbox] (m){$M$};
\draw[rounded corners] (m.east)--++(0.3,0)--++(0,-0.6)--++(-1.2,0)--++(0,0.6)--(m.west);
\end{tikzpicture}
\end{equation}
is possible as a tensor network, but not a string diagram of a symmetric monoidal category. In order to also make sense of such a diagram in a categorical language, we need to add extra structure. In a way, the two points where the ``flow of time'' from left to right is broken are the points where the line is changing its direction from towards the left to towards the right and vice versa. We can introduce a \emph{evaluation} morphisms with trivial source and a \emph{co-evaluation} morphism with trivial target, which represent the bend to the left and right.

\begin{mydef}
\label{cat:compact_closed}
A \emph{compact closed category} is a symmetric monoidal category $\catc$ with the following additional structure.
\begin{itemize}
\item For every object $a\in\ob_\catc$ there is a dual object $a^*$. $*$ is thus an example for a 1-function on objects.
\item For every object $a\in \ob$, there is a morphism
\begin{equation}
\eta_a\in \hom(1, a^*\otimes a)
\end{equation}
called the \emph{evaluation}, a 2-function with trivial domain.
\item For every $a\in \ob$, there is a morphism
\begin{equation}
\epsilon_a\in \hom(a\otimes a^*, 1)
\end{equation}
called the \emph{co-evaluation}, another 2-function with trivial domain.
\item There are two 2-axioms called the \emph{snake identities}. Those are most easily expressed in the diagrammatic calculus of symmetric monoidal categories. For all $a\in \ob$, we have
\begin{equation}
\begin{tikzpicture}
\node[roundbox, minimum height=1.25cm] (eta)at(0,0){$\eta_a$};
\node[roundbox, minimum height=1.25cm] (epsilon)at(1.8,-0.74){$\epsilon_a$};
\draw[<-] ([yshift=-0.37cm]epsilon.west)--++(-1.5,0);
\draw[->] ([yshift=0.37cm]epsilon.west)--([yshift=-0.37cm]eta.east);
\draw[->] ([yshift=0.37cm]eta.east)--++(1.5,0);
\end{tikzpicture}
=
\begin{tikzpicture}
\draw[->](0,0)--++(0.8,0);
\end{tikzpicture}
\end{equation}
and
\begin{equation}\label{compact_closed:dual_hexagon_morphism_equation}
\begin{tikzpicture}
\node[roundbox, minimum height=1.25cm] (eta)at(0,-0.74){$\eta_a$};

\node[roundbox, minimum height=1.25cm] (epsilon)at(1.8,0){$\epsilon_a$};

\draw[->]([yshift=-0.37cm]eta.east)--++(1.5,0);
\draw[<-]([yshift=0.37cm]epsilon.west)--++(-1.5,0);
\draw[->]([yshift=0.37cm]eta.east)--([yshift=-0.37cm]epsilon.west);
\end{tikzpicture}
=
\begin{tikzpicture}
\draw[->](0,0)--++(0.8,0);
\end{tikzpicture}\;.
\end{equation}
\end{itemize}
\end{mydef}

\begin{myrem}
If there exists a choice of dual, evaluation, and co-evaluation, then this choice is unique up to isomorphism. Therefore, often compact closed categories are defined as symmetric monoidal categories for which a dual, evaluation and co-evaluation exists, instead of explicitly including the extra structure.
\end{myrem}

As already mentioned, $\eta$ and $\epsilon$ correspond to ``bending lines backwards'',
\begin{equation}
\begin{tikzpicture}
\node[roundbox, minimum height=1.25cm] (u){$\eta_A$};

\draw[->] ([yshift=0.37cm]u.east)--++(0.5,0);
\draw[->] ([yshift=-0.37cm]u.east)--++(0.5,0);
\end{tikzpicture}
=
\begin{tikzpicture}
\draw[<->, rc](0,0)--++(-0.9,0)--++(0,0.7)--++(0.9,0);
\end{tikzpicture}
\end{equation}
and
\begin{equation}
\begin{tikzpicture}
\node[roundbox, minimum height=1.25cm] (u){$\epsilon_A$};
\draw ([yshift=0.37cm]u.west)--++(-0.5,0);
\draw ([yshift=-0.37cm]u.west)--++(-0.5,0);
\end{tikzpicture}
=
\begin{tikzpicture}
\draw[rc](0,0)--++(0.9,0)--++(0,0.7)--++(-0.9,0);
\end{tikzpicture}\;.
\end{equation}
This makes the significance of the snake equations for the consistency of the diagrammatic calculus apparent, as they turn into, e.g.,
\begin{equation}
\begin{tikzpicture}
\draw[->,rounded corners](0,0)--++(1.5,0)--++(0,0.8)--++(-1.5,0)--++(0,0.8)--++(1.5,0);
\end{tikzpicture}
=
\begin{tikzpicture}
\draw[->,rounded corners](0,0)--++(1.5,0);
\end{tikzpicture}\;.
\end{equation}

\begin{myexmp}
The category $\catmat_\times$ can be made compact closed in the following way:
\begin{itemize}
\item Every object is its own dual. I.e., $*$ is the identity function on objects.
\item For each object (i.e., number) $a$, $\eta_a$ is is obtained by taking the $a\times a$ identity matrix and reshaping it into a $1\times a^2$ matrix.
\item Analogously, $\epsilon_a$ is obtained by reshaping the $a\times a$ identity matrix into a $a^2\times 1$ matrix.

\end{itemize}
\end{myexmp}


\begin{myexmp}
In contrast to $\catmat_\times$ the category $\catset_\times$ cannot be made compact closed: No matter how we choose the duals, for every set $a$, there is only one unique function
\begin{equation}
a\times a^*\rightarrow \{\bullet\}\;,
\end{equation}
and hence only one possible choice of $\epsilon$. On the other hand, $\eta_a$ is a function
\begin{equation}
\{\bullet\}\rightarrow a^*\times a
\end{equation}
which is specified by the tuple
\begin{equation}
(x,y)=\eta_a(\bullet)\;.
\end{equation}
The left hand side of the snake diagrams yields a function which maps everything onto either $x$ or $y$, and not the identity function.
\end{myexmp}



\subsection{Summary}
In the previous sections we have seen how category theory can be used to define structures which fulfill graphical calculi. The more structure we add, the more powerful the diagrammatic calculus gets. Starting with pure categories we have 1-dimensional diagrams. After adding a monoidal structure the diagrams become 2-dimensional planar. The braiding allows us to consider lines crossing each other in a third dimension, and in the symmetric monoidal case the diagram is determined purely combinatorially by which morphisms are connected by lines. Finally, all diagrams in category theory have a flow of time, which can be reverted by introducing a compact closed category.

\section{Physics in terms of tensor networks}
In this section we sketch how tensor networks (and also category theory) can be used to formalize any sort of physics model which has a notion of space and locality.
\subsection{Tensor-network models}
As mentioned earlier, linear algebra and consequently array tensors appear as an important tool in many fields of physics. In this section we will see how tensor networks are not only an important tool, but can also be used as a systematic operational approach to theoretical physics with a notion of space or time, as described in Ref.~\cite{cstar_qmech}. To this end, it is important to generalize from array tensor networks to tensor networks of arbitrary tensor types.

First of all, let us introduce some vocabulary. Theoretical physics is all about mathematical \emph{models} which give \emph{predictions} for different \emph{setups}, which are supposed to describe some aspects of the real world.

The predictions can be any sort of data, whose type depends on the physical description we choose. For example, for the time evolution of a classical deterministic model, the predictions tell us which of a set of possible different events will happen, e.g., which measurement outcomes we will find. For a statistical or quantum theory, the predictions will instead consist of a probability distribution over all possible events or measurement outcomes. If we are studying electric circuits on the other hand, the predictions might be the electric properties of some chip, such as the resistance between two pins.

The setups are the different circumstances under which we probe the model. In a lab, those would be different experimental setups which we aim to describe by a single physical model. This can for example include different sizes of the experiment, different places where we take measurements. E.g., in an optics lab, different setups could be different ways to place mirrors or lenses on a table. Or in electric engineering, different setups could be different ways to connect transistors and resistors on a chip.

The models assign predictions to all the different setups. So, a model is some data $M$, and there is some prescription which associates to each model $M$ and each setup $S$ a prediction $P(M,S)$.

It is very important to be able to probe many different setups. In the worst case with only one single setup we can test, we could just write down the measurement outcomes for that specific setup, and use them as a model which per construction yields the correct predictions. Such a model would be useless, however, as we cannot learn anything from it about other setups. On the other hand, it is important for a model to be simple, such that it can be specified using only a small amount of data. If there is a lot of freedom in choosing the model, then we can find many models which make correct predictions for the probed setups. However, depending on which model we choose, we might get different predictions for unprobed setups.

A key aspect of theoretical physics is to find very simple models which yield correct predictions for a very large amount of different setups.

Basically any physical model has a notion of space(-time). Very roughly speaking, this means that there is an underlying $d$-dimensional manifold for some small usually fixed $d$, and a setup consists of information assigned to different points on that manifold. E.g., we could have mirrors and lenses at different points of the 2-dimensional surface of an optical table. Or, there are transistors and resistors at different points on an electrical chip. As another example, we could consider a condensed matter model with atoms at all different lattice sites in a $3$-dimensional crystal. Or, the transition cells of a $1$-dimensional cellular automaton that are placed at the vertices of a $2$-dimensional square lattice representing the discrete $2$-dimensional space-time. Usually, there is also a notion of locality. That is, the model is specified by some kind of ``interaction'' between the different constituents at points in space which are nearby. E.g., for a condensed-matter model, it is most common for them to be described by a local quantum Hamiltonian.

The notion of space and locality can be very elegantly formalized in terms of so-called \emph{tensor-network models}. For such a model, the setups correspond to different tensor-network diagrams. The latter naturally have a notion of locality and can be seen as combinatorial representations of the space(-time) of the setup. More precisely, the different tensor-networks contain copies of a fixed set of tensors, for example one 4-index tensor $A$ and one 3-index tensor $B$. A model is simply given by a choice of the tensors. In the mentioned example this would be a choice of the tensor $A$ and a choice of the tensor $B$. The predictions for a model $M$ and a setup $S$ are then obtained in the most simple fashion: $P(M,S)$ is the evaluation of the tensor-network diagram $S$ using the tensors $M$. Note that usually, $S$ contains a small number of open indices.

Usually, physical systems also have a flow of time, and a notion of causality with respect to that. That is, the space-time has light-cones, and predictions associated to space-like separated points are independent. In those cases it is natural to demand that also the network diagrams have a flow of time. The causality can be enforced demanding that ``dead ends'' of networks without open indices can be erased, such that we end up with light-cones going backwards in time starting at open indices. In this case it is natural to take the directed string diagrams of a symmetric monoidal category. The causality condition can be implemented by choosing a symmetric monoidal category whose unit object is \emph{terminal}. This means, that for any object there is exactly one morphism to the monoidal unit, which implies that this can pulled through backwards in time. Often, symmetric monoidal categories can be extended to tensor types, such that a symmetric-monoidal-category model can also be expressed as a tensor-network model. This tensor-network model however does not explicitly enforce causality. Additionally, there are physical models or theories which describe an equilibrium, such as thermal ensembles, which does not admit a flow of time. Such models are naturally expressed as tensor-network models.

The type of those tensors corresponds to the different kinds of matter, or the different kinds of physical descriptions of the latter. For example, deterministic physics corresponds to a certain tensor type, classical statistical physics to another. Yet another type is given by quantum physics. If we are dealing with fermions or fundamental symmetries are supposed to hold, this can be modeled by tensor types as well. If we do experiments with linear optics only and choose to work in the framework of Gaussian states, also this can be formalized as using a specific tensor type. A similar case of an efficiently simulable subset of quantum theory corresponding to a tensor type is the stabilizer framework for studying error correcting codes.

\subsection{Example: Classical statistical models}
Let us consider a simple example for how a type physical system can be described as a tensor-network model, namely classical statistical physics by conventional array tensors. To be more precise, we use tensors with non-negative real entries, a restriction which constitutes itself a tensor type. Consider a thermal classical statistical model with a set of degrees of freedom $i$ taking values $c_i$ distributed over some lattice, and Boltzmann weights
\begin{equation}
W_p(c_{p^{(0)}},c_{p^{(1)}},\ldots)=e^{-\beta H_p(c_{p^{(0)}},c_{p^{(1)}},\ldots)}
\end{equation}
at different places $p\in P_W$ depending on the values of some degrees of freedom $p^{(0)},p^{(1)},\ldots$ nearby. Furthermore, consider some conditional probabilities $O_p(o_p|c_{p^{(0)}},c_{p^{(1)}},\ldots)$ at some other places $p\in P_O$, again depending on the nearby degrees of freedom. Usually, $o$ is a deterministic function of $c$, that is, for any set of $c_{p^{(0)}},c_{p^{(1)}},\ldots$ there is exactly one value for $o$ with $O_p=1$.

A setup for such a model consists of a 1) a lattice (of some size) 2) a few places $P_O$ to which we associate observables. The probability distribution for such a setup can be obtained from the following partition function,
\begin{equation}
\begin{multlined}
Z(\mathbf{o})=\sum_{\mathbf{c}} e^{-\beta\sum_p H_p(\mathbf{c})} \mathbf{O}(\mathbf{o}|\mathbf{c})\\
=\sum_{\mathbf{c}} \prod_{p_W\in P_W} \prod_{p_O\in P_O} W_{p_W}(c_{{p_W}^{(0)}},c_{{p_W}^{(1)}},\ldots)\\
\cdot O_{p_O}(c_{{p_O}^{(0)}},c_{{p_O}^{(1)}},\ldots)\;.
\end{multlined}
\end{equation}
Here we used $\mathbf{c}$ and $\mathbf{o}$ for a collection of configurations or outputs.

In order to obtain a proper probability distribution we have to normalize it. The normalization is the partition function without any observables included,
\begin{equation}
P(\mathbf{o})=\frac{Z(\mathbf{o})}{Z(\{\})}\;.
\end{equation}

Such a function $Z(\mathbf{o})$ can be expressed as a tensor network with only positive-entry tensors as follows.
\begin{itemize}
\item Take for each degree of freedom a copy of the $\delta$-tensor, with one index for each Boltzmann weight and each measurement it is part of. The basis of the $\delta$-tensor is the set of configurations of the degree of freedom.
\item Replace each Boltzmann weight and each measurement by the corresponding tensor.
\item Contract all indices of all $\delta$-tensors with the corresponding index of a Boltzmann weight or measurement.
\item This yields a tensor network with one open index for each measurement.
\end{itemize}

Consider, for example the following kind of model: There are degrees of freedom on the vertices of a $2$-dimensional square lattice. One Boltzmann weight $x$ is associated to each plaquette, depending on the degrees of freedom in the corners of that plaquette. The measurements correspond to directly reading out a single degree of freedom. The following shows a patch of the representing tensor network with $3$ measurements (in red),
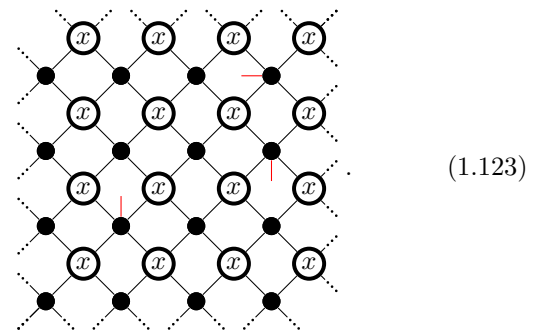
\begin{equation}
\begin{tikzpicture}
\foreach \x in {0,1,2,3}{
\foreach \y in {0,1,2,3}{
\atoms{labbox=$x$, bdastyle={circle,minimum width=0.4cm,minimum height=0.4cm,inner sep=1pt}}{b\x\y/p={\x+0.5,\y+0.5}}
\atoms{small,circ,all}{\x\y/p={\x,\y}}
}}
\foreach \x in {0,1,2,3}{
\foreach \y in {0,1,2,3}{
\draw (\x\y)--(b\x\y);
}}
\foreach[count=\xx from 1] \x in {0,1,2}{
\foreach \y in {0,1,2,3}{
\draw (\xx\y)--(b\x\y);
}}
\foreach[count=\xx from 1] \x in {0,1,2,3}{
\foreach[count=\yy from 1] \y in {0,1,2}{
\draw (\x\yy)--(b\x\y);
}}
\foreach[count=\xx from 1] \x in {0,1,2}{
\foreach[count=\yy from 1] \y in {0,1,2}{
\draw (\xx\yy)--(b\x\y);
}}
\foreach \x in {0,1,2,3}{
\draw (\x0)edge[enddots]++(-135:0.3);
\draw (\x0)edge[enddots]++(-45:0.3);
\draw (b\x3)edge[enddots]++(135:0.3);
\draw (b\x3)edge[enddots]++(45:0.3);
}
\foreach \y in {0,1,2,3}{
\draw (0\y)edge[enddots]++(-135:0.3);
\draw (0\y)edge[enddots]++(135:0.3);
\draw (b3\y)edge[enddots]++(-45:0.3);
\draw (b3\y)edge[enddots]++(45:0.3);
}
\draw[red] (11)--++(0,0.4);
\draw[red] (33)--++(-0.4,0);
\draw[red] (32)--++(0,-0.4);
\end{tikzpicture}\;.
\end{equation}

Or, consider another example where the degrees of freedom are on the vertices of a square lattice. The Boltzmann weights are associated to the edges, and only depend on the degrees of freedom at both ends of the edge. Measurements are some function depending on the configuration of two neighboring degrees of freedom. The following shows a patch of the representing tensor network with two measurements (in red),
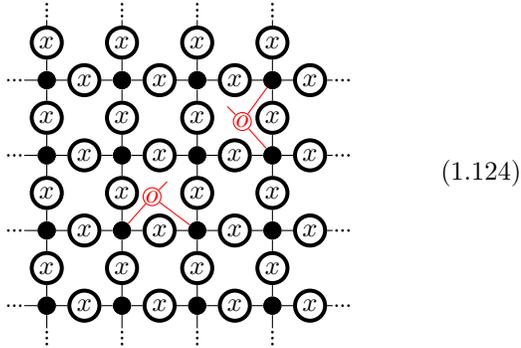
\begin{equation}
\begin{tikzpicture}
\foreach \x in {0,1,2,3}{
\foreach \y in {0,1,2,3}{
\atoms{labbox=$x$, bdastyle={draw,circle,minimum width=0.4cm,inner sep=1pt}}{a\x\y/p={\x+0.5,\y}, b\x\y/p={\x,\y+0.5}}
\atoms{small,circ,all}{\x\y/p={\x,\y}}
}}
\foreach \x in {0,1,2,3}{
\foreach \y in {0,1,2,3}{
\draw (\x\y)--(a\x\y);
\draw (\x\y)--(b\x\y);
}}
\foreach \x in {0,1,2,3}{
\foreach[count=\yy from 1] \y in {0,1,2}{
\draw (\x\yy)--(b\x\y);
}}
\foreach \y in {0,1,2,3}{
\foreach[count=\xx from 1] \x in {0,1,2}{
\draw (\xx\y)--(a\x\y);
}}
\foreach \x in {0,1,2,3}{
\draw (\x0)edge[enddots]++(-90:0.3);
\draw (0\x)edge[enddots]++(180:0.3);
\draw (b\x3)edge[enddots]++(90:0.3);
\draw (a3\x)edge[enddots]++(0:0.3);
}
\node[red,draw,circle,inner sep=0.1pt] (o1) at (1.4,1.45){$o$};
\draw[red] (11)--(o1) (21)--(o1) (o1)--++(0.2,0.2);
\node[red,draw,circle,inner sep=0.1pt] (o1) at (2.6,2.45){$o$};
\draw[red] (33)--(o1) (32)--(o1) (o1)--++(-0.2,0.2);
\end{tikzpicture}
\end{equation}

\chapter{Definition of tensor types}
\label{sec:ttypes_definition}
In this chapter, we introduce the general formalism of tensor types. In order to define tensor types, we'll use a new language which we'll call the \tdef{2-scheme language}{twoscheme_language}. This language will be explained in full generality elsewhere. Here we'll introduce the necessary concepts and terminology on the fly, and won't explicitly explain the full systematics. We do believe that large parts of the systematics will become apparent to the reader while reading this paper though.

The 2-scheme language borrows a few concepts from \emph{category theory}, most notably the idea of \emph{weakening}. In fact, tensor types have some similarity to \emph{monoidal categories}, more specifically \emph{compact closed categories}. Apart from various technical differences, we find that our approach also has conceptual advantages over similar categorical structures since it does not require a flow of time at any stage. For more details on the relation to category theory, see Section~\ref{sec:category}.

Tensor types come in different flavors, and for each flavor there are many different tensor types. In Section~\ref{sec:minimal_definition} we give what we think is the simplest useful flavor of tensor types. In Section~\ref{sec:effective_schemes}, we describe how to get from the axiomatic definition of tensor types to the effective diagrammatic calculus that describes calculations performed using a tensor type. In Section~\ref{sec:other_flavors}, we will discuss various extensions and generalizations yielding other flavors of tensor type, and look at how those alter the effective diagrammatic calculus. In Section~\ref{sec:tensor_mappings}, we define a very general way to map between different tensor types, which can also be used to establish the equivalence of different tensor types. In Section~\ref{sec:category}, we elaborate on the relation of tensor types to structures in monoidal category theory. In Section~\ref{sec:liquid}, we look at equations between different tensor networks which can be expressed using the diagrammatic calculus and which formalize algebraic structures.

\section{Rough idea}
The perspective we will take in this paper is to view a tensor network as a computation, whose input are the individual tensors, and whose output is the single tensor obtained by evaluating the tensor network. The computation itself corresponds to a sequence of tensor products and contractions applied to the inputs.

From the graphical representation of a tensor network, it is impossible to determine in which order different index pairs are contracted. Also, we cannot see in which order we take tensor products, and how we set the brackets in the sequence of tensor products. This is fine, because tensor product and contraction obey a few axioms which guarantee that different orderings/bracketings must yield the same result. E.g., contractions on different index pairs commute, which is easy to see to hold (for conventional tensors, where contraction is implemented by an Einstein summation), e.g.,
\begin{equation}
\label{eq:example_axiom1}
\sum_i \sum_j A_{ajixij} = \sum_j \sum_i A_{ajixij}\;.
\end{equation}
Or, tensor product commutes with contraction, e.g.,
\begin{equation}
\label{eq:example_axiom2}
\sum_i (A_{kl} B_{fii}) = A_{kl} (\sum_i B_{fii})\;.
\end{equation}

Different ways of ordering/bracketing the sequence of tensor products and contractions correspond to different concrete computations, and axioms like in Eq.~\eqref{eq:example_axiom1} or Eq.~\eqref{eq:example_axiom2} define equivalences between different concrete computations. Whereas the combinatorics needed to describe a concrete computation is relatively complex, the equivalence classes of computations have a more efficient (``effective'') representation. This effective representation has a graphical notation in terms of networks.

It is precisely the axioms that justify the graphical calculus. Thus, one might consider other data types (than arrays) and other contraction and tensor product (than Einstein summation and Kronecker product), which fulfill the same axioms. These other structures fulfill the same graphical calculus, and will be called tensor types.

In order to fully formalize our definition of tensor types, we do not require contractions between indices of different tensors, but only contractions of the form $\sum_i A_{xii}$. In order to obtain arbitrary contractions, we need to support this specific contraction with an index permutation operation which will be called the \emph{commutor}. Like in an actual computer numerics implementation, we think of arrays as being flattened into a vector when represented in storage. The flattening depends on an index ordering, and the commutor maps between the different flattened storage representations. For more general tensor types, the storage representations might not only depend on an index ordering, but on a process in which we combine the indices into a single one, including a bracketing. Then, there will be another operation called the \emph{associator} which maps between the storage representations of the different bracketings. For conventional tensors, the associator is trivial.

The axioms in the full definition are relatively complex (though the language we use to formulate them is very simple and systematic). It is not easy to find a ``minimal'' set of axioms that are needed to justify the graphical calculus. We spent quite a bit of time to think about what such a minimal set of axioms should look like, but we don't provide any detailed proofs, and cannot guarantee for sure that we haven't overlooked any axiom, or that there isn't any simpler set of axioms.

\section{Minimal useful flavor}
\label{sec:minimal_definition}
There are several ways to modify the graphical calculus of ``networks'' a little bit. For example, we could add a direction to every bond. Or, we could distinguish between ``in'' and ``out'' indices, and only allow bonds between an in-out pair. Yet another possibility is to demand that every tensor has at least one index, such that no scalars are allowed.

All these different flavors of networks correspond to different flavors of tensor types with slightly different operations and axioms. In this section we will introduce a flavor of tensor types which only contains the essential operations and axioms.

\subsection{1-networks and 1-axioms}
\label{sec:1networks}
The tensors of a tensor network are some kind of data, which we will refer to as \tdef{1-data}{1data}. For array tensors, the 1-data are arrays (or better, vectors) of real/complex numbers. The network itself represents a computation applied to the data, which is a sequence of functions which will be \tdef{2-functions}{2function} implementing individual tensor products, index contractions, index permutations, etc. E.g., for array tensors, the 2-functions are Kronecker product, Einstein summation, etc. For the pairing tensors mentioned in the introduction, the 2-functions are the disjoint union of diagrams and connecting lines/removing loops.

Also the individual indices have data associated to them, which is called \tdef{0-data}{0data} $a\in \dat_0$. A 1-data $A$ can be thought of as a tensor with a single index, and its data type is dependent on the 0-data of that index, $A\in \dat_1(a)$. E.g., for array tensors, the 0-data is the bond dimension of the index, so $\dat_0=\mathbb{N}$, and a 1-data $A$ is a vector with the number of entries given by the 0-data $a$, so $\dat_1(a)=\mathbb{R}^a$. For pairing tensors, the 0-data is the number of dots, and the 1-data is the pairing of those dots.

For a tensor with multiple indices, the individual 0-data need to be combined into a single 0-data, such that the resulting 0-data is the 0-data the 1-data depends on. This is implemented by a function combining two 0-data to a single one called \tdef{product}{0product},
\begin{equation}
\otimes: \dat_0\times \dat_0\rightarrow \dat_0\;.
\end{equation}
In general, functions whose domain and co-domain are (cartesian products of) 0-data will be called \tdef{1-functions}{1function}. E.g., for array tensors, the product is just the product of bond dimensions, corresponding to the fact that two indices of dimension $a$ and $b$ can be blocked into a single index of dimension $a\cdot b$. For pairing tensors, the product is the sum of the two numbers of dots, corresponding to the fact that two sets of $a$ and $b$ dots together have $a+b$ dots.

The product (or in general any 1-function) can be represented graphically by a circle (or a different shape) with three (or more, or less) lines sticking out. The two lines pointing up represent the two input 0-data, and the line pointing down is representing the output 0-data,
\begin{equation}
\label{eq:1product}
a\otimes b=c
\quad\rightarrow\quad
\begin{tikzpicture}
\atoms{prod}{p/}
\draw (p)edge[ind=a]++(135:0.5) (p)edge[ind=b]++(45:0.5) (p)edge[ind=c]++(-90:0.5);
\end{tikzpicture}\;.
\end{equation}

We can compose different products, e.g.,
\begin{equation}
x=(a\otimes b)\otimes (c\otimes d)\;.
\end{equation}
This can be represented graphically as, e.g.,
\begin{equation}
\begin{tikzpicture}
\atoms{prod}{0/, 1/p={-0.8,0.8}, 2/p={0.8,0.8}}
\draw (0)--(1) (0)--(2) (0)edge[ind=x]++(-90:0.5) (1)edge[ind=a]++(135:0.5) (1)edge[ind=b]++(45:0.5) (2)edge[ind=c]++(135:0.5) (2)edge[ind=d]++(45:0.5);
\end{tikzpicture}\;.
\end{equation}
The combinatorics of such a composition will be referred to as a \tdef{1-network}{1_network}. Note that networks of this kind are different from the tensor networks in the introduction in two ways: First, they have a flow of time, which we chose from the top to the bottom. Second, they have a fixed interpretation in terms of sets and functions, whereas the ``tensor networks'' can be interpreted in different tensor types.

Like every collection, the 0-data is automatically equipped with a $\cop$ function and a $\operatorname{Discard}$ function,
\begin{equation}
\begin{aligned}
\cop:\dat_0&\rightarrow \dat_0\times\dat_0\\
\cop(a)&=(a,a)\;,\\
\operatorname{Discard}:\dat_0&\rightarrow \{0\}\\
\operatorname{Discard}(a)&=0\;,
\end{aligned}
\end{equation}
where $\{0\}$ simply represents the empty cartesian product. We'll represent these functions by the following symbols,
\begin{equation}
\begin{gathered}
\cop(a)=(b,c) \quad \rightarrow\quad
\begin{tikzpicture}
\atoms{copy}{p/p={0,0}}
\draw (p)edge[ind=b]++(-0.4,-0.4) (p)edge[ind=c]++(0.4,-0.4) (p)edge[ind=a]++(0,0.4);
\end{tikzpicture}\;,\\
\operatorname{Discard}(a)=0 \quad \rightarrow\quad
\begin{tikzpicture}
\atoms{copy}{p/p={0,0}}
\draw (p)edge[ind=a]++(0,0.4);
\end{tikzpicture}\;.
\end{gathered}
\end{equation}
Those new functions can also be used in 1-networks. E.g., the combination
\begin{equation}
x=(a\otimes b)\otimes a
\end{equation}
is represented by
\begin{equation}
\begin{tikzpicture}
\atoms{prod}{0/, 1/p={-0.5,0.5}}
\atoms{copy}{c/p={-0.5,1.5}}
\draw[rc] (0)--(1) (0)edge[ind=x]++(-90:0.5) (1)--++(-0.5,0.5)--(c) (1)edge[ind=b]++(45:1.1) (0)--++(0.5,0.5)--(c) (c)edge[ind=a]++(90:0.5);
\end{tikzpicture}\;.
\end{equation}

Also a 1-data $A\in \dat_1(a)$ can be represented graphically, by the following shape,
\begin{equation}
\begin{tikzpicture}
\atoms{data}{0/}
\draw (0-t)edge[ind=a]++(0,0.4);
\end{tikzpicture}\;.
\end{equation}
Using this shape, we can also represent the 1-data for the result of a 1-network. E.g.,
\begin{equation}
\dat_1\left((a\otimes c)\otimes ((b\otimes c)\otimes c)\right)
\end{equation}
is represented by
\begin{equation}
\begin{tikzpicture}
\atoms{prod}{0/p={0,0}, 1/p={0.4,0.4}, 2/p={-0.4,0.4}, 3/p={-0.4,1.2}}
\atoms{copy}{x/p={0.4,1.2}, y/p={0.4,2}}
\atoms{data}{d/p={0,-0.4}}
\draw (0)--(1)--(3)--(y) (0)--(2)--(x)--(y) (2)edge[ind=a]++(-0.4,0.4) (3)edge[ind=b]++(-0.4,0.4) (y)edge[ind=c]++(0,0.4) (1)edge[bend right=50](x) (0)--(d);
\end{tikzpicture}\;.
\end{equation}
1-networks with multiple 1-data symbols are interpreted as the cartesian product of the individual data. E.g.,
\begin{equation}
\dat_1(a\otimes b)\times \dat_1(b\otimes c)
\end{equation}
is represented by
\begin{equation}
\begin{tikzpicture}
\atoms{prod}{0/p={-0.4,0}, 1/p={0.4,0}}
\atoms{copy}{c/p={0,0.4}}
\atoms{data}{d0/p={-0.4,-0.4}, d1/p={0.4,-0.4}}
\draw (d0)--(0) (d1)--(1) (0)--(c) (1)--(c) (c)edge[ind=b]++(90:0.5) (0)edge[ind=a]++(135:0.5) (1)edge[ind=c]++(45:0.5);
\end{tikzpicture}\;.
\end{equation}

There are different ways to combine the 0-data of the individual indices into a single 0-data. As the combination is not apparent from the tensor-network graphical calculus, it is natural to demand that all different combinations are equal. Equality of different ways of bracketing is ensured by the associativity axiom $\widetilde{\alpha}$,
\begin{equation}
\widetilde{\alpha}:\quad (a\otimes b)\otimes c = a\otimes (b\otimes c)\;.
\end{equation}
In graphical notation, this becomes
\begin{equation}
\label{eq:1associativity}
\widetilde{\alpha}:\quad
\begin{tikzpicture}
\atoms{prod}{0/, 1/p={-0.5,0.5}}
\draw (0)--(1) (0)edge[ind=x]++(-90:0.5) (1)edge[ind=a]++(135:0.5) (1)edge[ind=b]++(45:0.5) (0)edge[ind=c]++(45:0.5);
\end{tikzpicture}
=
\begin{tikzpicture}
\atoms{prod}{0/, 1/p={0.5,0.5}}
\draw (0)--(1) (0)edge[ind=x]++(-90:0.5) (0)edge[ind=a]++(135:0.5) (1)edge[ind=b]++(135:0.5) (1)edge[ind=c]++(45:0.5);
\end{tikzpicture}
\;.
\end{equation}
In general, we will call such an equation between two 1-networks a \tdef{1-axiom}{1axiom}. E.g., for array tensors, this 1-axiom is nothing but the associativity of integer multiplication, and for pairing tensors it corresponds to associativity of the addition.

The copy and discard functions have special properties which can be normalized as 1-axioms as well. Copying is ``co-associative'',
\begin{equation}
\begin{tikzpicture}
\atoms{copy}{p/p={0,0}, p0/p={0.4,-0.4}}
\draw (p)edge[ind=a]++(-0.4,-0.4) (p)--(p0) (p)edge[ind=d]++(0,0.4) (p0)edge[ind=b]++(-0.4,-0.4) (p0)edge[ind=c]++(0.4,-0.4);
\end{tikzpicture}
=
\begin{tikzpicture}
\atoms{copy}{p/p={0,0}, p0/p={-0.4,-0.4}}
\draw (p)edge[ind=c]++(0.4,-0.4) (p)--(p0) (p)edge[ind=d]++(0,0.4) (p0)edge[ind=a]++(-0.4,-0.4) (p0)edge[ind=b]++(0.4,-0.4);
\end{tikzpicture}
\end{equation}
and commutative,
\begin{equation}
\begin{tikzpicture}
\atoms{copy}{p/p={0,0}}
\draw (p)edge[ind=a]++(-0.4,-0.4) (p)edge[ind=b]++(0.4,-0.4) (p)edge[ind=c]++(0,0.4);
\end{tikzpicture}
=
\begin{tikzpicture}
\atoms{copy}{p/p={0,0}}
\draw (p)edge[ind=b]++(-0.4,-0.4) (p)edge[ind=a]++(0.4,-0.4) (p)edge[ind=c]++(0,0.4);
\end{tikzpicture}
\;.
\end{equation}
Also, making a copy and then discarding it has no effect,
\begin{equation}
\begin{tikzpicture}
\atoms{copy}{p/p={0,0}, p0/p={0.4,-0.4}}
\draw (p)edge[ind=a]++(-0.4,-0.4) (p)--(p0) (p)edge[ind=b]++(0,0.4);
\end{tikzpicture}
=
\begin{tikzpicture}
\draw[startind=a,ind=b] (0,0)--++(0,0.8);
\end{tikzpicture}
\;.
\end{equation}
Moreover, the copy function also satisfies the following property with respect to any other 1-function: Applying the 1-function and then copying the output is equal to copying all inputs and then applying the 1-function to all copies. For the product, the corresponding 1-axiom is
\begin{equation}
\begin{tikzpicture}
\atoms{copy}{p/p={0,0}}
\atoms{prod}{u/p={0,0.4}}
\draw (p)edge[ind=a]++(-0.4,-0.4) (p)edge[ind=b]++(0.4,-0.4) (u)edge[ind=c]++(-0.4,0.4) (u)edge[ind=d]++(0.4,0.4) (p)--(u);
\end{tikzpicture}
=
\begin{tikzpicture}
\atoms{copy}{p1/p={0,0}, p2/p={0.6,0}}
\atoms{prod}{u1/p={0,-0.6}, u2/p={0.6,-0.6}}
\draw (u1)edge[ind=a]++(0,-0.4) (u2)edge[ind=b]++(0,-0.4) (p1)edge[ind=c]++(0,0.4) (p2)edge[ind=d]++(0,0.4) (p1)--(u1) (p1)--(u2) (p2)--(u1) (p2)--(u2);
\end{tikzpicture}
\;.
\end{equation}
Similarly, applying any 1-function and then discarding the output is the same as directly discarding the input. For the product, we get the following 1-axiom,
\begin{equation}
\begin{tikzpicture}
\atoms{copy}{p/p={0,0}}
\atoms{prod}{u/p={0,0.4}}
\draw (u)edge[ind=c]++(-0.4,0.4) (u)edge[ind=d]++(0.4,0.4) (p)--(u);
\end{tikzpicture}
=
\begin{tikzpicture}
\atoms{copy}{0/, 1/p={0.8,0}}
\draw (0)edge[ind=c]++(0,0.4) (1)edge[ind=d]++(0,0.4);
\end{tikzpicture}
\;.
\end{equation}
Note that the term ``axiom'' for the equations involving copying and discarding might be misleading, as these are properties which hold automatically given the fixed interpretation of the black circles as copying and discarding functions. We will abbreviate all those ``axioms'' by $\widetilde{C}$.

\subsection{2-functions and weakening}
We have introduced associativity of the product in order to make the 1-data independent on how we set the brackets when we take the product of all the 0-data of the individual indices. In order to also make it independent of how we order the indices in the product, it would be straight forward to also make it commutative,
\begin{equation}
\begin{tikzpicture}
\atoms{prod}{p/}
\draw (p)edge[ind=a]++(135:0.5) (p)edge[ind=b]++(45:0.5) (p)edge[ind=c]++(-90:0.5);
\end{tikzpicture}
=
\begin{tikzpicture}
\atoms{prod}{p/}
\draw (p)edge[ind=b]++(135:0.5) (p)edge[ind=a]++(45:0.5) (p)edge[ind=c]++(-90:0.5);
\end{tikzpicture}\;.
\end{equation}
There are many tensor types for which this equation holds. However, imposing this as a 1-axiom within the formalism we're about to present has further consequences, namely that the 1-data
\begin{equation}
\dat_1(a\otimes b)= \dat_1(b\otimes a)\;
\end{equation}
are trivially identified. This is not the case for any non-trivial tensor type. E.g., for array tensors, the product of 0-data is indeed commutative. However, an element of $\dat_1(a\otimes b)$ is an length-$ab$ vector which we like to be reshaped into a $a\times b$ matrix. If we change the ordering of the indices, the reshaping of the vector should instead yield the transpose $b\times a$ matrix. For pairing tensors, $\dat_1(a\otimes b)$ is a pairing of $a+b$ dots. Even though $a+b=b+a$, if we change the first $a$ dots with the following $b$ dots this will generally change the pairing diagram.

In order to resolve this problem, we borrow a concept from \emph{category theory} called \emph{weakening}: Instead of the 0-data $a\otimes b$ and $b\otimes a$ being equal, we only demand the 1-data for the 0-data to be in one-to-one correspondence. This one-to-one correspondence is a function,
\begin{equation}
\sigma_0: \dat_1(a\otimes b)\rightarrow \dat_1(b\otimes a)\;.
\end{equation}
More precisely, $\sigma_0$ is a collection of functions $\sigma_0^{a,b}$, one of each choice of 0-data $a$ and $b$. We will however suppress such explicit superscripts as it will always be clear from the context what $a$ and $b$ are. In network notation we can write
\begin{equation}
\sigma_0:\quad
\begin{tikzpicture}
\atoms{prod}{p/}
\atoms{data}{d/p={0,-0.5}}
\draw (p)edge[ind=a]++(135:0.5) (p)edge[ind=b]++(45:0.5) (p)--(d);
\end{tikzpicture}
\quad\rightarrow\quad
\begin{tikzpicture}
\atoms{prod}{p/}
\atoms{data}{d/p={0,-0.5}}
\draw (p)edge[ind=b]++(135:0.5) (p)edge[ind=a]++(45:0.5) (p)--(d);
\end{tikzpicture}\;.
\end{equation}
Collections of functions like $\sigma_0$ are 2-functions, as mentioned above. The target and source 1-network of a 2-function together will be called a \tdef{1-move}{1-move}. Such moves can be applied to a part of a larger 1-network. The 1-axioms also give rise to 1-moves, in fact to two 1-moves each, as they can be applied in two directions.

E.g., for array tensors, a 1-data $A\in \dat_1(a\otimes b)$ is a vector of dimension $ab$ which can be reshaped into an $a\times b$ matrix. $\sigma_0$ consists of taking the transpose of this matrix, and flattening it into an dimension-$ab$ vector again. In other words, $\sigma_0$ maps between the two ways of flattening a $a\times b$ matrix into a $ab$-entry vector. For pairing tensors, $\sigma_0$ simply exchanges the first $a$ dots with the following $b$ dots in the pairing diagram.

Interchanging two indices with 0-data $a$ and $b$ corresponds to crossing the two corresponding index lines in the network diagrammatic calculus. We want that in the resulting tensor-network calculus crossing two indices twice is the same as doing nothing,
\begin{equation}
\begin{tikzpicture}
\draw (0,0)edge[ind=b,startind=a](1.5,0)  (0,0.5)edge[bend right=50,looseness=2,ind=d,startind=c](1.5,0.5);
\end{tikzpicture}
=
\begin{tikzpicture}
\draw (0,0)edge[ind=b,startind=a](1.5,0)  (0,0.5)edge[ind=d,startind=c](1.5,0.5);
\end{tikzpicture}
\;.
\end{equation}
So in order to justify the graphical calculus, we demand that applying $\sigma_0$ twice does nothing on the 1-data, i.e.,
\begin{equation}
\sigma_0\circ\sigma_0=\idop\;,
\end{equation}
as functions on $\dat_1(a,b)$ for all $a,b\in \dat_0$. It is very instructive to express such an equation as a commutative diagram, where we also show what happens to the corresponding 1-network,
\begin{equation}
\label{eq:2axiom_commutor0_involutive}
\begin{tikzpicture}[every node/.style={inner sep=0.2cm}]
\node (0) at (0,0){
\begin{tikzpicture}
\atoms{prod}{p/}
\atoms{data}{d/p={0,-0.5}}
\draw (p)edge[ind=a]++(135:0.5) (p)edge[ind=b]++(45:0.5) (p)--(d);
\end{tikzpicture}};
\node (1) at (4,0){
\begin{tikzpicture}
\atoms{prod}{p/}
\atoms{data}{d/p={0,-0.5}}
\draw (p)edge[ind=a]++(135:0.5) (p)edge[ind=b]++(45:0.5) (p)--(d);
\end{tikzpicture}};
\node (2) at (2,-3){
\begin{tikzpicture}
\atoms{prod}{p/}
\atoms{data}{d/p={0,-0.5}}
\draw (p)edge[ind=b]++(135:0.5) (p)edge[ind=a]++(45:0.5) (p)--(d);
\end{tikzpicture}};
\draw (0)edge[->]node[midway,below left]{$\sigma_0$}(2);
\draw (0)edge[->]node[midway,above]{$=$}(1);
\draw (2)edge[->]node[midway,below right]{$\sigma_0$}(1);
\end{tikzpicture}\;.
\end{equation}
The $=$ sign denotes the identity function on the 1-data ($\idop$ will be reserved for another 2-function). Equations of this kind will be called \tdef{2-axioms}{2axiom}. So, in our formalism, \emph{weakening} means replacing a 1-axiom by a 2-function together with some 2-axioms. It is very easy to see that the 2-axiom above holds for both array tensors and pairing tensors, e.g., transposing a matrix twice yields the original matrix.

Consider a 1-data for a product of many 0-data, say, $\dat_1(a\otimes b\otimes c\otimes d)$ (the bracketing doesn't matter due to the associativity 1-axiom). By applying $\sigma_0$ we can only reach cyclic permutations of $a,b,c,d$. In order to reach arbitrary permutations, we introduce another 2-function $\sigma$ (without the $0$ subscript) which exchanges two 0-data $b,c$ while keeping another one $a$ fixed,
\begin{equation}
\sigma:\quad
\begin{tikzpicture}
\atoms{prod}{0/p={0,0}, 1/p={0.4,0.4}}
\atoms{data}{d/p={0,-0.5}}
\draw (0)--(1) (0)edge[ind=a]++(-0.4,0.4) (0)--(d) (1)edge[ind=b]++(-0.4,0.4) (1)edge[ind=c]++(0.4,0.4);
\end{tikzpicture}
\quad \rightarrow \quad
\begin{tikzpicture}
\atoms{prod}{0/p={0,0}, 1/p={0.4,0.4}}
\atoms{data}{d/p={0,-0.5}}
\draw (0)--(1) (0)edge[ind=a]++(-0.4,0.4) (0)--(d) (1)edge[ind=c]++(-0.4,0.4) (1)edge[ind=b]++(0.4,0.4);
\end{tikzpicture}\;.
\end{equation}

$\sigma$ has to obey the same kind of 2-axiom as $\sigma_0$,
\begin{equation}
\label{eq:2axiom_commutor_involutive}
\begin{tikzpicture}[every node/.style={inner sep=0.2cm}]
\node (a) at (0,0){
\begin{tikzpicture}
\atoms{prod}{0/p={0,0}, 1/p={0.4,0.4}}
\atoms{data}{d/p={0,-0.5}}
\draw (0)--(d) (0)--(1) (0)edge[ind=a]++(-0.4,0.4) (1)edge[ind=c]++(0.4,0.4) (1)edge[ind=b]++(-0.4,0.4);
\end{tikzpicture}};
\node (b) at (4,0){
\begin{tikzpicture}
\atoms{prod}{0/p={0,0}, 1/p={0.4,0.4}}
\atoms{data}{d/p={0,-0.5}}
\draw (0)--(d) (0)--(1) (0)edge[ind=a]++(-0.4,0.4) (1)edge[ind=c]++(0.4,0.4) (1)edge[ind=b]++(-0.4,0.4);
\end{tikzpicture}};
\node (c) at (2,-3){
\begin{tikzpicture}
\atoms{prod}{0/p={0,0}, 1/p={0.4,0.4}}
\atoms{data}{d/p={0,-0.5}}
\draw (0)--(d) (0)--(1) (0)edge[ind=a]++(-0.4,0.4) (1)edge[ind=b]++(0.4,0.4) (1)edge[ind=c]++(-0.4,0.4);
\end{tikzpicture}};
\draw (a)edge[->]node[midway,below left]{$\sigma$}(c);
\draw (a)edge[->]node[midway,above]{$=$}(b);
\draw (c)edge[->]node[midway,below right]{$\sigma$}(b);
\end{tikzpicture}\;.
\end{equation}

There is one more relation that we would like to be fulfilled by $\sigma$. In the graphical calculus, it doesn't matter in which order three index lines cross each other,
\begin{equation}
\begin{tikzpicture}
\draw (0,0)edge[ind=b,startind=a](2,0.5) (0,0.5)edge[ind=d,startind=c](2,0) (0.5,-0.5)edge[ind=f,startind=e](0.5,1);
\end{tikzpicture}
=
\begin{tikzpicture}
\draw (0,0)edge[ind=b,startind=a](2,0.5) (0,0.5)edge[ind=d,startind=c](2,0) (1.5,-0.5)edge[ind=f,startind=e](1.5,1);
\end{tikzpicture}
\;.
\end{equation}
This is equivalent to the following 2-axiom (via $\sigma\circ\sigma=\idop$),
\begin{equation}
\label{eq:2axiom_hexagon}
\begin{tikzpicture}[every node/.style={inner sep=0.2cm}]
\node (x0) at (0,0){
\begin{tikzpicture}
\atoms{data}{d/p={-0.4,-0.9}}
\atoms{prod}{0/p={0,0}, 1/p={-0.4,0.4}, x/p={-0.4,-0.4}}
\draw (x)--(0)--(1) (0)edge[ind=d]++(0.4,0.4) (x)--(d) (1)edge[ind=b]++(-0.4,0.4) (1)edge[ind=c]++(0.4,0.4) (x)edge[ind=a]++(-0.4,0.4);
\end{tikzpicture}};
\node (x1) at (4,0){
\begin{tikzpicture}
\atoms{data}{d/p={-0.4,-0.9}}
\atoms{prod}{0/p={0,0}, 1/p={0.4,0.4}, x/p={-0.4,-0.4}}
\draw (x)--(0)--(1) (0)edge[ind=d]++(-0.4,0.4) (x)--(d) (1)edge[ind=b]++(-0.4,0.4) (1)edge[ind=c]++(0.4,0.4) (x)edge[ind=a]++(-0.4,0.4);
\end{tikzpicture}};
\node (x2) at (0,-4){
\begin{tikzpicture}
\atoms{data}{d/p={-0.4,-0.9}}
\atoms{prod}{0/p={0,0}, 1/p={0.4,0.4}, x/p={-0.4,-0.4}}
\draw (x)--(0)--(1) (0)edge[ind=b]++(-0.4,0.4) (x)--(d) (1)edge[ind=c]++(-0.4,0.4) (1)edge[ind=d]++(0.4,0.4) (x)edge[ind=a]++(-0.4,0.4);
\end{tikzpicture}};
\node (x3) at (4,-4){
\begin{tikzpicture}
\atoms{data}{d/p={-0.4,-0.9}}
\atoms{prod}{0/p={0,0}, 1/p={-0.4,0.4}, x/p={-0.4,-0.4}}
\draw (x)--(0)--(1) (0)edge[ind=c]++(0.4,0.4) (x)--(d) (1)edge[ind=d]++(-0.4,0.4) (1)edge[ind=b]++(0.4,0.4) (x)edge[ind=a]++(-0.4,0.4);
\end{tikzpicture}};
\draw (x0)edge[->]node[midway,above]{$\sigma$}(x1);
\draw (x0)edge[->]node[midway,left]{$\widetilde{\alpha}$}(x2);
\draw (x2)edge[<-]node[midway,below]{$\sigma\widetilde{\alpha}\sigma$}(x3);
\draw (x3)edge[<-]node[midway,right]{$\widetilde{\alpha}$}(x1);
\end{tikzpicture}
\;.
\end{equation}
Be aware that the index labels here and the diagram before have nothing to do with each other even though the letters are partly the same. The 2-axiom consists of $6$ steps. To save space, we combined three steps into a single $\sigma\widetilde{\alpha}\sigma$, which denotes composition of 2-functions or 1-axioms (for brevity we just drop the $\circ$ symbol). E.g., for array tensors this is easily seen to hold, as any sequence of index permutations of a $4$-index array acts trivially if the overall permutation is trivial. Also for pairing tensors it holds for rather trivial reasons, even though there are practically relevant tensor types for which this 2-axiom can be much more subtle.

We believe that the 2-axioms above generate equivalence of \emph{any} two sequences of $\widetilde{\alpha}$, $\sigma_0$ and $\sigma$ mapping between the same two tree-like 1-networks with a single 0-data element. In category theory, an analogous statement is known as \emph{coherence theorem} for strict symmetric monoidal categories.

In general, a 2-axiom is defined by two sequences of 2-functions and 1-axioms which are equated. Both 2-functions and (directed application of) 1-axioms in the sequence correspond to 1-moves, and the corresponding sequence of 1-moves transforms a \tdef{source 1-network}{source_1network} into a \tdef{target 1-network}{target_1network}. Two sequences are equivalent if they differ by exchanging the ordering of 2-functions/1-axioms whose 1-moves do not overlap. The equivalence classes are called \tdef{2-networks}{2network}. A 2-axiom thus consists of two 2-networks with the same source and target 1-network.

\subsection{Tensor product and contraction}
So far we've been defining the 1-data representing a tensor, as well as changes in its representations. The next step is to introduce operations that actually process the 1-data, namely tensor product and contraction. It may not come as a surprise that also those two operations will be formulated as 2-functions, just as the commutor.

The tensor product takes two tensors and combines them into a single one. The 0-data for the resulting tensor is the product of the original 0-data. I.e., we have a 2-function
\begin{equation}
\otimes:\quad
\begin{tikzpicture}
\atoms{data}{0/, 1/p={0.8,0}}
\draw (0-t)edge[ind=a]++(90:0.5) (1-t)edge[ind=b]++(90:0.5);
\end{tikzpicture}
\quad \rightarrow \quad
\begin{tikzpicture}
\atoms{prod}{0/}
\atoms{data}{d/p={0,-0.5}}
\draw (0)edge[ind=a]++(-0.4,0.4) (0)edge[ind=b]++(0.4,0.4) (0)--(d);
\end{tikzpicture}\;.
\end{equation}
We'd like to mention again that such a 2-functions is a collection of functions on 1-data, one for every choice of $a$ and $b$.

E.g., for array tensors we take the Kronecker product (or outer product) of the $a$-entry vector and $b$-entry vector which is a $a\times b$ matrix, and then flatten this matrix into a $ab$-entry vector. For pairing tensors, it is simply given by taking the disjoint union of two pairing diagrams, i.e., drawing them next to each other.

The contraction maps a tensor to a new tensor. The two contracted indices must have the same 0-data $b$. After the contraction, the two indices disappear, and only the 0-data $a$ of the remaining indices is left. Thus, contraction is a 2-function
\begin{equation}
\label{eq:2function_contraction}
[\cdot]:\quad
\begin{tikzpicture}
\atoms{prod}{0/, 1/p={0.4,0.4}}
\atoms{copy}{c/p={0.4,1.2}}
\atoms{data}{d/p={0,-0.4}}
\draw[rc] (0)--(1) (0)edge[ind=a]++(-0.4,0.4) (c)edge[ind=b]++(0,0.4) (1)--++(-0.4,0.4)--(c) (1)--++(0.4,0.4)--(c) (0)--(d);
\end{tikzpicture}
\quad \rightarrow \quad
\begin{tikzpicture}
\atoms{data}{d/}
\atoms{copy}{c/p={0.8,0}}
\draw (d)edge[ind=a]++(0,0.5) (c)edge[ind=b]++(0,0.5);
\end{tikzpicture}\;.
\end{equation}
For array tensors, we reshape the $ab^2$-entry vector $A$ into an $a\times b\times b$ array $\widetilde A$ and the contraction is given by the $a$-entry vector
\begin{equation}
[A]_i=\sum_{0\leq j<b} \widetilde A_{i,j,j}\;.
\end{equation}
For pairing tensors it consists in connecting the last $b$ dots with the $b$ dots before.

Those 2-functions need to satisfy a number of 2-axioms in order to justify the graphical calculus, which we'll list in the following.
\begin{itemize}
\item The tensor product is associative,
\begin{equation}
\label{eq:2axiom_tprod_associativity}
\begin{tikzpicture}[every node/.style={inner sep=0.2cm}]
\node (x0) at (0,0){
\begin{tikzpicture}
\atoms{data}{0/, 1/p={0.5,0}, 2/p={1,0}}
\draw[ind=a] (0-t)--++(0,0.4);
\draw[ind=b] (1-t)--++(0,0.4);
\draw[ind=c] (2-t)--++(0,0.4);
\end{tikzpicture}};
\node (x1) at (2.5,-2){
\begin{tikzpicture}
\atoms{prod}{0/p={1,0.5}}
\atoms{data}{d0/, d1/p={1,0}}
\draw (0)--(d1)  (d0-t)edge[ind=a]++(0,0.4) (0)edge[ind=c]++(0.4,0.4) (0)edge[ind=b]++(-0.4,0.4);
\end{tikzpicture}};
\node (x2) at (-2.5,-2){
\begin{tikzpicture}
\atoms{prod}{0/p={0,0.5}}
\atoms{data}{d1/, d0/p={1,0}}
\draw (0)--(d1)  (d0-t)edge[ind=c]++(0,0.4) (0)edge[ind=b]++(0.4,0.4) (0)edge[ind=a]++(-0.4,0.4);
\end{tikzpicture}};
\node (x3) at (2.5,-5){
\begin{tikzpicture}
\atoms{prod}{0/p={0,0}, 1/p={0.4,0.4}}
\atoms{data}{d/p={0,-0.5}}
\draw (0)--(d) (0)--(1) (0)edge[ind=a]++(-0.4,0.4) (1)edge[ind=c]++(0.4,0.4) (1)edge[ind=b]++(-0.4,0.4);
\end{tikzpicture}};
\node (x4) at (-2.5,-5){
\begin{tikzpicture}
\atoms{prod}{0/p={0,0}, 1/p={-0.4,0.4}}
\atoms{data}{d/p={0,-0.5}}
\draw (0)--(d) (0)--(1) (0)edge[ind=c]++(0.4,0.4) (1)edge[ind=b]++(0.4,0.4) (1)edge[ind=a]++(-0.4,0.4);
\end{tikzpicture}};
\draw (x0)edge[->]node[midway,above right]{$\otimes$}(x1);
\draw (x0)edge[->]node[midway,above left]{$\otimes$}(x2);
\draw (x1)edge[->]node[midway,right]{$\otimes$}(x3);
\draw (x2)edge[->]node[midway,left]{$\otimes$}(x4);
\draw (x3)edge[->]node[midway,below]{$\widetilde{\alpha}$}(x4);
\end{tikzpicture}\;.
\end{equation}
Note that in the first step, there are $6$ different ways in which we could apply the $\otimes$ 2-function. Which of the applications was chosen can be easily determined form the 0-data labels.
\item The tensor product is commutative,
\begin{equation}
\label{eq:2axiom_tprod_commutativity}
\begin{tikzpicture}[every node/.style={inner sep=0.2cm}]
\node (x0) at (0,0){
\begin{tikzpicture}
\atoms{data}{0/, 1/p={0.5,0}}
\draw[ind=a] (0-t)--++(0,0.4);
\draw[ind=b] (1-t)--++(0,0.4);
\end{tikzpicture}};
\node (x1) at (2.5,-2){
\begin{tikzpicture}
\atoms{prod}{0/p={0,0}}
\atoms{data}{d/p={0,-0.5}}
\draw (0)--(d) (0)edge[ind=b]++(0.4,0.4) (0)edge[ind=a]++(-0.4,0.4);
\end{tikzpicture}};
\node (x2) at (-2.5,-2){
\begin{tikzpicture}
\atoms{prod}{0/p={0,0}}
\atoms{data}{d/p={0,-0.5}}
\draw (0)--(d) (0)edge[ind=a]++(0.4,0.4) (0)edge[ind=b]++(-0.4,0.4);
\end{tikzpicture}};
\draw (x0)edge[->]node[midway,above right]{$\otimes$}(x1);
\draw (x0)edge[->]node[midway,above left]{$\otimes$}(x2);
\draw (x2)edge[->]node[midway,below]{$\sigma_0$}(x1);
\end{tikzpicture}
\;.
\end{equation}
As for the associativity, the two arrows from the start correspond to two different ways of applying the $\otimes$ 2-function.
\item The tensor product and the commutor are compatible,
\begin{equation}
\begin{tikzpicture}[every node/.style={inner sep=0.2cm}]
\node (x0) at (0,0){
\begin{tikzpicture}
\atoms{prod}{0/p={1,0.5}}
\atoms{data}{d0/, d1/p={1,0}}
\draw (0)--(d1)  (d0-t)edge[ind=a]++(0,0.4) (0)edge[ind=c]++(0.4,0.4) (0)edge[ind=b]++(-0.4,0.4);
\end{tikzpicture}};
\node (x1) at (4,0){
\begin{tikzpicture}
\atoms{prod}{0/p={0,0}, 1/p={0.4,0.4}}
\atoms{data}{d/p={0,-0.5}}
\draw (0)--(d) (0)--(1) (0)edge[ind=a]++(-0.4,0.4) (1)edge[ind=c]++(0.4,0.4) (1)edge[ind=b]++(-0.4,0.4);
\end{tikzpicture}};
\node (x2) at (0,-3){
\begin{tikzpicture}
\atoms{prod}{0/p={1,0.5}}
\atoms{data}{d0/, d1/p={1,0}}
\draw (0)--(d1)  (d0-t)edge[ind=a]++(0,0.4) (0)edge[ind=b]++(0.4,0.4) (0)edge[ind=c]++(-0.4,0.4);
\end{tikzpicture}};
\node (x3) at (4,-3){
\begin{tikzpicture}
\atoms{prod}{0/p={0,0}, 1/p={0.4,0.4}}
\atoms{data}{d/p={0,-0.5}}
\draw (0)--(d) (0)--(1) (0)edge[ind=a]++(-0.4,0.4) (1)edge[ind=b]++(0.4,0.4) (1)edge[ind=c]++(-0.4,0.4);
\end{tikzpicture}};
\draw (x0)edge[->]node[midway,above]{$\otimes$}(x1);
\draw (x0)edge[->]node[midway,left]{$\sigma_0$}(x2);
\draw (x1)edge[->]node[midway,right]{$\sigma$}(x3);
\draw (x2)edge[->]node[midway,below]{$\otimes$}(x3);
\end{tikzpicture}
\;.
\end{equation}
\item The contraction and the commutor are compatible,
\begin{equation}
\label{eq:2axiom_contraction_commutor}
\begin{tikzpicture}[every node/.style={inner sep=0.2cm}]
\node (x0) at (0,0){
\begin{tikzpicture}
\atoms{prod}{0/,1/p={0.4,0.4},2/p={-0.8,0.8}}
\atoms{data}{d/p={0,-0.5}}
\atoms{copy}{c/p={0.4,1.2}}
\draw[rc] (d)--(0) (0)--(1) (0)--(2) (1)--++(0.4,0.4)--(c) (1)--++(-0.4,0.4)--(c) (c)edge[ind=c]++(0,0.4) (2)edge[ind=b]++(0.4,0.4) (2)edge[ind=a]++(-0.4,0.4);
\end{tikzpicture}};
\node (x1) at (4,0){
\begin{tikzpicture}
\atoms{prod}{0/p={0,0.5}}
\atoms{data}{d/}
\atoms{copy}{c/p={1,0}}
\draw (0)--(d) (c)edge[ind=c]++(0,0.4) (0)edge[ind=b]++(0.4,0.4) (0)edge[ind=a]++(-0.4,0.4);
\end{tikzpicture}};
\node (x2) at (0,-4){
\begin{tikzpicture}
\atoms{prod}{0/,1/p={0.4,0.4},2/p={-0.8,0.8}}
\atoms{data}{d/p={0,-0.5}}
\atoms{copy}{c/p={0.4,1.2}}
\draw[rc] (d)--(0) (0)--(1) (0)--(2) (1)--++(0.4,0.4)--(c) (1)--++(-0.4,0.4)--(c) (c)edge[ind=c]++(0,0.4) (2)edge[ind=a]++(0.4,0.4) (2)edge[ind=b]++(-0.4,0.4);
\end{tikzpicture}};
\node (x3) at (4,-4){
\begin{tikzpicture}
\atoms{prod}{0/p={0,0.5}}
\atoms{data}{d/}
\atoms{copy}{c/p={1,0}}
\draw (0)--(d) (c)edge[ind=c]++(0,0.4) (0)edge[ind=a]++(0.4,0.4) (0)edge[ind=b]++(-0.4,0.4);
\end{tikzpicture}};
\draw (x0)edge[->]node[midway,above]{$[\cdot]$}(x1);
\draw (x0)edge[->]node[midway,left]{$\sigma_0 \sigma \sigma_0$}(x2);
\draw (x1)edge[->]node[midway,right]{$\sigma$}(x3);
\draw (x2)edge[->]node[midway,below]{$[\cdot]$}(x3);
\end{tikzpicture}
\;.
\end{equation}
Note that again, we combined $\sigma_0\sigma\sigma_0$ into one arrow to save space.
\item Contracting two index pairs successively is equal to contracting the blocked index pair, which we'll refer to as \mdef{block-compatibility} of the contraction,
\begin{equation}
\label{eq:2axiom_contraction_blocking}
\begin{tikzpicture}[every node/.style={inner sep=0.2cm}]
\node (x0) at (0,0){
\begin{tikzpicture}
\atoms{prod}{0/p={0,0}, 1/p={0.4,0.4}, 2/p={-0.8,0.8}, 3/p={-0.4,1.2}}
\atoms{copy}{c0/p={0.4,1.2}, c1/p={-0.4,2}}
\atoms{data}{f/p={0,-0.4}}
\draw[rc] (f-t)--(0) (0)--(1) (0)--(2) (2)--(3) (1)--++(0.4,0.4)--(c0) (3)--++(0.4,0.4)--(c1) (c0)edge[ind=c]++(0,0.4) (c1)edge[ind=b]++(0,0.4) (2)edge[ind=a]++(-0.4,0.4);
\draw[rounded corners] (1)--++(-0.4,0.4)--(c0);
\draw[rounded corners] (3)--++(-0.4,0.4)--(c1);
\end{tikzpicture}};
\node (x1) at (0,-4){
\begin{tikzpicture}
\atoms{copy}{c0/p={0.8,0}, c1/p={0.4,0}}
\atoms{data}{f/p={0,0}}
\draw (c0)edge[ind=c]++(0,0.4) (c1)edge[ind=b]++(0,0.4) (f-t)edge[ind=a]++(0,0.4);
\end{tikzpicture}};
\node (x2) at (4,0){
\begin{tikzpicture}
\atoms{prod}{0/p={0,0}, 1/p={0.4,0.4}, 2/p={0,0.8}, 3/p={0.8,0.8}}
\atoms{copy}{c0/p={0,1.6}, c1/p={0.8,1.6}}
\atoms{data}{f/p={0,-0.4}}
\draw[rc] (f-t)--(0) (0)--(1) (1)--(2) (1)--(3) (3)--(c0) (3)--(c1) (2)--(c0) (2)--(c1) (c1)edge[ind=c]++(90:0.4) (c0)edge[ind=b]++(90:0.4) (0)edge[ind=a]++(-0.4,0.4);
\end{tikzpicture}};
\node (x3) at (4,-4){
\begin{tikzpicture}
\atoms{prod}{0/p={0,0}, 1/p={0.4,0.4}, 2/p={0.4,1.6}}
\atoms{copy}{c0/p={0.4,1.2}}
\atoms{data}{f/p={0,-0.4}}
\draw[rc] (f-t)--(0) (0)--(1) (c0)--(2) (1)--++(0.4,0.4)--(c0) (2)edge[ind=c]++(45:0.4) (2)edge[ind=b]++(135:0.4) (0)edge[ind=a]++(-0.4,0.4) (1)--++(-0.4,0.4)--(c0);
\end{tikzpicture}};
\draw (x0)edge[->]node[midway,above]{$\widetilde{\alpha}\widetilde{\alpha}\sigma\widetilde{\alpha}\widetilde{\alpha}\sigma\widetilde{\alpha}$}(x2);
\draw (x0)edge[->]node[midway,left]{$[\cdot][\cdot]$}(x1);
\draw (x2)edge[->]node[midway,right]{$\widetilde{C}$}(x3);
\draw (x3)edge[->]node[midway,below]{$[\cdot]$}(x1);
\end{tikzpicture}
\;.
\end{equation}
The composition $\widetilde{\alpha}\widetilde{\alpha}\sigma\widetilde{\alpha}\widetilde{\alpha}\sigma\widetilde{\alpha}$ is ambiguous as we haven't specified where each $\widetilde{\alpha}$ is applied in the 1-network. This has to be done such that the first $\sigma$ exchanges the second copy of $b$ with both copies of $c$ and the second $\sigma$ exchanges the second copy of $b$ back with the second copy of $c$.
\item Contraction and tensor product commute,
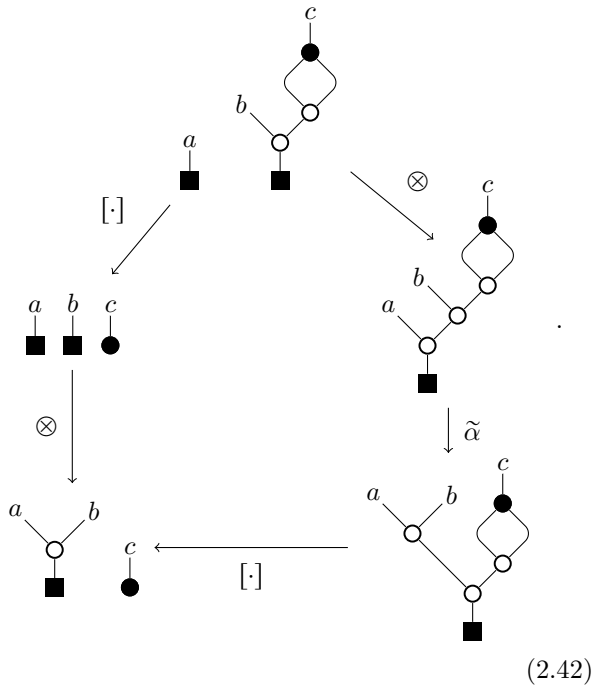
\begin{equation}
\label{eq:2axiom_contraction_tprod}
\begin{tikzpicture}[every node/.style={inner sep=0.2cm}]
\node (x0) at (0,0){
\begin{tikzpicture}
\atoms{prod}{0/p={0,0}, 1/p={0.4,0.4}}
\atoms{copy}{c/p={0.4,1.2}}
\atoms{data}{d0/p={0,-0.5}, d1/p={-1.2,-0.5}}
\draw[rc] (0)--(1) (0)edge[ind=b]++(-0.4,0.4) (c)edge[ind=c]++(0,0.4) (1)--++(-0.4,0.4)--(c) (1)--++(0.4,0.4)--(c) (0)--(d0) (d1)edge[ind=a]++(0,0.4);
\end{tikzpicture}};
\node (x1) at (2.5,-2.5){
\begin{tikzpicture}
\atoms{prod}{0/p={0,0}, 1/p={0.4,0.4}, x/p={-0.4,-0.4}}
\atoms{copy}{c/p={0.4,1.2}}
\atoms{data}{d/p={-0.4,-0.9}}
\draw[rc] (x)--(0)--(1) (0)edge[ind=b]++(-0.4,0.4) (c)edge[ind=c]++(0,0.4) (1)--++(-0.4,0.4)--(c) (1)--++(0.4,0.4)--(c) (x)--(d) (x)edge[ind=a]++(-0.4,0.4);
\end{tikzpicture}};
\node (x2) at (-2.5,-3){
\begin{tikzpicture}
\atoms{copy}{0/p={1,-0.5}}
\atoms{data}{d0/p={0,-0.5}, d1/p={0.5,-0.5}}
\draw (d0)edge[ind=a]++(0,0.4) (d1)edge[ind=b]++(0,0.4) (0)edge[ind=c]++(0,0.4);
\end{tikzpicture}};
\node (x3) at (2.5,-6){
\begin{tikzpicture}
\atoms{prod}{0/p={0,0}, 1/p={0.4,0.4}, x/p={-0.8,0.8}}
\atoms{copy}{c/p={0.4,1.2}}
\atoms{data}{d/p={0,-0.5}}
\draw[rc] (x)--(0)--(1) (x)edge[ind=a]++(-0.4,0.4) (x)edge[ind=b]++(0.4,0.4) (c)edge[ind=c]++(0,0.4) (1)--++(-0.4,0.4)--(c) (1)--++(0.4,0.4)--(c) (0)--(d);
\end{tikzpicture}};
\node (x4) at (-2.5,-6){
\begin{tikzpicture}
\atoms{prod}{0/p={0,0}}
\atoms{copy}{d/p={1,-0.5}}
\atoms{data}{r/p={0,-0.5}}
\draw[rc] (0)--(r) (0)edge[ind=b]++(0.4,0.4) (0)edge[ind=a]++(-0.4,0.4) (d)edge[ind=c]++(0,0.4);
\end{tikzpicture}};
\draw (1.2,-1)edge[->]node[midway,above right]{$\otimes$}(2.3,-1.9);
\draw (x0)edge[->]node[midway,above left]{$[\cdot]$}(x2);
\draw (x1.south)edge[->]node[midway,right]{$\widetilde{\alpha}$}++(0,-0.6);
\draw (x2)edge[->]node[midway,left]{$\otimes$}(x4);
\draw (x3)edge[->]node[midway,below]{$[\cdot]$}(x4);
\end{tikzpicture}
\;.
\end{equation}
\item Two successive contractions commute,
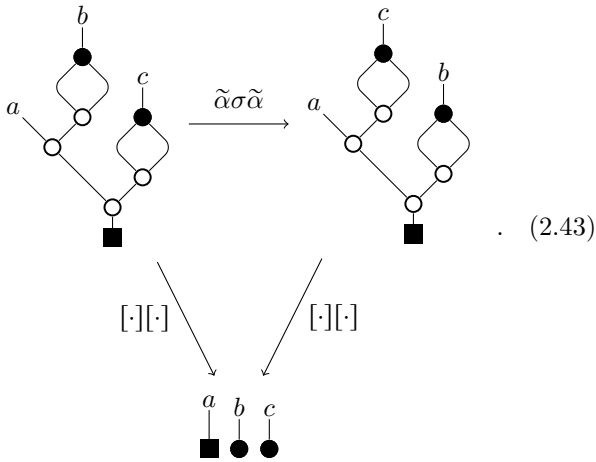
\begin{equation}
\label{eq:2axiom_contraction_commutative}
\begin{tikzpicture}[every node/.style={inner sep=0.2cm}]
\node (x0) at (0,0){
\begin{tikzpicture}
\atoms{prod}{0/p={0,0}, 1/p={0.4,0.4}, 2/p={-0.8,0.8}, 3/p={-0.4,1.2}}
\atoms{copy}{c0/p={0.4,1.2}, c1/p={-0.4,2}}
\atoms{data}{f/p={0,-0.4}}
\draw[rc] (f-t)--(0) (0)--(1) (0)--(2) (2)--(3) (1)--++(0.4,0.4)--(c0) (3)--++(0.4,0.4)--(c1) (c0)edge[ind=c]++(0,0.4) (c1)edge[ind=b]++(0,0.4) (2)edge[ind=a]++(-0.4,0.4);
\draw[rc] (1)--++(-0.4,0.4)--(c0) (3)--++(-0.4,0.4)--(c1);
\end{tikzpicture}};
\node (x1) at (2,-4){
\begin{tikzpicture}
\atoms{copy}{c0/p={0.8,0}, c1/p={0.4,0}}
\atoms{data}{f/p={0,0}}
\draw (c0)edge[ind=c]++(0,0.4) (c1)edge[ind=b]++(0,0.4) (f-t)edge[ind=a]++(0,0.4);
\end{tikzpicture}};
\node (x2) at (4,0){
\begin{tikzpicture}
\atoms{prod}{0/p={0,0}, 1/p={0.4,0.4}, 2/p={-0.8,0.8}, 3/p={-0.4,1.2}}
\atoms{copy}{c0/p={0.4,1.2}, c1/p={-0.4,2}}
\atoms{data}{f/p={0,-0.4}}
\draw[rc] (f-t)--(0) (0)--(1) (0)--(2) (2)--(3) (1)--++(0.4,0.4)--(c0) (3)--++(0.4,0.4)--(c1) (c0)edge[ind=b]++(0,0.4) (c1)edge[ind=c]++(0,0.4) (2)edge[ind=a]++(-0.4,0.4);
\draw[rc] (1)--++(-0.4,0.4)--(c0) (3)--++(-0.4,0.4)--(c1);
\end{tikzpicture}};
\draw (x0)edge[->]node[midway,above]{$\widetilde{\alpha}\sigma\widetilde{\alpha}$}(x2);
\draw (x0)edge[->]node[midway,left]{$[\cdot][\cdot]$}(x1);
\draw (x1)edge[<-]node[midway,right]{$[\cdot][\cdot]$}(x2);
\end{tikzpicture}
\;.
\end{equation}
\item The contraction is permutation invariant,
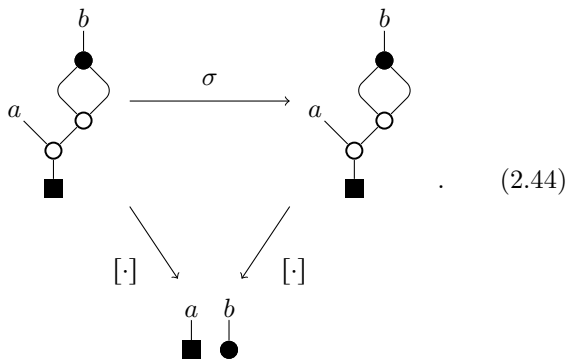
\begin{equation}
\label{eq:2ax_symmetric_contraction}
\begin{tikzpicture}[every node/.style={inner sep=0.2cm}]
\node (x0) at (0,0){
\begin{tikzpicture}
\atoms{prod}{0/p={0,0}, 1/p={0.4,0.4}}
\atoms{copy}{c/p={0.4,1.2}}
\atoms{data}{d/p={0,-0.5}}
\draw[rc] (0)--(1) (0)edge[ind=a]++(-0.4,0.4) (c)edge[ind=b]++(0,0.4) (1)--++(-0.4,0.4)--(c) (1)--++(0.4,0.4)--(c) (0)--(d);
\end{tikzpicture}};
\node (x1) at (4,0){
\begin{tikzpicture}
\atoms{prod}{0/p={0,0}, 1/p={0.4,0.4}}
\atoms{copy}{c/p={0.4,1.2}}
\atoms{data}{d/p={0,-0.5}}
\draw[rc] (0)--(1) (0)edge[ind=a]++(-0.4,0.4) (c)edge[ind=b]++(0,0.4) (1)--++(-0.4,0.4)--(c) (1)--++(0.4,0.4)--(c) (0)--(d);
\end{tikzpicture}};
\node (x2) at (2,-3){
\begin{tikzpicture}
\atoms{copy}{c/p={0.5,-0.5}}
\atoms{data}{d/p={0,-0.5}}
\draw (d)edge[ind=a]++(0,0.4);
\draw (c)edge[ind=b]++(0,0.4);
\end{tikzpicture}};
\draw (x0)edge[->]node[midway,above]{$\sigma$}(x1);
\draw (x0)edge[->]node[midway,below left]{$[\cdot]$}(x2);
\draw (x1)edge[->]node[midway,below right]{$[\cdot]$}(x2);
\end{tikzpicture}
\;.
\end{equation}
\end{itemize}

\subsection{Copying and discarding}
Just like 0-data, 1-data can be copied and discarded, yielding two additional 2-functions, namely
\begin{equation}
\operatorname{Discard}:\quad
\begin{tikzpicture}
\atoms{data}{0/}
\draw[ind=a] (0-t)--++(0,0.3);
\end{tikzpicture}
\quad\rightarrow\quad
\hspace{2cm}
\;,
\end{equation}
and
\begin{equation}
\cop:\quad
\begin{tikzpicture}
\atoms{data}{0/}
\draw[ind=a] (0-t)--++(0,0.3);
\end{tikzpicture}
\quad\rightarrow\quad
\begin{tikzpicture}
\atoms{copy}{0/}
\atoms{data}{d0/p={-0.3,-0.5}, d1/p={0.3,-0.5}}
\draw[rc] (d0-t)--++(0,0.3)--(0) (d1-t)--++(0,0.3)--(0) (0)edge[ind=a]++(90:0.4);
\end{tikzpicture}
\;.
\end{equation}
All the 1-axioms for the copying and discarding 1-functions shown in Section~\ref{sec:1networks} are 2-axioms for the copying and discarding 2-functions in the same way. That is, the copying is co-associative, discarding acts as a co-unit for copying, and copying and discarding can be ``pulled through'' any other 2-function.

The 2-networks we are interested in are 2-networks with the 2-functions from the sections above, but also copying and discarding.

\subsection{Schemes and Types}
Some of the introduced vocabulary refers only to the purely combinatorial structure represented by the diagrams, such as 1-network, 2-network, 1-axiom, 2-axiom. The other vocabulary refers to data structures associated to the combinatorial structure, such as 0-data, 1-data, 1-function, 2-function.

To make the text easier accessible we weren't always strict about separating the data side from the combinatorial side, and sometimes use terms of the data side to talk about the combinatorial structure. In general, a collection of 0-data, 1-data, 1-functions, 2-functions, fulfilling 1-axioms and 2-axioms will be called a \tdef{2-type}{2type}. The pure combinatorics of a 2-type (that is, the diagrams of all 1-axioms, 2-axioms, 1-functions, 2-functions without stating what they actually are) is called a \tdef{2-scheme}{2scheme}.

The product, associativity, commutor, contraction, associativity and commutativity of the tensor product, etc. define one specific 2-scheme called the \tdef{tensor 2-scheme}{tensor_2scheme}. In later sections we define other 2-schemes which are variants of this 2-scheme. We will refer to all those variants as \mdef{tensor 2-schemes}, and call the specific variants \tdef{flavors}{flavor}.

A specific choice for the 1-functions and 2-functions (i.e., product, contraction, commutor, etc.) of a tensor 2-scheme (of a certain flavor) will be called a \tdef{tensor 2-type}{tensor_2type} (or sometimes colloquially a ``tensor type'').




\section{Effective schemes}
\label{sec:effective_schemes}
In this section, we will describe how to go from the definition of tensor types to their effective graphical calculus in terms of tensor network diagrams. We will start by giving a very general definition of  effective graphical calculi for arbitrary 2-schemes and then investigate the effective graphical calculi for the tensor 2-scheme.

\subsection{General idea}
2-networks describe sequences of 2-functions which are interpreted as computations carried out on some 1-data. Some 2-functions are guaranteed to be invertible via 2-axioms, such as the commutor we introduced in the previous section, which is its own inverse. Those 2-functions can be interpreted as changes in the representation of the same 1-data. In contrast, other, non-invertible, 2-functions such as the contraction describe real processing of the 1-data.

Two different computations can be considered equivalent if they perform the same data processing in a different representation. Equivalence classes of computations might have a simple combinatorial representation. Such a combinatorial representation will be called an \tdef{effective scheme}{effective_scheme}. 

Entities of the combinatorial structure describing equivalence classes of 2-networks will be called \tdef{effective 2-networks}{effective_2network}. As we will see, tensor 2-schemes are designed such that effective 2-networks have an elegant diagrammatic representation, with diagrams looking like tensor networks. In other words, an effective scheme is nothing but a formalization of a graphical calculus.

Formally, two 2-networks are equivalent if
\begin{itemize}
\item they differ by cutting out the left hand side of an axiom and pasting the right hand side or vice versa, or,
\item they differ by pre- or post-composing with a 1-axiom or an invertible 2-function. More precisely, such a 2-function has to be right-invertible in the case of pre-composition, and left-invertible in the case of post-composition.
\end{itemize}
Note that the term ``invertible'' above does not describe the actual 2-function, but corresponds to the fact that invertibility follows from the 2-axioms, which is a purely combinatorial property. So while for a concrete 2-type, invertible 2-functions must always be implemented by invertible functions, the converse wouldn't have to be necessarily true. Note also that per definition, 2-networks themselves are already agnostic to the ordering in which 2-functions are applied that do not overlap.

Precomposing/post-composing with invertible 2-functions or 1-axioms, changes the source/target 1-network of a 2-network. Thus, effective 2-networks are not between a source- and target 1-network, but equivalence classes thereof under applying 1-axioms or the moves of invertible 2-functions, which we'll call \tdef{effective 1-networks}{effective_1network}.

Note that the concepts in this section can be applied to get effective graphical calculi for any 2-scheme. Specifically, for the tensor 2-scheme, the effective 2-networks are precisely tensor network diagrams. The effective graphical calculus applies to any tensor type. In the next two sections, we investigate the effective 2-scheme of the tensor 2-scheme presented in earlier sections.

\subsection{Effective 1-networks}
For the tensor 2-scheme, the only invertible 2-functions are the commutors $\sigma$ and $\sigma_0$. By applying the commutors and 1-axioms we can arbitrarily change the structure of any tree of products of 0-data. The only things that stay invariant and therefore describe effective 1-networks are the following:
\begin{itemize}
\item The number of different 1-data elements. The latter will be called \tdef{elements}{element}.
\item The number of different input 0-data. The latter will be called \tdef{bindings}{binding}.
\item For each different data 1-data element and input 0-data, the number of copies of the 0-data that end up as an argument of the 1-data. That is, the number of paths between that 0-data and 1-data. The latter will be called \tdef{indices}{index}.
\end{itemize}

We can use the following diagrammatic notation for effective 1-networks.
\begin{itemize}
\item Every element will be represented by a shape, such as a circle, triangle, or square. The shape might also be a rectangular box with a label inside, or an arbitrary shape with a label nearby.
\item Every binding will be represented by a line style, or a label.
\item Every index of an element will correspond to a point on the boundary of the corresponding shape. We will draw line segments sticking out of this point. The binding of that index will be either indicated by the line style of the line segment, or a label placed after the line end.
\end{itemize}
Surely, there isn't much point in drawing those diagrams alone. Their importance comes from the fact that they are used as part of the graphical calculus for effective 2-networks.

\begin{myexmp}
Let us consider a few examples for 1-networks and the corresponding effective 1-networks.
\begin{itemize}
\item Consider the following 1-network,
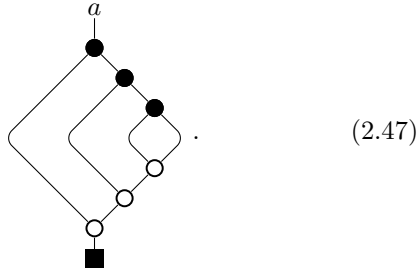
\begin{equation}
\begin{tikzpicture}
\atoms{data}{0/p={0,-0.4}}
\atoms{copy}{c0/p={0.8,1.6}, c1/p={0.4,2}, c2/p={0,2.4}}
\atoms{prod}{m0/p={0,0}, m1/p={0.4,0.4}, m2/p={0.8,0.8}}
\draw (0-t)--(m0) (m0)--(m1)--(m2) (c0)--(c1)--(c2);
\draw (c2)edge[ind=a]++(0,0.4);
\draw[rc] (m2)--++(0.4,0.4)--(c0) (m2)--++(-0.4,0.4)--(c0) (m1)--++(-0.8,0.8)--(c1) (m0)--++(-1.2,1.2)--(c2);
\end{tikzpicture}\;.
\end{equation}
The effective 1-network has one element with $4$ indices that all have the same binding. As shape we could choose a half black half white square
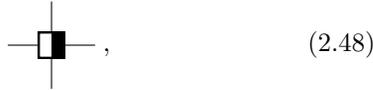
\begin{equation}
\begin{tikzpicture}
\atoms{square,rhalf}{t/}
\draw (t-t)--++(90:0.4) (t-l)--++(180:0.4) (t-r)--++(0:0.4) (t-b)--++(-90:0.4);
\end{tikzpicture}\;,
\end{equation}
or a rectangular box with a label,
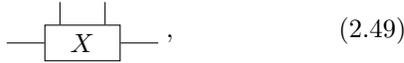
\begin{equation}
\begin{tikzpicture}
\node[box,wid=1] (t) {$X$};
\draw ([xshift=-0.3cm]t.north)--++(0,0.3) ([xshift=0.3cm]t.north)--++(0,0.3) (t.east)--++(0.5,0) (t.west)--++(-0.5,0);
\end{tikzpicture}\;,
\end{equation}
or, a circle with a label next to it,
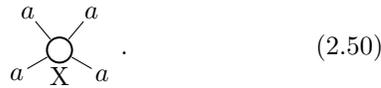
\begin{equation}
\begin{tikzpicture}
\atoms{lab={t=X,p=-90:0.35},circ}{t/p={0,0}}
\draw (t)edge[ind=a]++(-30:0.5) (t)edge[ind=a]++(50:0.5) (t)edge[ind=a]++(130:0.5) (t)edge[ind=a]++(-150:0.5);
\end{tikzpicture}\;.
\end{equation}
In the last representation we also explicitly labeled each index by its binding.
\item Consider the following 1-network,
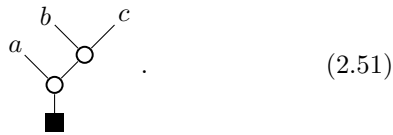
\begin{equation}
\begin{tikzpicture}
\atoms{prod}{0/p={0,0}, 1/p={0.4,0.4}}
\atoms{data}{d/p={0,-0.5}}
\draw (0)--(1) (0)edge[ind=a]++(-0.4,0.4) (0)--(d) (1)edge[ind=b]++(-0.4,0.4) (1)edge[ind=c]++(0.4,0.4);
\end{tikzpicture}
\;.
\end{equation}
The effective 1-network has $3$ bindings, one element, and $3$ indices (one for every binding). The different bindings of the indices can be denoted by labels,
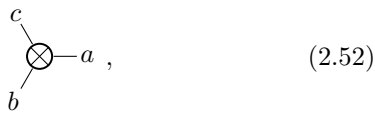
\begin{equation}
\begin{tikzpicture}
\atoms{circ,dcross}{0/p={0,0}}
\draw (0)edge[ind=a]++(0:0.5) (0)edge[ind=c]++(120:0.5) (0)edge[ind=b]++(-120:0.5);
\end{tikzpicture}\;,
\end{equation}
or, by using different line styles,
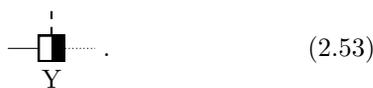
\begin{equation}
\begin{tikzpicture}
\atoms{lab={t=Y,p=-90:0.4},rhalf,square}{t/p={0,0}}
\draw (t-t)edge[thick,dashed]++(90:0.4) (t-l)--++(180:0.4) (t-r)edge[densely dotted]++(0:0.4);
\end{tikzpicture}\;.
\end{equation}
Labels and line styles might also be mixed,
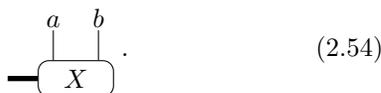
\begin{equation}
\begin{tikzpicture}
\node[roundbox,wid=1] (t) at (0,0){$X$};
\draw (t.west)edge[line width=2]++(180:0.4) ([xshift=-0.3cm]t.north)edge[ind=a]++(90:0.4) ([xshift=0.3cm]t.north)edge[ind=b]++(90:0.4);
\end{tikzpicture}\;.
\end{equation}
\item As an example with multiple 1-data elements, consider the following 1-network,
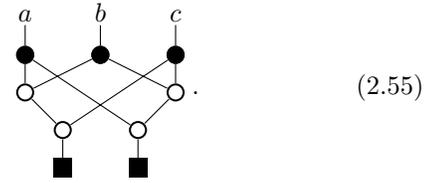
\begin{equation}
\begin{tikzpicture}
\atoms{data}{0/p={0,0}, 1/p={1,0}}
\atoms{copy}{c0/p={-0.5,1.5}, c1/p={0.5,1.5}, c2/p={1.5,1.5}}
\atoms{prod}{m0/p={-0.5,1}, m1/p={0,0.5}, m2/p={1,0.5}, m3/p={1.5,1}}
\draw (0-t)--(m1) (1-t)--(m2) (m0)--(m1) (m3)--(m2) (c0)--(m0) (c0)--(m2) (c1)--(m0) (c1)--(m3) (c2)--(m1) (c2)--(m3);
\draw (c0)edge[ind=a]++(0,0.4) (c1)edge[ind=b]++(0,0.4) (c2)edge[ind=c]++(0,0.4);
\end{tikzpicture}
\;.
\end{equation}
The effective 1-network has two elements, and $3$ bindings. Both elements have $3$ indices, one for each binding. In the graphical notation, we separate the two shapes by a comma (putting them next to each other without separator will be reserved for the tensor product in 2-networks),
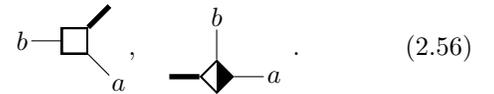
\begin{equation}
\label{eq:substrate_example1}
\begin{tikzpicture}
\atoms{square}{t/p={0,0}}
\draw (t-tr)edge[line width=2]++(45:0.4) (t-br)edge[ind=a]++(-45:0.4) (t-l)edge[ind=b]++(0:-0.4);
\end{tikzpicture}
,\quad
\begin{tikzpicture}
\atoms{diamond,rhalf}{0/p={0,0}}
\draw (0-l)edge[line width=2]++(180:0.4) (0-t)edge[ind=b]++(90:0.4) (0-r)edge[ind=a]++(0:0.4);
\end{tikzpicture}\;.
\end{equation}
\item Note that strictly speaking, the graphical representation does not fully specify the effective 1-network, as there might be bindings without any indices. E.g., consider the following 1-network,
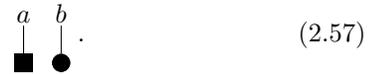
\begin{equation}
\begin{tikzpicture}
\atoms{data}{0/}
\atoms{copy}{1/p={0.5,0}}
\draw (0)edge[ind=a]++(0,0.5) (1)edge[ind=b]++(0,0.5);
\end{tikzpicture}\;.
\end{equation}
The effective 1-network has two bindings and one element with one index. There is no index of binding $B$. It would be denoted by, e.g.,
\begin{equation}
\begin{tikzpicture}
\atoms{square,dcross}{0/}
\draw (0-r)--++(0.4,0);
\end{tikzpicture}\;,
\end{equation}
which hides the binding $b$. Of course, such ``unused bindings'' are superficial, unless we are planning to ``extend'' the network by further elements.
\end{itemize}
\end{myexmp}

\subsection{Effective 2-networks}
Consider now effective 2-networks. First, we note that effective 2-networks do have a source and target effective 1-network, which is invariant.

Using the equivalence, every 2-network can be brought into a partial standard form as follows.
\begin{enumerate}
\item A 2-network is called \tdef{simple}{simple} if its target 1-network only has a single data 1-element. For a 2-network with $n$ target 1-data elements, we can use the pull-through 2-axioms for copying and discarding to bring it into a form where we start by making $n$ sets of copies of every input 1-data, and then apply a simple 2-network to each of the sets of copies. For each simple 2-network, we can proceed as follows.
\item Use the pull-through 2-axioms to bring all copying and discarding operations to the beginning.
\item Use the commutativity of the tensor product with the other 2-functions to move all tensor products to come next, right after copying/discarding.
\item We will say that a contraction in a 2-network \mdef{acts on individual indices}, if the 1-network it acts on does not continue where the 0-data $b$ in Eq.~\eqref{eq:2function_contraction} is. Using the block-compatibility of the contraction Eq.~\eqref{eq:2axiom_contraction_blocking}, we can replace a single contraction where the 0-data $b$ in Eq.~\eqref{eq:2function_contraction} is the output of a product by two contractions, one for each input of the product. Repeating this step we obtain a 2-network where all contractions act on individual indices.
\item Use the compatibility of contraction and commutor Eq.~\eqref{eq:2axiom_contraction_commutor} to move all contractions to the end of the 2-network.
\end{enumerate}

So the 2-network in its partial standard form consists of the following steps.
\begin{enumerate}
\item Make $n$ copies, one for every simple component. For each simple 2-network proceeds as follows.
\item Discard or make as many copies as needed.
\item Take the tensor product of all the copies.
\item Apply commutors and 1-axioms to get the 1-network ready for the final step.
\item Apply contractions which act on individual indices.
\end{enumerate}

This standard form is subject to further equivalences as follows.
\begin{itemize}
\item Which sequence of copying and discarding we take in steps 1 and 2 is irrelevant due to 2-axioms such as co-unitality and co-associativity of copying and discarding.
\item Which sequence of tensor products we take in step 3 is irrelevant due to the commutativity Eq.~\eqref{eq:2axiom_tprod_commutativity} and associativity Eq.~\eqref{eq:2axiom_tprod_associativity} of the tensor product.
\item Which sequence of commutors and 1-axioms we take for step 4 is irrelevant, as any two such sequences are equivalent due to the 2-axioms Eq.~\eqref{eq:2axiom_commutor0_involutive}, Eq.~\eqref{eq:2axiom_commutor_involutive}
\item The order of contractions in the last step is irrelevant, as two consecutive contractions can be swapped using the commutativity of the contraction Eq.~\eqref{eq:2axiom_contraction_commutative}.
\end{itemize}

Thus, an effective 2-network is specified by the following invariant data.
\begin{itemize}
\item The source effective 1-network.
\item For every simple component of the 2-network (i.e., every 1-data element of the target 1-network), the following information.
\item The target effective 1-network (or better, the part leading to the corresponding 1-data element). The indices will be referred to as \tdef{open indices}{open_index}.
\item For every element, the number of (non-discarded) copies made. The individual copies will be called \tdef{atoms}{atom} of the element.
\item A pair of an atom and an index (of the corresponding element) will be referred to as a \tdef{receptor}{receptor} of the corresponding network. A contraction acting on individual indices is specified by a pair of receptors. Such a pair will be referred to as a \tdef{bond}{bond}. The set of bonds is an invariant.
\end{itemize}

\begin{myrem}
Simple effective 2-networks have following graphical representation:
\begin{itemize}
\item For each atom, draw a copy of the shape representing its element. The position of those shapes is arbitrary. They can also be rotated by arbitrary angles, or even reflected.
\item Each receptor corresponds to a point on the boundary of the shape representing an atom. For each bond between two receptors, we connect the according points by a line. The line can run along an arbitrary path between the two points, and is also allowed to intersect with other lines, which has no effect.
\item More precisely, we distinguish two kinds of shapes: Shapes that have a label inside should not be drawn rotated or reflected. Shapes without label or with a label next to them can be drawn rotated or reflected arbitrary. If, in the latter case, the shape has any reflection or rotation symmetry, it will be impossible to uniquely identify which receptor corresponds to which index. We will see in Remark~\ref{rem:shape_symmetry} how this corresponds to certain constraints for the corresponding 1-data.
\end{itemize}
\end{myrem}

\begin{myexmp}
Let us give the a few examples for 2-networks and their effective 2-networks.
\begin{itemize}
\item Consider the following simple 2-network, consisting of a tensor product only,
\begin{equation}
\begin{tikzpicture}
\node(x0)at (0,0){
\begin{tikzpicture}
\atoms{data}{0/p={0,-0.5}, 1/p={1.2,0.7}}
\atoms{prod}{m0/p={0,0}, m1/p={0.4,0.4}, m2/p={1.2,1.2}}
\atoms{copy}{c0/p={0,0.8}, c1/p={0.4,1.2}, c2/p={0.8,1.6}, c3/p={1.2,2}}
\draw[rc] (0-t)--(m0) (1-t)--(m2) (m0)--(m1) (m0)--++(-0.4,0.4)--(c0) (m1)--(c0) (m1)--++(0.4,0.4)--(c1) (m2)--(c2) (m2)--++(0.4,0.4)--(c3) (c0)--(c1)--(c2)--(c3) (c3)edge[ind=a]++(90:0.4);
\end{tikzpicture}
};
\node(x1)at (4,0){
\begin{tikzpicture}
\atoms{data}{0/p={0.4,-0.9}}
\atoms{prod}{m0/p={0,0}, m1/p={0.4,0.4}, m2/p={1.2,1.2}, m3/p={0.4,-0.4}}
\atoms{copy}{c0/p={0,0.8}, c1/p={0.4,1.2}, c2/p={0.8,1.6}, c3/p={1.2,2}}
\draw[rc] (0)--(m3) (m0)--(m1) (m0)--++(-0.4,0.4)--(c0) (m1)--(c0) (m1)--++(0.4,0.4)--(c1) (m2)--(c2) (m2)--++(0.4,0.4)--(c3) (c0)--(c1)--(c2)--(c3) (c3)edge[ind=a]++(90:0.4) (m3)--(m0) (m2)--++(0,-0.8)--(m3);
\end{tikzpicture}
};
\draw (x0)edge[->]node[midway,above]{$\otimes$}(x1);
\end{tikzpicture}
\;.
\end{equation}
The effective network can be drawn as
\begin{equation}
\begin{tikzpicture}
\node[box] (a) at (0,0) {$X$};
\atoms{big,circ,all}{b/p={1.5,0}}
\draw (a.west)--++(180:0.3) (a.east)--++(0:0.3) (a.south)--++(-90:0.3) (b)--++(-90:0.6) (b)--++(90:0.6);
\end{tikzpicture}\;.
\end{equation}
Another way to draw the same network would be
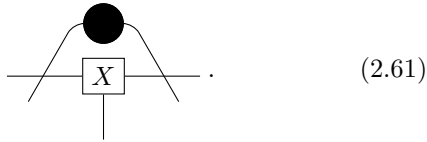
\begin{equation}
\begin{tikzpicture}
\node[box] (a) at (0,0) {$X$};
\atoms{big,circ,all}{b/p={0,0.7}}
\draw (a.west)--++(180:1) (a.east)--++(0:1) (a.south)--++(-90:0.6);
\draw[rounded corners] (b)--++(0.4,0)--++(-60:1.2);
\draw[rounded corners] (b)--++(-0.4,0)--++(-120:1.2);
\end{tikzpicture}\;.
\end{equation}
\item Consider the following simple 2-network consisting of a contraction only,
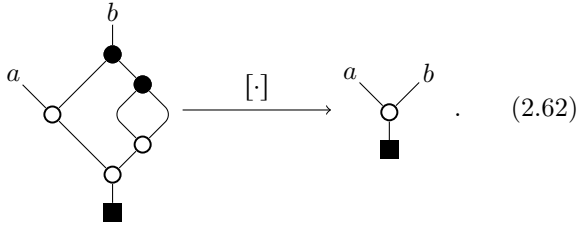
\begin{equation}
\begin{tikzpicture}
\node(x0)at(0,0){
\begin{tikzpicture}
\atoms{prod}{0/p={0,0}, 1/p={0.4,0.4}, 2/p={-0.8,0.8}}
\atoms{data}{d/p={0,-0.5}}
\atoms{copy}{c0/p={0.4,1.2}, c1/p={0,1.6}}
\draw[rc] (0)--(1) (0)--(2) (2)edge[ind=a]++(-0.4,0.4) (c1)edge[ind=b]++(0,0.4) (1)--++(-0.4,0.4)--(c0) (1)--++(0.4,0.4)--(c0) (0)--(d) (c0)--(c1) (2)--(c1);
\end{tikzpicture}
};
\node(x1)at(4,0){
\begin{tikzpicture}
\atoms{prod}{0/p={0,0}}
\atoms{data}{d/p={0,-0.5}}
\draw (0)edge[ind=a]++(-0.4,0.4) (0)edge[ind=b]++(0.4,0.4) (0)--(d);
\end{tikzpicture}
};
\draw (x0)edge[->]node[midway,above]{$[\cdot]$} (x1);
\end{tikzpicture}
\;.
\end{equation}
The effective 2-network can be drawn by
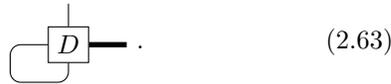
\begin{equation}
\begin{tikzpicture}
\node[box] (t) {$D$};
\draw[rc] (t.west)--++(-0.5,0)--++(0,-0.5)-|(t.south) (t.north)--++(0,0.3);
\draw[line width=2] (t.east)--++(0.5,0);
\end{tikzpicture}
\;.
\end{equation}
The fat line corresponds to the open index $b$.
\item Consider a 2-network which just consists of discarding one 1-data,
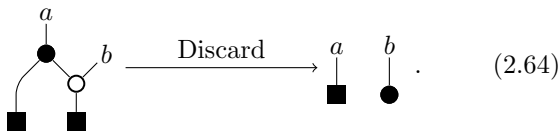
\begin{equation}
\begin{tikzpicture}
\node(x0)at(0,0){
\begin{tikzpicture}
\atoms{prod}{p/}
\atoms{copy}{c/p={-0.4,0.4}}
\atoms{data}{d0/p={0,-0.5}, d1/p={-0.8,-0.5}}
\draw[rc] (c)--++(-0.4,-0.4)--(d1) (d0)--(p) (p)--(c) (c)edge[ind=a]++(0,0.4) (p)edge[ind=b]++(45:0.4);
\end{tikzpicture}
};
\node(x1) at (4,0){
\begin{tikzpicture}
\atoms{data}{0/p={0,0}}
\draw (0)edge[ind=a]++(0,0.5);
\atoms{copy}{1/p={0.7,0}}
\draw (1)edge[ind=b]++(0,0.5);
\end{tikzpicture}
};
\draw (x0)edge[->]node[midway,above]{$\operatorname{Discard}$} (x1);
\end{tikzpicture}\;.
\end{equation}
The diagram for the effective 2-network consists of a single shape for the $a$-data only,
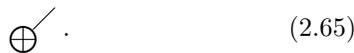
\begin{equation}
\begin{tikzpicture}
\atoms{circ,cross}{0/}
\draw (0)--++(45:0.6);
\end{tikzpicture}\;.
\end{equation}
The 2-index 1-data of the source 1-network is invisible in the diagram. A 2-network acting trivial on a 1-network with only one $a$ 1-data would be represented by the same diagram. We see that strictly speaking the diagram alone does not fully specify the effective 2-network if there are discarded elements.
\item Consider a more elaborate 2-network consisting of a tensor product and two contractions,
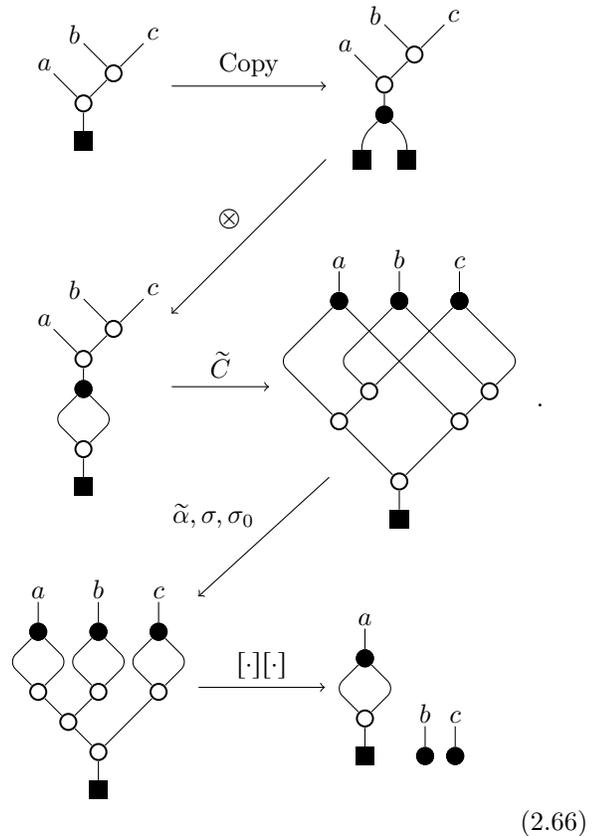
\begin{equation}
\begin{tikzpicture}
\node(x0) at (0,0){
\begin{tikzpicture}
\atoms{prod}{0/p={0,0}, 1/p={0.4,0.4}}
\atoms{data}{d/p={0,-0.5}}
\draw (0)--(1) (0)edge[ind=a]++(-0.4,0.4) (0)--(d) (1)edge[ind=b]++(-0.4,0.4) (1)edge[ind=c]++(0.4,0.4);
\end{tikzpicture}};
\node(x1) at (4,0){
\begin{tikzpicture}
\atoms{prod}{0/p={0,0}, 1/p={0.4,0.4}}
\atoms{data}{d0/p={-0.3,-1}, d1/p={0.3,-1}}
\atoms{copy}{c/p={0,-0.4}}
\draw[rc] (0)--(1) (0)edge[ind=a]++(-0.4,0.4) (0)--(c) (1)edge[ind=b]++(-0.4,0.4) (1)edge[ind=c]++(0.4,0.4) (c)--++(-0.3,-0.3)--(d0) (c)--++(0.3,-0.3)--(d1);
\end{tikzpicture}};
\node(x2) at (0,-4){
\begin{tikzpicture}
\atoms{prod}{0/p={0,0}, 1/p={0.4,0.4}, 2/p={0,-1.2}}
\atoms{data}{d/p={0,-1.7}}
\atoms{copy}{c/p={0,-0.4}}
\draw[rc] (0)--(1) (0)edge[ind=a]++(-0.4,0.4) (0)--(c) (1)edge[ind=b]++(-0.4,0.4) (1)edge[ind=c]++(0.4,0.4) (c)--++(-0.4,-0.4)--(2) (c)--++(0.4,-0.4)--(2) (2)--(d);
\end{tikzpicture}};
\node(x3) at (4,-4){
\begin{tikzpicture}
\atoms{prod}{0/p={0,0}, 1/p={0.8,0.8}, 2/p={1.2,1.2}, 3/p={-0.8,0.8}, 4/p={-0.4,1.2}}
\atoms{data}{d/p={0,-0.5}}
\atoms{copy}{c0/p={-0.8,2.4}, c1/p={0,2.4}, c2/p={0.8,2.4}}
\draw[rc] (0)--(1) (1)--(2) (0)--(3) (3)--(4) (3)--++(-0.8,0.8)--(c0) (1)--(c0) (4)--++(-0.4,0.4)--(c1) (2)--(c1) (2)--++(0.4,0.4)--(c2) (4)--(c2) (c0)edge[ind=a]++(0,0.4) (c1)edge[ind=b]++(0,0.4) (c2)edge[ind=c]++(0,0.4) (0)--(d);
\end{tikzpicture}};
\node(x4) at (0,-8){
\begin{tikzpicture}
\atoms{prod}{0/p={0,0}, 1/p={0.8,0.8}, 2/p={-0.4,0.4}, 3/p={0,0.8}, 4/p={-0.8,0.8}}
\atoms{data}{d/p={0,-0.5}}
\atoms{copy}{c0/p={-0.8,1.6}, c1/p={0,1.6}, c2/p={0.8,1.6}}
\draw[rc] (0)--(1) (0)--(2) (2)--(3) (2)--(4) (4)--++(-0.4,0.4)--(c0) (4)--++(0.4,0.4)--(c0) (3)--++(-0.4,0.4)--(c1) (3)--++(0.4,0.4)--(c1) (1)--++(-0.4,0.4)--(c2) (1)--++(0.4,0.4)--(c2) (c0)edge[ind=a]++(0,0.4) (c1)edge[ind=b]++(0,0.4) (c2)edge[ind=c]++(0,0.4) (0)--(d);
\end{tikzpicture}};
\node(x5) at (4,-8){
\begin{tikzpicture}
\atoms{prod}{1/p={0,0}}
\atoms{copy}{c/p={0,0.8}, c0/p={0.8,-0.5}, c1/p={1.2,-0.5}}
\atoms{data}{f/p={0,-0.5}}
\draw[rc] (1)--++(-0.4,0.4)--(c) (1)--++(0.4,0.4)--(c) (1)--(f-t) (c)edge[ind=a]++(0,0.4) (c0)edge[ind=b]++(0,0.4) (c1)edge[ind=c]++(0,0.4);
\end{tikzpicture}};
\draw (x0)edge[->] node[midway,above]{$\cop$}(x1);
\draw (x1)edge[->] node[midway,above left]{$\otimes$}(x2);
\draw (x2)edge[->] node[midway,above]{$\widetilde{C}$}(x3);
\draw ([xshift=0.8cm,yshift=0.8cm]x3.south west)edge[->] node[midway,above left]{$\widetilde{\alpha}, \sigma, \sigma_0$}(x4);
\draw (x4)edge[->] node[midway,above]{$[\cdot][\cdot]$}(x5);
\end{tikzpicture}
\;.
\end{equation}
Here, $\widetilde{\alpha}, \sigma, \sigma_0$ denotes some sequence of those 1-axioms and 2-functions performing the desired transformation. The effective 2-network would be denoted by
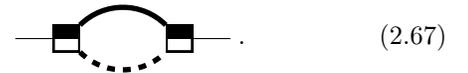
\begin{equation}
\begin{tikzpicture}
\atoms{square,dec={rotate=90}{rhalf}}{0/, 1/p={1.5,0}}
\draw[line width=2] (0-tr)to[bend left=45](1-tl);
\draw[line width=2,dashed] (0-br)to[bend right=45](1-bl);
\draw (0-l)--++(180:0.5) (1-r)--++(0:0.5);
\end{tikzpicture}\;.
\end{equation}
The 1-data which are input to the tensor product are copies of the same 1-data, so we use the same shape. Note that one of the shapes in the drawing has been reflected. The three different bindings are indicated by 3 line styles. Another way of drawing the same diagram is
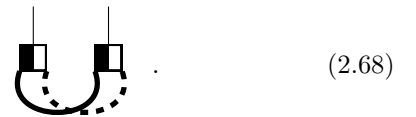
\begin{equation}
\begin{tikzpicture}
\atoms{rot=90,square,dec={rotate=90}{rhalf}}{0/, 1/p={1,0}}
\draw[line width=2] (0-tl)to[bend right=110,looseness=2](1-tl);
\draw[line width=2,dashed] (0-bl)to[bend right=110,looseness=2](1-bl);
\draw (0-r)--++(90:0.5) (1-r)--++(90:0.5);
\end{tikzpicture}\;.
\end{equation}
\item Consider the following (simple) effective 2-network (without giving the original 2-network),
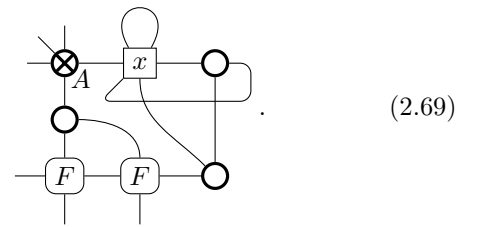
\begin{equation}
\begin{tikzpicture}
\node[draw,rectangle,rounded corners] (f1) at (0,0) {$F$};
\node[draw,rectangle,rounded corners] (f2) at (1,0) {$F$};
\node[draw,rectangle,minimum width=0.4cm,minimum height=0.4cm] (b) at (1,1.5) {$x$};
\node[draw,circle,minimum width=0.3cm,line width=1.2] (c1) at (0,0.75) {};
\node[draw,circle,minimum width=0.3cm,line width=1.2] (c2) at (2,0) {};
\node[draw,circle,minimum width=0.3cm,line width=1.2] (c3) at (2,1.5) {};
\node[draw,circle,minimum width=0.3cm,line width=1.2] (cr) at (0,1.5) {};
\begin{scope}
\clip (cr)circle(0.15);
\draw[line width=1.2] ($(cr)+(-0.15,0.15)$)--++(0.3,-0.3) ($(cr)+(0.15,0.15)$)--++(-0.3,-0.3);
\end{scope}
\node at ($(cr)+(-45:0.3)$){$A$};
\draw (f1)--(c1) (c1)--(cr) (cr)--(b) (b)--(c3) (c3)--(c2) (c2)--(f2) (f2)--(f1) (f1.south)--++(0,-0.4) (f1.west)--++(-0.4,0) (f2.south)--++(0,-0.4) (cr)--++(180:0.5) (cr)--++(135:0.5) (cr)--++(90:0.5) (b.south)to[out=-90,in=135] (c2) (c1)to[out=0,in=90](f2);
\draw[rc] (b.south west)--++(-0.3,-0.3)--++(2,0)|-(c3) ([xshift=-0.15cm]b.north)to[out=120,in=60,looseness=7]([xshift=0.15cm]b.north);
\end{tikzpicture}\;.
\end{equation}
The source effective 1-network has (at least) $4$ elements. The ``circle'' element is copied $3$ times, the $F$ element $2$ times, and there is only a single copy of the $A$ and $x$ elements. 
\end{itemize}
\end{myexmp}

As we saw in the examples, the diagrammatic calculus of the effective 2-scheme is an enormous simplification compared to specifying a full 2-network.

\section{Other flavors}
\label{sec:other_flavors}
We already gave a full definition of tensor types as 2-types of a certain 2-scheme. However, it makes sense to consider a few variations of this 2-scheme to be able to formalize more general structures as tensor types. In essence, we aim for 2-schemes which are as general as possible while maintaining roughly the same graphical calculus.

\subsection{Asymmetric  contraction}
We can do without the 2-axiom Eq.~\eqref{eq:2ax_symmetric_contraction}. We will say that a tensor type has a symmetric contraction, if this 2-axiom holds, and an asymmetric contraction otherwise. Tensor types with an asymmetric contraction only require a slight modification to the diagrammatic calculus: Every bond has to be equipped with a direction which tells us which index came first and which second in a contraction. Those bond directions will be indicated by little arrows, e.g.,
\begin{equation}
\begin{tikzpicture}
\atoms{circ}{0/, 1/p={1,0}, 2/p={0.5,0.5}}
\draw (0)edge[mark=arr](1) (1)edge[mark=arr](2) (0)edge[mark={arr,-}](2) (0)--++(-135:0.5) (0)--++(135:0.5) (1)--++(-45:0.5) (2)--++(90:0.5);
\end{tikzpicture}\;.
\end{equation}


Both array tensors and pairing tensors do have a symmetric contraction. E.g., the contraction of an array $A_{i,j,k}$ after reshaping only depends on the entries with $j=k$ which are invariant under permuting $j$ and $k$. Having an asymmetric contraction is very important for physics though, as it is needed to define $\zz_2$-twisted symmetric tensor which describe physical models with fermions. A 2-index tensor in this case is a matrix consisting of four blocks, corresponding to even-even, even-odd, odd-even, or odd-odd fermion parity (in fact only the even-even and odd-odd sectors are allowed to be non-zero as the total parity must be conserved). In addition to transposition, $\sigma_0$ negates the odd-odd sector corresponding to the fact that exchanging to fermions yields a factor of $-1$. This factor of $-1$ changes the outcome of the contraction.

\subsection{Adding a unit 1-function}
Conventional array tensors include scalars, i.e., numbers, which are tensors without indices and therefore without 0-data. In our current formulation, a 1-data is always with respect to some 0-data. In order to represent tensors without indices, we need a \tdef{unit}{unit} 0-data $1 \in \dat_0$ which represents the value of an ``empty product''. To keep our formalism as systematic as possible, we formalize this unit 0-data as a unit 1-function with a trivial 1-element set as domain,
\begin{equation}
1: \{0\} \rightarrow \dat_0
\end{equation}
such that $1(0)$ is the actual unit 0-data. E.g., for array tensors, the unit is the number $1$, as a scalar is a vector with one entry. For pairing tensors, the unit is the number $0$ and the only scalar is the unique empty pairing of no dots. In a 1-network, we will represent the unit by
\begin{equation}
\label{eq:1unit}
\begin{tikzpicture}
\atoms{prod}{p/}
\draw (p)edge[ind=a]++(0,-0.4);
\end{tikzpicture}
\;.
\end{equation}
So we use the same symbol as for the product, just that there are no inputs.

E.g., 1-data representing a scalar is drawn by
\begin{equation}
\begin{tikzpicture}
\atoms{prod}{0/}
\atoms{data}{d/p={0,-0.5}}
\draw (0)--(d);
\end{tikzpicture}\;.
\end{equation}

There are a few 1-axioms which it makes sense to assume. E.g., taking the product with the trivial 0-data shouldn't change anything,
\begin{equation}
\label{eq:1unitality}
\widetilde{U}:\quad
\begin{tikzpicture}
\atoms{prod}{0/, 1/p={-0.5,0.5}}
\draw (0)--(1) (0)edge[ind=b]++(0,-0.5) (0)edge[ind=a]++(45:0.5);
\end{tikzpicture}
=
\begin{tikzpicture}
\draw (0,0)edge[startind=b,ind=a]++(0,0.5);
\end{tikzpicture}
\;.
\end{equation}
The same holds for a product from the right, abbreviated by the same symbol $\widetilde{U}$,
\begin{equation}
\widetilde{U}:\quad
\begin{tikzpicture}
\atoms{prod}{0/, 1/p={0.5,0.5}}
\draw (0)--(1) (0)edge[ind=b]++(0,-0.5) (0)edge[ind=a]++(135:0.5);
\end{tikzpicture}
=
\begin{tikzpicture}
\draw (0,0)edge[startind=b,ind=a]++(0,0.5);
\end{tikzpicture}
\;.
\end{equation}
E.g., this works for array and pairing tensors, $1$ is the multiplicative unit and $0$ the additive unit of the natural numbers.

Also the following two ``1-axioms'' hold automatically due to the special properties of the copy and discard functions,
\begin{equation}
\begin{gathered}
\begin{tikzpicture}
\atoms{copy}{p/p={0,0}}
\atoms{prod}{u/p={0,0.4}}
\draw (p)edge[ind=a]++(-0.4,-0.4) (p)edge[ind=b]++(0.4,-0.4) (p)--(u);
\end{tikzpicture}
=
\begin{tikzpicture}
\atoms{prod}{0/, 1/p={0.5,0}}
\draw (0)edge[ind=a]++(0,-0.4) (1)edge[ind=b]++(0,-0.4);
\end{tikzpicture}\;,
\\
\begin{tikzpicture}
\atoms{copy}{p/}
\atoms{prod}{u/p={0,0.5}}
\draw (p)--(u);
\end{tikzpicture}
=
\hspace{1cm}\;.
\end{gathered}
\end{equation}

In the graphical calculus, we want be able to omit indices with unit 0-data. As a consequence, we demand that the 2-functions become trivial when one or more of the involved 0-data is the unit one. E.g., the commutor with one unit 0-data should be trivial,
\begin{equation}
\label{eq:2axiom_unit_commutor}
\begin{tikzpicture}[every node/.style={inner sep=0.2cm}]
\node (x0) at (0,0){
\begin{tikzpicture}
\atoms{data}{d/p={0,-0.5}}
\atoms{prod}{0/p={0,0}, 1/p={0.4,0.4}, 2/p={0.8,0.8}}
\draw (0)--(d) (0)--(1) (1)--(2) (1)edge[ind=b]++(135:0.4) (0)edge[ind=a]++(135:0.4);
\end{tikzpicture}};
\node (x1) at (4,0){
\begin{tikzpicture}
\atoms{data}{d/p={0,-0.5}}
\atoms{prod}{0/p={0,0}, 1/p={0.4,0.4}, 3/p={0,0.8}}
\draw (0)--(d) (0)--(1) (1)edge[ind=b]++(45:0.4) (1)--(3) (0)edge[ind=a]++(135:0.4);
\end{tikzpicture}};
\node (x2) at (2,-2.5){
\begin{tikzpicture}
\atoms{data}{d/p={0,-0.5}}
\atoms{prod}{0/p={0,0}}
\draw (0)--(d) (0)edge[ind=b]++(45:0.4) (0)edge[ind=a]++(135:0.4);
\end{tikzpicture}};
\draw (x0)edge[->]node[midway,above]{$\sigma$}(x1);
\draw (x0)edge[<->]node[midway,below left]{$\widetilde{U}$}(x2);
\draw (x1)edge[<->]node[midway,below right]{$\widetilde{U}$}(x2);
\end{tikzpicture}
\;.
\end{equation}
Also the contraction of two trivial indices should be trivial,
\begin{equation}
\label{eq:2axiom_unit_contraction}
\begin{tikzpicture}[every node/.style={inner sep=0.2cm}]
\node (x0) at (0,0){
\begin{tikzpicture}
\atoms{data}{d/p={0,-0.5}}
\atoms{prod}{0/p={0,0}, 1/p={0.4,0.4}, 2/p={0.4,1.6}}
\atoms{copy}{c/p={0.4,1.2}}
\draw[rc] (0)--(d) (0)edge[ind=a]++(135:0.4) (0)--(1) (1)--++(0.4,0.4)--(c) (1)--++(-0.4,0.4)--(c) (c)--++(2);
\end{tikzpicture}};
\node (x1) at (4,0){
\begin{tikzpicture}
\atoms{data}{d/p={0,-0.5}}
\atoms{prod}{1/p={0.6,0}}
\atoms{copy}{c/p={0.6,-0.5}}
\draw (d)edge[ind=a]++(0,0.5) (1)--(c);
\end{tikzpicture}};
\node (x2) at (2,-2.5){
\begin{tikzpicture}
\atoms{data}{d/p={0,-0.5}}
\draw (d)edge[ind=a]++(0,0.5);
\end{tikzpicture}};
\draw (x0)edge[->]node[midway,above]{$[\cdot]$}(x1);
\draw (x0)edge[<->]node[midway,below left]{$\widetilde{U}\widetilde{U}\widetilde{C}$}(x2);
\draw (x1)edge[<->]node[midway,below right]{$\widetilde{C}$}(x2);
\end{tikzpicture}
\;.
\end{equation}

Array tensors have the number $1$, which is a scalar such that the tensor product with this scalar leaves any tensor invariant. Consequently, the empty tensor network can be interpreted as this number $1$. In the general setting, this number $1$ is a 1-data with respect to the unit 0-data. This 1-data can be formalized as a 2-function with trivial domain called the \tdef{trivial tensor}{trivial_tensor},
\begin{equation}
\mathbf{1}:\quad
\hspace{2cm}
\quad \rightarrow \quad
\begin{tikzpicture}
\atoms{prod}{0/}
\atoms{data}{d/p={0,-0.5}}
\draw (0)--(d);
\end{tikzpicture}\;.
\end{equation}
Note that the empty left hand side has no 1-data elements, and thus denotes the empty cartesian product, which is a 1-element set, so the notation is consistent.

The trivial tensor has to obey the aforementioned 2-axiom, that it acts as a unit under tensor product,
\begin{equation}
\label{eq:2axiom_trivial_tensor}
\begin{tikzpicture}[every node/.style={inner sep=0.2cm}]
\node (x0) at (0,0){
\begin{tikzpicture}
\atoms{data}{0/}
\draw (0-t)edge[ind=a]++(0,0.4);
\end{tikzpicture}};
\node (x1) at (4,0){
\begin{tikzpicture}
\atoms{prod}{0/}
\atoms{data}{d0/p={0,-0.5}, d1/p={0.5,-0.5}}
\draw (d1-t)edge[ind=a]++(0,0.4) (0)--(d0);
\end{tikzpicture}};
\node (x2) at (0,-2.5){
\begin{tikzpicture}
\atoms{data}{0/}
\draw (0-t)edge[ind=a]++(0,0.4);
\end{tikzpicture}};
\node (x3) at (4,-2.5){
\begin{tikzpicture}
\atoms{data}{d/p={0,-0.5}}
\atoms{prod}{0/p={0,0}, 1/p={-0.4,0.4}}
\draw (0)--(d) (0)--(1) (0)edge[ind=a]++(0.4,0.4);
\end{tikzpicture}};
\draw (x0)edge[->]node[midway,above]{$\mathbf{1}$}(x1);
\draw (x0)edge[->]node[midway,left]{$=$}(x2);
\draw (x1)edge[->]node[midway,right]{$\otimes$}(x3);
\draw (x3)edge[->]node[midway,below]{$\widetilde{U}$}(x2);
\end{tikzpicture}
\;.
\end{equation}

Note that with the help of the unit 1-function and the trivial tensor, the commutor $\sigma_0$ without auxiliary index can be obtained from $\sigma$,
\begin{equation}
\begin{tikzpicture}[every node/.style={inner sep=0.2cm}]
\node (x0) at (0,0){
\begin{tikzpicture}
\atoms{prod}{0/p={0,0}}
\atoms{data}{d/p={0,-0.5}}
\draw (0)--(d) (0)edge[ind=b]++(0.4,0.4) (0)edge[ind=a]++(-0.4,0.4);
\end{tikzpicture}};
\node (x4) at (0,-6){
\begin{tikzpicture}
\atoms{prod}{0/p={0,0}}
\atoms{data}{d/p={0,-0.5}}
\draw (0)--(d) (0)edge[ind=a]++(0.4,0.4) (0)edge[ind=b]++(-0.4,0.4);
\end{tikzpicture}};
\node (x1) at (4,0){
\begin{tikzpicture}
\atoms{prod}{0/p={0,0}, 2/p={-0.8,0}}
\atoms{data}{d0/p={0,-0.5}, d1/p={-0.8,-0.5}}
\draw (0)--(d0) (0)edge[ind=b]++(0.4,0.4) (0)edge[ind=a]++(-0.4,0.4) (2)--(d1);
\end{tikzpicture}};
\node (x2) at (4,-3){
\begin{tikzpicture}
\atoms{prod}{0/p={0,0}, 1/p={0.4,0.4}, 2/p={-0.4,0.4}}
\atoms{data}{d/p={0,-0.5}}
\draw (0)--(d) (0)--(1) (0)--(2) (1)edge[ind=b]++(0.4,0.4) (1)edge[ind=a]++(-0.4,0.4);
\end{tikzpicture}};
\node (x3) at (4,-6){
\begin{tikzpicture}
\atoms{prod}{0/p={0,0}, 1/p={0.4,0.4}, 2/p={-0.4,0.4}}
\atoms{data}{d/p={0,-0.5}}
\draw (0)--(d) (0)--(1) (0)--(2) (1)edge[ind=a]++(0.4,0.4) (1)edge[ind=b]++(-0.4,0.4);
\end{tikzpicture}};
\draw (x0)edge[->]node[midway,above]{$\mathbf{1}$}(x1);
\draw (x1)edge[->]node[midway,right]{$\otimes$}(x2);
\draw (x2)edge[->]node[midway,right]{$\sigma$}(x3);
\draw (x3)edge[->]node[midway,below]{$\widetilde{U}$}(x4);
\draw (x0)edge[->]node[midway,left]{$\sigma_0$}(x4);
\end{tikzpicture}
\;.
\end{equation}
We can either view this as another 2-axiom involving $\sigma$ and $\sigma_0$, or delete $\sigma_0$ from the 2-functions by replacing it with the above definition in terms of $\sigma$ everywhere. We will choose to keep using $\sigma_0$ as sort of an auxiliary 2-function.

\paragraph{Effective scheme}
The way in which the unit 1-function and the trivial tensor enter the effective 2-scheme/graphical calculus is simple. Every occurrence of a unit in a 1-network can be removed using $\widetilde{U}$, except for units whose outputs are directly input of a 1-data element,
\begin{equation}
\begin{tikzpicture}
\atoms{data}{0/p={0,0}}
\atoms{prod}{1/p={0,0.5}}
\draw (0-t)--(1);
\end{tikzpicture}\;.
\end{equation}
Such a scalar 1-data (or a copy of it as part of a 2-network) is denoted like an ordinary 1-data, just that it has no lines sticking out, e.g.,
\begin{equation}
\label{eq:substrate_example2}
\begin{tikzpicture}
\atoms{circ,dcross}{0/p={0,0}}
\end{tikzpicture}\;,
\end{equation}
or
\begin{equation}
\begin{tikzpicture}
\node[box](0)at(0,0){$X$};
\end{tikzpicture}\;.
\end{equation}

\subsection{Weakening associativity and unitality}
In our formalism so far, we didn't demand commutativity of the product as a 1-axiom, but implemented it in a weakened fashion via the commutor 2-function. In this section, we'll also weaken the associativity and unitality 1-axioms of the product and unit. Note that the reason why we introduce this weakening only now is that there are many interesting examples with strict associativity and unitality, whereas examples with strict commutativity always seem to be trivial. Note that in principle it is possible to weaken only associativity while keeping unitality strict, or vice versa. However, we find that those often come together naturally.

Weakening associativity means replacing the associativity 1-axiom by an \tdef{associator}{associator} 2-function,
\begin{equation}
\alpha:\quad
\begin{tikzpicture}
\atoms{data}{d/p={-0.4,-0.9}}
\atoms{prod}{0/p={0,0}, 1/p={0.4,0.4}, x/p={-0.4,-0.4}}
\draw (x)--(0)--(1) (0)edge[ind=b]++(-0.4,0.4) (x)--(d) (1)edge[ind=c]++(-0.4,0.4) (1)edge[ind=d]++(0.4,0.4) (x)edge[ind=a]++(-0.4,0.4);
\end{tikzpicture}
\quad \rightarrow \quad
\begin{tikzpicture}
\atoms{data}{d/p={-0.4,-0.9}}
\atoms{prod}{0/p={0,0}, 1/p={-0.4,0.4}, x/p={-0.4,-0.4}}
\draw (x)--(0)--(1) (0)edge[ind=d]++(0.4,0.4) (x)--(d) (1)edge[ind=b]++(-0.4,0.4) (1)edge[ind=c]++(0.4,0.4) (x)edge[ind=a]++(-0.4,0.4);
\end{tikzpicture}\;.
\end{equation}
We notice that there is a 0-data $a$ that isn't changed, just as we have for $\sigma$ compared to $\sigma_0$. As we will see soon, this will be important to generate all possible reshapings of tree-like 1-networks. Analogously, the unitality 1-axiom is replaced by a \tdef{unitor}{unitor} 2-function,
\begin{equation}
U:\quad
\begin{tikzpicture}
\atoms{data}{d/p={0,-0.5}}
\atoms{prod}{0/p={0,0}, 1/p={-0.4,0.4}}
\draw (0)--(1) (0)edge[ind=a]++(0.4,0.4) (0)--(d);
\end{tikzpicture}
\quad \rightarrow \quad
\begin{tikzpicture}
\atoms{data}{d/p={0,-0.5}}
\draw (d)edge[ind=a]++(0,0.5);
\end{tikzpicture}\;.
\end{equation}
Surely, tensor types with strict associativity and unitality directly give rise to weak ones where $\alpha$ and $U$ are simply identity functions. E.g., for array tensors, consider flattening an $a\times b\times c\times d$ array into a vector. The result does not depend on whether we first block $c$ and $d$ and then block the result with $b$, or first $b$ with $c$ and then the result with $d$ (though it does depend on the ordering $a,b,c,d$). So, $\alpha$ is the identity.

More precisely, weakening means that we replace any occurrence of the associativity 1-axiom $\widetilde{\alpha}$ in every 2-axiom introduced above by the associator $\alpha$. However, associativity $\widetilde{\alpha}$ can be applied at arbitrary places in the 1-network, whereas $\alpha$ defined above only applies the effect of $\widetilde{\alpha}$ to the 1-network at one specific location relative to the 1-data element. Furthermore, $\widetilde{\alpha}$ can be applied forward and backward, whereas $\alpha$ is only defined in one direction. So, in order to be able to replace $\widetilde{\alpha}$ by a 2-function, we need to have different versions of $\alpha$ acting on different places of the 1-network, just as $\sigma_0$ is a different version of $\sigma$. Analogously, one might want to define different versions of $U$.

Luckily, all those different versions of $U$ and $\alpha$ can be generated by combining the ones already introduced. For example, a ``right unitor'' $U_r$ corresponding to applying the ``right unitality'' version of $\widetilde{U}$ can be obtained by combining the ``left'' unitor above with the commutor,
\begin{equation}
\begin{tikzpicture}
\node (x0) at (0,0){
\begin{tikzpicture}
\atoms{data}{d/p={0,-0.5}}
\atoms{prod}{0/p={0,0}, 1/p={0.4,0.4}}
\draw (0)--(1) (0)edge[ind=a]++(-0.4,0.4) (0)--(d);
\end{tikzpicture}};
\node (x1) at (4,0){
\begin{tikzpicture}
\atoms{data}{d/p={0,-0.5}}
\draw (d)edge[ind=a]++(0,0.5);
\end{tikzpicture}};
\node (x2) at (2,-2){
\begin{tikzpicture}
\atoms{data}{d/p={0,-0.5}}
\atoms{prod}{0/p={0,0}, 1/p={-0.4,0.4}}
\draw (0)--(1) (0)edge[ind=a]++(0.4,0.4) (0)--(d);
\end{tikzpicture}};
\draw (x0) edge[->]node[midway,above]{$U_r$}(x1);
\draw (x1)edge[<-]node[midway,below right]{$U$}(x2);
\draw (x0)edge[->]node[midway,below left]{$\sigma_0$}(x2);
\end{tikzpicture}
\;.
\end{equation}

Also, combining $\alpha$ with the trivial tensor, tensor product and unitor (or unitality if it is still strict) allows us to also define an associator $\alpha_0$ without auxiliary index,
\begin{equation}
\begin{tikzpicture}[every node/.style={inner sep=0.2cm}]
\node (x0) at (0,0){
\begin{tikzpicture}
\atoms{prod}{0/p={0,0}, 1/p={0.4,0.4}}
\atoms{data}{d0/p={0,-0.5}}
\draw (0)--(d0) (0)--(1) (0)edge[ind=a]++(-0.4,0.4) (1)edge[ind=c]++(0.4,0.4) (1)edge[ind=b]++(-0.4,0.4);
\end{tikzpicture}};
\node (x4) at (0,-6){
\begin{tikzpicture}
\atoms{prod}{0/p={0,0}, 1/p={-0.4,0.4}}
\atoms{data}{d0/p={0,-0.5}}
\draw (0)--(d0) (0)--(1) (0)edge[ind=c]++(0.4,0.4) (1)edge[ind=b]++(0.4,0.4) (1)edge[ind=a]++(-0.4,0.4);
\end{tikzpicture}};
\node (x1) at (4,0){
\begin{tikzpicture}
\atoms{prod}{0/p={0,0}, 1/p={0.4,0.4}, 2/p={-0.8,0}}
\atoms{data}{d0/p={0,-0.5}, d1/p={-0.8,-0.5}}
\draw (0)--(d0) (0)--(1) (0)edge[ind=a]++(-0.4,0.4) (1)edge[ind=c]++(0.4,0.4) (1)edge[ind=b]++(-0.4,0.4) (2)--(d1);
\end{tikzpicture}};
\node (x2) at (4,-3){
\begin{tikzpicture}
\atoms{prod}{0/p={0,0}, 1/p={0.4,0.4}, 2/p={0.8,0.8}, 3/p={-0.4,0.4}}
\atoms{data}{d0/p={0,-0.5}}
\draw (0)--(d0) (0)--(1) (1)--(2) (0)--(3) (1)edge[ind=a]++(-0.4,0.4) (2)edge[ind=c]++(0.4,0.4) (2)edge[ind=b]++(-0.4,0.4);
\end{tikzpicture}};
\node (x3) at (4,-6){
\begin{tikzpicture}
\atoms{prod}{0/p={0,0}, 1/p={0.4,0.4}, 2/p={0,0.8}, 3/p={-0.4,0.4}}
\atoms{data}{d0/p={0,-0.5}}
\draw (0)--(d0) (0)--(1) (1)--(2) (0)--(3) (2)edge[ind=a]++(-0.4,0.4) (2)edge[ind=b]++(0.4,0.4) (1)edge[ind=c]++(0.4,0.4);
\end{tikzpicture}};
\draw (x0)edge[->]node[midway,above]{$\mathbf{1}$}(x1);
\draw (x1)edge[->]node[midway,right]{$\otimes$}(x2);
\draw (x2)edge[->]node[midway,right]{$\alpha$}(x3);
\draw (x3)edge[->]node[midway,below]{$U$}(x4);
\draw (x0)edge[->]node[midway,left]{$\alpha_0$}(x4);
\end{tikzpicture}\;.
\end{equation}
Furthermore, we can obtain a 2-function $\alpha^{-1}$ performing the inverse move of $\alpha$, by a sequence of $\alpha$ and $\sigma$. To this end, we replace every occurrence of $\widetilde{\alpha}$ by $\alpha$ in Eq.~\eqref{eq:2axiom_hexagon}, except for the arrow from the top left to the bottom left which is replaced by $\alpha^{-1}$. The equation can then be seen as a definition for $\alpha^{-1}$ in terms of $\alpha$ and $\sigma$ instead of an axiom.

Analogously to obtaining $\alpha_0$ from $\alpha$, we can define $\alpha_0^{-1}$ from $\alpha^{-1}$. With the help of $\alpha_0$ and $\alpha_0^{-1}$, we can now also define a version $\alpha_2$ of the associator with two auxiliary indices,
\begin{equation}
\begin{tikzpicture}[every node/.style={inner sep=0.2cm}]
\node (x0) at (0,0){
\begin{tikzpicture}
\atoms{prod}{0/p={0,0}, 1/p={0.4,0.4}, 2/p={0.8,0.8}, 3/p={1.2,1.2}}
\atoms{data}{f/p={0,-0.4}}
\draw (f-t)--(0) (0)--(1) (1)--(2) (2)--(3) (0)edge[ind=a]++(-0.4,0.4) (1)edge[ind=b]++(-0.4,0.4) (2)edge[ind=c]++(-0.4,0.4) (3)edge[ind=d]++(-0.4,0.4) (3)edge[ind=e]++(0.4,0.4);
\end{tikzpicture}};
\node (x1) at (4,0){
\begin{tikzpicture}
\atoms{prod}{0/p={0,0}, 1/p={0.4,0.4}, 2/p={0.8,0.8}, 3/p={0.4,1.2}}
\atoms{data}{f/p={0,-0.4}}
\draw (f-t)--(0) (0)--(1) (1)--(2) (2)--(3) (0)edge[ind=a]++(-0.4,0.4) (1)edge[ind=b]++(-0.4,0.4) (2)edge[ind=e]++(0.4,0.4) (3)edge[ind=c]++(-0.4,0.4) (3)edge[ind=d]++(0.4,0.4);
\end{tikzpicture}};
\node (x2) at (0,-4){
\begin{tikzpicture}
\atoms{prod}{0/p={0,0}, 1/p={-0.4,0.4}, 2/p={0.8,0.8}, 3/p={1.2,1.2}}
\atoms{data}{f/p={0,-0.4}}
\draw (f-t)--(0) (0)--(1) (0)--(2) (2)--(3) (1)edge[ind=a]++(-0.4,0.4) (1)edge[ind=b]++(0.4,0.4) (2)edge[ind=c]++(-0.4,0.4) (3)edge[ind=d]++(-0.4,0.4) (3)edge[ind=e]++(0.4,0.4);
\end{tikzpicture}};
\node (x3) at (4,-4){
\begin{tikzpicture}
\atoms{prod}{0/p={0,0}, 1/p={-0.4,0.4}, 2/p={0.8,0.8}, 3/p={0.4,1.2}}
\atoms{data}{f/p={0,-0.4}}
\draw (f-t)--(0) (0)--(1) (0)--(2) (2)--(3) (1)edge[ind=a]++(-0.4,0.4) (1)edge[ind=b]++(0.4,0.4) (2)edge[ind=e]++(0.4,0.4) (3)edge[ind=c]++(-0.4,0.4) (3)edge[ind=d]++(0.4,0.4);
\end{tikzpicture}};
\draw (x0)edge[->]node[midway,above]{$\alpha_2$}(x1);
\draw (x0)edge[->]node[midway,left]{$\alpha_0$}(x2);
\draw (x2)edge[->]node[midway,below]{$\alpha$}(x3);
\draw (x3)edge[->]node[midway,right]{$\alpha_0^{-1}$}(x1);
\end{tikzpicture}
\;.
\end{equation}

As we already said, we need to replace every occurrence of $\widetilde{\alpha}$ with a version of the associator, depending on which place in the 1-network it is applied. Those versions could be $\alpha_0$, $\alpha$, $\alpha_2$, or something like $\sigma\alpha_2\sigma$ (recall that we drop the $\circ$ symbol in compositions of 2-functions). Specifically, in Eq.~\eqref{eq:2axiom_tprod_associativity} and Eq.~\eqref{eq:2axiom_contraction_tprod}, we replace $\widetilde{\alpha}$ by $\alpha_0$. In Eq.~\eqref{eq:2axiom_hexagon} it is $\alpha$ or $\alpha^{-1}$ as already said, and in Eq.~\eqref{eq:2axiom_contraction_commutative}, $\widetilde{\alpha}\sigma\widetilde{\alpha}$ is replaced by $\alpha_0\sigma\alpha_0^{-1}$.

The same thing happens for $\widetilde{U}$. In Eq.~\eqref{eq:2axiom_unit_commutor}, we replace one $\widetilde{U}$ by $\alpha_0^{-1}U_r\alpha_0$ and the other one by $\alpha_0^{-1}U_r\alpha_0\sigma$, which makes the 2-axiom a tautology and we can get rid of it. In Eq.~\eqref{eq:2axiom_unit_contraction}, we replace $\widetilde{U}\widetilde{U}\widetilde{C}$ by $U_rU_r\alpha_0\widetilde{C}$. In Eq.~\eqref{eq:2axiom_trivial_tensor}, we have to replace $\widetilde{U}$ by $U$.

It does not suffice to only replace the 1-axioms $\widetilde{U}$ and $\widetilde{\alpha}$ by the corresponding 2-functions in all the previously defined 2-axioms. We also need to introduce new 2-axioms that would have been trivial for $\widetilde{U}$ and $\widetilde{\alpha}$, but become now non-trivial. Those 2-axioms are needed to justify the same graphical calculus/effective 2-scheme as for the strict case. E.g., the associator should be consistent with the tensor product,
\begin{equation}
\label{eq:2axiom_tprod_commutor}
\begin{tikzpicture}[every node/.style={inner sep=0.2cm}]
\node (x0) at (0,0){
\begin{tikzpicture}
\atoms{prod}{0/p={1,0.5}, 1/p={1.4,0.9}}
\atoms{data}{d0/, d1/p={1,0}}
\draw (0)--(d1) (0)--(1) (d0-t)edge[ind=a]++(0,0.4) (1)edge[ind=d]++(0.4,0.4) (1)edge[ind=c]++(-0.4,0.4) (0)edge[ind=b]++(-0.4,0.4);
\end{tikzpicture}};
\node (x1) at (4,0){
\begin{tikzpicture}
\atoms{prod}{0/p={0,0}, 1/p={0.4,0.4}, 2/p={0.8,0.8}}
\atoms{data}{d/p={0,-0.5}}
\draw (0)--(d) (0)--(1)--(2) (0)edge[ind=a]++(-0.4,0.4) (1)edge[ind=b]++(-0.4,0.4) (2)edge[ind=c]++(-0.4,0.4) (2)edge[ind=d]++(0.4,0.4);
\end{tikzpicture}};
\node (x2) at (0,-3){
\begin{tikzpicture}
\atoms{prod}{0/p={1,0.5}, 1/p={0.6,0.9}}
\atoms{data}{d0/, d1/p={1,0}}
\draw (0)--(d1) (0)--(1) (d0-t)edge[ind=a]++(0,0.4) (0)edge[ind=d]++(0.4,0.4) (1)edge[ind=b]++(-0.4,0.4) (1)edge[ind=c]++(0.4,0.4);
\end{tikzpicture}};
\node (x3) at (4,-3){
\begin{tikzpicture}
\atoms{prod}{0/p={0,0}, 1/p={0.4,0.4}, 2/p={0,0.8}}
\atoms{data}{d/p={0,-0.5}}
\draw (0)--(d) (0)--(1)--(2) (0)edge[ind=a]++(-0.4,0.4) (1)edge[ind=d]++(0.4,0.4) (2)edge[ind=b]++(-0.4,0.4) (2)edge[ind=c]++(0.4,0.4);
\end{tikzpicture}};
\draw (x0)edge[->]node[midway,above]{$\otimes$}(x1);
\draw (x0)edge[->]node[midway,left]{$\alpha_0$}(x2);
\draw (x1)edge[->]node[midway,right]{$\alpha$}(x3);
\draw (x2)edge[->]node[midway,below]{$\otimes$}(x3);
\end{tikzpicture}
\;.
\end{equation}
If we would exchange $\alpha$ and $\alpha_0$ with $\widetilde{\alpha}$, this would be a triviality as $\widetilde{\alpha}$ and $\otimes$ wouldn't overlap. However $\alpha$ and $\alpha_0$ act on partly the same 1-data as $\otimes$ and thus this is a non-trivial 2-axiom. Similarly, the associator should be consistent with contraction,
\begin{equation}
\label{eq:2axiom_contraction_associator}
\begin{tikzpicture}[every node/.style={inner sep=0.2cm}]
\node (x0) at (0,0){
\begin{tikzpicture}
\atoms{prod}{0/,1/p={0.4,0.4},2/p={-0.8,0.8}, 3/p={-0.4,1.2}}
\atoms{data}{d/p={0,-0.5}}
\atoms{copy}{c/p={0.4,1.2}}
\draw[rc] (d)--(0) (0)--(1) (0)--(2)--(3) (1)--++(0.4,0.4)--(c) (1)--++(-0.4,0.4)--(c) (c)edge[ind=d]++(0,0.4) (3)edge[ind=c]++(0.4,0.4) (2)edge[ind=a]++(-0.4,0.4) (3)edge[ind=b]++(-0.4,0.4);
\end{tikzpicture}};
\node (x1) at (4,0){
\begin{tikzpicture}
\atoms{prod}{0/p={0,0.5}, 1/p={0.4,0.9}}
\atoms{data}{d/}
\atoms{copy}{c/p={1,0}}
\draw (0)--(d) (0)--(1) (c)edge[ind=d]++(0,0.4) (1)edge[ind=c]++(0.4,0.4) (0)edge[ind=a]++(-0.4,0.4) (1)edge[ind=b]++(-0.4,0.4);
\end{tikzpicture}};
\node (x2) at (0,-4){
\begin{tikzpicture}
\atoms{prod}{0/,1/p={0.4,0.4},2/p={-0.8,0.8},3/p={-1.2,1.2}}
\atoms{data}{d/p={0,-0.5}}
\atoms{copy}{c/p={0.4,1.2}}
\draw[rc] (d)--(0) (0)--(1) (0)--(2)--(3) (1)--++(0.4,0.4)--(c) (1)--++(-0.4,0.4)--(c) (c)edge[ind=d]++(0,0.4) (2)edge[ind=c]++(0.4,0.4) (3)edge[ind=a]++(-0.4,0.4) (3)edge[ind=b]++(0.4,0.4);
\end{tikzpicture}};
\node (x3) at (4,-4){
\begin{tikzpicture}
\atoms{prod}{0/p={0,0.5}, 1/p={-0.4,0.9}}
\atoms{data}{d/}
\atoms{copy}{c/p={1,0}}
\draw (0)--(d) (0)--(1) (c)edge[ind=d]++(0,0.4) (0)edge[ind=c]++(0.4,0.4) (1)edge[ind=a]++(-0.4,0.4) (1)edge[ind=b]++(0.4,0.4);
\end{tikzpicture}};
\draw (x0)edge[->]node[midway,above]{$[\cdot]$}(x1);
\draw (x0)edge[->]node[midway,left]{$\sigma_0 \alpha \sigma_0$}(x2);
\draw (x1)edge[->]node[midway,right]{$\alpha$}(x3);
\draw (x2)edge[->]node[midway,below]{$[\cdot]$}(x3);
\end{tikzpicture}
\;.
\end{equation}
Also the unitor should be compatible with the tensor product and contraction, and we're sure that the reader can by now work out by themselves what the corresponding commuting diagrams would look like.

Furthermore, the unitor should be consistent with the commutor,
\begin{equation}
\label{eq:2axiom_unit_braiding_weak}
\begin{tikzpicture}[every node/.style={inner sep=0.2cm}]
\node (x0) at (0,0){
\begin{tikzpicture}
\atoms{data}{d/p={0,-0.5}}
\atoms{prod}{0/p={0,0}, 1/p={0.4,0.4}, 2/p={0.8,0.8}, 3/p={0,0.8}}
\draw (0)--(d) (0)--(1) (1)--(2) (1)--(3) (0)edge[ind=a]++(135:0.4);
\end{tikzpicture}};
\node (x1) at (4,0){
\begin{tikzpicture}
\atoms{data}{d/p={0,-0.5}}
\atoms{prod}{0/p={0,0}, 1/p={0.4,0.4}, 2/p={0.8,0.8}, 3/p={0,0.8}}
\draw (0)--(d) (0)--(1) (1)--(2) (1)--(3) (0)edge[ind=a]++(135:0.4);
\end{tikzpicture}};
\node (x2) at (2,-2.5){
\begin{tikzpicture}
\atoms{data}{d/p={0,-0.5}}
\atoms{prod}{0/p={0,0}, 1/p={0.4,0.4}}
\draw (0)--(d) (0)--(1) (0)edge[ind=a]++(135:0.4);
\end{tikzpicture}};
\draw (x0)edge[->]node[midway,above]{$\sigma$}(x1);
\draw (x0)edge[->]node[midway,below left]{$U\sigma_0\alpha_0$}(x2);
\draw (x1)edge[->]node[midway,below right]{$U\sigma_0\alpha_0$}(x2);
\end{tikzpicture}
\;.
\end{equation}
The strict version of this 2-axiom with $U$ replaced by $\widetilde{U}$ follows from the 2-axiom Eq.~\eqref{eq:2axiom_unit_commutor}.

We are still missing the perhaps most important additional 2-axiom to introduce. As $\widetilde{\alpha}$ does not act on 1-data, it is not only automatically consistent with 2-functions such as contraction or tensor product, but also with itself. That is, it doesn't matter which of two sequences of $\widetilde{\alpha}$ at different places we apply to a 1-network, as long as the resulting 1-network is the same. However, the two corresponding sequences of associator 2-functions $\alpha$, $\alpha_0$, etc., do act non-trivially on the same 1-data and might have different results. In order to justify the same graphical calculus as for the strict case, we should add 2-axioms that ensure that all pairs of associator sequences with the same source and target 1-network correspond to the same operation.

We believe that it is sufficient to add a single 1-axiom. It shares some similarity with the \emph{pentagon equation} known from category theory, even though formally very different,
\begin{equation}
\begin{tikzpicture}[every node/.style={inner sep=0.2cm}]
\node (x0) at (0,0){
\begin{tikzpicture}
\atoms{prod}{0/p={0,0}, 1/p={0.4,0.4}, 2/p={0.8,0.8}, 3/p={1.2,1.2}}
\atoms{data}{f/p={0,-0.4}}
\draw (f-t)--(0) (0)--(1) (1)--(2) (2)--(3) (0)edge[ind=a]++(-0.4,0.4) (1)edge[ind=b]++(-0.4,0.4) (2)edge[ind=c]++(-0.4,0.4) (3)edge[ind=d]++(-0.4,0.4) (3)edge[ind=e]++(0.4,0.4);
\end{tikzpicture}};
\node (x4) at (0,-8){
\begin{tikzpicture}
\atoms{prod}{0/p={0,0}, 1/p={0.4,0.4}, 2/p={0,0.8}, 3/p={1.2,1.2}}
\atoms{data}{f/p={0,-0.4}}
\draw (f-t)--(0) (0)--(1) (1)--(3) (1)--(2) (0)edge[ind=a]++(-0.4,0.4) (2)edge[ind=b]++(-0.4,0.4) (2)edge[ind=c]++(0.4,0.4) (3)edge[ind=d]++(-0.4,0.4) (3)edge[ind=e]++(0.4,0.4);
\end{tikzpicture}};
\node (x1) at (4,0){
\begin{tikzpicture}
\atoms{prod}{0/p={0,0}, 1/p={0.4,0.4}, 2/p={0.8,0.8}, 3/p={0.4,1.2}}
\atoms{data}{f/p={0,-0.4}}
\draw (f-t)--(0) (0)--(1) (1)--(2) (2)--(3) (0)edge[ind=a]++(-0.4,0.4) (1)edge[ind=b]++(-0.4,0.4) (2)edge[ind=e]++(0.4,0.4) (3)edge[ind=c]++(-0.4,0.4) (3)edge[ind=d]++(0.4,0.4);
\end{tikzpicture}};
\node (x2) at (4,-4){
\begin{tikzpicture}
\atoms{prod}{0/p={0,0}, 1/p={0.4,0.4}, 2/p={0,0.8}, 3/p={0.4,1.2}}
\atoms{data}{f/p={0,-0.4}}
\draw (f-t)--(0) (0)--(1) (1)--(2) (2)--(3) (0)edge[ind=a]++(-0.4,0.4) (1)edge[ind=e]++(0.4,0.4) (2)edge[ind=b]++(-0.4,0.4) (3)edge[ind=c]++(-0.4,0.4) (3)edge[ind=d]++(0.4,0.4);
\end{tikzpicture}};
\node (x3) at (4,-8){
\begin{tikzpicture}
\atoms{prod}{0/p={0,0}, 1/p={0.4,0.4}, 2/p={0,0.8}, 3/p={-0.4,1.2}}
\atoms{data}{f/p={0,-0.4}}
\draw (f-t)--(0) (0)--(1) (1)--(2) (2)--(3) (0)edge[ind=a]++(-0.4,0.4) (1)edge[ind=e]++(0.4,0.4) (2)edge[ind=d]++(0.4,0.4) (3)edge[ind=b]++(-0.4,0.4) (3)edge[ind=c]++(0.4,0.4);
\end{tikzpicture}};
\draw (x0)edge[->]node[midway,above]{$\alpha_2$}(x1);
\draw (x1)edge[->]node[midway,right]{$\alpha$}(x2);
\draw (x2)edge[->]node[midway,right]{$\sigma\circ \alpha_2\circ \sigma$}(x3);
\draw (x3)edge[<-]node[midway,below]{$\alpha$}(x4);
\draw (x0)edge[->]node[midway,left]{$\alpha$}(x4);
\end{tikzpicture}
\;.
\end{equation}

\subsection{Adding an identity 2-function}
For array tensors, we have an $a\times a$ identity matrix for every bond dimension $a$. This is a tensor with two indices, with the special property that after contracting one of its indices with an index of another tensor (after performing the tensor product), we get the original tensor again. Thus, it is consistently represented by a ``free bond'' in the graphical language. In our framework, such an identity matrix is a 1-data $\idop\in \dat_1(a\otimes a)$, which can be formalized as a 2-function that spits out the corresponding 1-data from nothing,
\begin{equation}
\label{eq:2function_identity}
\operatorname{id}:\quad
\begin{tikzpicture}
\atoms{copy}{c/p={0.4,1.2}}
\draw (c)edge[ind=a]++(0,0.4);
\end{tikzpicture}
\quad \rightarrow \quad
\begin{tikzpicture}
\atoms{prod}{1/p={0,0}}
\atoms{copy}{c/p={0,0.8}}
\atoms{data}{f/p={0,-0.4}}
\draw[rc] (1)--++(-0.4,0.4)--(c) (1)--++(0.4,0.4)--(c) (1)--(f-t) (c)edge[ind=a]++(0,0.4);
\end{tikzpicture}\;.
\end{equation}
Also pairing tensors have an identity. It is the pairing of $a+a$ dots pairing the $i$th dot with the $i+a$th dot.

The above mentioned special property turns into the following \tdef{defining 2-axiom}{identity_defining_2axiom} of the identity,
\begin{equation}
\label{eq:2axiom_identity}
\begin{tikzpicture}
\node (x0) at (0,0){
\begin{tikzpicture}
\atoms{data}{f/p={0,0}}
\atoms{prod}{0/p={0,0.4}}
\draw (f-t)--(0) (0)edge[ind=b]++(45:0.4) (0)edge[ind=a]++(135:0.4);
\end{tikzpicture}};
\node (x1) at (4,0){
\begin{tikzpicture}
\atoms{data}{f0/p={0,0}, f1/p={1.2,-0.4}}
\atoms{prod}{0/p={0,0.4}, 1/p={1.2,0}}
\atoms{copy}{c0/p={1.2,0.8}, c1/p={0.8,1.2}}
\draw[rounded corners] (f0-t)--(0) (0)edge[ind=a]++(135:0.4) (0)--(c1) (f1-t)--(1) (1)--++(-0.4,0.4)--(c0) (1)--++(0.4,0.4)--(c0) (c0)--(c1) (c1)edge[ind=b]++(0,0.4);
\end{tikzpicture}};
\node (x2) at (4,-6){
\begin{tikzpicture}
\atoms{data}{f0/p={0,0}}
\atoms{prod}{0/p={0,0.4}, 1/p={0.4,0.8}, 2/p={-0.8,1.2}}
\atoms{copy}{c0/p={0.4,1.6}, c1/p={0,2}}
\draw[rounded corners] (f0-t)--(0) (0)--(1) (0)--(2) (2)edge[ind=a]++(135:0.4) (2)--(c1) (1)--++(-0.4,0.4)--(c0) (1)--++(0.4,0.4)--(c0) (c0)--(c1) (c1)edge[ind=b]++(0,0.4);
\end{tikzpicture}};
\node (x3) at (0,-6){
\begin{tikzpicture}
\atoms{data}{f0/p={0,0}}
\atoms{prod}{0/p={0,0.4}, 1/p={0.4,0.8}, 2/p={-0.8,1.2}}
\atoms{copy}{c0/p={0.4,1.6}, c1/p={-0.4,2.4}}
\draw[rounded corners] (f0-t)--(0) (0)--(1) (0)--(2) (2)edge[ind=a]++(45:0.4) (2)--++(-0.4,0.4)--(c1) (1)--++(0.4,0.4)--(c0) (1)--++(-0.4,0.4)--(c0) (c0)--(c1) (c1)edge[ind=b]++(0,0.4);
\end{tikzpicture}};
\node (x4) at (0,-3){
\begin{tikzpicture}
\atoms{data}{f/p={0,0}}
\atoms{prod}{0/p={0,0.4}}
\draw (f-t)--(0) (0)edge[ind=a]++(45:0.4) (0)edge[ind=b]++(135:0.4);
\end{tikzpicture}};
\draw (x0)edge[->]node[midway,above]{$\idop\widetilde{C}$}(x1);
\draw (x1)edge[->]node[midway,right]{$\otimes$}(x2);
\draw (x2)edge[->]node[midway,below]{$\widetilde{C}\alpha_0\alpha^{-1}\sigma_0\alpha_0$}(x3);
\draw (x3)edge[->]node[midway,left]{$\widetilde{C}[\cdot]$}(x4);
\draw (x0)edge[->]node[midway,left]{$\sigma_0$}(x4);
\end{tikzpicture}\;.
\end{equation}
The role of $\alpha_0\alpha^{-1}\sigma_0\alpha_0$ is to permute the indices such that the following $[\cdot]$ contracts the $b$-index of the original tensor with an index of the identity.

For array tensors, The $a\times a$ identity matrix is invariant under exchanging its two indices. We can implement this property as another 2-axiom,
\begin{equation}
\label{eq:2axiom_identity_symmetric}
\begin{tikzpicture}[every node/.style={inner sep=0.2cm}]
\node (x0) at (2,0){
\begin{tikzpicture}
\atoms{copy}{c/p={0.4,-0.4}}
\draw (c)edge[ind=a]++(0,0.4);
\end{tikzpicture}};
\node (x1) at (4,-2){
\begin{tikzpicture}
\atoms{prod}{1/p={0,0}}
\atoms{copy}{c/p={0,0.8}}
\atoms{data}{x/p={0,-0.5}}
\draw[rc] (1)--++(0.4,0.4)--(c) (1)--++(-0.4,0.4)--(c) (1)--(x) (c)edge[ind=a]++(0,0.4);
\end{tikzpicture}};
\node (x2) at (0,-2){
\begin{tikzpicture}
\atoms{prod}{1/p={0,0}}
\atoms{copy}{c/p={0,0.8}}
\atoms{data}{x/p={0,-0.5}}
\draw[rc] (1)--++(0.4,0.4)--(c) (1)--++(-0.4,0.4)--(c) (1)--(x) (c)edge[ind=a]++(0,0.4);
\end{tikzpicture}};
\draw (x0)edge[->]node[midway,above right]{$\operatorname{id}$}(x1);
\draw (x0)edge[->]node[midway,above left]{$\operatorname{id}$}(x2);
\draw (x1)edge[<->]node[midway,below]{$\sigma_0$}(x2);
\end{tikzpicture}\;.
\end{equation}
Identities fulfilling this 2-axioms will be called \tdef{symmetric}{id_symmetric} and \mdef{asymmetric} otherwise. Symmetry of the identity is naturally demanded together with symmetry of the contraction. For tensor types with asymmetric contraction and identity there are four different ways of contracting an index pair between an identity tensor and another arbitrary tensor. Either the first or second index of the identity tensor can be either the first or the second index in the contraction with the other tensor. The 2-axiom in Eq.~\eqref{eq:2axiom_identity} corresponds to the case where the first index of the identity tensor is the second index of the contraction. In this case we usually also demand the version of this 2-axiom where the second index of the identity is the first index of the contraction. Furthermore, we demand that the two other contractions with the identity are equal to each other. Moreover, if the arbitrary other tensor has only one single index which is contracted with an identity index, then all four contractions with the identity tensor yield the original tensor again. The motivation for imposing such a 2-axiom is that it holds for relevant tensor types with asymmetric contraction and identity such as $\zz_2$-twisted symmetric tensors. We will write them down explicitly in terms of the effective diagrammatic calculus in the next paragraph.

Another 2-axiom we usually add is that the identity is compatible with the 0-data product and tensor product,
\begin{equation}
\label{eq:2axiom_identity_blocking}
\begin{tikzpicture}
\node (x0) at (0,0){
\begin{tikzpicture}
\atoms{copy}{0/,1/p={0.5,0}}
\draw (0)edge[ind=a]++(0,0.4) (1)edge[ind=b]++(0,0.4);
\end{tikzpicture}};
\node (x1) at (4,0){
\begin{tikzpicture}
\atoms{data}{d0/p={0,-0.1}, d1/p={1.2,-0.1}}
\atoms{prod}{0/p={0,0.4}, 1/p={1.2,0.4}}
\atoms{copy}{c0/p={0,1.2}, c1/p={1.2,1.2}}
\draw[rc] (d0)--(0) (d1)--(1) (0)--++(-0.4,0.4)--(c0) (0)--++(0.4,0.4)--(c0) (1)--++(-0.4,0.4)--(c1) (1)--++(0.4,0.4)--(c1) (c0)edge[ind=a]++(0,0.4) (c1)edge[ind=b]++(0,0.4);
\end{tikzpicture}};
\node (x2) at (4,-3.3){
\begin{tikzpicture}
\atoms{data}{d0/p={0,-0.1}}
\atoms{prod}{0/p={0,0.4}, 1/p={0.4,0.8}, 2/p={-0.4,0.8}}
\atoms{copy}{c0/p={0.4,1.6}, c1/p={-0.4,1.6}}
\draw[rc] (d0)--(0) (0)--(1) (0)--(2) (1)--++(-0.4,0.4)--(c0) (1)--++(0.4,0.4)--(c0) (2)--++(-0.4,0.4)--(c1) (2)--++(0.4,0.4)--(c1) (c0)edge[ind=b]++(0,0.4) (c1)edge[ind=a]++(0,0.4);
\end{tikzpicture}};
\node (x3) at (4,-7){
\begin{tikzpicture}
\atoms{data}{d0/p={0,-0.1}}
\atoms{prod}{0/p={0,0.4}, 1/p={0.4,0.8}, 2/p={-0.4,0.8}}
\atoms{copy}{c0/p={0.4,1.6}, c1/p={-0.4,1.6}}
\draw[rc] (d0)--(0) (0)--(1) (0)--(2) (1)--(c0) (1)--(c1) (2)--(c0) (2)--(c1) (c0)edge[ind=b]++(0,0.4) (c1)edge[ind=a]++(0,0.4);
\end{tikzpicture}};
\node (x4) at (0,-7){
\begin{tikzpicture}
\atoms{data}{f/p={0,-0.1}}
\atoms{prod}{0/p={0,0.4}, 1/p={0,1.6}}
\atoms{copy}{c/p={0,1.2}}
\draw[rc] (0)--++(-0.4,0.4)--(c) (0)--++(0.4,0.4)--(c) (f-t)--(0) (c)--(1) (1)edge[ind=b]++(45:0.4) (1)edge[ind=a]++(135:0.4);
\end{tikzpicture}};
\node (x5) at (0,-3.3){
\begin{tikzpicture}
\atoms{prod}{0/p={0,0.4}}
\atoms{copy}{c/}
\draw[rc] (c)--(0) (0)edge[ind=b]++(45:0.4) (0)edge[ind=a]++(135:0.4);
\end{tikzpicture}};

\draw (x0)edge[->]node[midway,above]{$\idop\idop$}(x1);
\draw (x1)edge[->]node[midway,right]{$\otimes$}(x2);
\draw (x2)edge[->]node[midway,right]{$\sigma\alpha_0\alpha^{-1}\sigma\alpha_0^{-1}$}(x3);
\draw (x3)edge[<->]node[midway,below]{$\widetilde{C}$}(x4);
\draw (x0)edge[<->]node[midway,left]{$\widetilde{C}$}(x5);
\draw (x5)edge[->]node[midway,left]{$\idop$}(x4);
\end{tikzpicture}\;.
\end{equation}
Note that the effect of the series $\sigma\alpha_0\alpha^{-1}\sigma\alpha_0^{-1}$ is to exchange the second and third index. Identities fulfilling this 2-axiom will be called \tdef{block compatible}{block_compatible}. Block-compatibility of the identity is naturally demanded together with block-compatibility of the contraction.

\paragraph{Effective scheme}
The identity can be elegantly incorporated into the effective 2-scheme. Identities in a 2-network create new 1-data just as copying operations of the source 1-data. Therefore, identities should be represented by 2-index ``atoms'' in the effective 2-scheme,
\begin{equation}
\begin{tikzpicture}
\atoms{labbox=$\idop$}{0/}
\draw (0-l)edge[]++(180:0.4) (0-r)edge[]++(0:0.4);
\end{tikzpicture}\;.
\end{equation}
With this, the 2-axiom in Eq.~\eqref{eq:2axiom_identity} can be expressed in the graphical calculus,
\begin{equation}
\begin{tikzpicture}
\atoms{labbox=$\idop$}{0/}
\atoms{labbox=$A$}{a/p={-1,0}}
\draw (0-l)--(a-r) (a-l)edge[ind=a]++(180:0.4) (0-r)edge[ind=b]++(0:0.4);
\end{tikzpicture}
=
\begin{tikzpicture}
\atoms{labbox=$A$}{0/}
\draw (0-l)edge[ind=a]++(180:0.4) (0-r)edge[ind=b]++(0:0.4);
\end{tikzpicture}\;.
\end{equation}
The same for the symmetry axiom Eq.~\eqref{eq:2axiom_identity_symmetric},
\begin{equation}
\begin{tikzpicture}
\atoms{labbox=$\idop$}{0/}
\draw (0-l)edge[ind=a]++(180:0.4) (0-r)edge[ind=b]++(0:0.4);
\end{tikzpicture}
=
\begin{tikzpicture}
\atoms{labbox=$\idop$}{0/}
\draw (0-l)edge[ind=b]++(180:0.4) (0-r)edge[ind=a]++(0:0.4);
\end{tikzpicture}\;.
\end{equation}
We see that in the graphical calculus, it is natural to represent $\idop$ by a single line, such that the shape is not distinguishable from the two indices sticking out,
\begin{equation}
\begin{tikzpicture}
\atoms{labbox=$\idop$}{0/}
\draw (0-l)edge[ind=a]++(180:0.4) (0-r)edge[ind=b]++(0:0.4);
\end{tikzpicture}
\quad\rightarrow\quad
\begin{tikzpicture}
\draw (0,0)edge[ind=b,startind=a]++(1,0);
\end{tikzpicture}\;.
\end{equation}
This way, the two axioms above are automatically implied by the graphical calculus. In the graphical calculus, we will only see identities none of whose indices are involved in contractions, represented by disconnected ``free line segments''. An exception is that the two indices of the same $\idop$ might be contracted with each other, in which case we have a disconnected ``free loop'',
\begin{equation}
\begin{tikzpicture}
\atoms{labbox=$\idop$}{0/}
\draw[rc] (0-l)--++(-0.4,0)--++(0,-0.8)-|($(0-r)+(0.4,0)$)--(0-r);
\end{tikzpicture}
\quad\rightarrow\quad
\begin{tikzpicture}
\draw (0,0)circle(0.5);
\end{tikzpicture}\;.
\end{equation}

On a formal level, when we bring 2-networks in a partial standard form, we can proceed with the identities as follows. Using the block-compatibility of the identity Eq.~\eqref{eq:2axiom_identity_blocking} we can achieve that every identity always ``acts on individual indices'', such that the 0-data $a$ in Eq.~\eqref{eq:2function_identity} is always an input 0-data. Using the properties of the copying operation we can achieve that every identity is created from scratch and not by copying. We can use the fact that the identity does not act on any input 1-data to move all identities to after where all the copying happens and before where all the tensor products happen. Now, identities for which at least one index is later contracted to something else can be removed using the defining axiom of the identity Eq.~\eqref{eq:2axiom_identity}. We see that it suffices to specify the following additional invariants for effective 2-scheme with identity:
\begin{itemize}
\item The number of identities whose indices are not involved in any contraction. Those identities will be called \tdef{free bonds}{free_bond}.
\item The number of identities, where the first index is later contracted with the second index. Those identities will be called \tdef{loops}{loop}.
\item It also matters to which binding we apply the identity. Thus, we have to specify the above two numbers for every binding separately.
\end{itemize}

If the contraction is not symmetric, we have to slightly modify the graphical calculus: In order to distinguish the two indices of the identity, we have to equip the corresponding free line with a small arrow,
\begin{equation}
\begin{tikzpicture}
\atoms{labbox=$\idop$}{0/}
\draw (0-l)edge[ind=a]++(180:0.4) (0-r)edge[ind=b]++(0:0.4);
\end{tikzpicture}
\quad\rightarrow\quad
\begin{tikzpicture}
\draw (0,0)edge[arr=+,ind=b,startind=a]++(1,0);
\end{tikzpicture}\;.
\end{equation}
Note that this is analogous to the contraction. Also the latter is represented by a line in the graphical calculus, and we have to add arrows to these lines if contraction is incompatible with exchanging the contracted indices. The 2-axiom in Eq.~\eqref{eq:2axiom_identity} becomes
\begin{equation}
\begin{tikzpicture}
\atoms{labbox=$A$}{0/}
\draw (0-l)edge[ind=a]++(180:0.4) (0-r)edge[mark={arr,p=0.3},mark={arr,p=0.7},ind=b]++(0:1);
\end{tikzpicture}
=
\begin{tikzpicture}
\atoms{labbox=$A$}{0/}
\draw (0-l)edge[ind=a]++(180:0.4) (0-r)edge[ind=b]++(0:0.4);
\end{tikzpicture}\;.
\end{equation}
As described in words above we also demand
\begin{equation}
\begin{tikzpicture}
\atoms{labbox=$A$}{0/}
\draw (0-l)edge[ind=a]++(180:0.4) (0-r)edge[mark={arr,p=0.3,-},mark={arr,p=0.7,-},ind=b]++(0:1);
\end{tikzpicture}
=
\begin{tikzpicture}
\atoms{labbox=$A$}{0/}
\draw (0-l)edge[ind=a]++(180:0.4) (0-r)edge[ind=b]++(0:0.4);
\end{tikzpicture}\;.
\end{equation}
The analogous equations where the arrows of contraction and identity point in different directions do \emph{not} hold, however, we have
\begin{equation}
\begin{tikzpicture}
\atoms{labbox=$A$}{0/}
\draw (0-l)edge[ind=a]++(180:0.4) (0-r)edge[mark={arr,p=0.3,-},mark={arr,p=0.7},ind=b]++(0:1);
\end{tikzpicture}
=
\begin{tikzpicture}
\atoms{labbox=$A$}{0/}
\draw (0-l)edge[ind=a]++(180:0.4) (0-r)edge[mark={arr,p=0.3},mark={arr,p=0.7,-},ind=b]++(0:1);
\end{tikzpicture}\;.
\end{equation}
Using this, any occurrence of identity matrices in a network can be simplified, e.g.,
\begin{equation}
\label{eq:identity_opposite_square}
\begin{tikzpicture}
\draw[mark={arr,p=0.25},mark={arr,p=0.5,-},mark={arr,p=0.75}] (0,0)--(2,0);
\end{tikzpicture}
=
\begin{tikzpicture}
\draw[mark={arr,p=0.25},mark={arr,p=0.5},mark={arr,p=0.75,-}] (0,0)--(2,0);
\end{tikzpicture}
=
\begin{tikzpicture}
\draw[mark={arr,p=0.5,-}] (0,0)--(1,0);
\end{tikzpicture}
\end{equation}
such that the remaining ones are either free bonds, loops, or wrong-direction identities at open indices, which we'll denote by adding an ingoing arrow to those open indices,
\begin{equation}
\begin{tikzpicture}
\atoms{labbox=$A$}{0/}
\draw (0-l)edge[ind=a]++(180:0.4) (0-r)edge[mark={arr,p=0.3},mark={arr,p=0.7,-},ind=b]++(0:1);
\end{tikzpicture}
\quad\rightarrow\quad
\begin{tikzpicture}
\atoms{labbox=$A$}{0/}
\draw (0-l)edge[ind=a]++(180:0.4) (0-r)edge[mark={arr,-},ind=b]++(0:0.7);
\end{tikzpicture}\;.
\end{equation}
As we also described in text above, another 2-axiom is given by
\begin{equation}
\label{eq:inverse_identity}
\begin{tikzpicture}
\atoms{labbox=$A$}{0/}
\draw (0-r)edge[mark={arr,p=0.3,-},mark={arr,p=0.7},ind=b]++(0:1);
\end{tikzpicture}
=
\begin{tikzpicture}
\atoms{labbox=$A$}{0/}
\draw (0-r)edge[ind=b]++(0:0.4);
\end{tikzpicture}\;.
\end{equation}
So, for any network, we can simultaneously switch for all open indices whether they have a wrong-direction identity or not.

\subsection{Adding a dual 1-function}
In monoidal category theory, morphisms have \emph{source} a \emph{target} objects, and the effective graphical calculus has a global flow of time. By definition, tensor 2-schemes have an effective calculus without a flow of time. However, for a lot of relevant tensor types we still need the distinction between source and target indices (which we'll call input and output indices), without having a cycle-freeness condition implementing the flow of time.

This can be implemented by a \tdef{dual}{dual} 1-function
\begin{equation}
\label{eq:1dual}
*: \dat_0\rightarrow \dat_0
\end{equation}
which we'll represent graphically by
\begin{equation}
\begin{tikzpicture}
\atoms{dual}{0/}
\draw (0-b)edge[ind=a]++(-90:0.3) (0-t)edge[ind=b]++(90:0.3);
\end{tikzpicture}\;.
\end{equation}
For a given 1-network, indices which have an odd number of duals on the corresponding path are input, the others are output. It doesn't matter where on the path the duals are and how many there are apart from the whether this number is even or odd. This can be ensured by introducing additional 1-axioms. The first one is the \tdef{involutivity}{dual_involutivity} of the dual,
\begin{equation}
\widetilde{D_I}:\quad
\begin{tikzpicture}
\atoms{dual}{0/, 1/p={0,0.4}}
\draw (0-t)--(1-b) (0-b)edge[ind=a]++(-90:0.3) (1-t)edge[ind=b]++(90:0.3);
\end{tikzpicture}
=
\begin{tikzpicture}
\draw (0,0)edge[startind=b,ind=a]++(90:0.5);
\end{tikzpicture}\;.
\end{equation}
The second is the \tdef{product automorphicity}{product_automorphicity},
\begin{equation}
\widetilde{D_\otimes}:\quad
\begin{tikzpicture}
\atoms{dual}{0/p={-0.4,0.4}, 1/p={0.4,0.4}}
\atoms{prod}{p/}
\draw (p)--(0-b) (p)--(1-b) (0-t)edge[ind=a]++(90:0.3) (1-t)edge[ind=b]++(90:0.3) (p)edge[ind=c]++(-90:0.4);
\end{tikzpicture}
=
\begin{tikzpicture}
\atoms{dual}{0/p={0,-0.4}}
\atoms{prod}{p/}
\draw (p)--(0-t) (p)edge[ind=a]++(135:0.4) (p)edge[ind=b]++(45:0.4) (0-b)edge[ind=c]++(-90:0.3);
\end{tikzpicture}
\;.
\end{equation}
If we have a unit 1-data, we also want to demand \tdef{unit automorphicity}{unit_automorphicity},
\begin{equation}
\widetilde{D_1}:\quad
\begin{tikzpicture}
\atoms{prod}{p/}
\draw (p)edge[ind=a]++(-90:0.4);
\end{tikzpicture}
=
\begin{tikzpicture}
\atoms{dual}{0/p={0,-0.4}}
\atoms{prod}{p/}
\draw (p)--(0-t) (0-b)edge[ind=a]++(-90:0.3);
\end{tikzpicture}
\;.
\end{equation}

Apart from that, the dual 1-function automatically obeys the usual 1-axioms with respect to copying and discarding,
\begin{equation}
\label{eq:graph_equivalences}
\begin{gathered}
\begin{tikzpicture}
\atoms{copy}{p/}
\atoms{dual}{d0/p={-0.4,-0.4}, d1/p={0.4,-0.4}}
\draw (d0)--(p) (d1)--(p) (d0)edge[ind=b]++(0,-0.4) (d1)edge[ind=c]++(0,-0.4) (p)edge[ind=a]++(0,0.4);
\end{tikzpicture}
=
\begin{tikzpicture}
\atoms{copy}{p/}
\atoms{dual}{d0/p={0,0.4}}
\draw (p)edge[ind=b]++(-0.4,-0.4) (p)edge[ind=c]++(0.4,-0.4) (d0)edge[ind=a]++(0,0.4) (p)--(d0);
\end{tikzpicture}\;,\\
\begin{tikzpicture}
\atoms{copy}{p/}
\draw (p)edge[ind=a]++(0,0.4);
\end{tikzpicture}
=
\begin{tikzpicture}
\atoms{copy}{p/}
\atoms{dual}{d0/p={0,0.4}}
\draw (d0)edge[ind=a]++(0,0.4) (p)--(d0);
\end{tikzpicture}
\;.
\end{gathered}
\end{equation}
Note as usual that the latter two ``axioms'' do not actually impose any constraints on the dual 1-function. The point of the dual is that we restrict to contracting input with output indices. That is, the definition of the contraction gets changed such that it contracts one input and one output index,
\begin{equation}
[\cdot]:\quad
\begin{tikzpicture}
\atoms{prod}{0/, 1/p={0.4,0.4}}
\atoms{copy}{c/p={0.4,1.2}}
\atoms{data}{d/p={0,-0.4}}
\atoms{dual}{x/p={0.8,0.8}}
\draw[rc] (0)--(1) (0)edge[ind=a]++(-0.4,0.4) (c)edge[ind=b]++(0,0.4) (1)--++(-0.4,0.4)--(c) (1)--(x-b) (x-t)--(c) (0)--(d);
\end{tikzpicture}
\quad \rightarrow \quad
\begin{tikzpicture}
\atoms{data}{d/}
\atoms{copy}{c/p={0.8,0}}
\draw (d)edge[ind=a]++(0,0.5) (c)edge[ind=b]++(0,0.5);
\end{tikzpicture}\;.
\end{equation}
If there is an identity tensor, it is natural to impose that it also has one input and one output index,
\begin{equation}
\operatorname{id}:\quad
\begin{tikzpicture}
\atoms{copy}{c/p={0.4,1.2}}
\draw (c)edge[ind=b]++(0,0.4);
\end{tikzpicture}
\quad \rightarrow \quad
\begin{tikzpicture}
\atoms{prod}{1/p={0,0}}
\atoms{copy}{c/p={0,0.8}}
\atoms{dual}{d/p={-0.4,0.4}}
\atoms{data}{x/p={0,-0.5}}
\draw[rc] (1)--++(0.4,0.4)--(c) (1)--(d-b) (d-t)--(c) (1)--(x) (c)edge[ind=b]++(0,0.4);
\end{tikzpicture}\;.
\end{equation}
The fact that the dual is at the right for the contraction but on the left for the identity is pure convention and happens to be the more convenient choice.

All we need to do is to add duals to all the 2-axioms defined earlier, such that for every contraction and identity there is one dual involved. E.g., the defining property of the $\idop$ in Eq.~\eqref{eq:2axiom_identity} becomes
\begin{equation}
\begin{tikzpicture}
\node (x0) at (0,0){
\begin{tikzpicture}
\atoms{data}{f/p={0,0}}
\atoms{prod}{0/p={0,0.4}}
\draw (f-t)--(0) (0)edge[ind=b]++(45:0.4) (0)edge[ind=a]++(135:0.4);
\end{tikzpicture}};
\node (x1) at (4,0){
\begin{tikzpicture}
\atoms{data}{f0/p={0,0}, f1/p={1.2,-0.4}}
\atoms{prod}{0/p={0,0.4}, 1/p={1.2,0}}
\atoms{copy}{c0/p={1.2,0.8}, c1/p={0.8,1.2}}
\atoms{small,fflat,square}{d1/p={0.8,0.4}}
\draw[rounded corners] (f0-t)--(0) (0)edge[ind=a]++(135:0.4) (0)--(c1) (f1-t)--(1) (1)--(d1-b) (d1-t)--(c0) (1)--++(0.4,0.4)--(c0) (c0)--(c1) (c1)edge[ind=b]++(0,0.4);
\end{tikzpicture}};
\node (x2) at (4,-6){
\begin{tikzpicture}
\atoms{data}{f0/p={0,0}}
\atoms{prod}{0/p={0,0.4}, 1/p={0.4,0.8}, 2/p={-0.8,1.2}}
\atoms{copy}{c0/p={0.4,1.6}, c1/p={0,2}}
\atoms{small,fflat,square}{d/p={0,1.2}}
\draw[rounded corners] (f0-t)--(0) (0)--(1) (0)--(2) (2)edge[ind=a]++(135:0.4) (2)--(c1) (1)--(d-b) (d-t)--(c0) (1)--++(0.4,0.4)--(c0) (c0)--(c1) (c1)edge[ind=b]++(0,0.4);
\end{tikzpicture}};
\node (x3) at (0,-6){
\begin{tikzpicture}
\atoms{data}{f0/p={0,0}}
\atoms{prod}{0/p={0,0.4}, 1/p={0.4,0.8}, 2/p={-0.8,1.2}}
\atoms{copy}{c0/p={0.4,1.6}, c1/p={0,2}}
\atoms{small,fflat,square}{d0/p={0.8,1.2}}
\draw[rounded corners] (f0-t)--(0) (0)--(1) (0)--(2) (2)edge[ind=a]++(135:0.4) (2)--(c1) (1)--(d0-b) (d0-t)--(c0) (1)--++(-0.4,0.4)--(c0) (c0)--(c1) (c1)edge[ind=b]++(0,0.4);
\end{tikzpicture}};
\node (x4) at (0,-3){
\begin{tikzpicture}
\atoms{data}{f/p={0,0}}
\atoms{prod}{0/p={0,0.4}}
\draw (f-t)--(0) (0)edge[ind=b]++(45:0.4) (0)edge[ind=a]++(135:0.4);
\end{tikzpicture}};
\draw (x0)edge[->]node[midway,above]{$\idop\widetilde{C}$}(x1);
\draw (x1)edge[->]node[midway,right]{$\otimes$}(x2);
\draw (x2)edge[->]node[midway,below]{$\widetilde{C}\alpha_0\alpha^{-1}\sigma_0\alpha_0$}(x3);
\draw (x3)edge[->]node[midway,left]{$\widetilde{C}[\cdot]$}(x4);
\draw (x0)edge[->]node[midway,left]{$\sigma_0$}(x4);
\end{tikzpicture}\;.
\end{equation}

For some of the 2-axioms we also need to use the involutivity or automorphicity at some places. E.g., despite the diagram for the identity not looking symmetric anymore we can still define symmetry of the identity analogous to Eq.~\eqref{eq:2axiom_identity_symmetric},
\begin{equation}
\label{eq:2axiom_identity_symmetry_dual}
\begin{tikzpicture}[every node/.style={inner sep=0.2cm}]
\node (x0) at (0,0){
\begin{tikzpicture}
\atoms{copy}{c/p={0.4,-0.4}}
\draw (c)edge[ind=a]++(0,0.4);
\end{tikzpicture}};
\node (x1) at (4,0){
\begin{tikzpicture}
\atoms{prod}{1/p={0,0}}
\atoms{copy}{c/p={0,0.8}}
\atoms{dual}{d/p={-0.4,0.4}}
\atoms{data}{x/p={0,-0.5}}
\draw[rc] (1)--(d-b) (d-t)--(c) (1)--++(0.4,0.4)--(c) (1)--(x) (c)edge[ind=a]++(0,0.4);
\end{tikzpicture}};
\node (x2) at (0,-3){
\begin{tikzpicture}
\atoms{copy}{c/}
\atoms{dual}{d/p={0,0.4}}
\draw (c)--(d-b) (d)edge[ind=a]++(0,0.4);
\end{tikzpicture}};
\node (x3) at (4,-3){
\begin{tikzpicture}
\atoms{prod}{1/p={0,0}}
\atoms{copy}{c/p={0,1}}
\atoms{dual}{d0/p={0.4,0.35}, d1/p={0.4,0.65}, d2/p={-0.4,0.5}}
\atoms{data}{x/p={0,-0.5}}
\draw (1)--(d2-b) (d2-t)--(c) (1)--(d0-b) (d0-t)--(d1-b) (d1-t)--(c) (1)--(x) (c)edge[ind=a]++(0,0.4);
\end{tikzpicture}};
\node (x4) at (4,-6.5){
\begin{tikzpicture}
\atoms{prod}{1/p={0,0}}
\atoms{copy}{c/p={0,0.8}}
\atoms{dual}{d/p={0.4,0.4}, dx/p={0,1.2}}
\atoms{data}{x/p={0,-0.5}}
\draw[rc] (1)--++(-0.4,0.4)--(c) (1)--(d-b) (d-t)--(c) (1)--(x) (c)--(dx) (dx)edge[ind=a]++(0,0.4);
\end{tikzpicture}};
\node (x5) at (0,-6.5){
\begin{tikzpicture}
\atoms{prod}{1/p={0,0}}
\atoms{copy}{c/p={0,0.8}}
\atoms{dual}{d/p={-0.4,0.4}, dx/p={0,1.2}}
\atoms{data}{x/p={0,-0.5}}
\draw[rc] (1)--++(0.4,0.4)--(c) (1)--(d-b) (d-t)--(c) (1)--(x) (c)--(dx) (dx)edge[ind=a]++(0,0.4);
\end{tikzpicture}};
\draw (x0)edge[->]node[midway,above right]{$\idop$}(x1);
\draw (x0)edge[<->]node[midway,left]{$\widetilde{C}$}(x2);
\draw (x1)edge[<->]node[midway,right]{$\widetilde{D_I}$}(x3);
\draw (x4)edge[<->]node[midway,below]{$\sigma_0$}(x5);
\draw (x2)edge[->]node[midway,left]{$\operatorname{id}$}(x5);
\draw (x3)edge[<->]node[midway,right]{$\widetilde{C}$}(x4);
\end{tikzpicture}\;.
\end{equation}

Of course, we can weaken involutivity and automorphicities, and turn them into 2-functions. Involutivity is replaced by an \tdef{involutor}{dual_involutor} 2-function and its inverse,
\begin{equation}
\begin{tikzpicture}
\node(x0)at(0,0){
\begin{tikzpicture}
\atoms{prod}{0/p={0,0}}
\atoms{dual}{d0/p={0.4,0.4}, d1/p={0.4,0.7}}
\atoms{data}{x/p={0,-0.5}}
\draw (0)edge[ind=a]++(-0.4,0.4) (0)--(d0-b) (d0-t)--(d1-b) (d1-t)edge[ind=b]++(0,0.3) (0)--(x);
\end{tikzpicture}};
\node(x1)at(4,0){
\begin{tikzpicture}
\atoms{prod}{0/p={0,0}}
\atoms{data}{x/p={0,-0.5}}
\draw (0)edge[ind=a]++(-0.4,0.4) (0)edge[ind=b]++(0.4,0.4) (0)--(x);
\end{tikzpicture}};
\draw ([yshift=0.3cm]x0.east)edge[bend left,->]node[midway,above]{$D_I$} ([yshift=0.3cm]x1.west);
\draw ([yshift=-0.3cm]x0.east)edge[bend right,<-]node[midway,below]{$D_I^{-1}$} ([yshift=-0.3cm]x1.west);
\end{tikzpicture}\;.
\end{equation}
Product automorphicity is replaced by a \tdef{product automorphor}{product_automorphor} 2-function and its inverse,
\begin{equation}
\begin{tikzpicture}
\node(x0)at(0,0){
\begin{tikzpicture}
\atoms{prod}{0/p={0,0}, 1/p={0.4,0.7}}
\atoms{data}{x/p={0,-0.5}}
\atoms{dual}{d/p={0.4,0.3}}
\draw (0)--(d-b) (d-t)--(1) (0)edge[ind=a]++(-0.4,0.4) (0)--(x) (1)edge[ind=b]++(-0.4,0.4) (1)edge[ind=c]++(0.4,0.4);
\end{tikzpicture}};
\node(x1)at(4,0){
\begin{tikzpicture}
\atoms{prod}{0/p={0,0}, 1/p={0.4,0.4}}
\atoms{data}{x/p={0,-0.5}}
\atoms{dual}{d0/p={0.8,0.8}, d1/p={0,0.8}}
\draw (0)--(1) (0)edge[ind=a]++(-0.4,0.4) (0)--(x) (1)--(d0-b) (d0-t)edge[ind=c]++(0,0.3) (1)--(d1-b) (d1-t)edge[ind=b]++(0,0.3);
\end{tikzpicture}};
\draw ([yshift=0.3cm]x0.east)edge[bend left,->]node[midway,above]{$D_\otimes$} ([yshift=0.3cm]x1.west);
\draw ([yshift=-0.3cm]x0.east)edge[bend right,<-]node[midway,below]{$D_\otimes^{-1}$} ([yshift=-0.3cm]x1.west);
\end{tikzpicture}\;,
\end{equation}
and analogously there is a unit automorphor. The 3 pairs of 2-functions have to obey the obvious 2-axioms that they are actually inverses to each other. As for unitality and associativity earlier, weakening consists in three steps. First, we define versions of $D_I$, $D_\otimes$ or $D_1$ that act on different places relative to the 0-data, such as $D_{I0}\coloneqq UD_I\otimes \idop$ (as sequence composed from right to left). Then, we replace every occurrence of $\widetilde{D_I}$, $\widetilde{D_\otimes}$, or $\widetilde{D_1}$ by the corresponding version of $D_\otimes$, $D_I$, or $D_1$, such as $\widetilde{D_I}$ by $D_I$ in Eq.~\eqref{eq:2axiom_identity_symmetry_dual}. Then, we need to add additional 2-axioms that ensure the compatibility of $D_\otimes$, $D_I$, or $D_1$ with the other 2-functions, such as
\begin{equation}
\label{eq:2axiom_involutor_commutor}
\begin{tikzpicture}[every node/.style={inner sep=0.2cm}]
\node (x0) at (0,0){
\begin{tikzpicture}
\atoms{prod}{0/, 2/p={-0.8,0.8}}
\atoms{data}{d/p={0,-0.5}}
\atoms{dual}{c0/p={0.4,0.4}, c1/p={0.4,0.8}}
\draw[rc] (d)--(0) (0)--(2) (0)--(c0-b) (c0-t)--(c1-b) (c1-t)edge[ind=c]++(0,0.4) (2)edge[ind=a]++(-0.4,0.4) (2)edge[ind=b]++(0.4,0.4);
\end{tikzpicture}};
\node (x1) at (4,0){
\begin{tikzpicture}
\atoms{prod}{0/, 2/p={-0.8,0.8}}
\atoms{data}{d/p={0,-0.5}}
\atoms{dual}{c0/p={0.4,0.4}, c1/p={0.4,0.8}}
\draw[rc] (d)--(0) (0)--(2) (0)--(c0-b) (c0-t)--(c1-b) (c1-t)edge[ind=c]++(0,0.4) (2)edge[ind=b]++(-0.4,0.4) (2)edge[ind=a]++(0.4,0.4);
\end{tikzpicture}};
\node (x2) at (0,-4){
\begin{tikzpicture}
\atoms{prod}{0/, 2/p={-0.4,0.4}}
\atoms{data}{d/p={0,-0.5}}
\draw[rc] (d)--(0) (0)--(2) (0)edge[ind=c]++(0.4,0.4) (2)edge[ind=a]++(-0.4,0.4) (2)edge[ind=b]++(0.4,0.4);
\end{tikzpicture}};
\node (x3) at (4,-4){
\begin{tikzpicture}
\atoms{prod}{0/, 2/p={-0.4,0.4}}
\atoms{data}{d/p={0,-0.5}}
\draw[rc] (d)--(0) (0)--(2) (0)edge[ind=c]++(0.4,0.4) (2)edge[ind=b]++(-0.4,0.4) (2)edge[ind=a]++(0.4,0.4);
\end{tikzpicture}};
\draw (x0)edge[->]node[midway,above]{$\sigma_0 \sigma \sigma_0$}(x1);
\draw (x0)edge[->]node[midway,left]{$D_I$}(x2);
\draw (x1)edge[->]node[midway,right]{$D_I$}(x3);
\draw (x2)edge[->]node[midway,below]{$\sigma_0 \sigma \sigma_0$}(x3);
\end{tikzpicture}
\;.
\end{equation}

\paragraph{Effective scheme}
Introducing the dual 1-function yields one more invariant of effective 1-networks: For each index, number of duals on the corresponding path modulo 2 cannot be changed by any 2-function or 1-axiom. On the other hand, it is quite easy to see that all 1-networks where those numbers mod 2 are equal are related by invertible 2-functions. This allows us to distinguish between \tdef{output indices}{output_index} with an even number of duals and \tdef{input indices}{input_index} with an odd number.

In the graphical calculus we will mark the output indices with a small ingoing arrow at the corresponding line end. E.g., the following 1-network
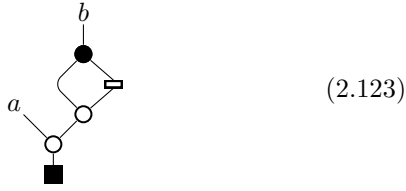
\begin{equation}
\begin{tikzpicture}
\atoms{prod}{0/p={0,0}, 1/p={0.4,0.4}}
\atoms{copy}{c/p={0.4,1.2}}
\atoms{data}{d/p={0,-0.4}}
\atoms{dual}{du/p={0.8,0.8}}
\draw[rc] (0)--(1) (0)edge[ind=a]++(-0.4,0.4) (c)edge[ind=b]++(0,0.4) (1)--++(-0.4,0.4)--(c) (1)--(du-b) (du-t)--(c) (0)--(d);
\end{tikzpicture}
\end{equation}
has 3 indices, one of which is an input index. The effective 1-network could be represented by
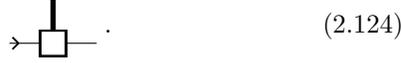
\begin{equation}
\begin{tikzpicture}
\atoms{square}{t/p={0,0}}
\draw (t-t)edge[line width=2]++(90:0.4) (t-l)edge[dualind]++(180:0.4) (t-r)--++(0:0.4);
\end{tikzpicture}\;.
\end{equation}

There is actually no visible difference in the graphical calculus of 2-networks. Merely, the input/output structure is a restriction to which network diagrams we can draw, namely only ones where we don't contract two outputs or two inputs.

\section{Tensor mappings}
\label{sec:tensor_mappings}
In this section we define a very powerful and general notion of one tensor type being able to emulate another tensor type: A \tdef{tensor mapping}{tensor_mapping} can be thought of as a way to embed one tensor type into another. If the mapping is invertible, then this embedding is faithful, which gives us a way for the target type to emulate the source tensor type. Tensor mappings are analogous to what is known as a \emph{(monoidal) functor} in category theory.

\subsection{Minimal useful flavor}
Technically, tensor mappings are nothing but 2-types of another 2-scheme. More precisely, there are different flavors of tensor mappings, just as there are different flavors of tensor types, and in this section we will describe the simplest such flavor. The 1-functions, 1-axioms, 2-functions, 2-axioms, etc. contain two copies of those of tensor types (of some flavor), along with a few additional 1-functions, 1-axioms, 2-functions, etc. connecting the two copies. The two different copies correspond to the source and the target tensor types, whereas the connecting components belong to the mapping itself. In order to distinguish the 1-functions, 2-functions, etc. of the different tensor types, we'll color those of the target tensor type in red. We'll work with an example of a mapping with prefactor pairing tensors as source and array tensors as target tensor type. Intuitively, this mapping corresponds to taking the pairing diagram and interpreting it as a tensor network consisting only of free bonds, which are $2\times 2$ identity matrices.

There is one additional 1-function which maps a 0-data of the source tensor type $\mathcal{A}$ to a 0-data of the target tensor type $\mathcal{B}$,
\begin{equation}
m:\quad \dat_0^{\mathcal{A}} \rightarrow \dat_0^{\mathcal{B}}\;.
\end{equation}
We'll depict it by a box whose lower half is red, according to the coloring of source and target,
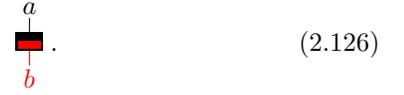
\begin{equation}
\begin{tikzpicture}
\atoms{mapping}{0/}
\draw (0-t)edge[ind=a]++(0,0.2);
\draw[red] (0-b)edge[ind=b]++(0,-0.2);
\end{tikzpicture}\;.
\end{equation}
For our example mapping from pairing to array tensors, $m$ maps a $a$ to $2^a$, i.e., the configurations of the resulting array tensors are length-$a$ bitstrings.

As an example for a 1-network, consider
\begin{equation}
\begin{tikzpicture}
\atoms{mapping second prod}{0/p={0,0}, 1/p={0.4,0.4}}
\atoms{mapping}{m/p={-0.4,0.4}}
\atoms{prod}{x0/p={-0.4,0.8}, x1/p={-0.8,1.2}}
\atoms{copy}{d0/p={-0.4,1.6}, d1/p={-1.2,1.6}}
\atoms{mapping second data}{f0/p={0,-0.4}}
\atoms{data}{f1/p={-1.6,0.8}}
\draw (m-t)--(x0) (x0)--(x1) (x1)--(d1) (x1)--(d0) (d0)edge[ind=b]++(0,0.4) (d1)edge[ind=a]++(0,0.4);
\draw[rc] (x0)--++(0.4,0.4)--(d0) (f1)--++(0,0.4)--(d1);
\draw[red] (0)--(m-b) (0)--(1) (0)--(f0);
\end{tikzpicture}
\end{equation}
representing an element of
\begin{equation}
\dat_1^{\mathcal{A}}(a) \times \dat_1^{\mathcal{B}}(m((a\otimes b)\otimes b)\otimes 1)\;.
\end{equation}

We also need one 2-function that maps 1-data of $\mathcal{A}$ to 1-data of $\mathcal{B}$,
\begin{equation}
M:\quad
\begin{tikzpicture}
\atoms{data}{d/}
\draw[ind=a] (d)--++(0,0.5);
\end{tikzpicture}
\quad\rightarrow\quad
\begin{tikzpicture}
\atoms{mapping}{0/p={0,0}}
\atoms{mapping second data}{d/p={0,-0.5}}
\draw[red] (0-b)--(d-t);
\draw (0-t)edge[ind=a]++(0,0.4);
\end{tikzpicture}
\;.
\end{equation}
For our example mapping, an $a$-dot pairing together with a prefactor $N$ is mapped to a vector with an entry for every length-$a$ bitstring. If the $i$th and $j$th dot are paired, the $i$th and $j$th entry in the bitstring must be the same. If this condition on the bitstring is fulfilled for all pairings, then the entry for this bitstring is $N$, otherwise it is $0$.

The mapping 1-function should be compatible with the product, i.e., we have the following 1-axiom which we'll call \tdef{product homomorphicity}{product_homomorphicity},
\begin{equation}
\widetilde{M_\otimes}:\quad
\begin{tikzpicture}
\atoms{prod}{0/}
\atoms{mapping}{m/p={0,-0.4}}
\draw[red] (m-b)edge[ind=c]++(0,-0.4);
\draw (0)--(m-t) (0)edge[ind=a]++(-0.4,0.4) (0)edge[ind=b]++(0.4,0.4);
\end{tikzpicture}
=
\begin{tikzpicture}
\atoms{mapping second prod}{0/}
\atoms{mapping}{m0/p={-0.4,0.4},m1/p={0.4,0.4}}
\draw[red] (0)--(m0-b) (0)--(m1-b) (0)edge[ind=c]++(0,-0.4);
\draw (m0-t)edge[ind=a]++(0,0.4) (m1-t)edge[ind=b]++(0,0.4);
\end{tikzpicture}
\;.
\end{equation}
This is true for our example mapping, as
\begin{equation}
2^{a+b}=2^a2^b\;.
\end{equation}

The 2-function needs to satisfy a number of 2-axioms. Basically, for every tensor 2-function $f$, there is a 2-axiom which makes sure that $f$ is compatible with the mapping 2-function $M$, schematically,
\begin{equation}
Mf^{\mathcal{A}}=f^{\mathcal{B}}M\;.
\end{equation}
E.g., for the tensor product 2-function we get,
\begin{equation}
\label{eq:2axiom_mapping_tprod}
\begin{tikzpicture}[every node/.style={inner sep=0.2cm}]
\node (x0) at (0,0){
\begin{tikzpicture}
\atoms{data}{0/, 1/p={0.5,0}}
\draw (0)edge[ind=a]++(0,0.4) (1)edge[ind=b]++(0,0.4);
\end{tikzpicture}};
\node (x1) at (4,0){
\begin{tikzpicture}
\atoms{prod}{0/}
\atoms{data}{d/p={0,-0.5}}
\draw (0)--(d) (0)edge[ind=a]++(-0.4,0.4) (0)edge[ind=b]++(0.4,0.4);
\end{tikzpicture}};
\node (x2) at (0,-3){
\begin{tikzpicture}
\atoms{mapping}{0/p={0,0}, 1/p={0.5,0}}
\atoms{mapping second data}{d0/p={0,-0.5}, d1/p={0.5,-0.5}}
\draw[red] (0-b)--(d0) (1-b)--(d1);
\draw (0-t)edge[ind=a]++(0,0.4) (1-t)edge[ind=b]++(0,0.4);
\end{tikzpicture}};
\node (x3) at (2,-6){
\begin{tikzpicture}
\atoms{mapping second prod}{0/p={0,0}}
\atoms{mapping}{m0/p={-0.4,0.4}, m1/p={0.4,0.4}}
\atoms{mapping second data}{d/p={0,-0.5}}
\draw[red] (0)--(d) (0)--(m0-b) (0)--(m1-b);
\draw (m0-t)edge[ind=a]++(0,0.4) (m1-t)edge[ind=b]++(0,0.4);
\end{tikzpicture}};
\node (x4) at (4,-3){
\begin{tikzpicture}
\atoms{prod}{0/p={0,0}}
\atoms{mapping}{m/p={0,-0.4}}
\atoms{mapping second data}{d/p={0,-0.9}}
\draw (m-t)--(0) (0)edge[ind=a]++(-0.4,0.4) (0)edge[ind=b]++(0.4,0.4);
\draw[mapping second 0dat] (m-b)--(d);
\end{tikzpicture}};
\draw (x0) edge[->] node[midway,left]{$M M$} (x2);
\draw (x0) edge[->] node[midway,above]{$\otimes$} (x1);
\draw (x2) edge[->] node[midway,below left,mapping second 0dat]{$\otimes$} (x3);
\draw (x1) edge[->] node[midway,right]{$M$} (x4);
\draw (x4)edge[->] node[midway,below right]{$\widetilde{M_\otimes}$} (x3);
\end{tikzpicture}\;.
\end{equation}

Or, for the commutor 2-function we get
\begin{equation}
\label{eq:2axiom_mapping_commutor}
\begin{tikzpicture}[every node/.style={inner sep=0.2cm}]
\node (x0) at (0,0){
\begin{tikzpicture}
\atoms{prod}{0/}
\atoms{data}{d/p={0,-0.5}}
\draw (0)--(d) (0)edge[ind=a]++(-0.4,0.4) (0)edge[ind=b]++(0.4,0.4);
\end{tikzpicture}};
\node (x1) at (4,0){
\begin{tikzpicture}
\atoms{prod}{0/}
\atoms{data}{d/p={0,-0.5}}
\draw (0)--(d) (0)edge[ind=b]++(-0.4,0.4) (0)edge[ind=a]++(0.4,0.4);
\end{tikzpicture}};
\node (x2) at (0,-3){
\begin{tikzpicture}
\atoms{prod}{0/p={0,0}}
\atoms{mapping}{m/p={0,-0.4}}
\atoms{mapping second data}{d/p={0,-0.9}}
\draw (m-t)--(0) (0)edge[ind=a]++(-0.4,0.4) (0)edge[ind=b]++(0.4,0.4);
\draw[mapping second 0dat] (m-b)--(d);
\end{tikzpicture}};
\node (x3) at (0,-6){
\begin{tikzpicture}
\atoms{mapping second prod}{0/p={0,0}}
\atoms{mapping}{m0/p={-0.4,0.4}, m1/p={0.4,0.4}}
\atoms{mapping second data}{d/p={0,-0.5}}
\draw[red] (0)--(d) (0)--(m0-b) (0)--(m1-b);
\draw (m0-t)edge[ind=a]++(0,0.4) (m1-t)edge[ind=b]++(0,0.4);
\end{tikzpicture}};
\node (x4) at (4,-3){
\begin{tikzpicture}
\atoms{prod}{0/p={0,0}}
\atoms{mapping}{m/p={0,-0.4}}
\atoms{mapping second data}{d/p={0,-0.9}}
\draw (m-t)--(0) (0)edge[ind=b]++(-0.4,0.4) (0)edge[ind=a]++(0.4,0.4);
\draw[mapping second 0dat] (m-b)--(d);
\end{tikzpicture}};
\node (x5) at (4,-6){
\begin{tikzpicture}
\atoms{mapping second prod}{0/p={0,0}}
\atoms{mapping}{m0/p={-0.4,0.4}, m1/p={0.4,0.4}}
\atoms{mapping second data}{d/p={0,-0.5}}
\draw[red] (0)--(d) (0)--(m0-b) (0)--(m1-b);
\draw (m0-t)edge[ind=b]++(0,0.4) (m1-t)edge[ind=a]++(0,0.4);
\end{tikzpicture}};
\draw (x0) edge[->] node[midway,left]{$\mathcal{M}$} (x2);
\draw (x0) edge[->] node[midway,above]{$\sigma_0$} (x1);
\draw (x2) edge[->] node[midway,left]{$\widetilde{M_\otimes}$} (x3);
\draw (x1) edge[->] node[midway,right]{$\mathcal{M}$} (x4);
\draw (x4)edge[->] node[midway,right]{$\widetilde{M_\otimes}$} (x5);
\draw (x3)edge[->] node[midway,below,mapping second 0dat]{$\sigma_0$} (x5);
\end{tikzpicture}
\;.
\end{equation}

It is easy to see that those compatibilities hold for our example mapping if we recall that it consists in interpreting the pairing diagram as a tensor product of $2\times 2$ identity matrices. E.g., the tensor product of identity matrices is compatible with the disjoint union of pairing lines. Or, the contraction connecting pairings of dots is compatible with carrying out all the corresponding contractions of identity matrices. At this point is becomes clear why the prefactor in the pairing tensors is needed: Closed loops represent traces over the $2\times 2$ identity matrix yielding the desired factor of $2$.

\subsection{Effective scheme}
\label{sec:mapping_effective_scheme}
Consider 2-networks that also involve mapping 1-functions and 2-functions. Consider equivalence classes of those 2-networks with respect to the 2-axioms and pre- and post-composing with invertible 2-functions. Due to the compatibility 2-axioms of the mapping 2-functions with all tensor 2-functions, it doesn't matter where in the 2-network we apply the mapping. All that matters is to which copies of the input source-type 1-data we apply it. Thus, in the graphical calculus, we can incorporate mappings by marking all elements to which the mapping is applied. As a marking we choose a semi-transparent overlay, e.g., in the following tensor product
\begin{equation}
\begin{tikzpicture}
\atoms{}{{0/circ,shadecirc}, {1/square,p={1,0}}}
\draw (0)--++(45:0.5) (0)--++(-45:0.5) (0)--++(135:0.5) (0)--++(-135:0.5) (1-t)--++(90:0.3) (1-b)--++(-90:0.3);
\end{tikzpicture}
\end{equation}
the circle-shaped element is a source-type 1-data to which the mapping is applied. Note that for any simple 2-network the target 1-data is either source-type in which case there cannot be any mappings, or it is target-type in which case the mapping must be applied to all source-type input 1-data. So in fact, the only invariant information is whether the target of each simple 2-network is source-type or target-type. The situation becomes a bit more interesting when we have variation of the mapping 2-scheme where the source and target tensor type are the same. In this case, the mapping can be applied to an arbitrary subset of the atoms in a network, e.g.,
\begin{equation}
\begin{tikzpicture}
\atoms{square,small}{{0/p={-0.4,-0.6},shadecirc}, 1/p={0.4,-0.6}}
\atoms{circ,small,all}{2/, {3/p={0,0.6},shadecirc}}
\draw[rc] (0-l)--++(180:0.4) (1-r)--++(0:0.4) (0-r)--(1-l) (0-t)--++(90:0.2)--(2) (1-t)--++(90:0.2)--(2) (2)--(3) (3)--++(180:0.5) (3)--++(0:0.5);
\end{tikzpicture}\;.
\end{equation}
We can even apply the mapping more than once.
\subsection{Other flavors}
In this section we consider different variants of tensor mappings. There are three main directions we can pursue. First, if we have different flavors of tensor types, we need different flavors of tensor mappings. Second, we can choose weaken the 1-axioms corresponding to the tensor mapping. Third, we can look at special kinds of mappings for which additional relations hold.

If $\mathcal{A}$ and $\mathcal{B}$ have a unit 0-data, it is natural to demand that it is compatible with the mapping as well, referred to as \tdef{unit homomorphicity}{unit_homomorphicity} of the mapping,
\begin{equation}
\widetilde{M_1}:\quad
\begin{tikzpicture}
\atoms{prod}{0/}
\atoms{mapping}{m/p={0,-0.4}}
\draw[mapping second 0dat] (m-b)edge[ind=a]++(0,-0.4);
\draw (0)--(m-t);
\end{tikzpicture}
=
\begin{tikzpicture}
\atoms{mapping second prod}{0/}
\draw[mapping second 0dat] (0)edge[ind=a]++(0,-0.4);
\end{tikzpicture}
\;.
\end{equation}

If $\mathcal{A}$ and $\mathcal{B}$ have a dual 1-function, we should impose \tdef{dual homomorphicity}{star_homomorphicity},
\begin{equation}
\widetilde{M_*}:\quad
\begin{tikzpicture}
\atoms{dual}{0/}
\atoms{mapping}{m/p={0,-0.4}}
\draw[mapping second 0dat] (m-b)edge[ind=b]++(0,-0.4);
\draw (0)--(m-t) (0)edge[ind=a]++(0,0.4);
\end{tikzpicture}
=
\begin{tikzpicture}
\atoms{mapping second dual}{0/}
\atoms{mapping}{m/p={0,0.4}}
\draw[mapping second 0dat] (0)edge[ind=b]++(0,-0.4) (m)--(0);
\draw (m)edge[ind=a]++(0,0.4);
\end{tikzpicture}\;.
\end{equation}
If $\mathcal{A}$ or $\mathcal{B}$ have 1-axioms weakened to 2-functions like unitor or product automorphor, or have additional 2-functions such as the trivial tensor or identity tensor, those need to be compatible with the mapping as well, yielding additional 2-axioms. E.g., for the identity, we get
\begin{equation}
\label{eq:2axiom_mapping_identity}
\begin{tikzpicture}[every node/.style={inner sep=0.2cm}]
\node (x0) at (0,0){
\begin{tikzpicture}
\atoms{copy}{0/}
\draw (0)edge[ind=a]++(0,0.4);
\end{tikzpicture}};
\node (x1) at (4,0){
\begin{tikzpicture}
\atoms{prod}{0/}
\atoms{data}{d/p={0,-0.5}}
\atoms{copy}{c/p={0,0.8}}
\draw[rc] (0)--(d) (0)--++(0.4,0.4)--(c) (0)--++(-0.4,0.4)--(c) (c)edge[ind=a]++(0,0.4);
\end{tikzpicture}};
\node (x2) at (0,-3){
\begin{tikzpicture}
\atoms{mapping second copy}{0/}
\atoms{mapping}{m/p={0,0.4}}
\draw[mapping second 0dat] (0)--(m);
\draw (m)edge[ind=a]++(0,0.4);
\end{tikzpicture}};
\node (x3) at (2,-6){
\begin{tikzpicture}
\atoms{mapping second prod}{0/}
\atoms{mapping}{m/p={0,1.2}}
\atoms{mapping second data}{d/p={0,-0.4}}
\atoms{mapping second copy}{c/p={0,0.8}}
\draw (m)edge[ind=a]++(0,0.4);
\draw[mapping second 0dat,rc] (d)--(0) (c)--(m-b) (0)--++(0.4,0.4)--(c) (0)--++(-0.4,0.4)--(c);
\end{tikzpicture}};
\node (x4) at (4,-3){
\begin{tikzpicture}
\atoms{prod}{0/}
\atoms{mapping}{m/p={0,-0.4}}
\atoms{mapping second data}{d/p={0,-0.8}}
\atoms{copy}{c/p={0,0.8}}
\draw[rc] (0)--(m-t) (0)--++(0.4,0.4)--(c) (0)--++(-0.4,0.4)--(c) (c)edge[ind=a]++(0,0.4);
\draw[mapping second 0dat] (d)--(m-b);
\end{tikzpicture}};
\draw (x0) edge[->] node[midway,left]{$\widetilde{C}$} (x2);
\draw (x0) edge[->] node[midway,above]{$\idop$} (x1);
\draw (x2) edge[->] node[midway,below left,mapping second 0dat]{$\idop$} (x3);
\draw (x1) edge[->] node[midway,right]{$M$} (x4);
\draw (x4)edge[->] node[midway,below right]{$\widetilde{C}\widetilde{M_\otimes}$} (x3);
\end{tikzpicture}\;.
\end{equation}

As already mentioned, the second possibility of generalization is to weaken the mapping 1-axioms themselves. We can replace the product homomorphicity by a \tdef{product homomorphor}{product_homomorphor} $M_\otimes$,
\begin{equation}
M_\otimes:\quad
\begin{tikzpicture}
\atoms{mapping second prod}{0/p={0,0}}
\atoms{prod}{1/p={0.4,0.8}}
\atoms{mapping}{m/p={0.4,0.4}}
\atoms{mapping second data}{d/p={0,-0.4}}
\draw[red] (0)--(m-b) (0)edge[ind=a]++(-0.4,0.4) (0)--(d);
\draw (m-t)--(1) (1)edge[ind=b]++(-0.4,0.4) (1)edge[ind=c]++(0.4,0.4);
\end{tikzpicture}
\quad\rightarrow\quad
\begin{tikzpicture}
\atoms{mapping second prod}{0/p={0,0}, 1/p={0.4,0.4}}
\atoms{mapping}{m0/p={0.8,0.8}, m1/p={0,0.8}}
\atoms{mapping second data}{d/p={0,-0.4}}
\draw[red] (0)--(1) (1)--(m0-b) (1)--(m1-b) (0)edge[ind=a]++(-0.4,0.4) (0)--(d);
\draw (m0-t)edge[ind=c]++(0,0.4) (m1-t)edge[ind=b]++(0,0.4);
\end{tikzpicture}\;.
\end{equation}

In the same way, the unit homomorphicity 1-axiom can be weakened to a \tdef{unit homomorphor}{unit_homomorphor} 2-function,
\begin{equation}
M_1:\quad
\begin{tikzpicture}
\atoms{mapping second prod}{0/p={0,0}}
\atoms{prod}{1/p={0.4,0.8}}
\atoms{mapping}{m/p={0.4,0.4}}
\atoms{mapping second data}{d/p={0,-0.5}}
\draw[mapping second 0dat] (0)--(m-b) (0)edge[ind=a]++(-0.4,0.4) (0)--(d);
\draw (m-t)--(1);
\end{tikzpicture}
\quad\rightarrow\quad
\begin{tikzpicture}
\atoms{mapping second prod}{0/p={0,0}, 1/p={0.4,0.4}}
\atoms{mapping second data}{d/p={0,-0.5}}
\draw[mapping second 0dat] (0)--(1) (0)edge[ind=a]++(-0.4,0.4) (0)--(d);
\end{tikzpicture}
\;.
\end{equation}

Analogously, the dual homomorphicity can be weakened to a \tdef{dual homomorphor}{dual_homomorphor},
\begin{equation}
\label{eq:dual_pull_through}
M_*:\quad
\begin{tikzpicture}
\atoms{mapping second prod}{0/p={0,0}}
\atoms{mapping}{m/p={0.4,0.4}}
\atoms{dual}{d/p={0.4,0.8}}
\atoms{mapping second data}{x/p={0,-0.5}}
\draw[mapping second 0dat] (0)--(m-b) (0)edge[ind=a]++(-0.4,0.4) (0)--(x);
\draw (m-t)--(d) (d)edge[ind=b]++(0,0.3);
\end{tikzpicture}
\quad\rightarrow\quad
\begin{tikzpicture}
\atoms{mapping second prod}{0/p={0,0}}
\atoms{mapping second dual}{d/p={0.4,0.4}}
\atoms{mapping second data}{x/p={0,-0.5}}
\atoms{mapping}{m/p={0.4,0.8}}
\draw[mapping second 0dat] (0)--(d-b) (0)edge[ind=a]++(-0.4,0.4) (0)--(x) (d)--(m);
\draw (m)edge[ind=b]++(0,0.3);
\end{tikzpicture}
\;.
\end{equation}

As usual, weakening consists in 3 steps. First, we need to define versions of the homomorphors acting at different places in a 1-network relative to the 0-data. E.g., a version of the product homomorphor without auxiliary index can be defined by
\begin{equation}
\begin{tikzpicture}[every node/.style={inner sep=0.2cm}]
\node (x0) at (0,0){
\begin{tikzpicture}
\atoms{prod}{1/p={0,0.4}}
\atoms{mapping}{m/}
\atoms{mapping second data}{d/p={0,-0.4}}
\draw[red] (d)--(m-b);
\draw (m-t)--(1) (1)edge[ind=a]++(-0.4,0.4) (1)edge[ind=b]++(0.4,0.4);
\end{tikzpicture}};
\node (x1) at (4,0){
\begin{tikzpicture}
\atoms{mapping second prod}{0/p={0,0}}
\atoms{mapping}{m0/p={0.4,0.4}, m1/p={-0.4,0.4}}
\atoms{mapping second data}{d/p={0,-0.4}}
\draw[mapping second 0dat] (0)--(m0-b) (0)--(m1-b) (0)--(d);
\draw (m0-t)edge[ind=b]++(0,0.4) (m1-t)edge[ind=a]++(0,0.4);
\end{tikzpicture}};
\node (x2) at (0,-3){
\begin{tikzpicture}
\atoms{mapping second prod}{0/p={0,0}, 2/p={-0.4,0.4}}
\atoms{prod}{1/p={0.4,0.8}}
\atoms{mapping}{m/p={0.4,0.4}}
\atoms{mapping second data}{d/p={0,-0.4}}
\draw[red] (0)--(m-b) (0)--(2) (0)--(d);
\draw (m-t)--(1) (1)edge[ind=a]++(-0.4,0.4) (1)edge[ind=b]++(0.4,0.4);
\end{tikzpicture}};
\node (x3) at (4,-3){
\begin{tikzpicture}
\atoms{mapping second prod}{0/p={0,0}, 1/p={0.4,0.4}, 2/p={-0.4,0.4}}
\atoms{mapping}{m0/p={0.8,0.8}, m1/p={0,0.8}}
\atoms{mapping second data}{d/p={0,-0.4}}
\draw[red] (0)--(1) (1)--(m0-b) (1)--(m1-b) (0)--(2) (0)--(d);
\draw (m0-t)edge[ind=b]++(0,0.4) (m1-t)edge[ind=a]++(0,0.4);
\end{tikzpicture}};
\draw (x0) edge[->] node[midway,above]{$M_{\otimes 0}$} (x1);
\draw (x0) edge[->] node[midway,left,mapping second 0dat]{$\otimes\idop$} (x2);
\draw (x2) edge[->] node[midway,below]{$M_\otimes$} (x3);
\draw (x3)edge[->] node[midway,right,mapping second 0dat]{$U$} (x1);
\end{tikzpicture}\;.
\end{equation}
If $\mathcal{A}$ and $\mathcal{B}$ do not have a unit 0-data and trivial 1-data, we need to explicitly add $M_{\otimes 0}$ as an additional 2-function.

Second, we have to replace every occurrence of $\widetilde{M_\otimes}$, $\widetilde{M_1}$, or $\widetilde{M_*}$ by the corresponding version of $M_\otimes$, $M_1$, or $M_*$. Specifically, in all of Eq.~\eqref{eq:2axiom_mapping_tprod}, Eq.~\eqref{eq:2axiom_mapping_commutor}, and Eq.~\eqref{eq:2axiom_mapping_identity} we have to replace $\widetilde{M_\otimes}$ by $M_{\otimes 0}$.

Third, we have to add a bunch of new 2-axioms which would be trivial in the strict case, to ensure compatibility with other 2-functions. E.g., compatibility with the commutor has to be ensured by
\begin{equation}
\begin{tikzpicture}[every node/.style={inner sep=0.2cm}]
\node (x0) at (0,0){
\begin{tikzpicture}
\atoms{mapping second prod}{0/p={0,0}, 2/p={-0.8,0.4}}
\atoms{prod}{1/p={0.4,0.8}}
\atoms{mapping}{m/p={0.4,0.4}}
\atoms{mapping second data}{d/p={0,-0.4}}
\draw[red] (0)--(m-b) (0)--(2) (0)--(d) (2)edge[ind=a]++(-0.4,0.4) (2)edge[ind=b]++(0.4,0.4);
\draw (m-t)--(1) (1)edge[ind=c]++(-0.4,0.4) (1)edge[ind=d]++(0.4,0.4);
\end{tikzpicture}};
\node (x1) at (4,0){
\begin{tikzpicture}
\atoms{mapping second prod}{0/p={0,0}, 1/p={0.4,0.4}, 2/p={-1.2,0.8}}
\atoms{mapping}{m0/p={0.8,0.8}, m1/p={0,0.8}}
\atoms{mapping second data}{d/p={0,-0.4}}
\draw[red] (0)--(1) (1)--(m0-b) (1)--(m1-b) (0)--(2) (0)--(d) (2)edge[ind=a]++(-0.4,0.4) (2)edge[ind=b]++(0.4,0.4);
\draw (m0)edge[ind=c]++(0,0.4) (m1)edge[ind=d]++(0,0.4);
\end{tikzpicture}};
\node (x2) at (0,-3){
\begin{tikzpicture}
\atoms{mapping second prod}{0/p={0,0}, 2/p={-0.8,0.4}}
\atoms{prod}{1/p={0.4,0.8}}
\atoms{mapping}{m/p={0.4,0.4}}
\atoms{mapping second data}{d/p={0,-0.4}}
\draw[red] (0)--(m-b) (0)--(2) (0)--(d) (2)edge[ind=b]++(-0.4,0.4) (2)edge[ind=a]++(0.4,0.4);
\draw (m-t)--(1) (1)edge[ind=c]++(-0.4,0.4) (1)edge[ind=d]++(0.4,0.4);
\end{tikzpicture}};
\node (x3) at (4,-3){
\begin{tikzpicture}
\atoms{mapping second prod}{0/p={0,0}, 1/p={0.4,0.4}, 2/p={-1.2,0.8}}
\atoms{mapping}{m0/p={0.8,0.8}, m1/p={0,0.8}}
\atoms{mapping second data}{d/p={0,-0.4}}
\draw[red] (0)--(1) (1)--(m0-b) (1)--(m1-b) (0)--(2) (0)--(d) (2)edge[ind=b]++(-0.4,0.4) (2)edge[ind=a]++(0.4,0.4);
\draw (m0)edge[ind=c]++(0,0.4) (m1)edge[ind=d]++(0,0.4);
\end{tikzpicture}};
\draw (x0) edge[->] node[midway,above]{$M_{\otimes}$} (x1);
\draw (x0) edge[->] node[midway,left,mapping second 0dat]{$\sigma_0\sigma\sigma_0$} (x2);
\draw (x2) edge[->] node[midway,below]{$M_\otimes$} (x3);
\draw (x3)edge[<-] node[midway,right,mapping second 0dat]{$\sigma_0\sigma\sigma_0$} (x1);
\end{tikzpicture}\;.
\end{equation}
Roughly, we get one such compatibility 2-axiom for every 2-function of $\mathcal{A}$ or $\mathcal{B}$.

A yet different version of the tensor mapping 2-scheme is when we let $\mathcal{A}$ and $\mathcal{B}$ be the same tensor type. The 2-scheme can be obtained by replacing all the red-colored 1-functions or 2-functions (i.e., the ones of $\mathcal{B}$) by black-colored ones (i.e., the ones of $\mathcal{A}$). We'll call such a tensor mapping a \mdef{self-mapping}.

As a special variant of self-mapping we can set the mapping 1-function to be the same as the dual 1-function of the tensor type. Additionally, in the weak case, the mapping homomorphors will be set equal to the dual automorphors.

Two mappings $\mathcal{M}$ from $\mathcal{A}$ to $\mathcal{B}$ and $\mathcal{N}$ from $\mathcal{B}$ to $\mathcal{A}$ can be inverses to each other, simply when their mapping 1-function and mapping 2-functions are inverses. This notion of inverse can be weakened: Instead of the 1-axiom
\begin{equation}
\begin{tikzpicture}
\atoms{mapping}{m1/p={0,0.4}, {m2/p={0,0.8},vflip}}
\draw[mapping second 0dat] (m1)edge[ind=b]++(-90:0.4) (m2)edge[ind=a]++(90:0.4);
\draw (m1)--(m2);
\end{tikzpicture}
=
\begin{tikzpicture}
\draw[mapping second 0dat, ind=a, startind=b] (0,0)--++(90:0.8);
\end{tikzpicture}
\end{equation}
stating that the mapping 1-functions are inverses, we can introduce a 2-function
\begin{equation}
\begin{tikzpicture}
\atoms{mapping second data}{d/}
\atoms{mapping}{m1/p={0,0.4}, {m2/p={0,0.8},vflip}}
\draw[mapping second 0dat] (d)--(m1) (m2)edge[ind=a]++(90:0.4);
\draw (m1)--(m2);
\end{tikzpicture}
\quad\rightarrow\quad
\begin{tikzpicture}
\atoms{mapping second data}{d/}
\draw[mapping second 0dat] (d)edge[ind=a]++(90:0.4);
\end{tikzpicture}\;.
\end{equation}
The notion of (weakly) inverse mappings can be used to define equivalence of tensor types.

\section{Relation to category theory}
\label{sec:category}
In this section we will elaborate on the relation between tensor types defined above and structures of (monoidal) category theory. In Section~\ref{sec:category_2scheme} we will see that basically all monoidal categorical structures such as symmetric monoidal categories, braided monoidal categories, or monoidal functors between categories can be formalized as 2-schemes. In Section~\ref{sec:monoidal_vs_ttypes} we will compare tensor type 2-schemes to categorical 2-schemes. In Section~\ref{sec:compact_category_to_tensor} and Section~\ref{sec:tensor_to_compact_category} we explicitly sketch the equivalence between compact closed categories and tensor types.

\subsection{Category theory in 2-scheme language}
\label{sec:category_2scheme}
In order to compare tensor types to categorical structures (such as symmetric monoidal categories), we will quickly sketch how the latter can be defined using the 2-scheme language. As we will see, \emph{composition}, \emph{identity morphisms}, \emph{functors} and \emph{natural isomorphisms} all become 2-functions. The fact that everything boils down to a 2-function shows that our 2-scheme language is simpler and more low level than the usual categorical language. It's therefore also more generally applicable, and allows the definition of other (non-categorical) types of structures such as tensor types. This comes at the expense that writing down concrete categorical structures in the 2-scheme language requires a little bit more paper than doing so in the conventional categorical language. The latter is more high-level and implicitly contains some of the axioms which would have to be spelled out explicitly in the 2-scheme language.

The 0-data in the context of category theory are the \emph{objects}, and the 1-data are the \emph{morphisms}. A 1-data (i.e., morphism) does not depend on a single 0-data (i.e., object), but on a pair of \emph{source} $a$ and \emph{target} object $b$. So diagrammatically, the 1-data has $2$ lines instead of one line sticking out,
\begin{equation}
\dat_1(a,b)
\quad\rightarrow\quad
\begin{tikzpicture}
\atoms{data}{0/p={0,0}}
\draw (0-t)edge[ind=a]++(-0.4,0.4) (0-t)edge[ind=b]++(0.4,0.4);
\end{tikzpicture}\;.
\end{equation}
If we have bare categories without any extra structure, the 1-networks consist only of this box as well as discarding and copying of objects. A \emph{category} consists of two 2-functions:
\begin{itemize}
\item The composition,
\begin{equation}
\circ:\quad
\begin{tikzpicture}
\atoms{data}{0/p={0,0}, 1/p={0.8,0}}
\atoms{copy}{c/p={0.4,0.4}}
\draw (0-t)edge[ind=a]++(-0.4,0.4) (1-t)edge[ind=c]++(0.4,0.4) (0-t)--(c) (1-t)--(c) (c)edge[ind=b]++(0,0.4);
\end{tikzpicture}
\quad\rightarrow\quad
\begin{tikzpicture}
\atoms{data}{0/p={0,0}}
\atoms{copy}{c/p={0,0.8}}
\draw (0-t)edge[ind=a]++(-0.4,0.4) (0-t)edge[ind=c]++(0.4,0.4) (c)edge[ind=b]++(0,0.4);
\end{tikzpicture}\;.
\end{equation}
\item The identity morphism,
\begin{equation}
\operatorname{id}:\quad
\begin{tikzpicture}
\atoms{copy}{c/p={0.4,0.4}}
\draw (c)edge[ind=a]++(0,0.4);
\end{tikzpicture}
\quad\rightarrow\quad
\begin{tikzpicture}
\atoms{data}{0/p={0,0}}
\atoms{copy}{c/p={0,0.8}}
\draw (c)edge[ind=a]++(0,0.4);
\draw[rounded corners] (0-t)--++(0.4,0.4)--(c);
\draw[rounded corners] (0-t)--++(-0.4,0.4)--(c);
\end{tikzpicture}\;.
\end{equation}
\end{itemize}
Those 2-functions have to obey the following 2-axioms:
\begin{itemize}
\item The composition is associative,
\begin{equation}
\begin{tikzpicture}[every node/.style={inner sep=0.2cm}]
\node (x0) at (0,0){
\begin{tikzpicture}
\atoms{data}{0/p={0,0}, 1/p={0.8,0}, 2/p={1.6,0}}
\atoms{copy}{c0/p={0.4,0.4}, c1/p={1.2,0.4}}
\draw (0-t)edge[ind=a]++(-0.4,0.4) (2-t)edge[ind=d]++(0.4,0.4) (0-t)--(c0) (1-t)--(c0) (1-t)--(c1) (2-t)--(c1) (c0)edge[ind=b]++(0,0.4) (c1)edge[ind=c]++(0,0.4);
\end{tikzpicture}};
\node (x1) at (4,0){
\begin{tikzpicture}
\atoms{data}{0/p={0,0}, 1/p={0.8,0}}
\atoms{copy}{c/p={0.4,0.4}}
\atoms{copy}{cx/p={0.8,0.8}}
\draw (0-t)edge[ind=a]++(-0.4,0.4) (1-t)edge[ind=d]++(0.4,0.4) (0-t)--(c) (1-t)--(c) (c)edge[ind=b]++(0,0.4) (cx)edge[ind=c]++(0,0.4);
\end{tikzpicture}};
\node (x2) at (0,-4){
\begin{tikzpicture}
\atoms{data}{0/p={0,0}, 1/p={0.8,0}}
\atoms{copy}{c/p={0.4,0.4}}
\atoms{copy}{cx/p={0,0.8}}
\draw (0-t)edge[ind=a]++(-0.4,0.4) (1-t)edge[ind=d]++(0.4,0.4) (0-t)--(c) (1-t)--(c) (c)edge[ind=c]++(0,0.4) (cx)edge[ind=b]++(0,0.4);
\end{tikzpicture}};
\node (x3) at (4,-4){
\begin{tikzpicture}
\atoms{data}{0/p={0,0}}
\atoms{copy}{c0/p={-0.3,1}}
\atoms{copy}{c1/p={0.3,1}}
\draw (0-t)edge[ind=a]++(-0.4,0.4) (0-t)edge[ind=d]++(0.4,0.4) (c0)edge[ind=b]++(0,0.4) (c1)edge[ind=c]++(0,0.4);
\end{tikzpicture}};
\draw (x0)edge[->]node[midway,above]{$\circ$}(x1);
\draw (x0)edge[->]node[midway,left]{$\circ$}(x2);
\draw (x1)edge[->]node[midway,right]{$\circ$}(x3);
\draw (x2)edge[->]node[midway,below]{$\circ$}(x3);
\end{tikzpicture}\;.
\end{equation}
\item Composition with the identity morphism yields the same morphism again.
\end{itemize}

For a \emph{monoidal category}, there is also a product and unit 1-function, just as for tensor types in Eq.~\eqref{eq:1product} and Eq.~\eqref{eq:1unit}.

Also, there is a \emph{tensor product} 2-function,
\begin{equation}
\otimes:\quad
\begin{tikzpicture}
\atoms{data}{f0/p={0,0}}
\atoms{data}{f1/p={1.6,0}}
\draw (f0-t)edge[ind=a]++(-0.4,0.4) (f0-t)edge[ind=b]++(0.4,0.4) (f1-t)edge[ind=c]++(-0.4,0.4) (f1-t)edge[ind=d]++(0.4,0.4);
\end{tikzpicture}
\quad\rightarrow\quad
\begin{tikzpicture}
\atoms{data}{f/p={0,0}}
\atoms{prod}{0/p={-0.8,0.8}, 1/p={0.8,0.8}}
\draw (f-t)--(0) (f-t)--(1) (0)edge[ind=a]++(-0.4,0.4) (0)edge[ind=c]++(0.4,0.4) (1)edge[ind=b]++(-0.4,0.4) (1)edge[ind=d]++(0.4,0.4);
\end{tikzpicture}\;.
\end{equation}

For a \emph{strictly monoidal category}, we impose associativity and unitality, i.e., we impose 1-axioms as in Eq.~\eqref{eq:1associativity} and Eq.~\eqref{eq:1unitality}. Those 1-axioms can be weakened. The categorical mechanism for weakening is very different from that for tensor types: For the latter, we replace the 1-axiom by a 2-function that performs the 1-axiom in a 1-network next to a 1-data element, possibly with an auxiliary index. For categories, we replace the 1-axiom by a natural transformation, which is a choice of morphism depending on objects. So in our language we have a 1-data depending on a choice of 0-data, which can be formalized as a 2-function which creates that 1-data from nothing, as we had for the identity tensor or trivial tensor for tensor types. The 1-network of this 1-data is such that its input side and its output side are related by applying the corresponding 1-axiom. The effect of the 1-axiom can then be performed by using the corresponding weakened 2-function creating a 1-data from nothing, and then composing with that 1-data, possibly after tensoring an identity.

So in order to weaken unitality, we replace it by a left unitor 2-function,
\begin{equation}
U_l:\quad
\begin{tikzpicture}
\atoms{copy}{0/p={0,0}}
\draw (0)edge[ind=a]++(0,0.4);
\end{tikzpicture}
\quad\rightarrow\quad
\begin{tikzpicture}
\atoms{data}{f/p={0,0}}
\atoms{prod}{0/p={-0.4,0.4}, 1/p={-0.8,0.8}}
\atoms{copy}{c0/p={0,0.8}}
\draw (c0)edge[ind=a]++(0,0.4) (f-t)--(0) (0)--(1) (0)--(c0);
\draw[rounded corners] (c0)--++(0.4,-0.4)--(f-t);
\end{tikzpicture}
\end{equation}
and a right unitor,
\begin{equation}
U_r:\quad
\begin{tikzpicture}
\atoms{copy}{0/p={0,0}}
\draw (0)edge[ind=a]++(0,0.4);
\end{tikzpicture}
\quad\rightarrow\quad
\begin{tikzpicture}
\atoms{data}{f/p={0,0}}
\atoms{prod}{0/p={-0.4,0.4}, 1/p={0,0.8}}
\atoms{copy}{c0/p={0,1.6}}
\draw (c0)edge[ind=a]++(0,0.4) (f-t)--(0) (0)--(1);
\draw[rounded corners] (c0)--++(0.8,-0.8)--(f-t);
\draw[rounded corners] (0)--++(-0.4,0.4)--(c0);
\end{tikzpicture}\;.
\end{equation}
Associativity gets replaced by an associator 2-function,
\begin{equation}
\alpha:\quad
\begin{tikzpicture}
\atoms{copy}{0/p={0,0}, 1/p={0.4,0}, 2/p={0.8,0}}
\draw (0)edge[ind=a]++(0,0.4) (1)edge[ind=b]++(0,0.4) (2)edge[ind=c]++(0,0.4);
\end{tikzpicture}
\quad\rightarrow\quad
\begin{tikzpicture}
\atoms{data}{f/p={0,0}}
\atoms{prod}{0/p={-0.4,0.4}, 1/p={0.4,0.4}, 2/p={-0.8,0.8}, 3/p={0.8,0.8}}
\atoms{copy}{c0/p={-0.8,1.6}, c1/p={0,1.6}, c2/p={0.8,1.6}}
\draw (c0)edge[ind=a]++(0,0.4) (c1)edge[ind=b]++(0,0.4) (c2)edge[ind=c]++(0,0.4) (f-t)--(0) (f-t)--(1) (0)--(2) (1)--(3) (0)--(c2) (1)--(c0) (2)--(c1) (3)--(c1);
\draw[rounded corners] (2)--++(-0.4,0.4)--(c0);
\draw[rounded corners] (3)--++(0.4,0.4)--(c2);
\end{tikzpicture}\;.
\end{equation}

Those come with a bunch of 2-axioms (which we won't all depict here):
\begin{itemize}
\item There is a 2-axiom known as \emph{pentagon equation} involving $5$ associators, $2$ tensor products, $2$ identity morphisms, and $3$ compositions.
\item There is a 2-axiom known as \emph{triangle diagram} involving $1$ associator, $2$ identity morphisms, $2$ tensor products, $1$ left unitor and $1$ right unitor.
\item In the usual categorical language, the tensor product is a \emph{bi-functor}, which in our language is a further 2-axiom,
\begin{equation}
\begin{tikzpicture}
\node (x0) at (2,0){
\begin{tikzpicture}
\atoms{data}{0/p={0,0}, 1/p={0.8,0}}
\atoms{copy}{c/p={0.4,0.4}}
\draw (0-t)edge[ind=a]++(-0.4,0.4) (1-t)edge[ind=c]++(0.4,0.4) (0-t)--(c) (1-t)--(c) (c)edge[ind=b]++(0,0.4);
\atoms{data}{0x/p={2.4,0}, 1x/p={3.2,0}}
\atoms{copy}{cx/p={2.8,0.4}}
\draw (0x-t)edge[ind=d]++(-0.4,0.4) (1x-t)edge[ind=f]++(0.4,0.4) (0x-t)--(cx) (1x-t)--(cx) (cx)edge[ind=e]++(0,0.4);
\end{tikzpicture}};
\node (x1) at (4,-3){
\begin{tikzpicture}
\atoms{data}{f0/p={0,0}, f1/p={1.6,0}}
\atoms{copy}{c0/p={0.4,1.2}, c1/p={1.2,1.2}}
\atoms{prod}{0/p={-0.8,0.8}, 1/p={0.4,0.4}, 2/p={1.2,0.4}, 3/p={2.4,0.8}}
\draw (c0)edge[ind=b]++(0,0.4) (c1)edge[ind=e]++(0,0.4) (0)edge[ind=a]++(-0.4,0.4) (0)edge[ind=d]++(0.4,0.4) (3)edge[ind=c]++(-0.4,0.4) (3)edge[ind=f]++(0.4,0.4);
\draw (f0-t)--(0) (f0-t)--(1) (f1-t)--(2) (f1-t)--(3) (1)--(c1) (2)--(c0);
\draw[rounded corners] (1)--++(-0.4,0.4)--(c0);
\draw[rounded corners] (2)--++(0.4,0.4)--(c1);
\end{tikzpicture}};
\node (x2) at (0,-4){
\begin{tikzpicture}
\atoms{data}{0/p={0,0}}
\atoms{copy}{c/p={0,0.8}}
\draw (0-t)edge[ind=a]++(-0.4,0.4) (0-t)edge[ind=c]++(0.4,0.4) (c)edge[ind=b]++(0,0.4);
\atoms{data}{0/p={1.6,0}}
\atoms{copy}{c/p={1.6,0.8}}
\draw (0-t)edge[ind=d]++(-0.4,0.4) (0-t)edge[ind=e]++(0.4,0.4) (c)edge[ind=f]++(0,0.4);
\end{tikzpicture}};
\node (x3) at (4,-7){
\begin{tikzpicture}
\atoms{data}{f0/p={0,0}, f1/p={1.6,0}}
\atoms{copy}{c0/p={0.8,0.8}}
\atoms{prod}{0/p={-0.8,0.8}, x/p={0.8,1.2}, 3/p={2.4,0.8}}
\draw (x)edge[ind=b]++(-0.4,0.4) (x)edge[ind=e]++(0.4,0.4) (0)edge[ind=a]++(-0.4,0.4) (0)edge[ind=d]++(0.4,0.4) (3)edge[ind=c]++(-0.4,0.4) (3)edge[ind=f]++(0.4,0.4);
\draw (f0-t)--(0) (f0-t)--(c0) (f1-t)--(c0) (f1-t)--(3) (x)--(c0);
\end{tikzpicture}};
\node (x4) at (2,-11){
\begin{tikzpicture}
\atoms{data}{f/p={0,0}}
\atoms{prod}{0/p={-0.8,0.8}, 1/p={0.8,0.8}}
\draw (f-t)--(0) (f-t)--(1) (0)edge[ind=a]++(-0.4,0.4) (0)edge[ind=d]++(0.4,0.4) (1)edge[ind=c]++(-0.4,0.4) (1)edge[ind=f]++(0.4,0.4);
\atoms{copy}{c/p={2,0}}
\atoms{prod}{x/p={2,0.4}}
\draw (x)--(c) (x)edge[ind=b]++(-0.4,0.4) (x)edge[ind=e]++(0.4,0.4);
\end{tikzpicture}};
\node (x5) at (0,-8){
\begin{tikzpicture}
\atoms{data}{f/p={0,0}}
\atoms{prod}{0/p={-0.8,0.8}, 1/p={0.8,0.8}}
\draw (f-t)--(0) (f-t)--(1) (0)edge[ind=a]++(-0.4,0.4) (0)edge[ind=d]++(0.4,0.4) (1)edge[ind=c]++(-0.4,0.4) (1)edge[ind=f]++(0.4,0.4);
\atoms{copy}{c0/p={1.6,0}}
\atoms{copy}{c1/p={2,0}}
\draw (c0)edge[ind=b]++(0,0.4) (c1)edge[ind=e]++(0,0.4);
\end{tikzpicture}};
\draw (x0)edge[->]node[midway,above right]{$\otimes\otimes$}(x1);
\draw (x0)edge[->]node[midway,above left]{$\circ\circ$}(x2);
\draw (x1)edge[<->]node[midway,right]{$\widetilde{C}$}(x3);
\draw (x4)edge[<->]node[midway,below left]{$\widetilde{C}$}(x5);
\draw (x2)edge[->]node[midway,left]{$\otimes$}(x5);
\draw (x3)edge[->]node[midway,below right]{$\circ$}(x4);
\end{tikzpicture}\;.
\end{equation}
\end{itemize}

For the monoidal category to be \emph{braided} we need one further 2-function, called braiding (the categorical analogue of what we called the commutor for tensor types),
\begin{equation}
R:\quad
\begin{tikzpicture}
\atoms{copy}{0/p={0,0}, 1/p={0.4,0}}
\draw (0)edge[ind=a]++(0,0.4) (1)edge[ind=b]++(0,0.4);
\end{tikzpicture}
\quad\rightarrow\quad
\begin{tikzpicture}
\atoms{data}{f/p={0,0}}
\atoms{prod}{0/p={-0.4,0.4}, 1/p={0.4,0.4}}
\atoms{copy}{c0/p={0,0.8}, c1/p={0,1.8}}
\draw (c0)edge[ind=a]++(0,0.3) (c1)edge[ind=b]++(0,0.3) (f-t)--(0) (f-t)--(1) (0)--(c0) (1)--(c0);
\draw[rounded corners] (0)--++(-0.5,0.5)--(c1);
\draw[rounded corners] (1)--++(0.5,0.5)--(c1);
\end{tikzpicture}\;.
\end{equation}
It comes along with one further 2-axiom known as the \emph{hexagon equation}, involving the $3$ associators, $3$ braidings, $4$ compositions, and lots of 1-axioms.

A braided monoidal category is called \emph{symmetric} if one more 2-axiom holds,
\begin{equation}
\begin{tikzpicture}
\node (x0) at (0,0){
\begin{tikzpicture}
\atoms{copy}{0/p={0,0}, 1/p={0.4,0}}
\draw (0)edge[ind=a]++(0,0.4) (1)edge[ind=b]++(0,0.4);
\end{tikzpicture}};
\node (x1) at (4,0){
\begin{tikzpicture}
\atoms{data}{f0/p={0,0}, f1/p={1.6,0}}
\atoms{copy}{0/p={0,0.8}, 1/p={0,1.6}, 2/p={1.6,0.8}, 3/p={1.6,1.6}, 4/p={0.4,2}, 5/p={1.2,2}}
\atoms{prod}{p0/p={-0.4,0.4}, p1/p={0.4,0.4}, p2/p={1.2,0.4}, p3/p={2,0.4}}
\draw (4)edge[ind=a]++(0,0.4) (5)edge[ind=b]++(0,0.4);
\draw[rounded corners] (f0-t)--(p0) (f0-t)--(p1) (f1-t)--(p2) (f1-t)--(p3) (p0)--(0) (p1)--(0) (p2)--(2) (p3)--(2) (p0)--++(-0.4,0.4)--(1) (p1)--++(0.4,0.4)--(1) (p2)--++(-0.4,0.4)--(3) (p3)--++(0.4,0.4)--(3) (0)--(5) (1)--(4) (2)--(4) (3)--(5);
\end{tikzpicture}};
\node (x2) at (0,-4){
\begin{tikzpicture}
\atoms{data}{0/p={0,0}}
\atoms{copy}{c/p={0,0.8}}
\draw[rounded corners] (0-t)--++(0.4,0.4)--(c) (0-t)--++(-0.4,0.4)--(c) (c)edge[ind=a]++(0,0.4);
\atoms{data}{1/p={0.8,0}}
\atoms{copy}{c1/p={0.8,0.8}}
\draw[rounded corners] (1-t)--++(0.4,0.4)--(c1) (1-t)--++(-0.4,0.4)--(c1) (c1)edge[ind=b]++(0,0.4);
\end{tikzpicture}};
\node (x3) at (4,-4){
\begin{tikzpicture}
\atoms{data}{f/p={0,0}}
\atoms{prod}{0/p={-0.4,0.4}, 1/p={0.4,0.4}}
\atoms{copy}{c0/p={-0.4,1.2}, c1/p={0.4,1.2}}
\draw[rounded corners] (c0)edge[ind=a]++(0,0.3) (c1)edge[ind=b]++(0,0.3) (f-t)--(0) (f-t)--(1) (0)--(c1) (1)--(c0) (0)--++(-0.4,0.4)--(c0) (1)--++(0.4,0.4)--(c1);
\end{tikzpicture}};
\draw (x0)edge[->]node[midway,above]{$RR$}(x1);
\draw (x0)edge[->]node[midway,left]{$\idop\idop$}(x2);
\draw (x1)edge[->]node[midway,right]{$\widetilde{C}\circ$}(x3);
\draw (x2)edge[->]node[midway,below]{$\otimes$}(x3);
\end{tikzpicture}
\;.
\end{equation}

A dual of an object $A$ (in a symmetric monoidal category) is an object $A^*$, together with two morphisms called \emph{evaluation} and \emph{coevaluation} satisfying certain axioms. A \emph{compact closed category} is a symmetric monoidal category where for each object a dual exists. One can introduce a very similar notion of a ``compact structure'', which refers to an actual choice of a dual, evaluation and coevaluation for each object. Symmetric monoidal categories with a compact structure are a 2-scheme. The dual becomes a 1-function, just as the dual for tensor types in Eq.~\eqref{eq:1dual}.

The evaluation and co-evaluation become 2-functions. The evaluation is denoted by
\begin{equation}
\operatorname{ev}:\quad
\begin{tikzpicture}
\atoms{copy}{0/p={0,0}}
\draw (0)edge[ind=a]++(0,0.4);
\end{tikzpicture}
\quad\rightarrow\quad
\begin{tikzpicture}
\atoms{data}{f/p={0,0}}
\atoms{prod}{0/p={-0.4,0.4}, 1/p={0.4,0.4}}
\atoms{copy}{c/p={-0.4,1.2}}
\atoms{dual}{d/p={0,0.8}}
\draw (c)edge[ind=a]++(0,0.4) (f-t)--(0) (f-t)--(1) (0)--(d-b) (d-t)--(c);
\draw[rounded corners] (0)--++(-0.4,0.4)--(c);
\end{tikzpicture}\;.
\end{equation}
The co-evaluation has the following source and target 1-network,
\begin{equation}
\operatorname{cev}:\quad
\begin{tikzpicture}
\atoms{copy}{0/p={0,0}}
\draw (0)edge[ind=A]++(0,0.4);
\end{tikzpicture}
\quad\rightarrow\quad
\begin{tikzpicture}
\atoms{data}{f/p={0,0}}
\atoms{prod}{0/p={-0.4,0.4}, 1/p={0.4,0.4}}
\atoms{copy}{c/p={0.4,1.2}}
\atoms{dual}{d/p={0,0.8}}
\draw (c)edge[ind=A]++(0,0.4) (f-t)--(0) (f-t)--(1) (1)--(d-b) (d-t)--(c);
\draw[rounded corners] (1)--++(0.4,0.4)--(c);
\end{tikzpicture}\;.
\end{equation}

Those 2-functions are subject to some 2-axioms, which we won't specify here.

\subsection{Monoidal categories versus tensor types}
\label{sec:monoidal_vs_ttypes}
We already mentioned in the introduction that tensor types share a great deal of similarity with compact closed categories. In this section, we will elaborate more on the relation between the two.

First of all, as we mentioned in the previous section, compact closedness is conventionally formulated as a property and not as a structure. Such a property is incompatible with our notion of a 2-scheme. Instead, we have to consider what we call \emph{categories with compact closed structure}, where we explicitly make a fixed choice of a dual, evaluation and coevaluation for each object.

Categories with compact closed structure are one specific categorical structure corresponding to one specific 2-scheme (apart from possibly strict versions). In contrast, tensor types come in many different flavors. We claim that one specific flavor of tensor type, i.e., one specific tensor 2-scheme is equivalent to the compact closed 2-scheme, as we'll sketch in the next two sections. Note that despite being equivalent, categorical 2-schemes and tensorial 2-schemes use completely different mechanisms to define a monoidal structure.

The flavor of tensor types which we claim to be equivalent to categories with compact closed structure has a unit 1-function, trivial tensor, dual, identity, but does neither have symmetric nor block-compatible contraction and identity. Also, all 1-axioms are weakened (in principle, one could also consider the equivalence of strict versions of compact closed categories to strict versions of the specified tensor type).

Tensor mappings between two tensor types also have a great deal of similarity with a categorical notion: They are roughly equivalent to monoidal functors between the corresponding compact closed categories.

While for some flavors of tensor types (like strict versions) we can define analogue versions of compact closed categories in a categorical language, this is not the case in general. One important instance of this is the identity. While we can define tensor types without identity, this is not possible in the categorical language. This is because the composition for categories always composes the whole target of the first morphism with the whole source of the second morphism. So in order to perform an associativity move in a part of the target 1-network of a morphism, e.g.,
\begin{equation}
\begin{multlined}
\dat_1(a,b \otimes(c \otimes (d\otimes (e\otimes f))))
\\\rightarrow\dat_1(a,b\otimes(c\otimes ((d\otimes e)\otimes f)))\;,
\end{multlined}
\end{equation}
we do this by composing with, e.g.,
\begin{equation}
\idop\otimes (\idop\otimes \alpha_{d,e,f})\;.
\end{equation}
So the identity morphism is essential for the definition of, e.g., the pentagon equation. In contrast, an identity tensor is totally optional for tensor types.

Conversely, there are flavors of monoidal categories that are impossible to define via a variant of the tensor 2-scheme. This includes everything that does not have a rigid structure given by a evaluation and coevaluation, via which we could identify the source and target level. On the other hand, we could also introduce non-symmetric and non-braided versions of tensor types. For non-symmetric, we simply drop the 2-axiom $\sigma\sigma=\idop$ (with $\idop$ the identity function not the identity tensor). For non-braided, we drop the commutor $\sigma$ as a whole. Of course, in both cases, we would also have to introduce some new 2-axioms or adapt some of the existing ones.

So we see that more generally, we can distinguish between a categorical mechanism and a tensor types mechanism for formulating 2-schemes whose 2-types have some network-like diagrammatic calculus. In the categorical mechanism, we have 1-data (morphisms) with respect to source and target 0-data (objects). We have a composition  which always combines \emph{all} target ``indices'' of one morphism with \emph{all} source indices of another morphism, and a tensor product. Weakening means replacing a 1-axiom by a morphism for any choice of 0-data (i.e., a natural transformation), and instead of applying the 1-axiom composing with that morphism. In the tensor types mechanism, we have 1-data with respect to a single 0-data. We have a contraction of individual indices, and a tensor product. Weakening means replacing a 1-axiom by a 2-function, and instead of applying the 1-axiom, applying this 2-function.

Due to the distinction of source and target in (monoidal) categorical 2-schemes, the corresponding diagrammatic calculi (in terms of ``string diagrams'') have a flow of time inherently inbuilt, which doesn't exist for tensor 2-schemes. This flow of time can be undone by adding a rigid structure, i.e., and evaluation and coevaluation. For tensor types, there is no notion of a flow of time in the string diagram from the very beginning on. One could argue that this is a bit more natural than starting with a source/target structure and then undoing it afterwards. Vaguely, tensor 2-schemes are more simplified than the rigid monoidal categories in the sense that they involve less structural degrees of freedom subject to less constraints. E.g., in the categorical language we can write down collections of morphisms with 2-functions that treat the source- and target level on manifestly unequal footing such as the category of matrices with the direct sum as product, and then realize that such a definition cannot be compatible with any rigid structure. In the tensor language, it wouldn't be possible to write down such a definition in the first place, as 1-data is defined for a single 0-data.

Surely, most of the tensor types in Section~\ref{sec:enzyclopedia} could have been equivalently formulated in the language of compact closed categories. Readers who prefer the categorical mechanism for 2-schemes are welcome to substitute the title of the document by ``Compact closed categories and their use in physics'', and to translate all tensor types in Section~\ref{sec:enzyclopedia} into compact closed categories as described in the following Section~\ref{sec:compact_category_to_tensor}.

\subsection{From compact closed categories to tensor types}
\label{sec:compact_category_to_tensor}
In this section we will sketch how any category with compact closed structure $\mathcal{C}$ yields a tensor type $\calg$ of the flavor specified in the section before. The identification goes as follows:
\begin{itemize}
\item The 0-data of $\calg$ are the objects (0-data) of $\mathcal{C}$,
\begin{equation}
\dat_0^\calg \coloneqq \dat_0^{\mathcal{C}}\;.
\end{equation}
\item Product, dual, and unit of 0-data in $\calg$ are the tensor product, dual and unit for objects/0-data in $\mathcal{C}$.
\item The 1-data of $\calg$ are the morphisms of $\mathcal{C}$ with the unit object in the source,
\begin{equation}
\dat_1^\calg(a)\coloneqq \dat_1^{\mathcal{C}}(1,a)\;,
\end{equation}
or diagrammatically,
\begin{equation}
\begin{tikzpicture}
\atoms{data}{0/p={0,0}}
\draw (0-t)edge[ind=a]++(0,0.4);
\end{tikzpicture}
\coloneqq
\begin{tikzpicture}
\atoms{category data}{0/p={0,0}}
\atoms{prod}{u/p={-0.4,0.4}}
\draw (0-t)--(u) (0-t)edge[ind=a]++(0.4,0.4);
\end{tikzpicture}\;.
\end{equation}
To avoid confusion, we color the categorical 1-data in blue.
\item The commutor $\sigma$ is given by the composition with the tensor product of the identity and the braiding ($\sigma_0$ is the composition with the braiding alone).
\item The associator $\alpha$ is given by the composition with the tensor product of the identity and the associator of $\mathcal{C}$.
\item The unitor is given by the composition with the left unitor.
\item The dual involutor is given by the tensor product with the identity, followed by the composition with the tensor product of the identity and the evaluation, followed by the pre-composition with the tensor product of the coevaluation and the identity.
\item The contraction is given by the composition with the tensor product of the identity and the evaluation, followed by a composition with the unitor.
\item The identity tensor is given by the coevaluation,
\begin{equation}
\begin{tikzpicture}
\node(x0)at(0,0){
\begin{tikzpicture}
\atoms{copy}{0/p={0,0}}
\draw (0)edge[ind=a]++(0,0.4);
\end{tikzpicture}};
\node(x1)at(4,0){
\begin{tikzpicture}
\atoms{category data}{f/p={0,0}}
\atoms{prod}{0/p={-0.4,0.4}, 1/p={0.4,0.4}}
\atoms{copy}{c/p={0.4,1.2}}
\atoms{dual}{d/p={0,0.8}}
\draw (c)edge[ind=a]++(0,0.4) (f-t)--(0) (f-t)--(1) (1)--(d-b) (d-t)--(c);
\draw[rounded corners] (1)--++(0.4,0.4)--(c);
\end{tikzpicture}};
\node(x2)at(2,-3){
\begin{tikzpicture}
\atoms{prod}{1/p={0,0}}
\atoms{copy}{c/p={0,0.8}}
\atoms{data}{d/p={0,-0.4}}
\atoms{dual}{s/p={-0.4,0.4}}
\draw[rc] (1)--++(0.4,0.4)--(c) (1)--(s-b) (s-t)--(c) (1)--(d) (c)edge[ind=a]++(0,0.4);
\end{tikzpicture}};
\draw (x0)edge[->]node[midway,below left]{$\idop$}(x2);
\draw (x2)edge[<->]node[midway,below right]{$=$}(x1);
\draw (x0)edge[->]node[midway,above]{$\textcolor{blue}{\operatorname{cev}}$}(x1);
\end{tikzpicture}
\;.
\end{equation}
\item The tensor product is given by the tensor product of $\mathcal{C}$ plus pre-composition with the unitor for the trivial object,
\begin{equation}
\begin{tikzpicture}
\node(x0)at(0,0){
\begin{tikzpicture}
\atoms{data}{0/p={0,0}, 1/p={0.5,0}}
\draw (0-t)edge[ind=a]++(0,0.4) (1-t)edge[ind=b]++(0,0.4);
\end{tikzpicture}};
\node(x1)at(4,0){
\begin{tikzpicture}
\atoms{category data}{0/p={0,0}, 1/p={1.5,0}}
\atoms{prod}{m0/p={-0.5,0.5}, m1/p={1,0.5}}
\draw (0-t)edge[ind=a]++(45:0.4) (1-t)edge[ind=b]++(45:0.4) (0-t)--(m0) (1-t)--(m1);
\end{tikzpicture}};
\node(x2)at(0,-3){
\begin{tikzpicture}
\atoms{data}{f/p={0,0}}
\atoms{prod}{0/p={0,0.4}}
\draw (f-t)--(0) (0)edge[ind=a]++(135:0.4) (0)edge[ind=b]++(45:0.4);
\end{tikzpicture}};
\node(x3)at(4,-3){
\begin{tikzpicture}
\atoms{category data}{f/p={0,0}}
\atoms{prod}{0/p={-0.4,0.4}, 1/p={0.8,0.8}, 2/p={-0.8,0.8}, 3/p={0,0.8}}
\draw (1)edge[ind=a]++(135:0.4) (1)edge[ind=b]++(45:0.4) (f-t)--(0) (f-t)--(1) (0)--(2) (0)--(3);
\end{tikzpicture}};
\node(x4)at(4,-6){
\begin{tikzpicture}
\atoms{category data}{f/p={0,0}}
\atoms{prod}{0/p={-0.4,0.4}, 1/p={0.4,0.4}, 2/p={-0.8,2}}
\atoms{copy}{c0/p={-0.4,1.2}, c1/p={-0.8,1.6}, c2/p={-1.2,1.2}}
\draw (1)edge[ind=a]++(135:0.4) (1)edge[ind=b]++(45:0.4) (f-t)--(0) (f-t)--(1) (0)edge[bend left=45](c0) (0)edge[bend right=45](c0) (c0)--(c1) (c1)--(c2) (c1)--(2);
\end{tikzpicture}};
\node(x5)at(4,-9){
\begin{tikzpicture}
\atoms{category data}{f/p={0,0}, fx/p={-2,0}}
\atoms{prod}{0/p={-0.4,0.4}, 1/p={0.4,0.4}, 2/p={-0.8,2}, 0x/p={-1.6,0.4}, 1x/p={-1.2,0.8}}
\atoms{copy}{c0/p={-0.4,1.2}, c1/p={-0.8,1.6}, c2/p={-2,0.8}}
\draw (1)edge[ind=a]++(135:0.4) (1)edge[ind=b]++(45:0.4) (f-t)--(0) (f-t)--(1) (0)edge[bend left=45](c0) (0)edge[bend right=45](c0) (c0)--(c1) (c1)--(c2) (c1)--(2) (fx-t)--(0x) (0x)--(1x) (0x)--(c2);
\draw[rounded corners] (c2)--++(-0.4,-0.4)--(fx-t);
\end{tikzpicture}};
\node(x6)at(0,-9){
\begin{tikzpicture}
\atoms{category data}{f/p={0,0}, fx/p={-0.8,0}}
\atoms{prod}{1/p={0.4,0.4}, 2/p={-0.4,0.8}, 3/p={-0.8,1.2}, 4/p={0,1.2}, 5/p={-1.2,0.4}}
\atoms{copy}{c0/p={-0.4,0.4}}
\draw (1)edge[ind=a]++(135:0.4) (1)edge[ind=b]++(45:0.4) (f-t)--(1) (f-t)--(c0) (fx-t)--(c0) (fx-t)--(5) (c0)--(2) (2)--(3) (2)--(4);
\end{tikzpicture}};
\node(x7)at(0,-6){
\begin{tikzpicture}
\atoms{category data}{f/p={0,0}}
\atoms{prod}{1/p={0.4,0.4}, 2/p={-0.4,0.4}}
\draw (1)edge[ind=a]++(135:0.4) (1)edge[ind=b]++(45:0.4) (f-t)--(1) (f-t)--(2);
\end{tikzpicture}};
\draw (x0)edge[<->]node[midway,above]{$=$}(x1);
\draw (x0)edge[->]node[midway,left]{$\otimes$}(x2);
\draw (x1)edge[->]node[midway,right]{$\color{blue}{\otimes}$}(x3);
\draw (x3)edge[->]node[midway,right]{$\widetilde{C}$}(x4);
\draw (x4)edge[->]node[midway,right]{$\color{blue}{U_l^{-1}}$}(x5);
\draw (x5)edge[->]node[midway,below]{$\widetilde{C}$}(x6);
\draw (x6)edge[->]node[midway,left]{$\widetilde{C}\color{blue}{\circ}$}(x7);
\draw (x2)edge[<->]node[midway,left]{$=$}(x7);
\end{tikzpicture}\;.
\end{equation}
\item The trivial tensor is obtained from the identity applied to discarding the trivial object.
\end{itemize}

\subsection{From tensor types to compact closed categories}
\label{sec:tensor_to_compact_category}
In this section, we sketch the mapping back from a tensor type $\calg$ to a category with compact closed structure $\mathcal{C}$:
\begin{itemize}
\item As for the converse identification, 0-data and product, dual and unit 1-functions of $\mathcal{C}$ are those of $\mathcal{G}$.
\item The morphisms (1-data) of $\mathcal{C}$ between two objects (0-data) $a$ and $b$ are the 1-data of $\calg$ with 0-data $a^*\otimes b$,
\begin{equation}
\dat_1^{\mathcal{C}}(a,b)\coloneqq \dat_1^\calg(a^*\otimes b)\;,
\end{equation}
or diagrammatically,
\begin{equation}
\begin{tikzpicture}
\atoms{category data}{0/p={0,0}}
\draw (0-t)edge[ind=a]++(135:0.4);
\draw (0-t)edge[ind=b]++(45:0.4);
\end{tikzpicture}
\coloneqq
\begin{tikzpicture}
\atoms{data}{0/p={0,0}}
\atoms{prod}{u/p={0,0.4}}
\atoms{dual}{d/p={-0.4,0.8}}
\draw (0-t)--(u) (u)edge[ind=b]++(45:0.4) (u)--(d-b) (d-t)edge[ind=a]++(0,0.3);
\end{tikzpicture}
\;.
\end{equation}
\item The braiding is given by the identity tensor applied to the product of two 0-data, followed by the commutor.
\item The associator is obtained by the identity tensor applied the product of three 0-data, followed by the associator.
\item The unitor is given by the identity tensor applied to the product of a 0-data and the unit 0-data, followed by the unitor.
\item The composition is given by the tensor product, followed by the sequence $\sigma\alpha\alpha_0^{-1}$ of invertible 2-functions, and finally a contraction.
\item The tensor product is given by the tensor product, followed by many invertible 2-functions.
\end{itemize}

\section{Liquids}
\label{sec:liquid}
Fulfilling a 2-axiom of a 2-scheme (such as the different flavors of the tensor 2-scheme) means that, for every choice of 0-data and input 1-data, the output 1-data for the two sides of the 2-axiom are equal. We can give a different interpretation to a 2-axiom as a constraint on the 0-data and input 1-data rather than the 2-functions. I.e., we look at the 0-data and input 1-data for which the two output 1-data are equal. The 0-data fulfilling this are what will be called \emph{models} of a \emph{liquid}.

\begin{mydef}
A \tdef{liquid}{liquid} for a 2-scheme consists of an additional 2-axiom called the \tdef{liquid 2-axiom}{liquid_2axiom}.

Two liquids are equivalent if their liquid 2-axioms differ by applying the other 2-axioms, or pre-/post-composing both 2-networks with the same invertible 2-function.

A liquid 2-axiom is called \tdef{simple}{liquid_2axiom_simple} if the target 1-network has only one 1-data element. Any liquid 2-axiom can be split into a set of simple liquid axioms, just as any 2-network can be split into simple ones. Those simple 2-axioms will be called \tdef{moves}{move} of the liquid.
\end{mydef}

So far it might seem like a liquid is nothing but an extension of a 2-scheme by further 2-axioms. However, the liquid moves have a different interpretation when it comes to defining models.

\begin{mydef}
A \tdef{model}{liquid_model} of a liquid in a 2-type (which is a model of the corresponding 2-scheme), consists of a 0-data and 1-data for the source 1-network of the liquid. The two 1-data resulting from applying the liquid 2-axiom have to be equal.

Equivalent liquids also have equivalent models. Models remain models when applying 2-axioms and post-composing with invertible 2-functions. When two liquids are equivalent under pre-composing the liquid 2-axiom by an invertible 2-function, the models of the equivalent liquids are related by applying the corresponding (inverse) 2-function to the 1-data.
\end{mydef}

\begin{myrem}
Liquids can be denoted using the effective graphical calculus of the corresponding 2-scheme. We simply draw the graphical representation of all the pairs of 2-networks for all the simple liquid 2-axioms. We should keep in mind though, that an effective 2-axiom is not quite the same as a pair of effective 2-networks, as pre- and post-composition with invertible 2-functions has to be applied to both 2-networks simultaneously.

For tensor types in particular, when we pre- or post-compose with the commutor to swap to indices, we can only do this at the same time for both 2-networks. Thus, the correspondence between the open indices of the two 2-networks is another invariant. In order to memorize which open indices on the left and right correspond to each other, we we put matching labels at the corresponding line ends. Sometimes we might omit them and instead use the position of the endpoints and/or the directions they are pointing at to indicate the latter.
\end{myrem}

\begin{myexmp}
Consider the following examples for moves:
\begin{itemize}
\item Consider an effective 1-network with one element and one binding,
\begin{equation}
\label{eq:substrate_example3}
\begin{tikzpicture}
\atoms{top half}{0/}
\draw (0-tr)edge[]++(45:0.5) (0-tl)edge[]++(135:0.5) (0-br)edge[dualind]++(-45:0.5) (0-bl)edge[dualind]++(-135:0.5);
\end{tikzpicture}\;.
\end{equation}
An example of a move with this effective 1-network as source is given by
\begin{equation}
\begin{tikzpicture}
\atoms{rot=-150,top half}{0/p={-150:0.6}}
\atoms{rot=-30,top half}{1/p={-30:0.6}}
\atoms{rot=90,top half}{2/p={90:0.6}}
\draw (0-tl)edge[cdir=+](1-bl) (1-tl)edge[cdir=+](2-bl) (2-tl)edge[cdir=+](0-bl) (0-br)edge[ind=a]++(180:0.4) (0-tr)edge[ind=b]++(-120:0.4) (2-br)edge[ind=e]++(60:0.4) (2-tr)edge[ind=f]++(120:0.4) (1-tr)edge[ind=d]++(0:0.4) (1-br)edge[ind=c]++(-60:0.4);
\end{tikzpicture}
=
\begin{tikzpicture}
\atoms{rot=-90,vflip,top half}{0/p={-90:0.6}}
\atoms{rot=30,vflip,top half}{1/p={30:0.6}}
\atoms{rot=150,vflip,top half}{2/p={150:0.6}}
\draw (0-bl)edge[cdir=-](1-tl) (1-bl)edge[cdir=-](2-tl) (2-bl)edge[cdir=-](0-tl) (0-tr)edge[ind=b]++(-120:0.4) (0-br)edge[ind=c]++(-60:0.4) (2-tr)edge[ind=f]++(120:0.4) (2-br)edge[ind=a]++(180:0.4) (1-br)edge[ind=e]++(60:0.4) (1-tr)edge[ind=d]++(0:0.4);
\end{tikzpicture}\;.
\end{equation}
\item Another move with the same source 1-network is
\begin{equation}
\label{eq:move_example1}
\begin{tikzpicture}
\atoms{top half}{0/p={0,0}}
\draw (0-br)edge[ind=a]++(-45:0.4) (0-tr)edge[ind=b]++(45:0.4) (0-tl)edge[ind=c]++(135:0.4) (0-bl)edge[ind=d]++(-135:0.4);
\end{tikzpicture}
=
\begin{tikzpicture}
\atoms{top half}{0/p={0,0}, 1/p={1.5,0}}
\draw (1-br)edge[ind=a]++(-45:0.4) (1-tr)edge[ind=b]++(45:0.4) (0-tl)edge[ind=c]++(135:0.4) (0-bl)edge[ind=d]++(-135:0.4);
\draw[cdir=-] (0-tr)to[out=45,in=90,looseness=2](0.6,0)to[out=-90,in=-45,looseness=2](0-br);
\draw[cdir=-] (1-tl)to[out=135,in=90,looseness=2](0.9,0)to[out=-90,in=-135,looseness=2](1-bl);
\end{tikzpicture}\;.
\end{equation}
\item For a source effective 1-network with $2$ elements and $2$ bindings,
\begin{equation}
\label{eq:substrate_example4}
\begin{tikzpicture}
\atoms{square,rhalf}{0/p={0,0}}
\draw (0-r)edge[]++(0:0.4) (0-t)edge[dualind,line width=1.8]++(90:0.4) (0-l)edge[dualind]++(180:0.4);
\end{tikzpicture}
,\quad
\begin{tikzpicture}
\atoms{triang,sect6}{0/p={0,0}}
\draw (0-mb)edge[line width=1.8]++(-90:0.4) (0-mr)edge[line width=1.8]++(30:0.4) (0-ml)edge[line width=1.8]++(150:0.4);
\end{tikzpicture}\;,
\end{equation}
an example of a move is given by,
\begin{equation}
\begin{tikzpicture}
\atoms{triang,sect6}{0/p={0,0}}
\atoms{square,rhalf}{1/p={0,-0.6}}
\draw (0-mb)edge[cdir=-,line width=1.8](1-t) (0-mr)edge[line width=1.8,ind=a]++(30:0.3) (0-ml)edge[line width=1.8,ind=b]++(150:0.3) (1-l)edge[ind=c]++(180:0.3) (1-r)edge[ind=d]++(0:0.3);
\end{tikzpicture}
=
\begin{tikzpicture}
\atoms{square,rhalf}{1/p={0,-1.1}, x0/p={-90:0.5}}
\atoms{rot=120,square,rhalf}{x1/p={30:0.5}}
\atoms{rot=-120,square,rhalf}{x2/p={150:0.5}}
\draw (x0-r)edge[cdir=+,out=0,in=-60](x1-l) (x1-r)edge[cdir=-,out=120,in=60](x2-l) (x2-r)edge[cdir=+,out=-120,in=180](x0-l) (x0-b)edge[cdir=-,line width=1.8](1-t) (x1-b)edge[line width=1.8,ind=a]++(30:0.3) (x2-b)edge[line width=1.8,ind=b]++(150:0.3) (1-l)edge[ind=c]++(180:0.3) (1-r)edge[ind=d]++(0:0.3);
\end{tikzpicture}\;.
\end{equation}
\end{itemize}
\end{myexmp}

\begin{myrem}
\label{rem:shape_symmetry}
Note that the element shapes in an effective 2-network can be arbitrarily rotated and reflected. Sometimes we choose shapes for element in a 1-network which have rotation/reflection symmetries, which makes it impossible to distinguish its indices, such as
\begin{equation}
\begin{tikzpicture}
\atoms{square}{0/p={0,0}}
\draw (0-l)--++(-0.3,0) (0-r)--++(0.3,0);
\end{tikzpicture}\;.
\end{equation}
We can now give an elegant interpretation to this: Symmetries that correspond to index permutations are an implicit notation for moves that hold for the corresponding element. E.g., if we had choose a shape without symmetry in the above example,
\begin{equation}
\begin{tikzpicture}
\atoms{square,rhalf}{0/p={0,0}}
\draw (0-l)--++(-0.3,0) (0-r)--++(0.3,0);
\end{tikzpicture}\;,
\end{equation}
then the symmetry of the shape corresponds to the following index-permuting,
\begin{equation}
\begin{tikzpicture}
\atoms{square,rhalf}{0/p={0,0}}
\draw (0-l)edge[ind=a]++(-0.3,0) (0-r)edge[ind=b]++(0.3,0);
\end{tikzpicture}
=
\begin{tikzpicture}
\atoms{square,rhalf}{0/p={0,0}}
\draw (0-l)edge[ind=b]++(-0.3,0) (0-r)edge[ind=a]++(0.3,0);
\end{tikzpicture}\;.
\end{equation}
If we express the same move using the symmetric shape, it becomes indistinguishable from a trivial move where one side is the same as the other but rotated.
\end{myrem}

\begin{myrem}
If we really want, we can make a liquid into a 2-scheme extending the 2-scheme it is defined in, such that 2-types of this 2-scheme are the liquid models:
\begin{itemize}
\item For every binding of the liquid 2-axiom, we add a 1-function which spits out a 0-data from nothing (i.e., a 1-element set).
\item For every 1-data element of the source 1-network of the liquid 2-axiom, add a 2-function, which spits out a 1-data from nothing.
\item Pre-compose the liquid 2-axiom with the 2-function generating 1-data above, and the corresponding 1-networks with the 1-function generating 0-data above. A model of the so extended 2-scheme will be a model of the 2-scheme together with a model of the liquid. In particular, the unique output of the 1-functions and 2-functions from the previous points define the liquid model alone.
\end{itemize}
\end{myrem}

\chapter{Concrete tensor types}
\label{sec:enzyclopedia}
This chapter will systematically present a large collection of different tensor types. The individual sections are each devoted to families of tensor types, parametrized by a different choices of \emph{dependencies}. Those dependencies can be things like natural numbers, groups, or rings. In fact, many of the presented families have a dependency another tensor type which we'll call the \emph{ground tensor type}. So those families can be understood as very general constructions which build new tensor types out of that ground tensor type. Every section will roughly have the following contents:
\begin{enumerate}
\item Definition of the (family of) tensor type, consisting of the dependency, the flavor, the 0-data, 1-data, 1-functions and 2-functions, together with an argument for why the 2-axioms hold.
\item Discussion of what the tensor type looks like for specific choices of dependencies.
\item Mappings to and from other (previously defined) tensor types.
\item Relations to established data structures/formalisms, discussion of their use in physics.
\end{enumerate}
\section{Overview}
In this section we will give an overview of the most relevant tensor types presented in this paper, by giving a short summary of their properties.

\subsection{Definitions}
\paragraph{Array tensors}
\begin{description}
\item[Dependencies:] Commutative semiring
\item[0-data:] Finite sets with cartesian product
\item[1-data:] Array of ring elements
\item[Tensor product:] Kronecker product
\item[Contraction:] Einstein summation
\item[Invertible 2-functions:] Canonical
\end{description}

\paragraph{Multi-mode infinite array tensors}
\begin{description}
\item[Dependencies:] Commutative semiring
\item[0-data:] Finite sets of ``infinity-modes'' with disjoint union
\item[1-data:] Multi-dimensionally infinite array with whose entries vanish for large coefficient of the modes
\item[Tensor product:] Kronecker product
\item[Contraction:] Einstein summation
\item[Invertible 2-functions:] Canonical
\end{description}

\paragraph{Rectangular Schur-complement tensors}
\begin{description}
\item[Dependencies:] Division ring
\item[0-data:] Pair of finite sets. Product is component-wise disjoint union, unit is twice empty set, dual interchanges the two sets in the pair.
\item[1-data:] Rectangular matrix, where the 0-data corresponds to the set of columns and the set of rows.
\item[Tensor product:] Direct sum of matrices
\item[Contraction:] Reshaping into a block matrix, then addition of identity matrices to certain blocks, then Schur complement of the block matrix.
\item[Invertible 2-functions:] Canonical
\end{description}

\paragraph{Square Schur-complement tensors}
\begin{description}
\item[Dependencies:] Division ring, optionally either symmetry or anti-symmetry constraint for the matrices.
\item[0-data:] Finite sets with disjoint union
\item[1-data:] Square matrix, 0-data corresponds to the set of rows which equals the set of columns
\item[Tensor product:] Direct sum of matrices
\item[Contraction:] Reshaping into a block matrix, then addition of identity matrices to certain blocks, then Schur complement of the block matrix.
\item[Invertible 2-functions:] Canonical
\end{description}

\paragraph{Level tensors}
\begin{description}
\item[Dependencies:] Ground tensor type, commutative unital associative algebra
\item[0-data:] Ground type 0-data
\item[1-data:] Ground type tensor with one additional ``level index'' whose 0-data that of the algebra
\item[Tensor product:] Ground type tensor product, followed by fusion of level indices with the algebra product
\item[Contraction:] Ground type contraction
\item[Invertible 2-functions:] Canonical via those of ground type
\end{description}

\paragraph{Projective tensors}
\begin{description}
\item[Dependencies:] Ground tensor type, subgroup of scalars
\item[0-data:] Ground type 0-data
\item[1-data:] Equivalence class of ground type 1-data modulo products with scalars from subgroup
\item[Tensor product:] Ground type tensor product applied to equivalence classes
\item[Contraction:] Ground type contraction applied to equivalence classes
\item[Invertible 2-functions:] Those of ground type
\end{description}

\paragraph{Hopf-symmetric tensors}
\begin{description}
\item[Dependencies:] Ground tensor type, co-commutative Hopf algebra
\item[0-data:] Algebra representation, containing a ground type 0-data
\item[1-data:] Ground type 1-data, subject to the ``symmetry constraint'' of being invariant under the algebra representation
\item[Tensor product:] Ground type tensor product
\item[Contraction:] Ground type contraction
\item[Invertible 2-functions:] Those of ground type
\end{description}

\paragraph{Twisted Hopf-symmetric tensors}
\begin{description}
\item[Dependencies:] Ground tensor type, triangular Hopf algebra
\item[0-data:] Algebra representation
\item[1-data:] Ground type 1-data, subject to the ``symmetry constraint''
\item[Tensor product:] Concatenation of orderings, ground tensor product
\item[Contraction:] Ground contraction, after putting contracted indices in consecutive ordering
\item[Invertible 2-functions:] Commutor involves the braiding of the triangular Hopf algebra. Dual involutor involves conjugation with the ``Drinfeld element''. Other 2-functions are canonical.
\end{description}

\paragraph{Frobenius-positive tensors}
\begin{description}
\item[Dependencies:] Ground tensor type, self-mapping in the case of $\dagger$-Frobenius algebras
\item[0-data:] $T$-Frobenius algebra, including ground type 0-data. Alternatively, $\dagger$-Frobenius algebra
\item[1-data:] Ground type 1-data fulfilling a ``positivity constraint'', being the ``square'' of a root tensor
\item[Tensor product:] Ground type tensor product
\item[Contraction:] Ground type contraction
\item[Invertible 2-functions:] Those of ground type
\end{description}

\subsection{Use in physics}
The following table lists a few physical theories, or ways to describe physical systems. Mathematically, their description can be formalized using tensor networks of certain tensor types, which are listed in the column on the right. This is surely not a complete list.
\begin{tabularx}{\linewidth}{X|X}
\bf{Physical theory} & \bf{Tensor type}\\
\hline
Classical deterministic many-body physics (mechanics and statics) & (Infinite/continuous versions of) Relation array tensors \\
Classical statistical physics (processes and thermal physics) & Non-negative-real array tensors \\
Pure-state real or imaginary time evolution and in quantum spin systems & Invertible-projective complex array tensors\\
Quantum spin systems including measurements or dissipation & Invertible-projective $\dagger$-Frobenius positive complex array tensors\\
Quantum many-body systems with fermions & $\zz_2$-twisted graded tensors\\
Many-body physics with on-site symmetries & Group-symmetric or Hopf-symmetric tensors (for a suitable ground type)\\
Quantum spin systems with time-reversal symmetry & ($T$-Frobenius positive) real array tensors\\
Quantum single-particle physics & Schur-complement tensors\\
Free-fermion models with particle-number conservation & Rectangular complex Schur-complement tensors with $u=i\sigma_y$\\
Free-fermion models without particle-number conservation & Anti-symmetric square complex Schur-complement tensors with $u=i\sigma_y$\\
\end{tabularx}

\section{Product types and mappings}
In this section, we give a few very simple general constructions for tensor types and mappings between them.

\begin{mydef}
Given two tensor types $\calg_1$ and $\calg_2$ (of the same flavor), the \tdef{product tensor type}{product_tensor_type} is the following tensor type $\calg_1\times \calg_2$:
\begin{itemize}
\item The 0-data are pairs of 0-data from the two tensor types,
\begin{equation}
\dat_0=\dat_0^{\calg_1}\times \dat_0^{\calg_2}\;.
\end{equation}
\item The 1-data are pairs of 1-data from the two tensor types,
\begin{equation}
\dat_1((a,b))=\dat_1^{\calg_1}(a)\times \dat_1^{\calg_2}(b)\;.
\end{equation}
\item The 1-functions are cartesian products of the 1-functions of the two 2-types, e.g.,
\begin{equation}
(a,b)\otimes (c,d)=(a\otimes b,c\otimes d)\;.
\end{equation}
\item The 2-functions are cartesian products of the 2-functions of the two 2-types, e.g.,
\begin{equation}
[(A,B)]=([A],[B])\;.
\end{equation}
\end{itemize}
The 1-axioms and 2-axioms of $\calg_1\times \calg_2$ hold as an immediate consequence of the 1-axioms and 2-axioms of $\calg_1$ and $\calg_2$. Note that the product type is not only defined for tensor types, but for 2-types of arbitrary 2-schemes.
\end{mydef}

\begin{mydef}
For every tensor type $\calg$ and tensor type $\mathcal{H}$ with unit and trivial tensor, the \tdef{trivial tensor mapping}{trivial_tensor_mapping} the following mapping from $\calg$ to $\mathcal{H}$:
\begin{itemize}
\item Every 0-data $a\in \dat_0^\calg$ is mapped to the unit 0-data of $\mathcal{H}$,
\begin{equation}
m(a)=1^{\mathcal{H}}\;.
\end{equation}
Diagrammatically, we can write
\begin{equation}
\label{eq:trivial_mapping0}
\begin{tikzpicture}
\atoms{mapping}{0/p={0,0}}
\draw (0-t)edge[mark={lab={$a$},a}]++(0,0.4) (0-b)edge[mapping second 0dat,ind=b]++(0,-0.4);
\end{tikzpicture}
=
\begin{tikzpicture}
\atoms{mapping second prod}{0/p={0,0}}
\atoms{copy}{c/p={0,0.4}}
\draw[mapping second 0dat] (0)edge[ind=b]++(0,-0.4);
\draw (c)edge[ind=a]++(0,0.4);
\end{tikzpicture}\;.
\end{equation}
\item Every 1-data $A\in \dat_1^\calg(a)$ is mapped to the trivial tensor,
\begin{equation}
M(A)=\mathbf{1}^{\mathcal{H}}\;.
\end{equation}
Diagrammatically, we have
\begin{equation}
\begin{tikzpicture}[every node/.style={inner sep=0.2cm}]
\node (xa) at (0,0){
\begin{tikzpicture}
\atoms{data}{d/}
\draw[mark={lab={$a$},a}] (d)--++(0,0.4);
\end{tikzpicture}};
\node (xb) at (4,0){
\begin{tikzpicture}
\atoms{mapping second data}{d/p={0,-0.5}}
\atoms{mapping}{m/}
\draw (m-t)edge[mark={lab={$a$},a}]++(0,0.4);
\draw[mapping second 0dat] (m-b)--(d);
\end{tikzpicture}};
\node (xc) at (0,-3){
\begin{tikzpicture}
\atoms{mapping second copy}{c/}
\atoms{mapping second prod}{p/p={0,0.5}}
\atoms{copy}{c1/p={-0.5,0}}
\draw[mapping second 0dat] (c)--(p);
\draw (c1)edge[ind=a]++(0,0.4);
\end{tikzpicture}};
\node (xe) at (4,-3){
\begin{tikzpicture}
\atoms{mapping second prod}{0/}
\atoms{mapping second data}{d/p={0,-0.5}}
\atoms{copy}{c/p={0,0.4}}
\draw[mapping second 0dat] (0)--(d);
\draw (c)edge[ind=a]++(0,0.4);
\end{tikzpicture}};
\draw (xa) edge[->] node[midway,left]{$\textcolor{red}{\widetilde{C}}\text{Discard}$} (xc);
\draw (xa) edge[->] node[midway,above]{$M$} (xb);
\draw (xb) edge[<->] node[midway,right]{$=$} (xe);
\draw (xe)edge[<-] node[midway,below,mapping second 0dat]{$\mathbf{1}$} (xc);
\end{tikzpicture}
\;.
\end{equation}
\item The product homomorphor is given by
\begin{equation}
\begin{tikzpicture}[every node/.style={inner sep=0.2cm}]
\node (a) at (0,0){
\begin{tikzpicture}
\atoms{mapping second prod}{0/p={0,0}}
\atoms{prod}{1/p={0.4,0.8}}
\atoms{mapping}{m/p={0.4,0.4}}
\atoms{mapping second data}{d/p={0,-0.5}}
\draw (m-t)--(1) (1)edge[ind=b]++(-0.4,0.4) (1)edge[ind=c]++(0.4,0.4);
\draw[mapping second 0dat] (0)--(d) (0)--(m-b) (0)edge[ind=a]++(-0.4,0.4);
\end{tikzpicture}};
\node (b) at (4,0){
\begin{tikzpicture}
\atoms{mapping second prod}{0/p={0,0}, 1/p={0.4,0.4}}
\atoms{mapping}{m0/p={0,0.8}, m1/p={0.8,0.8}}
\atoms{mapping second data}{d/p={0,-0.5}}
\draw (m0-t)edge[ind=b]++(0,0.4) (m1-t)edge[ind=c]++(0,0.4);
\draw[mapping second 0dat] (0)--(d) (0)--(1) (1)--(m0-b) (1)--(m1-b) (0)edge[ind=a]++(-0.4,0.4);
\end{tikzpicture}};
\node (c) at (0,-4){
\begin{tikzpicture}
\atoms{mapping second prod}{0/p={0,0}, 1/p={0.4,0.4}}
\atoms{mapping second data}{d/p={0,-0.5}}
\atoms{prod}{p0/p={1.4,0}}
\atoms{copy}{c0/p={1.4,-0.5}}
\draw[mapping second 0dat, rc] (0)--(d) (0)--(1) (0)edge[ind=a]++(-0.4,0.4);
\draw (p0)--(c0) (p0)edge[ind=b]++(-0.4,0.4) (p0)edge[ind=c]++(0.4,0.4);
\end{tikzpicture}};
\node (e) at (4,-4){
\begin{tikzpicture}
\atoms{mapping second prod}{0/p={0,0}, 1/p={0.4,0.4}, 2/p={0.2,1}, 3/p={0.6,1}}
\atoms{mapping second data}{d/p={0,-0.5}}
\atoms{copy}{c1/p={0.2,1.4}, c2/p={0.6,1.4}}
\draw[mapping second 0dat, rc] (0)--(d) (0)--(1) (2)--++(0,-0.3)--(1) (3)--++(0,-0.3)--(1) (0)edge[ind=a]++(-0.4,0.4);
\draw (c1)edge[ind=b]++(0,0.4) (c2)edge[ind=c]++(0,0.4);
\end{tikzpicture}};
\draw (a) edge[<->] node[midway,left]{$=$} (c) (a) edge[->] node[midway,above]{$M_\otimes$} (b) (b) edge[<->] node[midway,right]{$=$} (e) (e)edge[<-] node[midway,below]{$\textcolor{red}{\alpha_0^{-1}\otimes\mathbf{1}}\widetilde{C}$} (c);
\end{tikzpicture}
\;.
\end{equation}
\item It has strict unit homomorphicity (if $\calg$ has a unit), as we can see from
\begin{equation}
\begin{tikzpicture}[every node/.style={inner sep=0.2cm}]
\node (a) at (0,0){
\begin{tikzpicture}
\atoms{mapping second prod}{0/p={0,0}}
\atoms{prod}{1/p={0.4,0.8}}
\atoms{mapping}{m/p={0.4,0.4}}
\atoms{mapping second data}{d/p={0,-0.5}}
\draw (m-t)--(1);
\draw[mapping second 0dat] (0)--(d) (0)--(m-b) (0)edge[ind=a]++(-0.4,0.4);
\end{tikzpicture}};
\node (b) at (4,0){
\begin{tikzpicture}
\atoms{mapping second prod}{0/p={0,0}, 1/p={0.4,0.4}}
\atoms{mapping second data}{d/p={0,-0.5}}
\draw[mapping second 0dat] (0)--(d) (0)--(1) (0)edge[ind=a]++(-0.4,0.4);
\end{tikzpicture}};
\node (c) at (2,-3){
\begin{tikzpicture}
\atoms{mapping second prod}{0/p={0,0}, 1/p={0.4,0.4}}
\atoms{mapping second data}{d/p={0,-0.5}}
\draw[mapping second 0dat, rc] (0)--(d) (0)--(1) (0)edge[ind=a]++(-0.4,0.4);
\atoms{prod}{p/p={1,0}}
\atoms{copy}{c/p={1,-0.5}}
\draw (p)--(c);
\end{tikzpicture}};
\draw (a) edge[<->] node[midway,below left]{$=$} (c) (a) edge[->] node[midway,above]{$M_1$} (b) (c) edge[->] node[midway,right]{$\widetilde{C}$} (b);
\end{tikzpicture}
\;.
\end{equation}
\item The dual homomorphor (if $\calg$ has a dual) can be defined analogously and consists of the unit automorphor.
\end{itemize}
\end{mydef}

\begin{mydef}
Consider a mapping $\mathcal{M}$ from $\mathcal{A}$ to $\mathcal{B}$, and a mapping $\mathcal{N}$ from $\mathcal{A}$ to $\mathcal{C}$. The \tdef{product}{mapping_product} of $\mathcal{M}\times\mathcal{N}$ is a mapping from $\mathcal{A}$ to $\mathcal{B}\times \mathcal{C}$. It's 1-functions and 2-functions are simply tensor products of the 1-functions and 2-functions of $\mathcal{M}$ and $\mathcal{N}$, e.g.,
\begin{equation}
m^{\mathcal{M}\times\mathcal{N}}(a)=(m^{\mathcal{M}}(a), m^{\mathcal{N}}(a))
\end{equation}
for every $a\in \dat_0^{\mathcal{A}}$, or,
\begin{equation}
M_\otimes^{\mathcal{M}\times\mathcal{N}}(a)=(M_\otimes^{\mathcal{M}}(a), M_\otimes^{\mathcal{N}}(a))\;.
\end{equation}
\end{mydef}

\begin{mydef}
For every tensor type $\calg$, the \tdef{tensor product mapping}{tensor_product_mapping} is the following tensor mapping from $\calg\times\calg$ to $\calg$:
\begin{itemize}
\item The flavor depends on the flavor of $\calg$. In any case, it does not have strict product homomorphicity.
\item The mapping 1-function is the same as the product 1-function of $\calg$,
\begin{equation}
m((a,b))=a\otimes b\;.
\end{equation}
Diagrammatically, we can write
\begin{equation}
\label{eq:tensor_product_mapping_basis}
\begin{tikzpicture}
\atoms{mapping}{0/p={0,0}}
\draw (0-t)edge[mark={lab={$(a,b)$},a}]++(0,0.4) (0-b)edge[mapping second 0dat,ind=c]++(0,-0.4);
\end{tikzpicture}
=
\begin{tikzpicture}
\atoms{mapping second prod}{0/p={0,0}}
\draw[mapping second 0dat, rc, ind=a] (0)--++(-0.2,0.3)--++(0,0.3);
\draw[mapping second 0dat, rc, ind=b] (0)--++(0.2,0.3)--++(0,0.3);
\draw[mapping second 0dat] (0)edge[ind=c]++(0,-0.4);
\end{tikzpicture}\;.
\end{equation}
\item The mapping 2-function is the same as the tensor product,
\begin{equation}
M((A,B))=A\otimes B\;.
\end{equation}
Diagrammatically, we have
\begin{equation}
\begin{tikzpicture}[every node/.style={inner sep=0.2cm}]
\node (a) at (0,0){
\begin{tikzpicture}
\atoms{data}{d/}
\draw[mark={lab={$(a,b)$},a}] (d)--++(0,0.4);
\end{tikzpicture}};
\node (b) at (4,0){
\begin{tikzpicture}
\atoms{mapping second data}{d/p={0,-0.5}}
\atoms{mapping}{m/}
\draw (m-t)edge[mark={lab={$(a,b)$},a}]++(0,0.4);
\draw[mapping second 0dat] (m-b)--(d);
\end{tikzpicture}};
\node (c) at (0,-3){
\begin{tikzpicture}
\atoms{mapping second data}{0/, 1/p={0.4,0}}
\draw[mapping second 0dat] (0)edge[ind=a]++(0,0.4) (1)edge[ind=b]++(0,0.4);
\end{tikzpicture}};
\node (e) at (4,-3){
\begin{tikzpicture}
\atoms{mapping second prod}{0/}
\atoms{mapping second data}{d/p={0,-0.5}}
\draw[mapping second 0dat] (0)--(d) (0)edge[ind=a]++(-0.4,0.4) (0)edge[ind=b]++(0.4,0.4);
\end{tikzpicture}};
\draw (a) edge[<->] node[midway,left]{$=$} (c) (a) edge[->] node[midway,above]{$M$} (b) (b) edge[<->] node[midway,right]{$=$} (e) (e)edge[<-] node[midway,below,mapping second 0dat]{$\otimes$} (c);
\end{tikzpicture}
\;.
\end{equation}
\item The product homomorphor is given by
\begin{equation}
\begin{tikzpicture}[every node/.style={inner sep=0.2cm}]
\node (a) at (0,0){
\begin{tikzpicture}
\atoms{mapping second prod}{0/p={0,0}}
\atoms{prod}{1/p={0.4,0.8}}
\atoms{mapping}{m/p={0.4,0.4}}
\atoms{mapping second data}{d/p={0,-0.5}}
\draw (m-t)--(1) (1)edge[mark={lab={$(b,c)$},a}]++(-0.4,0.4) (1)edge[mark={lab={$(d,e)$},a}]++(0.4,0.4);
\draw[mapping second 0dat] (0)--(d) (0)--(m-b) (0)edge[ind=a]++(-0.4,0.4);
\end{tikzpicture}};
\node (b) at (4,0){
\begin{tikzpicture}
\atoms{mapping second prod}{0/p={0,0}, 1/p={0.4,0.4}}
\atoms{mapping}{m0/p={0,0.8}, m1/p={0.8,0.8}}
\atoms{mapping second data}{d/p={0,-0.5}}
\draw (m0-t)edge[mark={lab=${(b,c)}$,a}]++(0,0.4) (m1-t)edge[mark={lab={$(d,e)$},a}]++(0,0.4);
\draw[mapping second 0dat] (0)--(d) (0)--(1) (1)--(m0-b) (1)--(m1-b) (0)edge[ind=a]++(-0.4,0.4);
\end{tikzpicture}};
\node (c) at (0,-4){
\begin{tikzpicture}[blue]
\atoms{mapping second prod}{0/p={0,0}, 1/p={0.4,0.4}, 2/p={0.2,1}, 3/p={0.6,1}}
\atoms{mapping second data}{d/p={0,-0.5}}
\draw[mapping second 0dat, rc] (0)--(d) (0)--(1) (2)--++(0,-0.3)--(1) (3)--++(0,-0.3)--(1) (2)edge[ind=b]++(-0.4,0.4) (3)edge[ind=c]++(-0.4,0.4) (2)edge[ind=d]++(0.4,0.4) (3)edge[ind=e]++(0.4,0.4) (0)edge[ind=a]++(-0.4,0.4);
\end{tikzpicture}};
\node (e) at (4,-4){
\begin{tikzpicture}
\atoms{mapping second prod}{0/p={0,0}, 1/p={0.4,0.4}, 2/p={0,0.8}, 3/p={0.8,0.8}}
\atoms{mapping second data}{d/p={0,-0.5}}
\draw[mapping second 0dat, rc] (0)--(d) (0)--(1) (2)--(1) (3)--(1) (0)edge[ind=a]++(-0.4,0.4);
\draw[mapping second 0dat, rc,ind=b] (2)--++(-0.2,0.2)--++(0,0.3);
\draw[mapping second 0dat, rc,ind=c] (2)--++(0.2,0.2)--++(0,0.3);
\draw[mapping second 0dat, rc,ind=d] (3)--++(-0.2,0.2)--++(0,0.3);
\draw[mapping second 0dat, rc,ind=e] (3)--++(0.2,0.2)--++(0,0.3);
\end{tikzpicture}};
\draw (a) edge[<->] node[midway,left]{$=$} (c) (a) edge[->] node[midway,above]{$M_\otimes$} (b) (b) edge[<->] node[midway,right]{$=$} (e) (e)edge[<-] node[mapping second 0dat, midway,below]{$F$} (c);
\end{tikzpicture}
\;,
\end{equation}
where
\begin{equation}
F=\alpha\alpha_0^{-1}\sigma_2\alpha^{-1}\sigma\alpha_0\alpha^{-1}
\end{equation}
is a sequence of invertible 2-functions of $\calg$ whose job it is to exchange the indices $c$ and $d$.
\item The unit homomorphor (if $\calg$ has a unit) is given by
\begin{equation}
\begin{tikzpicture}[every node/.style={inner sep=0.2cm}]
\node (a) at (0,0){
\begin{tikzpicture}
\atoms{mapping second prod}{0/p={0,0}}
\atoms{prod}{1/p={0.4,0.8}}
\atoms{mapping}{m/p={0.4,0.4}}
\atoms{mapping second data}{d/p={0,-0.5}}
\draw (m-t)--(1);
\draw[mapping second 0dat] (0)--(d) (0)--(m-b) (0)edge[ind=a]++(-0.4,0.4);
\end{tikzpicture}};
\node (b) at (4,0){
\begin{tikzpicture}
\atoms{mapping second prod}{0/p={0,0}, 1/p={0.4,0.4}}
\atoms{mapping second data}{d/p={0,-0.5}}
\draw[mapping second 0dat] (0)--(d) (0)--(1) (0)edge[ind=a]++(-0.4,0.4);
\end{tikzpicture}};
\node (c) at (2,-3){
\begin{tikzpicture}
\atoms{mapping second prod}{0/p={0,0}, 1/p={0.4,0.4}, 2/p={0.2,1}, 3/p={0.6,1}}
\atoms{mapping second data}{d/p={0,-0.5}}
\draw[mapping second 0dat, rc] (0)--(d) (0)--(1) (2)--++(0,-0.3)--(1) (3)--++(0,-0.3)--(1) (0)edge[ind=a]++(-0.4,0.4);
\end{tikzpicture}};
\draw (a) edge[<->] node[midway,below left]{$=$} (c) (a) edge[->] node[midway,above]{$M_1$} (b) (c) edge[->] node[midway,right]{$U_r\alpha_0$} (b);
\end{tikzpicture}
\;.
\end{equation}
\item The dual homomorphor (if $\calg$ has a dual) is defined analogously.
\end{itemize}
\end{mydef}

\begin{myrem}
Note that the tensor product mapping has weak product automorphicity even if $\calg$ is strictly associative, as $F$ involves commutors. On the other hand, the mapping does have strict unit homomorphicity and dual homomorphicity if $\calg$ has strict associativity, unitality, and dual involutivity and automorphicity.
\end{myrem}

\section{Array tensors}
\label{sec:array_tensor}
\subsection{Motivation}
Array tensors are probably the most natural 2-type of the tensor 2-scheme. Array tensors for real or complex numbers are what is conventionally called ``tensors'' in the context of ``tensor networks''. The tensor types necessary for a first-principle formulation of both quantum mechanics and classical statistics are based on these array tensors.

\subsection{Definition}
Here and in all other sections of this chapter, we will not define only an individual tensor type but a whole family of tensor types. This family is parametrized by some algebraic data, which we'll introduce first. For array tensors, this algebraic data will be a commutative semiring, which generalizes the field of real or complex numbers that the array entries live in.

\begin{mydef}
A \tdef{semigroup}{semigroup} is a set $B$ together with a product operation,
\begin{equation}
\otimes: B\times B\rightarrow B\;.
\end{equation}
It has to obey the \emph{associativity axiom},
\begin{equation}
\label{eq:associativity_axiom}
a\otimes (b\otimes c)=(a\otimes b)\otimes c\;,
\end{equation}
\end{mydef}

\begin{mydef}
A \tdef{monoid}{monoid} is a semigroup $(B,\otimes)$ together with a unit element,
\begin{equation}
\mathbf{1} \in B\;.
\end{equation}
It has to obey the following \emph{unit axiom},
\begin{equation}
\label{eq:unit_axiom}
a\otimes \mathbf{1}=\mathbf{1} \otimes a=a\;.
\end{equation}
\end{mydef}

\begin{mydef}
A semigroup (or monoid) is \tdef{commutative}{semigroup_commutative} if
\begin{equation}
a\otimes b=b \otimes a\;.
\end{equation}
\end{mydef}

\begin{mydef}
A \tdef{commutative semiring}{commutative_semiring} is a set $K$ together with structures
\begin{equation}
\begin{gathered}
+: K\times K \rightarrow K\;,\\
1\in K\;,\\
\cdot: K\times K \rightarrow K\;,\\
\end{gathered}
\end{equation}
such that
\begin{itemize}
\item $+$ forms a commutative semigroup,
\item $1$ and $\cdot$ form a commutative monoid,
\item the following \emph{distributivity law} holds,
\begin{equation}
x\cdot (y+z)=x\cdot y+x\cdot z\;.
\end{equation}
\end{itemize}
\end{mydef}

The 0-data of array tensors will be finite sets with cartesian product as product. There are different ways to take cartesian products of the same sets, such as
\begin{equation}
a\times (b\times c)\neq (a\times b)\times c
\end{equation}
for three finite sets $a$, $b$, $c$. Even though those are different as sets, their elements can be identified canonically. In order to define the 2-functions of array tensors we will make use of such canonical bijections.

\begin{mydef}
Let us define a canonical bijection for associativity,
\begin{equation}
\begin{aligned}
\Phi_\times^\alpha:a\times (b\times c)&\rightarrow (a\times b)\times c\\
\Phi_\times^\alpha((x,(y,z)))&= ((x,y),z)\;,
\end{aligned}
\end{equation}
commutativity,
\begin{equation}
\begin{aligned}
\Phi_\times^\sigma:a\times b&\rightarrow b\times a\\
\Phi_\times^\sigma((x,y))&= (y,x)\;,
\end{aligned}
\end{equation}
and unitality,
\begin{equation}
\begin{aligned}
\Phi_\times^U:\{0\}\times a&\rightarrow a\\
\Phi_\times^U((0,x))&= x\;.
\end{aligned}
\end{equation}
Here, $a$, $b$, $c$ are finite sets, $x$, $y$, $z$ elements of those, and $0$ is the single element of an arbitrarily chosen 1-element set. We will denote the inverse of each $\Phi$ by $\bar{\Phi}$.
\end{mydef}

We are now ready to define array tensors in full generality.
\begin{mydef}
\tdef{Array tensors}{array_tensor} are the following tensor type.
\begin{itemize}
\item They can be defined for any commutative semiring $K$.
\item They allow for the definition of a unit and a trivial tensor. They have a weakened associator and unitor. They don't need a dual. They have symmetric contraction.
\item The 0-data is given by non-empty finite sets.
\item The product of 0-data is the cartesian product.
\item The unit 0-data is an arbitrary 1-element set, say
\begin{equation}
1=\{0\}\;.
\end{equation}
\item A \tdef{configuration}{array_configuration} refers to an element of a 0-data set $a$. A 1-data $A\in \dat_1(a)$ is a map that associates to each configuration an element of $K$, referred to as the \tdef{entry}{array_entry} for that configuration,
\begin{equation}
A: a \rightarrow K\;.
\end{equation}
\item The commutor, associator and unitor are defined via the canonical bijections $\Phi_\times^{\alpha/\sigma/U}$. That is, for a 1-data $A\in \dat_1(a\times (b\times (c\times d)))$, we have
\begin{equation}
\alpha(A)((x,y))=A((x,\bar{\Phi}_\times^\alpha(y)))\;,
\end{equation}
for $x\in a$ and $y\in (b\times c)\times d$. Recall that the bar over $\Phi$ denotes its inverse. More explicitly, we can write
\begin{equation}
\alpha(A)((w,((x,y),z)))=\alpha(A)((w,(x,(y,z))))\;.
\end{equation}
For a 1-data $A\in \dat_1(a\times (b\times c))$, we have
\begin{equation}
\sigma(A)((x,y))=A((x,\bar{\Phi}_\times^\sigma(y)))\;,
\end{equation}
and for $A\in \dat_1(\{0\}\times a)$,
\begin{equation}
U(A)(x)=A(\bar{\Phi}_\times^U(x))\;.
\end{equation}
\item The trivial tensor $\mathbf{1}\in \dat_1(\{0\})$ has only one entry which is the multiplicative unit $1$ of $K$,
\begin{equation}
\mathbf{1}(0)=1\;.
\end{equation}
\item The configurations after a tensor product of two 1-data $A\in \dat_1(a)$ and $B\in \dat_1(b)$ are given by pairs of configurations before the tensor product. The entry for such a configuration is the product of the entries for the two original configurations,
\begin{equation}
(A\otimes B)((i,j))=
A(i)\cdot B(j)
\end{equation}
for all $(i,j)\in a\times b$.
\item The configurations of a 1-data $A\in \dat_1(a\otimes (b\otimes b))$ before contraction are of the form $(i,(j,j'))$. The entry for the configuration $i$ after the contraction is the sum over all entries before the contraction for which $j=j'$,
\begin{equation}
\label{eq:array_contraction}
[A](i)=\sum_{j\in B} A((i,(j,j)))
\end{equation}
for all $i\in a$.
\end{itemize}
\end{mydef}

\begin{myobs}
All the 2-axioms are easily observed to hold, as associator, commutor and unitor are all canonical. E.g., for the block-compatibility of the contraction, we find
\begin{equation}
\begin{multlined}
[[A]](i)=\sum_{j\in b} \sum_{k\in c} A(((i,(j,j)),(k,k))) \\= \sum_{(j,k)\in b\times c} X(A)((i,((j,k),(j,k)))) = [X(A)](i)\;,
\end{multlined}
\end{equation}
where $X(A)$ denotes the appropriate sequence of associator and commutor 2-functions applied to $A$.
\end{myobs}

The real and complex numbers contain a $0$ element, which allows us to define the identity matrix and arrays of bond dimension $0$. If we want those, we have to equip $K$ with such a $0$.
\begin{mydef}
A \tdef{commutative semiring with zero}{cs_with_zero} is a commutative semiring together with an additive unit
\begin{equation}
0\in K
\end{equation}
such that $(K,+,0)$ form a commutative monoid. Note that sometimes in the literature already the term ``semiring'' assumes an additive unit.
\end{mydef}

\begin{mydef}
\tdef{Array tensors with zero}{at_with_zero} are the following variant of array tensors:
\begin{itemize}
\item The dependency $R$ is a commutative semiring with zero (instead of just a commutative semiring).
\item The 0-data are finite sets (including the empty one, instead of non-empty finite sets).
\item The additive unit over of the semiring is needed to define the outcome of the summation over an empty set in the contraction of a data $A\in \dat_1(a\times(\{\}\times\{\}))$,
\begin{equation}
[A](i)=0 \stackrel{!}{=}\sum_{j\in\{\}}A((i,(j,j)))\;.
\end{equation}
\item We can define an identity tensor $\idop \in \dat_1(a\otimes a)$ given by:
\begin{equation}
\idop((i,i'))=
\begin{cases}
1 &\text{if}\quad i=i'\\
0 &\text{otherwise}
\end{cases}
\end{equation}
for all $(i,i')\in a\times a$.
\end{itemize}
\end{mydef}

We promised to follow a data-centered approach which is very near to actual concrete computations. However, the 1-data that we chose, ``finite sets'' is not a very concrete data type that we would want so store on a computer. Luckily, all we need to know to store a 1-data is the number of elements of the corresponding 0-data set. The following alternative definition uses integers as 0-data, and could be directly implemented on a computer.

\begin{mydef}
Array tensors in their \emph{skeletal formulation} are the following tensor type.
\begin{itemize}
\item As in the finite-set formulation, they are defined for a commutative semiring $K$, and there is a version with and one without zero.
\item They have strict associativity and unitality 1-axioms.
\item The 0-data is given by non-negative (for the version with $0$) or positive (for the version without $0$) integers. Product and unit are those of the integers,
\begin{equation}
1=1,\qquad a\otimes b=ab\;.
\end{equation}
\item A 1-data $A\in \dat_1(a)$ is just an array of length $a$, that is, a function
\begin{equation}
A: \{0,\ldots,a-1\}\rightarrow K\;.
\end{equation}
\item In order to define the 2-functions it is useful to define a bijection
\begin{equation}
\label{eq:product_bijection}
\begin{gathered}
\begin{multlined}
\Phi_\cdot^{a,b}(i,j): \{0,\ldots,a-1\}\times \{0,\ldots,b-1\}\\\rightarrow \{0,\ldots,ab-1\}
\end{multlined}\\
\Phi_\cdot^{a,b}(i,j) \coloneqq bi+j\;,\\
\bar{\Phi}_\cdot^{a,b}(i) \coloneqq (\bar{\Phi}_{\cdot 0}^{a,b}(i), \bar{\Phi}_{\cdot 1}^{a,b}(i)) \coloneqq (\Phi_\cdot^{a,b})^{-1}(i)\;.
\end{gathered}
\end{equation}
\item The commutor $\sigma_0$ is given by acting on a 1-data $A\in \dat_1(ab)$ is given by
\begin{equation}
\sigma_0(A)(i)=A(\Phi_\cdot^{b,a}(\bar{\Phi}_{\cdot 1}^{a,b}(i), \bar{\Phi}_{\cdot 0}^{a,b}(i)))\;,
\end{equation}
where $0\leq i<ab$, or, more explicitly
\begin{equation}
\sigma_0(A)(ak+j)=A(bj+k)\;,
\end{equation}
where $0\leq j<a$ and $0\leq k<b$.
\item The tensor product of two 1-data $A\in\dat_1(a)$ and $B\in\dat_1(b)$ is given by
\begin{equation}
(A\otimes B)(i)=A(\bar{\Phi}_{\cdot 0}^{a,b}(i)) \cdot B(\bar{\Phi}_{\cdot 1}^{a,b}(i))\;.
\end{equation}
\item The contraction of a 1-data $A\in \dat_1(abb)$ is given by
\begin{equation}
[A](i)=\sum_{0\leq j<b} A(\Phi_\cdot^{a,bb}(i,\Phi_\cdot^{b,b}(j,j)))\;.
\end{equation}
\end{itemize}
\end{mydef}

\begin{myobs}
It may not come as a surprise that associativity and unitality are strict in the skeletal formulation. Everyone who has worked with multi-dimensional arrays in a modern programming language knows that we need to keep track of an ordering of the different axes/indices, but we don't need to book-keep a bracketing. In contrast to the finite-set formulation, we have associativity due to
\begin{equation}
\begin{multlined}
\psi^{ab,c}(\psi^{a,b}(i,j),k)=
c(bi+j)+k\\
=bci+(cj+k)=
\psi^{a,bc}(i,\psi^{b,c}(j,k))
\;.
\end{multlined}
\end{equation}
\end{myobs}

\begin{myrem}
The name ``skeletal'' is borrowed from category theory, whose construction of a \emph{skeleton} we can directly take over into our tensor types language. First, we define a 2-index tensor, i.e., a 1-data $G\in \dat_1(a\otimes b)$, to be \mdef{invertible} if there exists another 1-data $G^{-1}\in \dat_1(b\otimes a)$ such that
\begin{equation}
\begin{gathered}
\begin{tikzpicture}
\atoms{square,rhalf}{{0/lab={p=-90:0.35,t=$G$}}, {1/p={0.8,0},lab={p=-90:0.35,t=$G^{-1}$}}}
\draw (0)--(1) (0-l)edge[ind=x]++(180:0.4) (1-r)edge[ind=y]++(0:0.4);
\end{tikzpicture}
=
\begin{tikzpicture}
\draw (0,0)edge[ind=y,startind=x](1,0);
\end{tikzpicture}
\;,\\
\begin{tikzpicture}
\atoms{square,rhalf}{{0/lab={p=-90:0.35,t=$G^{-1}$}}, {1/p={0.8,0},lab={p=-90:0.35,t=$G$}}}
\draw (0)--(1) (0-l)edge[ind=x]++(180:0.4) (1-r)edge[ind=y]++(0:0.4);
\end{tikzpicture}
=
\begin{tikzpicture}
\draw (0,0)edge[ind=y,startind=x](1,0);
\end{tikzpicture}
\;.
\end{gathered}
\end{equation}
Note that this requires an identity tensor. Now, taking the skeleton means considering two 0-data $a$ and $b$ equivalent if there exists is an invertible 1-data $G\in\dat_1(a\otimes b)$, and taking the equivalence classes as the new 0-data. Then we choose a representative of every equivalence class, together with invertible 1-data which relate the representatives of a product with the product of representatives (and the analogous for other 1-functions). The new 1-data is then the 1-data for the representative.

For array tensors, an invertible 1-data is simply an invertible $a\times b$ matrix, thus two 0-data finite sets $|a|$ and $|b|$ are equivalent exactly if $|a|=|b|$. So the 0-data of the skeletonized version are numbers, corresponding to the number of elements of the sets.

We will make the relation between the finite-set formulation and skeletal formulation formal by giving a corresponding tensor mapping in Section~\ref{sec:array_mapping}.
\end{myrem}

\begin{myrem}
In general, taking the skeleton will yield a formulation of a tensor type which is more practically oriented, for, e.g., explicitly carrying out computations. In general, we aim for such a ``more skeletal'' description, where the 1-data contains the actual data, whereas the 0-data is rather a bookkeeping of some simple information. E.g., for the skeletal version of array tensors, the actual data processing consists of Einstein summations and Kronecker products applied to arrays, whereas the 0-data only corresponds to a bookkeeping of the different dimensions.

Let as consider yet a different formulation of array tensors, which are even ``less skeletal'' than the finite-set formulation. The 0-data are finite-dimensional vector spaces, with the product by the tensor product of vector spaces. The 1-data is simply a vector in the corresponding vector space. This is perfectly fine as formal definition, but a nightmare if we directly want to store things on a computer. First, vector spaces aren't a sensible data type as they form a class rather than a set. But even if the set $V$ underlying the vector space itself was fixed, the vector addition and a scalar multiplication are a data structure
\begin{equation}
(\cdot,+)\in V^{V\times \mathbb{R}}\times V^{V\times V}\;.
\end{equation}
To directly represent an instance of such a structure in storage, we would first need to pick some sensible discretization of the continuous sets $\mathbb{R}$ and $V$, and even then the storage size would scale exponentially in the vector space dimension $n$. Note that a good rule of thumb for practically sensible tensor types is that the storage/processing of the 0-data should only give rise to a small overhead to the storage/processing of the 1-data. This is clearly not the case for the presented example.

All vector spaces of the same dimension $n$ are isomorphic, and equivalent as 0-data. So after taking the skeleton, the 0-data is given by non-negative integer $n$, and the 1-data is defined with respect to the standard representative $\mathbb{R}^n$. So when taking the skeleton we again end up with the skeletal formulation of array tensors.
\end{myrem}

\subsection{Specific commutative semirings}
\begin{myobs}
Examples of commutative semirings are:
\begin{itemize}
\item Every field is a commutative ring. Every commutative ring is a commutative semiring.
\item Real numbers, complex numbers.
\item Positive, or non-negative real numbers.
\item Prime number fields, or more generally $\mathbb{Z}_n$ rings with modulo $n$ addition and multiplication.
\item (Non-negative or positive) integers with addition and multiplication.
\item Boolean variables with $\mathop{AND}$ and $\mathop{OR}$.
\item The cartesian product of any two commutative semirings is a commutative semiring again. E.g., two complex numbers with component-wise addition and multiplications form a commutative semiring.
\end{itemize}
Every of these commutative semirings yields an instance of array tensors. Surely, arrays with real or complex entries are the most useful ones for common practical purposes. But also $\mathbb{Z}_2$-valued arrays may have some applications. Boolean operations have a special interpretation that we'll comment on below.
\end{myobs}

\begin{mydef}
Let us take a closer look at the commutative semiring of booleans with $\mathop{AND}$ and $\mathop{OR}$. We will call array tensors for this semiring \tdef{relation array tensors}{relation_array_tensor}. This commutative semiring written out explicitly over the set $\{0,1\}$ is,
\begin{equation}
\begin{tabular}{l|l|l}
$+$ & $0$ & $1$\\
\hline
$0$ & $0$ & $1$\\
\hline
$1$ & $1$ & $1$
\end{tabular}
,\qquad
\begin{tabular}{l|l|l}
$\cdot$ & $0$ & $1$\\
\hline
$0$ & $0$ & $0$\\
\hline
$1$ & $0$ & $1$
\end{tabular}\;.
\end{equation}
Recall that this is different from the finite field $\mathbb{Z}_2$.
\end{mydef}

\begin{myrem}
A data $A\in \dat_1(a)$ of a relation array tensor is a function
\begin{equation}
A: a\rightarrow \{0,1\}\;,
\end{equation}
which can be interpreted as a subset of ``allowed'' configurations with entry $1$. 1-data for a 0-data $a\otimes b$ are subsets of the cartesian product $a\times b$, which are sometimes referred to as \emph{binary relations} between the sets $a$ and $b$. Consider taking the tensor product of two relation array tensors $A\in \dat_1(a\times b)$ and $B\in \dat_1(b\times c)$ and then contracting over the $b$ components. A configuration $(i,j)\in a\times c$ of the result is allowed if there exists a configuration $k\in b$ such that both the configuration $(i,k)$ of $A$ and $(k,j)$ of $B$ are allowed. This operation is known as the \emph{composition of binary relations} $A$ and $B$.

This indicates that relation array tensors are equivalent to the compact closed category of sets and relations.
\end{myrem}

\subsection{Mappings}
\label{sec:array_mapping}
\begin{mydef}
For any homomorphism $\mathcal{H}$ between two commutative semirings, the \tdef{entry-wise mapping}{entry_wise_mapping} is the following tensor mapping between the corresponding array tensors:
\begin{itemize}
\item It has strict homomorphor 1-axioms.
\item The mapping 1-function is the identity.
\item The mapping 2-function acting on a 1-data $A\in \dat_1(a)$ is given by
\begin{equation}
\mathcal{M}(A)(i)=\mathcal{H}(A(i))
\end{equation}
for all $i\in a$.
\end{itemize}
\end{mydef}

\begin{myobs}
Let's consider a few examples of semiring homomorphisms and the corresponding entry-wise mappings:
\begin{itemize}
\item Complex conjugation is a homomorphism from the commutative semiring of complex numbers to itself. The corresponding tensor mapping acts by complex conjugating every entry.
\item The canonical embedding of real numbers into complex ones is a homomorphism. The corresponding tensor mapping takes a real array tensor and interprets its entries as complex numbers.
\item The canonical embedding of integer numbers into real numbers.
\item The embedding of positive reals into general reals.
\item Complex numbers with integer real and imaginary part are a sub commutative semiring of complex numbers, yielding a homomorphism from the former to the latter.
\item There is a homomorphism from $\mathbb{Z}$ to $\mathbb{Z}_n$ by taking integers mod $n$.
\end{itemize}
\end{myobs}

Next we'll introduce the direct sum mapping, which is based on the disjoint union of sets.
\begin{mydef}
The \mdef{disjoint union} of two sets $a$ and $b$ is the set that contains one element for each element of $a$ and one for each element of $b$. It is the natural notion of union when we view the sets as data types where it doesn't make sense for two different types to contain the same elements. In terms of the conventional set union, we can define the disjoint union as
\begin{equation}
\label{eq:disjoint_union_definition}
a\sqcup b = \{(0,i)\ \text{for}\ i\in a\} \cup \{(1,i)\ \text{for}\ i\in b\}\;.
\end{equation}
So, we use $(0,i)$ to denote elements $i$ coming from the first set, and $(1,i)$ for elements coming from the second set.
\end{mydef}

\begin{mydef}
The direct sum is the following tensor mapping from the product of two times array tensors to array tensors again:
\begin{itemize}
\item The two input tensor type and output tensor types need to share the same commutative semi-ring $K$.
\item The mapping is not a full mapping as it is \emph{not} compatible with the tensor product. It is however compatible with the contraction, and with a sequence of tensor product and contraction of an index pair between the two components.
\item The mapping applied to two 0-data sets $a$ and $b$ is their disjoint union
\begin{equation}
m((a,b))=a\sqcup b\;.
\end{equation}
\item The mapping applied to two 1-data $A\in \dat_1(a)$ and $B\in \dat_1(b)$ is given by
\begin{equation}
\begin{aligned}
M((A,B))((0,i))&= A(i)\;,\\
M((A,B))((1,i))&=B(i)\;.
\end{aligned}
\end{equation}
\item The product homomorphor is a map
\begin{equation}
M_\otimes: \dat_1(a\times c\sqcup b\times d)\rightarrow \dat_1((a\sqcup b)\times(c\sqcup d))
\end{equation}
given by
\begin{equation}
\label{eq:direct_sum_homomorphor}
\begin{aligned}
M_\otimes(A)(((0,i),(0,j)))&= A((0,(i,j)))\;,\\
M_\otimes(A)(((1,i),(0,j)))&= 0\;,\\
M_\otimes(A)(((0,i),(1,j)))&= 0\;,\\
M_\otimes(A)(((1,i),(1,j)))&= A((1,(i,j)))\;.
\end{aligned}
\end{equation}
\item The unit homomorphor is a map
\begin{equation}
M_1: \dat_1(a\times (\{0\}\sqcup\{0\}))\rightarrow \dat_1(a)
\end{equation}
given by
\begin{equation}
M_1(A)(i)=A((i,(0,0)))+A((i,(1,0)))\;.
\end{equation}
Note that the second $0$ in every summand denotes the element of $\{0\}$, the first $0$ in the first summand refers to the first component of the direct sum, whereas the $0$ on the right side of Eq.~\eqref{eq:direct_sum_homomorphor} is the zero ring element.
\end{itemize}
\end{mydef}

\begin{myobs}
It is a straight-forward computation to see that the direct sum mapping is compatible with the 2-functions of array tensors. As an example, let us consider compatibility with the contraction. For every two 1-data $A\in \dat_1(a\times (b\times b))$ and $B\in \dat_1(c\times (d\times d))$, we find
\begin{equation}
\begin{multlined}
[M_\otimes M_\otimes M((A,B))]((0,i))
\\=\sum_{l\in b\sqcup d}M_\otimes M_\otimes M((A,B))(((0,i),(l,l)))
\\=\sum_{j\in b}M_\otimes M_\otimes M((A,B))(((0,i),((0,j),(0,j))))
\\+\sum_{j\in d}M_\otimes M_\otimes M((A,B))(((0,i),((1,j),(1,j))))
\\=\sum_{j\in b}A((i,(j,j))
 = [A](i) = M(([A],[B]))((0,i))
\\=M([(A,B)])((0,i))\;,
\end{multlined}
\end{equation}
for $i\in a$, and analogously for an argument $(1,i)$ with $i\in c$.

It is easy to see that the direct sum is \emph{not} compatible with the tensor product alone. For $4$ 1-data $A\in \dat_1(a)$, $B\in \dat_1(b)$, $C\in \dat_1(c)$, and $D\in \dat_1(d)$, we find in non-trivial cases that
\begin{equation}
\begin{multlined}
M_\otimes M((A,B)\otimes (C,D))(((0,i), (1,j))\\
= 0 \neq A(i) D(j)\\
= M((A,B))\otimes M((C,D))(((0,i),(1,j)))\;.
\end{multlined}
\end{equation}
\end{myobs}

We already mentioned that array tensors in their finite-set and their skeletal formulation are equivalent. This is formalized by the following mapping and its reverse.
\begin{mydef}
The \emph{skeleton embedding mapping} is the following mapping from array tensors in their skeletal formulation to array tensors in their finite-set formulation:
\begin{itemize}
\item A 0-data integer $a$ mapped to a standard representative set with $a$ elements,
\begin{equation}
m(a)=\{0,\ldots,a-1\}\;.
\end{equation}
\item The mapping of a 1-data $A\in \dat_1(a)$ is trivial,
\begin{equation}
M(A)=A\;.
\end{equation}
\item The homomorphors are defined canonically via the bijections $\Phi_\cdot$. E.g., the product homomorphor $M_{\otimes 0}$ acting on a 1-data $A\in\dat_1(a\times b)$ is a map
\begin{equation}
\begin{multlined}
M_{\otimes0}:  \dat_1(\{0,\ldots,ab-1\}) \\\rightarrow \dat_1(\{0,\ldots,a-1\}\times\{0,\ldots,b-1\})
\end{multlined}
\end{equation}
given by
\begin{equation}
M_{\otimes0}(A)((i,j))=A(\Phi_\cdot^{a,b}(i,j))\;.
\end{equation}
\end{itemize}
\end{mydef}

\begin{mydef}
The \emph{skeleton projection mapping} is the following mapping from array tensors in their finite-set formulation to array tensors in their skeletal formulation:
\begin{itemize}
\item The mapping of a 0-data finite set $a$ is cardinality,
\begin{equation}
m(a)=|a|\;.
\end{equation}
\item It has strict homomorphicity, which is possible as
\begin{equation}
|a\times b|=|a||b|\;,\quad |\{0\}|=1\;.
\end{equation}
\item For every finite set $a$, choose a standard bijection
\begin{equation}
\rho^a:  \{0,\ldots,|a|-1\} \rightarrow a\;.
\end{equation}
The mapping applied to a 1-data $A\in \dat_1(a)$ uses this bijection,
\begin{equation}
M(A)(i)=A(\rho^a(i))\;.
\end{equation}
\end{itemize}
\end{mydef}

\subsection{Use in physics}
\label{sec:array_physics}
Array tensors with non-negative entries are the tensor type we have to use to formulate classical statistical physics. It has been known for some time that the partition function of a classical statistical model with a local Hamiltonian at a fixed temperature can be written as a tensor network consisting of delta tensors and Boltzmann weight tensors. As pointed out in Ref.~\cite{cstar_qmech}, this can be turned into a full tensor-network model. That is, for any choice of system size and local measurements at certain points, we can construct a tensor network with open indices for the measurement outcomes. The probabilities for the different measurement outcomes are simply obtained by evaluating the tensor network. The type of all the tensors in the network, namely the delta tensors, the Boltzmann weights, as well as the measurement tensors, are array tensors with non-negative real entries.

Not only thermal statistical physics, but also any ``discrete stochastic time evolution'', or ``probabilistic cellular automaton'', is a circuit of stochastic maps, and therefore a tensor network of non-negative real array tensors. In that case, the networks are better described as string diagrams for the terminal symmetric monoidal category of stochastic maps.

Array tensors with complex entries are the tensor type needed to formulate pure-state quantum spin (i.e. non-fermionic qu-$d$-it) many-body systems. A ``discrete time evolution'' by a quantum circuit is directly a tensor network. But also a continuous time evolution by a local Hamiltonian can be Trotterized and approximated by a tensor network, as described, e.g., in Ref.~\cite{liquid_intro}. As pointed out in Ref.~\cite{cstar_qmech}, we can construct a proper tensor-network model by taking two layers of this pure-state tensor network, one of them complex conjugated, corresponding to a ket- and a bra level. We then insert measurement tensors (i.e., POVMs) with open indices corresponding to classical measurement outcomes, which couple the ket- and bra layer. As usual, the probability distribution over measurement outcomes for a specific system size, initial state, and local measurements at different places, can be obtained by simply contracting the tensor network.

The imaginary-time evolution of a local spin Hamiltonian can be Trotterized into a tensor network as well. If we take the trace of this tensor network, i.e., we compactify it in time direction, we get a tensor network representing a thermal quantum state. The inverse temperature is the compactified imaginary time interval. As usual, we include measurement tensors into this tensor network, to obtain a tensor-network model yielding the probabilities of measurement outcomes in the thermal state. In the limit of infinite imaginary time we get a tensor-network model describing the ground state properties.

Conventionally, ``tensor networks'' in quantum physics known as \emph{MPS} or \emph{PEPS} refer to tensor-network representations of a state vector (usually approximately representing the ground state of a local Hamiltonian) in a local quantum spin model. Also such a conventional tensor network ansatzes can be considered a tensor-network model for array tensors: First, we construct the overlap of the MPS or PEPS with a complex conjugated copy. Then we insert, or ``sandwich'' measurement tensors at the desired positions. The contraction of such a tensor network yields the probability distribution over measurement outcomes of the represented state, which in many cases is (an approximation of) the ground state of some local Hamiltonian.

Array tensors with \emph{real} entries are the tensor type we need to formalize quantum spin systems with a \emph{time-reversal symmetry}, i.e. a local anti-unitary representation of $\zz_2$. Any such representation is equal to complex conjugation in some local basis, such that after a basis change to that local basis we end up with a model invariant under complex conjugation in the chosen bases. So, if the model is formalized by a tensor network in that basis, all tensor entries must be real.

A totally different physical interpretation have tensor-network models based on relation-array tensors. Those formalize classical local combinatorial static physics, that is, systems with discrete classical deterministic degrees of freedom that must satisfy local constraint equations. For example, the set of \emph{dimer coverings} on a square lattice can be formalized as a tensor-network of relation-array tensors. When we contract this tensor network on a patch of lattice with boundaries, the configurations of open indices correspond to different boundary conditions of the dimer coverings. The entry of the evaluation of the tensor network tells us whether there exists a dimer covering for the given boundary condition or not.

Very often, classical degrees of freedom can take values in a continuous set. In contrast to array tensors for real or complex numbers, relation array tensors can be easily generalized to allow for infinite/continuous sets as 0-data. This generalized version of relation array tensors can formalize classical local static physics with continuous degrees of freedom. E.g., consider a system with little magnets at all vertices of a square lattice which can be rotated at arbitrary continuous angles, with an energy function containing a some magnetic repulsion term between neighboring magnets and a global magnetic field. For every single magnet, we ca figure out what its equilibrium angles (i.e. energy local extrema) are depending on the angles of the four neighboring magnets. This defines a 5-index continuous relation-array tensor. The model can be formalized as a tensor network by taking one delta tensor and one 5-index lowest-energy tensor for every vertex, and contracting indices accordingly. We can now put this tensor network on some square lattice with periodic or open boundary conditions, and add open indices to some of the delta-tensors at some places. Contracting the tensor network will result in a relation-array tensor which tells us which combinations of angles at the chosen places are compatible with a global equilibrium configuration.

Another example are classical field theories such as electrostatics. In order to formalize those as tensor-network models, we need to choose some discretization, i.e., some tiling of the continuum space into ``small'' cubic blocks. We then obtain a model in discrete space, whose local configurations at a site are the field configurations within a block. Again, we can construct a constraint tensor corresponding to the equilibrium configurations of the field within a block depending on the field within the surrounding blocks.

\section{Infinite array tensors}
\label{sec:infinite_array}
\subsection{Motivation}
Physical systems often have continuous degrees of freedom such as spacial coordinates, or local field values. The paradigmatic example for a Hamiltonian on such a continuous degree of freedom is the harmonic oscillator, whose eigenfunctions in terms of Legendre polynomials give rise to an occupation number basis of a single bosonic mode. The interesting states are the low-energy ones, whose amplitude rapidly decays for large occupation numbers. This can be formalized by infinite array tensors which have an infinite bond dimension, but a decay condition which ensures that contraction is still well defined.

\subsection{Definition: Via modes}
The goal of this section is to define array tensors with countably infinite bond dimension, that is, where the 0-data is a set like $\mathbb{N}$ rather than a finite set. While is possible to define array tensors where some bond dimensions are finite and some infinite, it is easier to only allow infinite bond dimensions. The $0$-data of an index is then simply given by the number of $\mathbb{N}$-components at that index.

The 0-data of multi-mode infinite array tensors tensors will be finite sets, as for array tensors, but their product will be given by the disjoint union introduced in Eq.~\eqref{eq:disjoint_union_definition}. Similar to the cartesian product, there are inequivalent ways to build disjoint unions from the same sets, such as,
\begin{equation}
a\sqcup (b\sqcup c)\neq (a\sqcup b)\sqcup c\;.
\end{equation}
However, the elements of those sets can be canonically identified, which is important for the definition of the 2-functions.

\begin{mydef}
\label{def:disjoint_union_bijection}
We define a canonical bijection for associativity,
\begin{equation}
\label{eq:disjoint_union_alpha_bijection}
\begin{aligned}
\Phi_\sqcup^\alpha:a\sqcup (b\sqcup c)&\rightarrow (a\sqcup b)\sqcup c\\
\Phi_\sqcup^\alpha((0,x))&= (0,(0,x))\;,\\
\Phi_\sqcup^\alpha(1,(0,y))&= (0,(1,y))\;,\\
\Phi_\sqcup^\alpha(1,(1,z))&= (1,z)\;,
\end{aligned}
\end{equation}
where we used the notation in Eq.~\eqref{eq:disjoint_union_definition}. We can also define a canonical bijection for commutativity,
\begin{equation}
\begin{aligned}
\Phi_\sqcup^\sigma:a\sqcup b&\rightarrow b\sqcup a\\
\Phi_\sqcup^\sigma((0,x))&= (1,x)\;,\\
\Phi_\sqcup^\sigma((1,y))&= (0,y)\;.
\end{aligned}
\end{equation}
and one for unitality,
\begin{equation}
\label{eq:disjoint_union_u_bijection}
\begin{aligned}
\Phi_\sqcup^U:a\sqcup \{\}&\rightarrow a\\
\Phi_\sqcup^U((0,x))&= x\;.
\end{aligned}
\end{equation}
We also want to introduce the notation $x^a$ for the set of functions from a set $a$ to a set $x$. For two functions $\vec\alpha\in x^a$, $\vec\beta\in x^b$, we define their concatenation to be the function
\begin{equation}
(\vec\alpha\sqcup \vec\beta) \in x^{a\sqcup b}
\end{equation}
defined by
\begin{equation}
\label{eq:function_concatenation}
\begin{gathered}
(\vec{\alpha}\sqcup \vec{\beta})((0,i)) = \vec\alpha(i)\;,\\
(\vec{\alpha}\sqcup \vec{\beta})((1,i)) = \vec\beta(i)\\
\end{gathered}
\end{equation}
for $i\in a$ or $i \in b$. We can also concatenate two functions $\vec\alpha\in x^a$, $\vec\beta\in y^b$, to a function
\begin{equation}
(\vec\alpha\hat\sqcup \vec\beta) \in (x\sqcup y)^{a\sqcup b}
\end{equation}
defined by
\begin{equation}
\label{eq:function_concatenation_diffout}
\begin{gathered}
(\vec{\alpha}\hat\sqcup \vec{\beta})((0,i)) = (0,\vec\alpha(i))\;,\\
(\vec{\alpha}\hat\sqcup \vec{\beta})((1,i)) = (1,\vec\beta(i))\\
\end{gathered}
\end{equation}
Moreover, for $\vec\gamma\in x^{a\sqcup b}$ we define $\vec\gamma\rvert_0\in x^a$ and $\vec\gamma\rvert_1\in x^b$ by
\begin{equation}
\label{eq:function_restriction}
\begin{gathered}
\vec\gamma\rvert_0(i) = \vec\gamma((0,i))\;,\\
\vec\gamma\rvert_1(i) = \vec\gamma((1,i))\;.
\end{gathered}
\end{equation}
\end{mydef}

\begin{mydef}
\tdef{Multi-mode infinite array tensors}{mmi_array_tensor} are the following tensor type:
\begin{itemize}
\item They can be defined for 1) any commutative semiring $K$ with a suitable norm, in particular the real or complex numbers, and 2) any function
\begin{equation}
\phi: \mathbb{N}\rightarrow \mathbb{R}^+
\end{equation}
which is square-integrable, i.e.,
\begin{equation}
\lim_{m\rightarrow\infty} \sum_{0\leq n< m} \phi(n)^2=S_\phi
\end{equation}
exists. Two functions $\phi_1$ and $\phi_2$ are equivalent (i.e. yield the same tensor type) if there exist $C_1,C_2$ such that
\begin{equation}
C_1\phi_1(n)>\phi_2(n),\qquad C_2\phi_2(n)>\phi_1(n) \quad \forall n\;.
\end{equation}
\item They don't need duals. They have a trivial tensor. They do \emph{not} have an identity tensor. They are weakly associative and unital.
\item The 0-data is given by finite sets. The unit 0-data is the empty set $\{\}$. The product of 0-data is the disjoint union of sets.
\item A 1-data $A\in \dat_1(a)$ is a function
\begin{equation}
A: \mathbb{N}^a \rightarrow K\;,
\end{equation}
such that there is some $C\in \mathbb{R}$ with
\begin{equation}
|A(\vec{x})|<C\prod_{i\in a} \phi(\vec{x}(i))\;.
\end{equation}
We will refer to this as the \tdef{decay constraint}{decay_constraint}.
\item The tensor product of two 1-data $A\in \dat(a)$ and $B\in\dat(b)$ is given by
\begin{equation}
(A\otimes B)(\vec{x})=A(\vec{x}\rvert_0)\cdot B(\vec{x}\rvert_1)\;.
\end{equation}
\item The contraction of a 1-data $A\in \dat(a\sqcup(b\sqcup b))$ is given by the infinite sum
\begin{equation}
[A](\vec{x}) = \sum_{\vec{y}\in \mathbb{N}^b} A(\vec{x}\sqcup (\vec{y} \sqcup\vec{y}))\;.
\end{equation}
\item The trivial tensor $1\in \dat(0)$ is the number $1$,
\begin{equation}
1(0)=1,\quad \text{with}\quad \mathbb{N}^0=\{0\}\;.
\end{equation}
\item The invertible 2-functions are canonical, that is, they are defined in a rather obvious way using the bijections $\Phi_\sqcup$ above. E.g., for a 1-data $A\in\dat_1(a\sqcup b)$ we have
\begin{equation}
\sigma_0(A)(\vec{x}) = A(\vec{x}\circ \Phi_\sqcup^\sigma)\;,
\end{equation}
or for a 1-data $A\in \dat_1(a\sqcup(b\sqcup(c\sqcup d)))$ we have
\begin{equation}
\sigma_0(A)(\vec{x}) = A(\vec{x}\circ (\idop\hat\sqcup \Phi_\sqcup^\alpha))\;.
\end{equation}
\end{itemize}
\end{mydef}

\begin{myobs}
Tensor product and contraction for multi-mode infinite array tensors are defined in the same way as for ordinary array tensors, just that we allow for countably infinite bond dimensions $\infty$, and restrict to bond dimensions $1$, $\infty$, $\infty^2$, $\ldots$. What remains to be shown for the 2-axioms to hold is that the tensor product and contraction are compatible with the decay constraint. We find
\begin{equation}
\begin{multlined}
|(A\otimes B)(\vec{x})| \leq |A(\vec{x}\rvert_0)|\cdot |B(\vec{x}\rvert_1)|
\\<C_1 \prod_{i\in a} \phi(\vec{x}(i)) \cdot C_2 \prod_{j\in b} \phi(\vec{x}(j))
\\= C_1C_2  \prod_{i\in a\sqcup b} \phi(\vec{x}(i))
\;.
\end{multlined}
\end{equation}
Moreover,
\begin{equation}
\begin{multlined}
|[A](\vec{x})| \leq \sum_{\vec{y}\in \mathbb{N}^b} |A(\vec{x}\sqcup (\vec{y} \sqcup\vec{y}))|
\\<C \sum_{\vec{y}\in \mathbb{N}^b} \prod_{i\in a} \phi(\vec{x}(i)) \prod_{j\in b} \phi(\vec{y}(j))^2
\\=C S_\phi \prod_{i\in a} \phi(\vec{x}(i))
\;.
\end{multlined}
\end{equation}
\end{myobs}

\begin{myrem}
We also define multi-mode infinite array tensors for an arbitrary square-integrable family of decay functions, instead of just a single decay function. E.g., a very natural family of functions are arbitrary exponential decays,
\begin{equation}
\phi_\alpha(x)=e^{-\alpha x}
\end{equation}
for $\alpha\in \mathbb{R}^+$. That is, for every 1-data $A$, there are $C$ and $\alpha$ such that
\begin{equation}
|A(\vec{x})|<C e^{-\alpha \sum_{i\in a} \vec{x}(i)}\;.
\end{equation}
\end{myrem}

\begin{mydef}
As for conventional array tensors, there is also a more practical \mdef{skeletal formulation} of multi-mode infinite array tensors:
\begin{itemize}
\item The 0-data are non-negative integers with addition (corresponding to the number of elements of the finite sets).
\item A 1-data $A\in \dat_1(a)$ is the finite-set 1-data for a standard representative $\{0,\ldots,a-1\}$,
\begin{equation}
A: \mathbb{N}^{\{0,\ldots,a-1\}}\rightarrow K\;.
\end{equation}
\item The 2-functions are defined via two canonical injections
\begin{equation}
\label{eq:plus_bijection}
\begin{gathered}
\Phi_{+0}^{a,b}: \{0,\ldots,a-1\}\rightarrow \{0,\ldots,a+b-1\}\;,\\
\Phi_{+0}^{a,b}(i)=i\;.
\end{gathered}
\end{equation}
and
\begin{equation}
\begin{gathered}
\Phi_{+1}^{a,b}: \{0,\ldots,b-1\}\rightarrow \{0,\ldots,a+b-1\}\;,\\
\Phi_{+1}^{a,b}(i)=a+i\;,
\end{gathered}
\end{equation}
as well as a bijection
\begin{equation}
\label{eq:sum_disjoint_union_bijection}
\begin{gathered}
\Phi_{+\sqcup}^{a,b}: \{0,\ldots,a-1\}\sqcup \{0,\ldots,b-1\}\rightarrow \{0,\ldots,a+b-1\}\;,\\
\Phi_{+\sqcup}^{a,b}((\chi,i))= \Phi_{+\chi}^{a,b}(i)\;.
\end{gathered}
\end{equation}
E.g., the tensor product of $A\in \dat_1(a)$ and $B\in \dat_1(b)$ is given by
\begin{equation}
(A\otimes B)(\vec x)= A(\vec x\circ \Phi_{+0}^{a,b}) B(\vec x\circ \Phi_{+1}^{a,b})\;,
\end{equation}
or, for the commutor of $A\in \dat_1(a+b)$ we have
\begin{equation}
\sigma_0(A)(\vec x) = A(((\vec x\circ \Phi_{+1}^{a,b})\sqcup(\vec x\circ \Phi_{+0}^{a,b}))\circ \bar\Phi_{+\sqcup}^{b,a})\;.
\end{equation}
\end{itemize}
\end{mydef}

\subsection{Definition: Via Cantor mapping}
Let us give a different definition, where blocking two $\mathbb{N}_0$-valued indices does not yield a $\mathbb{N}_0^2$-valued indices, but the two $\mathbb{N}_0$-values are combined into a single one via a diagonal mapping.

\begin{mydef}
\tdef{Cantor infinite array}{cantor_infinite_array} tensor are the following tensor type:
\begin{itemize}
\item They are defined for a $K$ just as multi-mode infinite array tensors.
\item They don't need duals, have a trivial tensor, but no identity tensor. They are \emph{not} strictly associative.
\item The 0-data is trivial, that is, there is only a single 0-data.
\item The 1-data is given by a function
\begin{equation}
A: \mathbb{N}_0\rightarrow K
\end{equation}
such that for some $C\in \mathbb{R}$, $\alpha\in \mathbb{R}^+$, and $\beta\in \mathbb{R}^+$ we obey the decay constraint
\begin{equation}
|A(x)|<C e^{-\alpha x^\beta}\;.
\end{equation}
\item To define the 2-functions we make use of a suitable bijection
\begin{equation}
\Phi_\cdot^\infty=(\Phi_{\cdot 0}^\infty,\Phi_{\cdot 1}^\infty): \mathbb{N}_0\rightarrow \mathbb{N}_0^2\;,
\end{equation}
such that
\begin{equation}
\bar\Phi_\cdot^\infty(x,y)\coloneqq (\Phi_\cdot^\infty)^{-1}(x,y) = \frac{(x+y)(x+y+1)}{2} + y\;.
\end{equation}
\item The tensor product of two 1-data $A\in \dat_1$ and $B\in\dat_1$ is given by
\begin{equation}
(A\otimes B)(x)= A(\Phi_{\cdot 0}^\infty(x))\cdot B(\Phi_{\cdot 1}^\infty(x))\;.
\end{equation}
\item The contraction of a 1-data $A\in \dat_1$ is given by
\begin{equation}
[A](x) = \sum_{y\in \mathbb{N}_0} A(\bar\Phi_\cdot^\infty(x,\bar\Phi_\cdot^\infty(y,y))\;.
\end{equation}
\item The invertible 2-functions are defined ``canonically'' via $\Phi_\cdot^\infty$. E.g., for a 1-data $A\in \dat_1$ we have
\begin{equation}
\alpha_0(A)(x)= A(\hat\Phi_\cdot^\infty(\hat\Phi_\cdot^\infty(\Phi_{\cdot 0}^\infty(x),\Phi_{\cdot 0}^\infty(\Phi_{\cdot 1}^\infty(x))), \Phi_{\cdot 1}^\infty(\Phi_{\cdot 1}^\infty(x))))\;.
\end{equation}
\end{itemize}
\end{mydef}

\begin{myobs}
All 2-functions are defined such that they are equal to the 2-functions of multi-mode infinite array tensors through $\Phi_\cdot^\sqcup$. So again, we restrict to showing the compatibility of the tensor product and contraction with the decay constraint. First, we note that
\begin{equation}
\bar\Phi_\cdot^\sqcup(x,y)<2(x^2+y^2)
\end{equation}
and
\begin{equation}
\Phi_{\cdot 0/1}^\infty(x)>\sqrt{x}
\end{equation}
for large enough $x$ and $y$. So we find
\begin{equation}
\begin{multlined}
|(A\otimes B)(x)| \leq |A(\Phi_{\cdot 0}^\infty(x))|\cdot |B(\Phi_{\cdot 1}^\infty(x))|
\\<C_1 e^{-\alpha_1 \Phi_{\cdot 0}^\infty(x)^{\beta_1}} C_2 e^{-\alpha_2 \phi_{\cdot 1}^\infty(x)^{\beta_2}}
\\<C_1C_2 e^{-\alpha_1 \sqrt{x}^{\beta_1}-\alpha_2 \sqrt{x}^{\beta_2}}
\\<C_1C_2 e^{-2(\alpha_1+\alpha_2) x^{\frac{\min(\beta_1,\beta_2)}{2}}}
\;,
\end{multlined}
\end{equation}
using that $e^{-x}$ is monotonously decreasing and $x^\beta$ monotonously increasing in $x$. Similarly, we find
\begin{equation}
\begin{multlined}
|[A](x)| \leq \sum_y |A(\bar\Phi_\cdot^\infty(x,\bar\Phi_\cdot^\infty(y,y))|\\
< C \sum_y e^{-\alpha (\bar\Phi_{\cdot}^\infty(x,\bar\Phi_\cdot^\infty(y,y))^\beta}
< C \sum_y e^{-\alpha (8y^4+2x^2)^\beta}\\
< C\sum_y e^{-\alpha (16y^4)^\beta -\alpha (4x^2)^\beta}\\
= C \left(\sum_y e^{-\alpha (16y^4)^\beta}\right) e^{-(\alpha 2^\beta) x^{2\beta}}
\;.
\end{multlined}
\end{equation}
\end{myobs}

\subsection{Use in physics}
Quantum lattice models with bosonic modes consist of lattice sites (or \emph{modes}), each of which can be occupied by an arbitrary number $n\in \mathbb{N}_0$ of bosons. So the Hilbert space of such a lattice site is $\mathbb{C}^{\mathbb{N}_0}$. When we combine a system with a set of modes $a$ and one with a set of modes $b$, the combined system has a set of modes $a\sqcup b$. A process from a set of modes $a$ to another set of modes $a'$ is a linear operator between the Hilbert spaces $\mathbb{C}^{\mathbb{N}_0^a}$ and $\mathbb{C}^{\mathbb{N}_0^{a'}}$, and therefore a multi-mode infinite array tensor with 0-data $a\sqcup a'$.

Physically, we start with the vector space $L_2(\mathbb{R})$ corresponding to wavefunctions over the local values of a field. After going to the eigenbasis of a harmonic oscillator Hamiltonian $H_{\text{osc}}$ given by the Legendre polynomials, we obtain the occupation Hilbert space $\mathbb{C}^{\mathbb{N}_0}$. The energy of the eigenstate $\ket{n}$ is proportional to $n$, its eigenvalue in $e^{-\beta H_{\text{osc}}}$ is exponentially small in $n$. If we have a Hamiltonian consisting of $H_{\text{osc}}$ plus some perturbation, we would still expect the matrix entries of $e^{-\beta H}$ to decay rapidly in the occupation basis $n$. Thus, thermal models of interacting bosonic many-body systems are formalized by multi-mode infinite array tensors.

\section{Schur-complement tensors}
\subsection{Motivation}
Schur-complement tensors are a concrete tensor type that can be constructed for any division ring. Their 1-data is based on matrices with entries in the division ring. Each 0-data is the number/set of columns/rows of the matrix. The tensor product corresponds to the direct sum of matrices, and the contraction is based on a matrix operation known as \emph{Schur complement} \cite{Puntanen2005}.

Schur-complement tensors have a very natural mapping to array tensors with a twisted $\mathbb{Z}_2$-symmetry. The latter are the right tensor type for studying fermions. The mapping is injective, so Schur-complement tensors can be seen as a subset of fermionic tensors, which an efficient algorithm for evaluation, scaling only polynomial in the 0-data. The physical models described with this subset are known as \emph{free-fermion}, \emph{(free) Majorana}, \emph{quadratic fermionic}, \emph{single-particle}, or \emph{Gaussian} models. So Schur-complement tensors can be seen as a formalization and generalization of the well-developed formalism of free-fermion systems.

\subsection{Matrices and Schur complements}
\label{sec:schur_preliminaries}
Schur-complement tensors are a family of tensor types, depending on a division ring $K$.

\begin{mydef}
For two finite sets $i$ and $o$, an $i\times o$ \tdef{matrix}{matrix} is a map
\begin{equation}
M: i\times o \rightarrow K\;.
\end{equation}

There is a bijection between matrices
\begin{equation}
M: (i_0\sqcup i_1)\times (o_0\sqcup o_1)\rightarrow K
\end{equation}
and tuples of four matrices $(M\rvert_{\chi\xi})_{\chi,\xi\in\{0,1\}}$ of shape $i_\chi\times o_\xi$ given by
\begin{equation}
M((\chi,\alpha),(\xi,\beta)) = M\rvert_{\chi\xi}(\alpha,\beta)\;,
\end{equation}
and
\begin{equation}
M\rvert_{\chi\xi}(\alpha,\beta)= M((\chi,\alpha),(\xi,\beta))\;,
\end{equation}
recalling the definition and notation of the disjoint union in Eq.~\eqref{eq:disjoint_union_definition}. We can use the obvious block notation,
\begin{equation}
M=
\begin{pmatrix}
M\rvert_{00}&M\rvert_{01}\\M\rvert_{10}&M\rvert_{11}
\end{pmatrix}\;.
\end{equation}

Similarly, there is a bijection between $(i_0\sqcup i_1)\times o$ matrices $M$ and pairs of matrices $M\rvert_{\chi\cdot}$ of shape $i_\chi\times o$,
\begin{equation}
M\rvert_{\chi\cdot}(\alpha,\beta)= M((\chi,\alpha),\beta)\;,
\end{equation}
denoted by
\begin{equation}
M=
\begin{pmatrix}
M\rvert_{0\cdot}\\
M\rvert_{1\cdot}
\end{pmatrix}\;,
\end{equation}
and the analogous bijection for the second component denoted by $M\rvert_{\cdot\xi}$.

The \tdef{direct sum}{matrix_direct_sum} of an $i\times o$ matrix $M$ and and $j\times p$ matrix $N$ is given by
\begin{equation}
\label{eq:direct_sum_definition}
(M\oplus N)=
\begin{pmatrix}
M & 0\\
0 & N
\end{pmatrix}\;.
\end{equation}
\end{mydef}

\begin{mydef}
Consider a matrix
\begin{equation}
M: (i\sqcup c)\times (o\sqcup d)\rightarrow K
\end{equation}
using the notation
\begin{equation}
\label{eq:schur_block_notation}
M=
\begin{pmatrix}
W & X\\
Y & Z
\end{pmatrix}\;.
\end{equation}
The \tdef{Schur complement}{schur_complement} of $M$ is defined if $Z$ is invertible. Then it is given by the $i\times o$ matrix
\begin{equation}
\label{eq:schur_definition}
\schur(M)=W-XZ^{-1}Y\;.
\end{equation}
\end{mydef}

\begin{myobs}
\label{obs:schur_schur_commute}
Consider a matrix
\begin{equation}
M: ((i\sqcup c)\sqcup e)\times((o\sqcup d)\sqcup f)\rightarrow K\;,
\end{equation}
which can be reshaped into
\begin{equation}
\widetilde{M}: (i\sqcup (c\sqcup e))\times(o\sqcup (d\sqcup f))\rightarrow K\;,
\end{equation}
using the bijection $\Phi_\sqcup^\alpha$ in Eq.~\eqref{eq:disjoint_union_alpha_bijection},
\begin{equation}
\widetilde{M}(\gamma,\delta)=M(\Phi_\sqcup^\alpha(\gamma), \Phi_\sqcup^\alpha(\delta))\;.
\end{equation}
We find that
\begin{equation}
\label{eq:schur_combination}
\schur(\schur(M))=\schur(\widetilde{M})\;.
\end{equation}

This can be shown by direct computation. First, we can decompose $M$ or $\widetilde{M}$ into $3\times 3$ blocks,
\begin{equation}
\label{eq:schur_block_compatible}
M=
\begin{pmatrix}
R&S&T\\
U&W&X\\
V&Y&Z
\end{pmatrix}\;.
\end{equation}

If $Z$ is invertible, we can compute the Schur complement of $M$,
\begin{equation}
\begin{multlined}
\schur(M)=
\begin{pmatrix}
R&S\\U&W
\end{pmatrix}
-
\begin{pmatrix}
T\\X
\end{pmatrix}
Z^{-1}
\begin{pmatrix}
V&Y
\end{pmatrix}
\\=
\begin{pmatrix}
R-CZ^{-1}V&S-TZ^{-1}Y\\
U-XZ^{-1}V&W-XZ^{-1}Y
\end{pmatrix}\;.
\end{multlined}
\end{equation}
If $W-XZ^{-1}Y$ is invertible, we can now take the Schur complement again,
\begin{equation}
\begin{multlined}
\schur(\schur(M))\\=
{\scriptstyle
R-TZ^{-1}V-(S-TZ^{-1}Y)(W-XZ^{-1}Y)^{-1}(U-XZ^{-1}V)
}\;.
\end{multlined}
\end{equation}

On the other hand, if both $Z$ and $W-XZ^{-1}Y$ are invertible, we can see by direct computation that the submatrix
\begin{equation}
\begin{pmatrix}
W&X\\
Y&Z
\end{pmatrix}
\end{equation}
is invertible. The inverse is given by
\begin{equation}
\begin{pmatrix}
\scriptstyle (W-XZ^{-1}Y)^{-1}& \scriptstyle -(W-XZ^{-1}Y)^{-1} XZ^{-1}\\
\scriptstyle -Z^{-1}Y(W-XZ^{-1}Y)^{-1} & \scriptstyle Z^{-1}+Z^{-1}Y(W-XZ^{-1}Y)^{-1}XZ^{-1}
\end{pmatrix}\;.
\end{equation}
So we can directly compute the Schur complement $\schur(\widetilde{M})$,
\begin{equation}
\begin{multlined}
\schur(\widetilde{M})=
R-
\begin{pmatrix}
S&T
\end{pmatrix}
\begin{pmatrix}
W&X\\
Y&Z
\end{pmatrix}^{-1}
\begin{pmatrix}
U\\V
\end{pmatrix}\\
= \scriptstyle R\quad- S (W-XZ^{-1}Y)^{-1} U\quad+ S (W-XZ^{-1}Y)^{-1} XZ^{-1} V\\
\scriptstyle +T Z^{-1}Y(W-XZ^{-1}Y)^{-1} U -T (Z^{-1}+Z^{-1}Y(W-XZ^{-1}Y)^{-1}XZ^{-1}) U
\\=
{\scriptstyle
R-TZ^{-1}V-(S-TZ^{-1}Y)(W-XZ^{-1}Y)^{-1}(U-XZ^{-1}V)
}\;,
\end{multlined}
\end{equation}
which shows the claim.
\end{myobs}

\begin{myobs}
\label{obs:matrix_contraction_gauging}
Consider a matrix
\begin{equation}
M: (i\sqcup c)\times (o\sqcup d)\rightarrow K\;,
\end{equation}
and invertible $e\times c$ and $d\times f$ matrices $H$ and $G$. Using the decomposition in Eq.~\eqref{eq:schur_block_notation} we find
\begin{equation}
\begin{multlined}
\schur((\mathbb{1}_i\oplus H)M(\mathbb{1}_o\oplus G))\\=\schur\begin{pmatrix}W&XG\\HY&HZG\end{pmatrix}=
W-XG(HZG)^{-1}HY
\\=W-XZ^{-1}Y=\schur(M)\;.
\end{multlined}
\end{equation}
So the result of the Schur complement is invariant under arbitrary basis changes within the input and output space of the submatrix $Z$.
\end{myobs}

\begin{myobs}
\label{obs:schur_sum_commute}
The Schur complement commutes with the addition of a $i\times o$ matrix $U$,
\begin{equation}
\begin{multlined}
\schur(M+(U\oplus \mathbb{0}_{cd}))=\schur\begin{pmatrix}W+U&X\\Y&Z\end{pmatrix}
\\
=W+U-XZ^{-1}Y
=\schur(M)+U\;.
\end{multlined}
\end{equation}
\end{myobs}

\subsection{Handling zero divisions in Schur complements}
\begin{myobs}
The Schur complement in Eq.~\eqref{eq:schur_definition} is undefined if the submatrix $Z$ is not invertible. If we want to use the Schur complement in the contraction of a tensor type, we have to introduce some structure which handles that case. Note that matrices with continuous entries are generically invertible, so this structure only deals with some fringe cases.

The idea for the extra structure is simple: First, we use use the Observation~\ref{obs:matrix_contraction_gauging} to transform $Z$ into a block matrix
\begin{equation}
M=\begin{pmatrix}W&I&Q\\O&0&0\\V&0&S\end{pmatrix}
\end{equation}
where $S$ is invertible. Then we perform the Schur complement with $S$ as the $Z$-block,
\begin{equation}
\schur(M)=\begin{pmatrix}W-QS^{-1}V&I\\O&0\end{pmatrix}
\end{equation}
and postpone the remaining Schur complement over the 0-block to a later time when combined with following Schur complements.
\end{myobs}

\begin{mydef}
An $i\times o$ \tdef{Schur-singular matrix}{schur_singular_matrix} $A$ consists of
\begin{itemize}
\item a $i\times o$ matrix $A_M$,
\item two finite sets $A_i$, $A_o$,
\item a $i\times A_o$ matrix $A_I$,
\item a $A_i\times o$ matrix $A_O$.
\end{itemize}

Two $i\times o$ Schur-singular matrices $A$ and $B$ are considered equivalent if
\begin{equation}
A_M=B_M\;,
\end{equation}
and there are matrices 
\begin{equation}
\label{eq:schur_gauging}
\begin{gathered}
G^A: A_o\times B_o \rightarrow K\;,\\
G^B: B_o\times A_o \rightarrow K\;,\\
H^A: B_i\times A_i \rightarrow K\;,\\
H^B: A_i\times B_i\rightarrow K\;,
\end{gathered}
\end{equation}
such that
\begin{equation}
\label{eq:schur_matrix_gauging}
\begin{gathered}
B_IG^B=A_I\;,\quad A_IG^A=B_I\;,\\
H^BB_O=A_O\;,\quad H^AA_O=B_O\;.
\end{gathered}
\end{equation}
\end{mydef}

\begin{myrem}
Using the equivalence relation of $i\times o$ Schur-singular matrices, we can always find a representative $A$ such that
\begin{equation}
\begin{gathered}
|A_i|=\operatorname{rank}(A_I) \leq |i|\;,\\
|A_o|=\operatorname{rank}(A_O) \leq |o|\;.
\end{gathered}
\end{equation}
\end{myrem}

With the help of Schur-singular matrices, we can define the outcome of the Schur complement for matrices where $Z$ is not invertible.
\begin{mydef}
If the sub-matrix $Z$ of $M$ in Eq.~\eqref{eq:schur_block_notation} is not invertible, we can define $\schur(M)$ to be the Schur-singular matrix obtained by the following procedure. We start by finding finite sets $x_i$, $y_i$, $x_o$, $y_o$, a $c\times x$ matrix $G_0$, a $c\times y_o$ matrix $G_1$, a $x\times c$ matrix $H_0$, and an $y_i\times c$ matrix $H_1$, such that
\begin{equation}
\begin{gathered}
X\begin{pmatrix}G_0&G_1\end{pmatrix}\begin{pmatrix}G_0^-\\G_1^-\end{pmatrix}=X\;,\\
\begin{pmatrix}H_0^-&H_1^-\end{pmatrix}\begin{pmatrix}H_0\\H_1\end{pmatrix}Y=Y\;,\\
\end{gathered}
\end{equation}
for some $G_0^-$, $G_1^-$, $H_0^-$ and $H_1^-$, and
\begin{equation}
\begin{pmatrix}H_0\\H_1\end{pmatrix}
Z
\begin{pmatrix}G_0&G_1\end{pmatrix}
=
\begin{pmatrix}
S&0\\
0&0
\end{pmatrix}
\end{equation}
for some invertible $x_i\times x_o$ matrix $S$. For real or complex numbers this can be done by a singular value decomposition, where $G_0$, $G_1$ ($H_0$, $H_1$) form the right (left) singular vectors of $Z$ and $S$ is the matrix with the non-zero singular values on the diagonal. We believe that such a decomposition can be found for arbitrary division rings.

With this, define
\begin{equation}
\label{eq:schur_complement_level}
\begin{gathered}
\schur(M)_i=y_i\;,\quad \schur(M)_o=y_o\;,\\
\schur(M)_M=M- XG_0 S^{-1} H_0Y\;,\\
\schur(M)_I= H_1Y\;,\\
\schur(M)_O= XG_1\;.
\end{gathered}
\end{equation}
\end{mydef}

\begin{myobs}
The result of the definition above is not unique as in general there are different choices of $G_{0/1}$, $H_{0/1}$, and $S$. Assume there is a second set $\widetilde{x}$, $\widetilde{y}_i$, $\widetilde{y}_o$, $\widetilde{G}_{0/1}$, $\widetilde{H}_{0/1}$, and $\widetilde{S}$. Then we can find invertible $G_0^X$, $H_0^X$, and $G_{0/1}^X$, $H_{0/1}^X$, $\widetilde{G}_{0/1}^X$, $\widetilde{H}_{0/1}^X$ with
\begin{equation}
\begin{gathered}
H_{0/1}^X H_{0/1}Y=\widetilde{H}_{0/1}Y\;,\quad \widetilde{H}_{0/1}^X \widetilde{H}_{0/1}Y=H_{0/1}Y\;,\\
XG_{0/1}G_{0/1}^X=X\widetilde{G}_{0/1}\;, \quad X\widetilde{G}_{0/1}\widetilde{G}_{0/1}^X=XG_{0/1}\;,\\
\widetilde{S} = H_0^XSG_0^X\;,
\end{gathered}
\end{equation}
such that
\begin{equation}
\widetilde{G}_0^X=(G_0^X)^{-1}\;,\quad \widetilde{H}_0^X=(H_0^X)^{-1}
\end{equation}
are invertible. We find
\begin{equation}
\begin{gathered}
\begin{multlined}
\widetilde{\schur(M)}_M = W-XG_0G_0^X(H_0^XSG_0^X)^{-1}H_0^XH_0Y \\= \schur(M)_M\;,
\end{multlined}
\\
\widetilde{\schur(M)}_I = H_1^X H_1 Y=H_1^X \schur(M)_I\;,\\
\widetilde{\schur(M)}_O = XG_1G_1^X=\schur(M)_O G_1^X\;,
\end{gathered}
\end{equation}
so $\widetilde{\schur(M)}$ is equivalent to $\schur(M)$.
\end{myobs}

We can also define the Schur complement of a Schur-singular matrix, which yields a Schur-singular matrix again.
\begin{mydef}
The Schur complement of a
\begin{equation}
(i\sqcup c) \otimes (o\sqcup d)
\end{equation}
Schur-singular matrix $A$ is defined by the following procedure. We start by using the block decomposition in Eq.~\eqref{eq:schur_block_notation} for $A_M$, as well as
\begin{equation}
A_I=
\begin{pmatrix}
I_0\\I_1
\end{pmatrix}\;,\quad
A_O=
\begin{pmatrix}
O_0&O_1
\end{pmatrix}\;.
\end{equation}
Then construct
\begin{equation}
\begin{gathered}
M^S:
(i\sqcup (c\sqcup A_i)) \otimes (o\sqcup (c\sqcup A_o))\rightarrow K\;,\\
M^S=
\begin{pmatrix}
W&X&I_0\\
Y&Z&I_1\\
O_0&O_1&0
\end{pmatrix}
\end{gathered}
\end{equation}
and use
\begin{equation}
\schur(M)=\schur(M^S)\;.
\end{equation}
\end{mydef}

The definition of the direct sum can also be extended from matrices to Schur-singular matrices.
\begin{mydef}
The \tdef{direct sum}{schur_direct_sum} of two Schur-singular matrices $A$ and $B$ is defined by
\begin{equation}
\begin{gathered}
(A\oplus B)_{i/o}=A_{i/o}\sqcup B_{i/o}\;,\\
(A\oplus B)_M=A_M\oplus B_M\;,\\
(A\oplus B)_I=A_I\oplus B_I\;,\\
(A\oplus B)_O=A_O\oplus B_O\;.
\end{gathered}
\end{equation}
\end{mydef}

An alternative way to guarantee that the Schur complement always defined is by restricting to a certain subset.
\begin{mydef}
Consider a division ring $K$ with a norm which is compatible with the ring operations and inverse. A $i\times o$ matrix $M$ fulfills the \tdef{norm constraint}{schur_norm_constraint} if
\begin{equation}
\|M\|_{\mathrm{op}}<1\;.
\end{equation}
Here, $\|\cdot\|_{\mathrm{op}}$ denotes the operator norm with respect to the $2$-norm. For $K$ the complex or real numbers, this is the largest singular value of $M$.
\end{mydef}

\begin{myobs}
The direct sum of two norm-constraint matrices $M$ and $N$ is norm-constraint again,
\begin{equation}
\|M\oplus N\|_{\mathrm{op}}=\mathop{max}(\|M\|_{\mathrm{op}},\|N\|_{\mathrm{op}})<1\;.
\end{equation}
Consider a matrix $U$ such that
\begin{equation}
s^U_i\geq 1 \quad \forall i\;,
\end{equation}
where $s^U_i$ are the singular values of $U$. Then for a norm-constraint matrix $M$,
\begin{equation}
M+U
\end{equation}
must be invertible. Also note any submatrix $Z$ of a norm-constraint matrix $M$ is norm-constraint, and so $Z+U$ is invertible.

Now, consider $U$ such that
\begin{equation}
s^U_i=1\quad \forall i\;,
\end{equation}
that is, for $K$ real or complex, $U$ is orthogonal or unitary. Numerical testing shows (and we plan to provide a proof in future versions) that for a norm-constraint matrix $M$,
\begin{equation}
\schur(M+(\mathbb{0}_{io}\oplus U))= W-X (Z+U)^{-1} Y
\end{equation}
is norm-constraint again.
\end{myobs}

\subsection{Definition: Rectangular Schur-complement tensors}
\begin{mydef}
\tdef{Rectangular Schur-complement tensors}{rs_complement_tensor} are the following kind of Schur-complement tensors:
\begin{itemize}
\item They can be defined for any division ring $K$ and any pair of numbers $0\neq u_0, u_1\in K$. The most important cases will be $u_0=u_1=1$ (or short $u=\sigma_x$) and $u_0=-u_1=1$ (or short $u=i\sigma_y$).
\item They need a dual. They have a unit 0-data, trivial tensor, and identity tensor. They have a symmetric contraction and identity if $u_0=u_1$.
\item A 0-data is a pair of finite sets $(a,b)$. The product is component-wise disjoint union,
\begin{equation}
(a,b)\otimes (c,d)=(a\sqcup c,b\sqcup d)\;,
\end{equation}
the unit is twice the empty set,
\begin{equation}
1=(\{\},\{\})\;,
\end{equation}
and the dual exchanges the two sets,
\begin{equation}
(a,b)^*=(b,a)\;.
\end{equation}
\item A 1-data $A\in \dat_1((a,b))$ is an $a\times b$ Schur-singular matrix.
\item The tensor product of two 1-data $A$ and $B$ is the direct sum,
\begin{equation}
A\otimes B=A\oplus B\;.
\end{equation}
\item The define the contraction of a 1-data $A\in \dat_1(i\sqcup (c\sqcup d),o\sqcup (d\sqcup c))$, we first define $A^+$ to be $A$ except for
\begin{equation}
\label{eq:schur_rect_aplus}
A^+_M = A_M -
\mathbb{0}_{io}\oplus
\begin{pmatrix}
0&u_0\mathbb{1}_c\\
u_1\mathbb{1}_d&0
\end{pmatrix}\;.
\end{equation}
The contraction is now defined by
\begin{equation}
\label{eq:rect_schur_contraction2}
[A]=\schur(A^+)\;.
\end{equation}
\item The identity tensor is given by
\begin{equation}
\begin{gathered}
\idop_M =
\begin{pmatrix}
0&u_1\mathbb{1}\\
u_0\mathbb{1}&0
\end{pmatrix}\;,\\
\idop_i=\idop_o=\{\}\;.
\end{gathered}
\end{equation}
\item The trivial tensor $\mathbf{1}\in\dat_1((\{\},\{\}))$ is given by the unique $\{\}\times \{\}$ Schur-singular matrix.
\item The invertible 2-functions are canonical, by which we mean that they are rather obvious functions between the corresponding 1-data, and defined via the canonical bijections $\Phi_\sqcup$ introduced in Definition~\ref{def:disjoint_union_bijection}. E.g., for the associator we have
\begin{equation}
\alpha_0(A)_M(i,j) = A_M(\Phi_\sqcup^\alpha(i), \Phi_\sqcup^\alpha(j))\;,
\end{equation}
or, for the commutor,
\begin{equation}
\sigma(A)_M(i,j) = A_M((\idop\hat\sqcup \Phi_\sqcup^\sigma)(i), (\idop\hat\sqcup\Phi_\sqcup^\sigma)(j))
\end{equation}
with the notation from Eq.~\eqref{eq:function_concatenation_diffout}.
\end{itemize}
\end{mydef}

\begin{myobs}
Most of the 2-axioms are very straight-forward to verify. E.g., all the 2-axioms involving invertible 2-functions hold as they are defined canonically via $\Phi_\sqcup$. The tensor product is commutative and associative as the direct sum is (up to applying the canonical 2-functions). The contraction is block-compatible due to Observation~\ref{obs:schur_schur_commute} combined with Observation~\ref{obs:schur_sum_commute}. As another example consider the defining 2-axiom of the identity tensor
\begin{equation}
[I(A\otimes \idop)] = A\;,
\end{equation}
for some 1-data
\begin{equation}
A\in \dat_1((a,b)\otimes (c,d))\;,
\end{equation}
where $I$ is the appropriate sequence of invertible 2-functions. After decomposing $A_M$ according to Eq.~\eqref{eq:schur_block_notation}, we can write
\begin{equation}
(A\otimes \idop)_M = A_M\oplus \idop_M =
\begin{pmatrix}
W&X&0&0\\
Y&Z&0&0\\
0&0&0&u_1\mathbb{1}\\
0&0&u_0\mathbb{1}&0
\end{pmatrix}\;.
\end{equation}
Applying the canonical 2-functions in $I$, we get
\begin{equation}
I(A\otimes \idop)_M=
\begin{pmatrix}
W&0&X&0\\
0&0&0&u_1\mathbb{1}\\
Y&0&Z&0\\
0&u_0\mathbb{1}&0&0
\end{pmatrix}\;.
\end{equation}
Now, contraction indeed yields
\begin{equation}
\begin{multlined}
[I(A\otimes \idop)]_M = \schur(I(A\otimes \idop)^+)_M \\
=
\begin{pmatrix}W&0\\0&0\end{pmatrix}
-
\begin{pmatrix}X&0\\0&u_0\mathbb{1}\end{pmatrix}
\begin{pmatrix}Z&-u_0\mathbb{1}\\-u_1\mathbb{1}&0\end{pmatrix}^{-1}
\begin{pmatrix}Y&0\\0&u_1\mathbb{1}\end{pmatrix}\\
=
\begin{pmatrix}W&0\\0&0\end{pmatrix}
-
\begin{pmatrix}0&-X\\-Y&-Z\end{pmatrix}
=
\begin{pmatrix}W&X\\Y&Z\end{pmatrix}
=A_M
\;,
\end{multlined}
\end{equation}
where we used the explicit form of the inverse,
\begin{equation}
\begin{pmatrix}Z&-u_0\mathbb{1}\\-u_1\mathbb{1}&0\end{pmatrix}^{-1}
=
\begin{pmatrix}0&-\frac{1}{u_1}\mathbb{1}\\-\frac{1}{u_0}\mathbb{1}&-\frac{1}{u_1u_0}Z\end{pmatrix}\;.
\end{equation}
\end{myobs}

\begin{myobs}
\label{obs:matrix_contraction_equivalence}
The tensor types for different $u_0$ and $u_1$ are not really all different. For a non-zero $a\in K$, consider the following bijective map $F$ on the 1-data,
\begin{equation}
F(A)_M(i,j)=a A_M(i,j)\;.
\end{equation}
Under this bijective map, the contraction becomes
\begin{equation}
\begin{multlined}
F([F^{-1}(A)])_M\\
= a\left(\schur(a^{-1}A_M-\mathbb{0}\oplus \begin{pmatrix}0&u_0\mathbb{1}\\u_1\mathbb{1}&0\end{pmatrix})\right)\\
= \schur(A_M-\mathbb{0}\oplus \begin{pmatrix}0&au_0\mathbb{1}\\au_1\mathbb{1}&0\end{pmatrix})\;.
\end{multlined}
\end{equation}
So, the tensor type for $au_0$ and $au_1$ is equivalent to the tensor type for $u_0$ and $u_1$.

In fact, we can find other bijections from which we see that $u=\sigma_x$ and $u=i\sigma_y$ are essentially the only different cases.
\end{myobs}

\begin{myobs}
The only formal problem with the case $u_0=u_1=0$ is that we get problems defining an identity. However, as we saw in Observation~\ref{obs:matrix_contraction_gauging}, the Schur complement is invariant under arbitrary permutations of the contracted rows and columns, and so is the contraction for $u_0=u_1=0$. So, the result of the evaluation of a tensor network does depend only on which indices are contracted at all, but not on how the indices are paired up with bonds at the contraction.
\end{myobs}

Alternatively, we could make use of norm-constraint matrices.
\begin{mydef}
\tdef{Norm-constraint rectangular Schur-complement tensors}{ncrs_complement_tensor} are the following variant of rectangular Schur-complement tensors:
\begin{itemize}
\item They can be defined if $|u_0|\geq 1$ and $|u_1|\geq 1$.
\item They do not have an identity tensor. This is because the operator norm of the identity matrix is not strictly smaller than $1$.
\item A 1-data $A\in \dat_1((i,o))$ is not a Schur-singular matrix, but an $i\times o$ norm-constraint matrix.
\item The contraction is the same as in Eq.~\eqref{eq:rect_schur_contraction2} where the Schur complement is just an ordinary Schur complement.
\end{itemize}
\end{mydef}

\begin{mydef}
Again, there also is a skeletal formulation of rectangular Schur-complement tensors:
\begin{itemize}
\item As 0-data we take pairs of non-negative integers, the product being component-wise sum, unit $(0,0)$, and dual exchanging the numbers.
\item A 1-data $A\in \dat_1((a,b))$ is the 1-data of the finite-set formulation for standard representatives $\{0,\ldots,a-1\}$ and $\{0,\ldots,b-1\}$.
\item The 2-functions are defined via the bijections $\Phi_+$ and $\Phi_{+\sqcup}$ defined in Eq.~\eqref{eq:plus_bijection} and Eq.~\eqref{eq:sum_disjoint_union_bijection}.
\end{itemize}
\end{mydef}

\subsection{Definition: Square Schur-complement tensors}
\begin{mydef}
\tdef{Square Schur-complement tensors}{ss_complement_tensor} the following kind of Schur-complement tensors:
\begin{itemize}
\item They can be defined for any $2\times 2$ matrix $u$.
\item They do not need a dual. They have a unit 1-function, trivial tensor and identity tensor. They have a symmetric contraction and identity if
\begin{equation}
u_{00}=u_{11}\;,\quad u_{01}=u_{10}\;.
\end{equation}
\item The 0-data $a$ are finite sets. The product and unit are given by the disjoint union and empty set.
\item A 1-data $A\in \dat_1(a)$ is an $a\times a$ Schur-singular matrix.
\item The tensor product is the direct sum of Schur-singular matrices.
\item The contraction of a 1-data $A\in \dat_1(a\sqcup(b\sqcup b))$ is given by the following procedure. First, we construct $A^+$ like $A$ except for
\begin{equation}
\label{eq:schur_square_aplus}
A^+_M = A_M -
\mathbb{0}\oplus
\begin{pmatrix}
u_{00}\mathbb{1}&u_{01}\mathbb{1}\\
u_{10}\mathbb{1}&u_{11}\mathbb{1}
\end{pmatrix}\;.
\end{equation}
Then, the contraction is given by
\begin{equation}
[A] = \schur(A^+)\;.
\end{equation}
\item The trivial tensor $\mathbf{1}\in \dat_1(\{\})$ is given the unique empty $\{\}\times \{\}$ Schur-singular matrix.
\item The identity tensor $\idop\in \dat_1(a\sqcup a)$ is given by
\begin{equation}
\begin{multlined}
\idop_M =
\begin{pmatrix}
u_{00}\mathbb{1} &u_{01}\mathbb{1}\\
u_{10}\mathbb{1}&u_{11}\mathbb{1}
\end{pmatrix}
\;.
\end{multlined}
\end{equation}
\end{itemize}
\end{mydef}

\begin{myobs}
Analogous to Observation~\ref{obs:matrix_contraction_equivalence}, conjugating by a bijection $F$ shows that for $0\neq a\in K$, the tensor types for $u$ and $au$ are equivalent.

Another bijection consists in
\begin{equation}
\label{eq:square_contraction_equivalence2}
F(A)_M=A_M+a\mathbb{1}\;,
\end{equation}
under which the contraction becomes
\begin{equation}
\begin{multlined}
F([F^{-1}(A)])_M\\
= \schur(A_M-a\mathbb{1}-\mathbb{0}\oplus \begin{pmatrix}
u_{00}\mathbb{1} &u_{01}\mathbb{1}\\
u_{10}\mathbb{1}&u_{11}\mathbb{1}
\end{pmatrix})
+a\mathbb{1}\\
= \schur(A_M-\mathbb{0}\oplus 
\begin{pmatrix}
(u_{00}+a)\mathbb{1} &u_{01}\mathbb{1}\\
u_{10}\mathbb{1}&(u_{11}+a)\mathbb{1}
\end{pmatrix})
\;.
\end{multlined}
\end{equation}
Thus, the tensor types with $u+a\mathbb{1}$ and $u$ are equivalent.

It is possible to find further bijections which show that the cases $u=\sigma_x$ and $u=i\sigma_y$ are essentially the only inequivalent ones.
\end{myobs}

\subsection{Definition: (Anti-)symmetric subset}
Square matrices can be symmetric or anti-symmetric, a constraint which is compatible with the Schur complement and which we can use to restrict the 1-data of square Schur-complement tensors.

\begin{mydef}
\tdef{Symmetric Schur-complement tensors}{sys_complement_tensor} are the following variant of square Schur-complement tensors:
\begin{itemize}
\item We restrict to 1-data that is symmetric,
\begin{equation}
\begin{gathered}
A_M(i,j)=A_M(j,i)\;,\\
A_i=A_o\;,\ A_I(i,l)=A_O(l,i)\;,
\end{gathered}
\end{equation}
or in other words,
\begin{equation}
A_X=A_X^T\;,\ A_I=A_O^T\;.
\end{equation}
\item Consequently, the matrix $u$ that appears in the contraction and identity is symmetric as well,
\begin{equation}
u_{01}=u_{10}\;,
\end{equation}
which leaves us with $u=\sigma_x$ as essentially the only choice.
\end{itemize}
\end{mydef}

\begin{myobs}
It is easy to see that the symmetry condition is invariant under the tensor product and contraction. E.g., with the decomposition Eq.~\eqref{eq:schur_block_notation}, we have
\begin{equation}
\begin{multlined}
\schur(A_M^T) = W^T-Y^T(Z^{-1})^TX^T\\= (W-XZ^{-1}Y)^T = \schur(A_M)^T\;.
\end{multlined}
\end{equation}
\end{myobs}

\begin{mydef}
\tdef{Anti-symmetric Schur-complement tensors}{ass_complement_tensor} are the following variant of square Schur-complement tensors:
\begin{itemize}
\item We restrict to 1-data that is anti-symmetric,
\begin{equation}
\begin{gathered}
A_X(i,j)=-A_X(j,i)\\
A_i=A_o\;,\ A_I(i,l)=-A_O(l,i)\;,
\end{gathered}
\end{equation}
or,
\begin{equation}
A_X=-A_X^T\;,\ A_I=-A_O^T\;.
\end{equation}
\item The matrix $u$ has to be anti-symmetric, $u_{00}=u_{11}=0$, $u_{01}=-u_{10}$, thus $u=i\sigma_y$ essentially being the only choice.
\end{itemize}
\end{mydef}

\begin{myobs}
It is easy to see that both the tensor product and contraction are compatible with the anti-symmetry constraint, due to, e.g.,
\begin{equation}
\begin{multlined}
\schur(A_X^T) = -W^T- (-Y^T)(-Z^{-1})^T(-X^T)\\=-\schur(A)_X^T\;.
\end{multlined}
\end{equation}
\end{myobs}

\subsection{Definition: Adding prefactors}
For Schur-complement tensors to be physically useful we need a sensible mapping to array tensors. However, Schur-complement tensors only have a single scalar, as there is only one empty matrix. As such, there only exists a mapping to projective array tensors. If we want a mapping to array tensors which correctly handles prefactors, we need to add some extra structure.

\begin{mydef}
A \tdef{prefactor-matrix}{prefactor_matrix} $A$ consists of a matrix $A_M$ and a prefactor $A_P\in C(K)$, where $C(K)$ is the centralizer of the division ring $K$, i.e., the subset of ring elements that multiplicatively commute with all other ring elements. The Schur complement of a prefactor matrix is the Schur complement of $A_M$, with
\begin{equation}
\label{eq:prefactor_schur}
\schur(A)_P=\det(Z) A_P\;,
\end{equation}
using the decomposition in Eq.~\eqref{eq:schur_block_notation}.

Similarly, a \tdef{prefactor Schur-singular matrix}{ps_singular_matrix} is a Schur-singular matrix $A$ with additional prefactor $A_P$. When taking the Schur complement with a non-invertible sub-matrix $Z$ as in Eq.~\eqref{eq:schur_complement_level}, we need to take
\begin{equation}
\label{eq:prefactor_schur_singular}
\schur(A)_P=\det(H_0^-SG_0^-) A_P\;,
\end{equation}

At the direct sum, the prefactors multiply,
\begin{equation}
\label{eq:prefactor_direct_sum}
(A\oplus B)_P = A_P\cdot B_P\;.
\end{equation}
\end{mydef}

\begin{mydef}
\tdef{Prefactor Schur-complement tensors}{ps_complement_tensor} is the following variant of Schur-complement tensors:
\begin{itemize}
\item Prefactors can be added for both rectangular and square Schur-complement tensors, and both norm-constraint and Schur-singular versions.
\item Instead a (Schur-singular) matrix, the 1-data is given by a prefactor (Schur-singular) matrix.
\item The tensor product is the direct sum for prefactor (Schur-singular) matrices as in Eq.~\eqref{eq:prefactor_direct_sum}, and the contraction is the Schur complement for prefactor (Schur-singular) matrices as in Eq.~\eqref{eq:prefactor_schur} or Eq.~\eqref{eq:prefactor_schur_singular}.
\end{itemize}
\end{mydef}

\begin{myobs}
Let us quickly motivate why the prefactor is compatible with the 2-axioms. Most of those relations are rather trivial, e.g., associativity and commutativity of the tensor product follows from associativity of the ring and the fact that the prefactors are from the centralizer. The most non-trivial consistency is the block-compatibility of the contraction. For a prefactor-matrix $A$, using the block decomposition from Eq.~\eqref{eq:schur_block_compatible} for $A_M$, we find
\begin{equation}
\begin{multlined}
\schur(\schur(A_M))_P = A_P \det(Z)\det(\schur\begin{pmatrix}W&X\\Y&Z\end{pmatrix})\\
= A_P \det\begin{pmatrix}W&X\\Y&Z\end{pmatrix} = \schur(\widetilde{A}_M)_P\;,
\end{multlined}
\end{equation}
where we used the well-known relation
\begin{equation}
\label{eq:schur_determinant}
\begin{multlined}
\det\begin{pmatrix}W&X\\Y&Z\end{pmatrix} \\=
\det\left(
\begin{pmatrix}\mathbb{1}&0\\0&Z\end{pmatrix}
\begin{pmatrix}\mathbb{1}&X\\0&\mathbb{1}\end{pmatrix}
\begin{pmatrix}W-XY^{-1}Z&0\\Z^{-1}Y&\mathbb{1}\end{pmatrix}
\right)
\\=
\det(Z) \det(W-XY^{-1}Z)\\
=
\det(Z) \det(\schur
\begin{pmatrix}W&X\\Y&Z\end{pmatrix}
)\;.
\end{multlined}
\end{equation}
Together with the fact that for $A^+$ with
\begin{equation}
A^+_M=A_M+(U\oplus \mathbb{0})
\end{equation}
we obviously have
\begin{equation}
\schur(A^+)_P=\schur(A)_P\;,
\end{equation}
It follows that
\begin{equation}
[[A]]_P=[\widetilde{A}]_P\;.
\end{equation}
\end{myobs}

\begin{myobs}
Analogously to Eq.~\eqref{eq:schur_determinant}, we can arrive at a formula for the \emph{Pfaffian} of an anti-symmetric $2n\times 2n$ matrix,
\begin{equation}
\pfaff\begin{pmatrix}W&X\\-X^T&Z\end{pmatrix}
=\pfaff(W) \pfaff(\schur\begin{pmatrix}W&X\\-X^T&Z\end{pmatrix})\;.
\end{equation}
Thus, we can define a variant of prefactor anti-symmetric square Schur-complement tensors where we take the Pfaffian instead of the determinant in Eq.~\eqref{eq:prefactor_schur} or Eq.~\eqref{eq:prefactor_schur_singular}.

Note that we have $\pfaff(M)^2=\det(M)$, so using the Pfaffian instead of determinant prefactor contains information about an additional $\pm 1$.
\end{myobs}

\subsection{Mappings}
In this section we define various mappings involving Schur-complement tensors. Those can be grouped into 1) mappings between the different variants of Schur-complement tensors and 2) mappings from Schur-complement tensors to array tensors.
\paragraph{Mappings between matrix tensors}
\begin{mydef}
For every homomorphism $\mathcal{H}$ from a division ring $K_1$ to a division ring $K_2$, the \tdef{entry-wise mapping}{schur_ew_mapping} is the following tensor mapping between $K_1$ Schur-complement tensors to $K_2$ Schur-complement tensors:
\begin{itemize}
\item It works for any variant of Schur-complement tensors.
\item It is strictly homomorphic. The mapping of 0-data is trivial.
\item The mapping of 1-data $A$ consists in applying $\mathcal{H}$ entry-wise,
\begin{equation}
M(A)_M(i,j)=\mathcal{H}(A_M(i,j))\;,
\end{equation}
and the same for the other components of the Schur-singular matrices and prefactors.
\end{itemize}
\end{mydef}

\begin{myexmp}
Consider the following examples of division ring homomorphisms and the corresponding entry-wise mappings:
\begin{itemize}
\item Complex conjugation of the complex numbers.
\item Embedding of real numbers into complex numbers.
\item Embedding of complex numbers or real numbers into quaternions.
\item There are many automorphisms of the quaternions. These include, e.g., cyclic permutations of $i,j,k$, $i\leftrightarrow j, k\rightarrow -k$, or $i\rightarrow -i, j\rightarrow -j$.
\end{itemize}
\end{myexmp}

\begin{mydef}
Consider a ring homomorphism from a division ring $R_1$ to the matrix ring $M_n(R_2)$ over another division ring $R_2$, that is, for every $r\in R_1$ there is a matrix $\mathcal{H}_{ab}(r)\in R_2$. The \tdef{matrix-embedding mapping}{matrix_embedding_mapping} is a tensor mapping from $R_1$ square Schur-complement tensors to $R_2$ square Schur-complement tensors:
\begin{itemize}
\item The mapping of a 0-data set $a$ is
\begin{equation}
m(a)=a\times\{0,\ldots n-1\}\;.
\end{equation}
\item The mapping of a 1-data $A\in \dat_1(a)$ is given by
\begin{equation}
M(A)_M((i,\alpha),(j,\beta))=\mathcal{H}_{\alpha\beta}(A_M(i,j))\;.
\end{equation}
\item The product homomorphor is given canonically with the help of the bijection
\begin{equation}
\begin{gathered}
\Phi^\delta: (a\sqcup b) \times c\rightarrow (a\times c)\sqcup (b\times c)\;,\\
\Phi^\delta((\chi,\beta),\gamma))=(\chi,(\beta,\gamma))\;.
\end{gathered}
\end{equation}
\end{itemize}
We can define an analogous mapping for rectangular Schur-complement tensors.
\end{mydef}

\begin{myexmp}
Consider the following examples for matrix representations of division rings, and the corresponding matrix-embedding mappings:
\begin{itemize}
\item The real representation of complex numbers,
\begin{equation}
\mathcal{H}(a+\mathbf{i}b)=
\begin{pmatrix}
a&b\\
-b&a
\end{pmatrix}\;.
\end{equation}
\item The complex representation of quaternions,
\begin{equation}
\begin{gathered}
\mathcal{H}(a+\mathbf{i}b+\mathbf{j}c+\mathbf{k}d)=
\begin{pmatrix}
x&y\\
-y^*&x^*
\end{pmatrix}\;,\\
x=a+\mathbf{i}b \qquad y=c+\mathbf{i}d\;.
\end{gathered}
\end{equation}
\item The real representation of the quaternions,
\begin{equation}
\mathcal{H}(a+\mathbf{i}b+\mathbf{j}c+\mathbf{k}d)=
\begin{pmatrix}
a & -b & -c & -d\\
b & a & -d & c\\
c & d & a & -b\\
d & -c & b & a
\end{pmatrix}\;.
\end{equation}
\end{itemize}
\end{myexmp}

\begin{mydef}
The \tdef{anti-symmetrization mapping}{anti_symmetrization_mapping} is the following tensor mapping from rectangular Schur-complement tensors with $u=i\sigma_Y$ to anti-symmetric square Schur-complement tensors with $u=i\sigma_Y$:
\begin{itemize}
\item The mapping of a 0-data $(a_i,a_o)$ is given by
\begin{equation}
m((a_i,a_o))=a_i\sqcup a_o\;.
\end{equation}
\item The mapping of a 1-data $A\in \dat_1(a)$ is given by
\begin{equation}
M(A)_M=
\begin{pmatrix}
0 & A_M\\
-A_M^T & 0
\end{pmatrix}\;.
\end{equation}
\item The product homomorphor is given via the canonical bijection between
\begin{equation}
(a_i\sqcup a_o)\sqcup (b_i\sqcup b_o) \leftrightarrow (a_i\sqcup b_i)\sqcup (a_o\sqcup b_o)\;.
\end{equation}
\item The dual homomorphor is given via the canonical bijection $\Phi_\sqcup^\sigma$ applied to $a_i\sqcup a_o$.
\end{itemize}
\end{mydef}


\begin{mydef}
The \tdef{in-out-pair mapping}{io_pair_mapping} is the following mapping from square Schur-complement tensors to rectangular Schur complement tensors:
\begin{itemize}
\item If we call $\widetilde{u}$ the $2\times 2$ matrix $u$ for the square Schur-complement tensors and $u$ the numbers $u$ for the rectangular Schur-complement tensors, then the mapping works for any $u_0=\widetilde{u}_{01}$, $u_1=\widetilde{u}_{10}$, and $\widetilde{u}_{00}=\widetilde{u}_{11}=0$.
\item It has strict product, dual and unit homomorphicity.
\item The mapping applied to a 0-data $a$ is
\begin{equation}
m(a)=(a,a)\;.
\end{equation}
\item The mapping of a 1-data $A\in \dat_1(a)$ is
\begin{equation}
M(A)_M(i,j)=A_M(i,j)\;.
\end{equation}
In other words, we interpret the square matrix as a special rectangular matrix.
\end{itemize}
\end{mydef}

\begin{myobs}
Obviously, we can forget about the symmetry- or anti-symmetry constraint, which formally is a tensor mapping from symmetric or anti-symmetric square Schur-complement tensors to general square Schur-complement tensors.
\end{myobs}

\paragraph{Mappings to array tensors}

\begin{mydef}
The \tdef{determinant mapping}{determinant_mapping} is the following tensor mapping from prefactor $u=i\sigma_y$ rectangular Schur-complement tensors to $\zz$-twisted-symmetric tensors from Definition~\ref{def:ztwisted_symmetric_tensors}:
\begin{itemize}
\item The division ring of matrix tensors and the commutative semiring of array tensors must be given by the same field $K$. We here give the mapping only for norm-constraint Schur-complement tensors, but it can also be adapted to Schur-singular ones.
\item The mapping of a 0-data $(a_i,a_o)$ is
\begin{equation}
m((a_i,a_o))=\{0,-1\}^{a_i}\times \{0,1\}^{a_o}\;.
\end{equation}
Recall that we use $a^b$ to denote the set of functions from a set $b$ to a set $a$. The $\zz$-grading of some
\begin{equation}
(\vec\alpha_i,\vec\alpha_o) \in \{0,-1\}^{a_i}\times \{0,1\}^{a_o}
\end{equation}
is given by
\begin{equation}
|(\vec\alpha_i,\vec\alpha_o)|=|\vec\alpha_i| + |\vec\alpha_o| = \sum_{x\in a_i} \vec\alpha_i(x) +  \sum_{y\in a_o} \vec\alpha_o(y)\;.
\end{equation}
\item For every finite set $a$, choose an ordering $O_a$ of its elements, such that the ordering of a disjoint union is the concatenation of the individual orderings,
\begin{equation}
O_{a\sqcup b}=O_aO_b\;.
\end{equation}
This way we can define the matrix $A_M\rvert_{\vec{\alpha}_i,\vec{\alpha}_o}$ as $A_M$ restricted to the rows $x\in a_i$ with $\vec{\alpha}_i(x)=-1$ and columns $y\in a_o$ with $\vec{\alpha}_o(x)=1$, ordered according to $O_{a_i}$ and $O_{a_o}$. The mapping of a 1-data $A\in \dat_1((a_i,a_o))$ is given by
\begin{equation}
M(A)((\vec{\alpha}_i, \vec{\alpha}_o)) = A_P\cdot \det(A_M\rvert_{\vec{\alpha}_i, \vec{\alpha}_o})\;.
\end{equation}
The determinant of a non-square matrix is defined to be $0$. Note that the determinant is only defined for matrices with a fixed ordering of columns and rows (and not for our abstract notion of $a_i\times a_o$ ``matrix'' with sets $a_i$ and $a_o$), which is why in the definition we need the ordering of all different $a_i$ and $a_o$.
\item The product homomorphor $M_{\otimes 0}$ is a map
\begin{equation}
\begin{multlined}
M_{\otimes 0}: \dat_1^{\text{array}}(\{0,-1\}^{a_i\sqcup b_i}\times \{0,1\}^{a_o\sqcup b_o})
\\\rightarrow
\dat_1^{\text{array}}((\{0,-1\}^{a_i}\times \{0,1\}^{a_o})\\\times (\{0,-1\}^{b_i}\times \{0,1\}^{b_o}))
\end{multlined}
\end{equation}
given by
\begin{equation}
\begin{multlined}
M_{\otimes 0}(A)((\vec{\alpha}_i,\vec{\alpha}_o),(\vec{\beta}_i,\vec{\beta}_o))
\\=
(-1)^{|\vec\alpha_o||\vec\beta_i|} A((\vec\alpha_i\sqcup\vec\beta_i, \vec\alpha_o\sqcup\vec\beta_o))\;,
\end{multlined}
\end{equation}
where $\vec{\alpha}\sqcup \vec{\beta}$ is the concatenation of functions as defined in Eq.~\eqref{eq:function_concatenation}.
\item The unit and dual homomorphor are canonical.
\end{itemize}
\end{mydef}

\begin{myobs}
Let us check whether the definition above actually defines a tensor mapping.
\begin{itemize}
\item The outcome array 1-data $M(A)$ should fulfill the $\zz$-grading. This is true because $A_M\rvert_{\vec{\alpha}_i,\vec{\alpha}_o}$ is only square if
\begin{equation}
|(\vec\alpha_i,\vec\alpha_o)| = |\vec{\alpha}_i|+|\vec{\alpha}_o| = 0\;.
\end{equation}
\item The mapping should be consistent with the commutor. E.g., for a 1-data $A\in \dat_1(a_i\sqcup b_i,a_o\sqcup b_o)$ we find
\begin{equation}
\begin{multlined}
\sigma_0 M_{\otimes 0}M(A)(((\vec{\alpha}_i,\vec{\alpha}_o),(\vec{\beta}_i,\vec{\beta}_o)))\\
=(-1)^{(|\vec{\alpha}_i|+|\vec{\alpha}_o|)(|\vec{\beta}_i|+|\vec{\beta}_o|)}M_{\otimes 0}M(A)(((\vec{\beta}_i,\vec{\beta}_o),(\vec{\alpha}_i,\vec{\alpha}_o)))\\
=(-1)^{(|\vec{\alpha}_i|+|\vec{\alpha}_o|)(|\vec{\beta}_i|+|\vec{\beta}_o|)}(-1)^{|\vec\beta_o||\vec\alpha_i|}\\M(A)((\vec{\beta}_i\sqcup\vec{\alpha}_i, \vec{\beta}_o\sqcup\vec{\alpha}_o))\\
=(-1)^{(|\vec{\alpha}_i|+|\vec{\alpha}_o|)(|\vec{\beta}_i|+|\vec{\beta}_o|)+|\vec\beta_o||\vec\alpha_i|} A_P \det(A_M\rvert_{\vec{\beta}_i\sqcup\vec{\alpha}_i, \vec{\beta}_o\sqcup\vec{\alpha}_o})\\
=(-1)^{|\vec\beta_i||\vec\alpha_o|}A_P \det(A_M\rvert_{\vec{\alpha}_i\sqcup\vec{\beta}_i, \vec{\alpha}_o\sqcup\vec{\beta}_o})\\
=(-1)^{|\vec\beta_i||\vec\alpha_o|}M\sigma_0(A)((\vec{\alpha}_i\sqcup\vec{\beta}_i, \vec{\alpha}_o\sqcup\vec{\beta}_o))\\
= M_{\otimes 0}M \sigma_0(A)(((\vec{\alpha}_i,\vec{\alpha}_o),(\vec{\beta}_i,\vec{\beta}_o)))\;,
\end{multlined}
\end{equation}
where we used the property of the determinant that exchanging two odd-size blocks of rows/columns yields a factor of $-1$.
\item The mapping should be consistent with the tensor product. This holds because the determinant is multiplicative under direct sums.
\item The mapping should be consistent with the contraction. In order to calculate the entry
\begin{equation}
M([A])((\vec{\alpha}_i, \vec{\alpha}_o))
\end{equation}
for $A\in \dat_1((a_i\sqcup b_i\sqcup b_o,a_o\sqcup b_o\sqcup b_i))$, we consider
\begin{equation}
A' \coloneqq A_X\rvert_{\vec{\alpha}_i\sqcup (-1_{b_i}\sqcup -1_{b_o}), \vec{\alpha}_o\sqcup (1_{b_o}\sqcup -1_{b_i})}
\end{equation}
where e.g., $1_{b_o}$ denotes the function which yields $1$ for all $x\in b_o$. We decompose it as a block matrix,
\begin{equation}
A'=
\begin{pmatrix}
R&S&T\\
U&W&X\\
V&Y&Z
\end{pmatrix}\;.
\end{equation}
Now we find
\begin{equation}
\begin{multlined}
M([A])((\vec{\alpha}_i, \vec{\alpha}_o)) = [A]_P \cdot \det([A]_M\rvert_{\vec\alpha_i, \vec\alpha_o})\\
=A_P \cdot \det
\begin{pmatrix}
W&X+\mathbb{1}\\
Y-\mathbb{1}&Z
\end{pmatrix}
\\\cdot \det(\schur
\begin{pmatrix}
R&S&T\\
U&W&X+\mathbb{1}\\
V&Z-\mathbb{1}&Z
\end{pmatrix}
)\\
=A_P \det\begin{pmatrix}
R&S&T\\
U&W&X+\mathbb{1}\\
V&Y-\mathbb{1}&Z
\end{pmatrix}\\
=A_P \sum_{\vec\beta_i, \vec\beta_o} \det(A_X\rvert_{\vec\alpha_i\sqcup(\vec\beta_i\sqcup \vec\beta_o), \vec\alpha_o\sqcup(\vec\beta_o\sqcup \vec\beta_i)})\\
=\sum_{\vec\beta_i, \vec\beta_o} M(A)((\vec\alpha_i\sqcup(\vec\beta_i\sqcup \vec\beta_o), \vec\alpha_o\sqcup(\vec\beta_o\sqcup \vec\beta_i)))\\
=\sum_{\vec\beta_i, \vec\beta_o} M_{*2} M_\otimes M_{\otimes 0}M(A)((\vec\alpha_i, \vec\alpha_o), ((\vec\beta_i, \vec\beta_o),(\vec\beta_i, \vec\beta_o)))\\
=[M_{*2} M_\otimes M_{\otimes 0}M(A)]((\vec{\alpha}_i, \vec{\alpha}_o))\;,
\end{multlined}
\end{equation}
In the third equation we used the relation Eq.~\eqref{eq:schur_determinant}. The fourth equation is obtained from writing the determinant as polynomial and expanding. The terms containing the added $1$s or $-1$s at fixed positions are the determinant of a submatrix.
\end{itemize}
\end{myobs}

\begin{mydef}
The \tdef{pfaffian mapping}{pfaffian_mapping} is the following tensor mapping from prefactor anti-symmetric Schur-complement tensors to $\zz_2$-twisted symmetric tensors:
\begin{itemize}
\item The mapping of a 0-data $a$ is
\begin{equation}
m(a)=\{0,1\}^a\;,
\end{equation}
where $|0|=0$, $|1|=1$.
\item The mapping of a 1-data $A\in \dat_1(a)$ is given by
\begin{equation}
M(A)(\vec{\alpha})= A_P \operatorname{Pf}\left(A_X\rvert_{\vec\alpha,\vec\alpha}\right)\;,
\end{equation}
where $\operatorname{Pf}$ denotes the \emph{Pfaffian} of an anti-symmetric matrix, and $A_X\rvert_{\vec{\alpha}}$ is the matrix $A_X$ restricted to the columns and rows $x\in a$ for which $\alpha(x)=1$. We again need an ordering of the elements of any set $a$ to define matrix $A_X\rvert_{\vec\alpha,\vec\alpha}$ of which we take the Pfaffian.
\end{itemize}
\end{mydef}

\begin{myobs}
We will omit the proof that the pfaffian mapping is indeed is a tensor mapping, but it is analogous to the determinant case. We only remark that the Pfaffian is only defined for matrices with an even number of rows and columns, which is why the result of the mapping is consistent with the $\zz_2$ symmetry constraint.
\end{myobs}

\begin{myrem}
We have presented the determinant/pfaffian mappings from rectangular/square $u=i\sigma_y$ Schur-complement tensors to $\zz_2$-twisted symmetric array tensors. Analogously, there are mappings based on the \emph{permanent}/\emph{Hafnian} of matrices from rectangular/symmetric square $u=\sigma_x$ Schur-complement tensors to $\zz$/$\zz_2$-graded multi-mode infinite array tensors as described in Section~\ref{sec:infinite_array}. Note that in those cases we need to take norm-constraint Schur-complement tensors, as then the resulting array has exponentially decaying entries.
\end{myrem}

\subsection{Use in physics}
Schur-complement tensors are a formalization of the fact that we can calculate quantities of a free-fermion model with a single-particle formalism. The determinant mapping (and its analogues) formalize the second quantization that allows us to go from a single-particle model to a many-body model.

In general, rectangular $u=i\sigma_y$ Schur-complement tensors describe particle-number conserving free-fermion physics, whereas anti-symmetric square Schur-complement tensors are only fermion parity conserving, in which case the single-particle formalism is often referred to as \emph{Majorana}. The anti-symmetrization mapping corresponds to going from a standard-single-particle description to a Majorana description by simply forgetting about the particle-number conservation.

In single-particle quantum physics there is one particle which can be located at some set of possible places, called \emph{modes}. When we combine two systems with sets of modes $P_1$ and $P_1$, the combined system has a set of modes $P_1\sqcup P_2$. A ``process'' acting on a system with set of modes $P$ is a $P\times P$ matrix, whose entries are the ``hopping amplitudes'' of the particle to go from one mode to another. The typical example for such a process is the time-evolution $U=e^{iHt}$ under a single-particle Hamiltonian $H$. More generally, it's possible to change the set of modes during the process to another set of modes $P'$, which would be formalized by a $P\times P'$ matrix. Combining different processes corresponds to the product of matrices, and performing two processes $P_1\rightarrow P_1'$ and $P_2\rightarrow P_2'$ simultaneously corresponds to taking the direct sum of the $P_1\times P_1'$ and $P_2\times P_2'$ matrices yielding a $(P_1\sqcup P_2)\times (P_1'\sqcup P_2')$ matrix. So in that sense, single-particle physics corresponds to the symmetric monoidal category of matrices with direct sum.

This symmetric monoidal category does not directly admit a compact closed structure, and therefore does not directly yield a tensor type. For this we would need a one-to-one correspondence between the $P\times P'$ matrices and the $(P\sqcup (P')^*)\times \{\}$ matrices, the latter ones consisting only of the empty matrix no matter how we define $(P')^*$. Physically speaking, $P$ and $P'$ are the modes where the particle can hop \emph{from} and hop \emph{to}, which play a totally different roles due to the flow of time.

The problem can be solved by explicitly remembering which modes are input and which are output modes. When we combine a set of input modes $P$ and a output set $P'$ we keep them both,
\begin{equation}
P\otimes P'=(P,P')\;,
\end{equation}
whereas if $P$ and $P'$ both consist of input modes or both of output modes, we still take their disjoint union. Physically, this corresponds to not only having \emph{particle modes}, but also \emph{anti-particle modes} present. Now, a process from particle modes $P_1$ and anti-particle models $P_1'$ to particle models $P_2$ and anti-particle modes $P_2'$ is given by a $(P_1\sqcup P_2')\times(P_1'\sqcup P_2)$ matrix. That is, we have hopping of particles from $P_1$ to $P_2$, as well as \emph{backwards} hopping of anti-particles from $P_2'$ to $P_1'$, but also \emph{annihilation} of a particle in $P_1$ with an anti-particle in $P_1'$ and \emph{creation} of a particle-anti-particle pair from $P_2$ and $P_2'$. The new category is agnostic of the flow of time, as we can simply exchange a particle with an anti-particle traveling backwards in time, defining the dual
\begin{equation}
(P,P')^*=(P',P)\;.
\end{equation}

There are also free-fermionic systems without particle-number conservation, which are often referred to as \emph{superconducting}, \emph{Majorana}, or \emph{Bogoliubov-de Gennes} models. In the single-particle picture of those we automatically have particles and anti-particles around. However, particle- and anti-particle modes are not independent but come in pairs, such that compared to the (anti-)particle-number conserving case we always have $P=P'$. In other words, each particle is its own anti-particle. Moreover, hopping from $a$ to $b$ as a particle is the same as hopping backwards from $b$ to $a$ as an anti-particle, and we thus demand that those amplitudes are the same up to a minus sign. This yields anti-symmetric square Schur-complement tensors.

We still haven't explained where the Schur complement is coming from. To this end, imagine a string diagram for the symmetric monoidal category of matrices with direct sum, and let each index correspond to a single mode. The amplitude of the string diagram between to modes $a$ and $b$ is the sum over all paths in the diagram between $a$ and $b$, multiplied by all the amplitudes of the traversed matrices. There is always a finite number of such paths. Now, apply the same reasoning to tensor networks with closed time-like loops, which therefore aren't string diagrams. E.g., consider a tensor $M$ with two in modes $b$ and $d$ and two out modes $a$ and $c$,
\begin{equation}
\begin{tikzpicture}
\atoms{labbox=$M$,bdastyle={wid=1}}{0/}
\draw ([sx=-0.3]0-t)edge[ind=a]++(90:0.4) ([sx=-0.3]0-b)edge[ind=b]++(-90:0.4) ([sx=0.3]0-t)edge[ind=c]++(90:0.4) ([sx=0.3]0-b)edge[ind=d]++(-90:0.4);
\end{tikzpicture}
\end{equation}
and contract
\begin{equation}
\begin{tikzpicture}
\atoms{labbox=$[M]$}{0/}
\draw[rc] (0-t)edge[ind=a]++(90:0.4) (0-b)edge[ind=b]++(-90:0.4);
\end{tikzpicture}
\coloneqq
\begin{tikzpicture}
\atoms{labbox=$M$,bdastyle={wid=1}}{0/}
\draw[rc] ([sx=-0.3]0-t)edge[ind=a]++(90:0.4) ([sx=-0.3]0-b)edge[ind=b]++(-90:0.4) ([sx=0.3]0-t)|-++(0.6,0.4)|-([sx=0.3,sy=-0.4]0-b)--([sx=0.3]0-b);
\end{tikzpicture}\;.
\end{equation}
In order to get from $b$ to $a$ we can go directly, but we can also go from $b$ to $c$ ending up in $d$ and then going to $a$. Instead of directly going to $a$ we can also go to $c$ again and repeat this loop from $d$ to $c$ an arbitrary number of times. Summing up the infinite number of paths yields
\begin{equation}
\begin{multlined}
[M](b,a) = M(b,a) + \sum_{i=0}^\infty M(b,c) M(d,c)^i M(d,a)\\ = M(b,a) + M(b,c) (1-M(d,c))^{-1} M(d,a)\\
= M(b,a) - M(b,c) (M(d,c)-1)^{-1} M(d,a)\;.
\end{multlined}
\end{equation}
This is precisely the contraction we introduced via the Schur complement.

Note that circuits of free-fermionic/single-particle gates have been known under the name \emph{match-gate formalism} \cite{Bravyi2008}.

If we want to have more realistic models, we have go from a single-particle picture to a many-body picture such that we can potentially add interactions, known as \emph{second quantization}. This second quantization is a functor from the category of matrices with direct sum to the symmetric monoidal category of super-vectorspaces. The latter category is the one describing many-body fermionic physics. Conventionally, a single-particle Hamiltonian $H$ can be second-quantized to a many-body operator
\begin{equation}
\hat{H}=\sum_{i,j} H(i,j) c_i^\dagger c_j\;.
\end{equation}
This operator is a morphism in the category of super-vectorspaces (or $\zz_2$-twisted symmetric tensor), however none that would occur in the tensor network/string diagram describing a physical setup. The latter is given by the time evolution
\begin{equation}
\hat{U}=e^{it\hat{H}}\;.
\end{equation}
For the matrix entries of $\hat{U}$, we find
\begin{equation}
\begin{multlined}
\bra{\vec\alpha}\hat{U}\ket{\vec\beta}=\bra{\vec\alpha}\hat U \prod_{i\in\vec\beta} c_i^\dagger \ket{0}=
\bra{\vec\alpha} \prod_{i\in\vec\beta} (\sum_j U_{ij}c_j^\dagger) \ket{0}\\
=\sum_{\Pi(\vec\alpha,\vec\beta)} \sigma(\Pi) \prod_{i\in\vec\beta} U_{i,\Pi(i)}
=\det(U\rvert_{\vec\alpha,\vec\beta})\;.
\end{multlined}
\end{equation}
Here, we made use of familiar formulas for the time-evolution of a single creation operator,
\begin{equation}
\hat U c^\dagger_i \hat U^\dagger = \sum_j U_{ij} c^\dagger_j\;,
\end{equation}
as well as
\begin{equation}
\hat U\ket{0}=\ket{0}\;.
\end{equation}
Also, we identified $\vec\alpha$ and $\vec\beta$ with the subsets of occupied modes, and $\Pi(\vec\alpha,\vec\beta)$ are permutations between those sets.

If there isn't a flow of time, we need to use Schur-complement tensors in the single-particle case and $\zz_2$-twisted symmetric tensors for the many-body case. The determinant mapping is the tensor-type analogue of the second quantization functor.

Single-particle quantum physics as such is described by Schur-complement tensors whose division ring $K$ is given by the complex numbers. But also other division rings have interpretations in terms of symmetries. Different variants of Schur-complement tensors are closely related to the different \emph{Altland-Zirnbauer classes}.

\section{Level tensors}
\subsection{Motivation}
Level tensors are a way to generate a new tensor type from an arbitrary ground tensor type. The basic idea is to add one additional index to each tensor. As after a tensor product we would end up with two instead of one additional index, we need some structure to fuse these into one. This is done by a unital commutative associative algebra.

Level tensors are helpful for the understanding of the role of complex numbers in quantum mechanics, and the connection between unitarity, complex conjugation, and orientation reversal. They can also be used to formalize global constraints such as a fixed particle number which would otherwise be incompatible with a local tensor-network model description.

\subsection{Definition}
As for the tensor types in the following sections, we'll use models of some liquid in the ground type $\calg$ with the help of which we'll construct a new tensor type. If $\calg$ is real or complex array tensors, then the models of the liquid we'll introduce below are equivalent unital associative algebras, hence the name.

\begin{mydef}
\tdef{Unital associative algebras}{unital_associative_algebra} are the following liquid:
\begin{itemize}
\item The flavor of tensor type is allowed to have duals and asymmetric contraction. It needs to have an identity.
\item There is one binding. Recall that this means that a model of this liquid is determined by a single 0-data (together with 1-data).
\item There are two elements. The one is the \emph{product},
\begin{equation}
\label{eq:algebra_multiplication}
\begin{tikzpicture}
\atoms{algebra}{0/p={0,0}}
\draw (0)edge[mark={ar,s}]++(-120:0.6) (0)edge[mark={ar,s}]++(120:0.6) (0)--++(0:0.6);
\end{tikzpicture}\;.
\end{equation}
We add little arrows to some indices and a ``swirl'' in order to distinguish the three indices. Indices with ingoing arrows will be input indices, and the other ones are output indices, so we don't need to indicate that separately.
\item The other element is the \emph{unit} with one output index,
\begin{equation}
\label{eq:algebra_unit}
\begin{tikzpicture}
\atoms{algebra}{0/p={0,0}}
\draw (0)edge[]++(0:0.5);
\end{tikzpicture}
\;.
\end{equation}
\item The first move is \emph{associativity},
\begin{equation}
\label{eq:algebra_associativity}
\begin{tikzpicture}
\atoms{algebra}{0/p={0,0}, 1/p={0.5,0.5}}
\draw (0) edge[mark={ar,e}](1) (0)edge[mark={ar,s},ind=a]++(-135:0.5)(0)edge[mark={ar,s},ind=b]++(135:0.5)(1)edge[mark={ar,s},ind=c]++(135:0.8)(1)edge[ind=d]++(0:0.5);
\end{tikzpicture}=
\begin{tikzpicture}
\atoms{algebra}{0/p={0,0}, 1/p={0.5,-0.5}}
\draw (0) edge[mark={ar,e}](1) (0)edge[mark={ar,s},ind=b]++(-135:0.5)(0)edge[mark={ar,s},ind=c]++(135:0.5)(1)edge[mark={ar,s},ind=a]++(-135:0.8)(1)edge[ind=d]++(0:0.5);
\end{tikzpicture}\;.
\end{equation}
Note that the contraction direction in all diagrams will be from output to input.
\item Then, there is left unitality,
\begin{equation}
\begin{tikzpicture}
\atoms{algebra}{0/p={0,0}, 1/p={0.5,0.5}}
\draw (0) edge[mark={ar,e}](1)(1)edge[mark={ar,s},ind=a]++(135:0.5)(1)edge[ind=b]++(0:0.5);
\end{tikzpicture}=
\begin{tikzpicture}
\draw (0,0)edge[arr=+,startind=a,ind=b] (0.7,0);
\end{tikzpicture}\;,
\end{equation}
\item and right unitality,
\begin{equation}
\begin{tikzpicture}
\atoms{algebra}{0/p={0,0}, 1/p={0.5,-0.5}}
\draw (0) edge[mark={ar,e}](1)(1)edge[mark={ar,s},ind=a]++(-135:0.5)(1)edge[ind=b]++(0:0.5);
\end{tikzpicture}=
\begin{tikzpicture}
\draw (0,0)edge[arr=+,startind=a,ind=b] (0.7,0);
\end{tikzpicture}\;.
\end{equation}
\end{itemize}
\end{mydef}

\begin{mydef}
\tdef{Commutative}{algebra_commutative} unital associative algebras are the liquid obtained by extending unital associative algebras with the following \emph{commutativity} move,
\begin{equation}
\label{eq:product_commutativity}
\begin{tikzpicture}
\atoms{algebra}{0/p={0,0}}
\draw (0)edge[mark={ar,s},ind=a]++(-120:0.5) (0)edge[mark={ar,s},ind=b]++(120:0.5)(0)edge[ind=c]++(0:0.5);
\end{tikzpicture}=
\begin{tikzpicture}
\atoms{algebra}{0/p={0,0}}
\draw (0)edge[mark={ar,s},ind=b]++(-120:0.5) (0)edge[mark={ar,s},ind=a]++(120:0.5)(0)edge[ind=c]++(0:0.5);
\end{tikzpicture}\;.
\end{equation}
We will use a smaller circle as shape denote to denote the elements of commutative unital associative algebras, e.g., for the product,
\begin{equation}
\begin{tikzpicture}
\atoms{prod}{0/p={0,0}}
\draw (0)edge[mark={ar,s}]++(-120:0.5) (0)edge[mark={ar,s}]++(120:0.5)(0)edge[]++(0:0.5);
\end{tikzpicture}\;.
\end{equation}
\end{mydef}

With the help of this liquid, we can define level tensors over an arbitrary ground tensor type.
\begin{mydef}
\tdef{Level tensors}{level_tensor} are the following tensor type:
\begin{itemize}
\item They can be defined for any ground tensor type $\calg$ and any commutative unital associative algebra $X$ in $\calg$, whose 0-data we denote by $a_X$.
\item The flavor is the flavor of $\calg$, that is, level tensors are strictly associative if $\calg$ is strictly associative, and so on.
\item The 0-data is that of the ground tensor type $\calg$.
\item A 1-data $A\in \dat_1(a)$ consists of a 1-data of $\calg$ with 0-data $a_X\otimes a$,
\begin{equation}
A\in \dat_1^{\mathcal{G}}(a_X\otimes a)\;.
\end{equation}
We can think of $A$ as a $2$-index $\calg$-tensor, and express it within the graphical calculus of $\calg$,
\begin{equation}
\label{eq:level_tensor}
\begin{tikzpicture}
\atoms{labbox=$A$}{t/}
\draw (t.west)edge[ind=a]++(-0.4,0);
\draw[level] (t.east)edge[ind=l]++(0.4,0);
\end{tikzpicture}\;,
\end{equation}
where the dotted index is the one with 0-data $a_X$ and will be called \tdef{level index}{level_index}.
\item The invertible 2-functions are given by the invertible 2-functions of $\mathcal{G}$ with one additional auxiliary index, in order to accommodate for the $a_X\otimes $ in front. They are trivial when expressed within the graphical calculus of $\calg$.
\item If $\calg$ has a trivial tensor, we can define a trivial tensor $\mathbf{1}\in \dat_1^{\mathcal{G}}(a_x\otimes 1^{\mathcal{G}})$ given by the unit of $X$,
\begin{equation}
\mathbf{1}= 1_X\otimes \mathbf{1}^{\calg}\;.
\end{equation}
Or, within the graphical calculus of $\calg$,
\begin{equation}
\begin{tikzpicture}
\atoms{labbox=$\mathbf{1}$}{t/}
\draw[level] (t.east)edge[ind=l]++(0.4,0);
\end{tikzpicture}=
\begin{tikzpicture}
\atoms{prod}{0/p={0,0}}
\draw[level] (0)edge[ind=l]++(0.5,0);
\end{tikzpicture}\;.
\end{equation}
\item The tensor product of two 1-data $A\in\dat_1(a)$ and $B\in\dat_1(b)$ is given by the following procedure: 1) Take the tensor product in $\mathcal{G}$. 2) After the tensor product we end up with two level indices. 3) Use the multiplication tensor of $X$ to fuse these two level indices into a single one,
\begin{equation}
\begin{tikzpicture}
\atoms{labbox=$A\otimes B$}{t1/}
\draw (t1.west)edge[ind=ab]++(-0.4,0) (t1.east)edge[level, ind=l]++(0.5,0);
\end{tikzpicture}=
\begin{tikzpicture}
\atoms{prod}{c/p={1.2,0.5}}
\atoms{labbox=$A$}{t1/}
\draw (t1.west)edge[ind=b]++(-0.4,0);
\atoms{labbox=$B$}{t2/p={0,1}}
\draw (t2.west)edge[ind=a]++(-0.4,0);
\draw[level] (t1.east)edge[mark={ar,e},arr=+](c) (t2.east)edge[mark={ar,e},arr=+](c) (c)edge[ind=l]++(0.5,0);
\end{tikzpicture}
\;.
\end{equation}
\item The contraction of a 1-data $A\in \dat_1(a\otimes(b\otimes b^*))$ is the contraction of $\mathcal{G}$,
\begin{equation}
\begin{tikzpicture}
\atoms{labbox=$[A]$}{t/}
\draw (t.west)edge[ind=a]++(-0.4,0) (t.east)edge[level, ind=l]++(0.5,0);
\end{tikzpicture}
=
\begin{tikzpicture}
\atoms{labbox=$A$,bdastyle={wid=1.6}}{t/}
\draw ([xshift=-0.5cm]t.north)edge[ind=a]++(0,0.4);
\draw[rounded corners,arr=+] (t.north)--++(0,0.4)-|([xshift=0.5cm]t.north);
\draw[level] (t.east)edge[ind=l]++(0.4,0);
\end{tikzpicture}
\;.
\end{equation}
\item The identity tensor $\idop \in \dat_1(a\otimes a^*)$ (if $\calg$ has one) is the tensor product of the unit of $X$ and the identity tensor of $\mathcal{G}$,
\begin{equation}
\begin{tikzpicture}
\atoms{labbox=$\idop$,bdastyle={wid=0.8}}{t/}
\draw ([xshift=-0.3cm]t.north)edge[ind=a]++(0,0.4) ([xshift=0.3cm]t.north)edge[ind=b]++(0,0.4);
\draw[level] (t.east)edge[ind=l]++(0.4,0);
\end{tikzpicture}=
\begin{tikzpicture}
\atoms{prod}{0/p={0,0}}
\draw[level] (0)edge[ind=l]++(0.5,0);
\draw [startind=a,ind=b,arr=+] (0,0.5)--(1,0.5);
\end{tikzpicture}\;.
\end{equation}
\end{itemize}
\end{mydef}

\begin{myrem}
The 2-axioms follow very directly from the 2-axioms of $\calg$ and the liquid axioms of the algebra. E.g., 2-axioms involving the associator or commutor follow from 2-axioms of $\mathcal{G}$ together with the associativity and commutativity of $X$.
\end{myrem}

\subsection{Specific dependencies}
\label{sec:level_special_dependencies}
Physically, level tensors with real or complex array tensors as ground type are the most relevant. Thus, let us give a few examples for unital associative algebras in array tensors.

\begin{mydef}
For every finite group $G$, the \tdef{group algebra}{group_algebra} is the following unital associative algebra:
\begin{itemize}
\item The 0-data is given by the set of group elements.
\item The product is
\begin{equation}
\begin{tikzpicture}
\atoms{algebra}{c/p={0,0}}
\draw[]  (c)edge[ind=c]++(-90:0.5) (c)edge[mark={ar,s},ind=b]++(30:0.5) (c)edge[mark={ar,s},ind=a]++(150:0.5);
\end{tikzpicture}=
\delta_{ab,c}\;.
\end{equation}
\item The unit is
\begin{equation}
\begin{tikzpicture}
\atoms{algebra}{x/p={0,0}}
\draw (x)edge[ind=a]++(0.5,0);
\end{tikzpicture}
=
\delta_{a,1}\;.
\end{equation}
\item It is commutative for abelian groups.
\end{itemize}
\end{mydef}

\begin{mydef}
For each $n$, the \tdef{delta algebra}{delta_algebra} is the following unital associative algebra:
\begin{itemize}
\item It is commutative.
\item The 0-data is the set $\{0,\ldots,n-1\}$.
\item The product is
\begin{equation}
\begin{tikzpicture}
\atoms{comalg}{c/p={0,0}}
\draw[]  (c)edge[ind=c]++(-90:0.5) (c)edge[mark={ar,s},ind=b]++(30:0.5) (c)edge[mark={ar,s},ind=a]++(150:0.5);
\end{tikzpicture}
=
\begin{cases}
1 &\text{if}\quad a=b=c\\
0 &\text{otherwise}
\end{cases}\;.
\end{equation}
\item The unit is
\begin{equation}
\begin{tikzpicture}
\atoms{comalg}{x/p={0,0}}
\draw (x)edge[ind=a]++(0.5,0);
\end{tikzpicture}
=
1 \quad\forall a\;.
\end{equation}
\end{itemize}
\end{mydef}

\begin{mydef}
The \tdef{complex number algebra}{complex_number_algebra} is the following unital associative algebra:
\begin{itemize}
\item It is commutative.
\item The 0-data is the set $\{\mathbf{1},\mathbf{i}\}$.
\item The product is given by
\begin{equation}
\begin{tikzpicture}
\atoms{comalg}{c/p={0,0}}
\draw[]  (c)edge[ind=c]++(-90:0.5) (c)edge[mark={ar,s},ind=b]++(30:0.5) (c)edge[mark={ar,s},ind=a]++(150:0.5);
\end{tikzpicture}
=
\left(\begin{pmatrix}1&0\\0&-1\end{pmatrix}, \begin{pmatrix}0&1\\1&0\end{pmatrix}\right)
\;.
\end{equation}
where on the right side $a, b, c$ correspond to row, column, and block, respectively, and $\mathbf{1}$ and $\mathbf{i}$ correspond the first and second entry.
\item The unit is given by
\begin{equation}
\begin{tikzpicture}
\atoms{comalg}{x/p={0,0}}
\draw (x)edge[ind=a]++(0.5,0);
\end{tikzpicture}
=
\begin{pmatrix}
1&0
\end{pmatrix}
\;.
\end{equation}
\item As an algebra in complex array tensors it is isomorphic to the delta algebra for $n=2$. As an algebra in real array tensors it is not isomorphic to any delta algebra.
\end{itemize}
\end{mydef}

\begin{mydef}
For every number $n$, the \tdef{counting algebra}{counting_algebra} is a unital associative algebra:
\begin{itemize}
\item It is commutative.
\item The 0-data is the set $\{0,\ldots,n-1\}$.
\item The product is
\begin{equation}
\begin{tikzpicture}
\atoms{comalg}{c/p={0,0}}
\draw[]  (c)edge[ind=c]++(-90:0.5) (c)edge[mark={ar,s},ind=b]++(30:0.5) (c)edge[mark={ar,s},ind=a]++(150:0.5);
\end{tikzpicture}=
\delta_{a+b,c}\;.
\end{equation}
Note that the addition is not modulo $n$, i.e., if $a+b\geq n$, then the tensor entry is $0$.
\item The unit is
\begin{equation}
\begin{tikzpicture}
\atoms{comalg}{x/p={0,0}}
\draw (x)edge[ind=a]++(0.5,0);
\end{tikzpicture}
=
\delta_{a,1}\;.
\end{equation}
\end{itemize}
\end{mydef}

Let us also give two examples for a algebras in a tensor types other than array tensors.
\begin{mydef}
The \tdef{Schur-complement counting algebra}{sc_counting_algebra} is the following unital associative algebra in rectangular Schur-complement tensors:
\begin{itemize}
\item It is commutative.
\item The 0-data is the set $\{0\}$, i.e., it consists of a single mode.
\item The product is
\begin{equation}
\begin{tikzpicture}
\atoms{comalg}{c/p={0,0}}
\draw[]  (c)edge[ind=c]++(-90:0.5) (c)edge[mark={ar,s},ind=b]++(30:0.5) (c)edge[mark={ar,s},ind=a]++(150:0.5);
\end{tikzpicture}=
\begin{vmatrix}
&a&b\\
c&1&1
\end{vmatrix}\;.
\end{equation}
\item The unit is the empty matrix with one row but no columns,
\begin{equation}
\begin{tikzpicture}
\atoms{comalg}{x/p={0,0}}
\draw (x)edge[ind=a]++(0.5,0);
\end{tikzpicture}
=
\begin{vmatrix}
\\
c
\end{vmatrix}\;.
\end{equation}
\end{itemize}
\end{mydef}

\begin{mydef}
The \tdef{matrix $\mathbb{Z}_2$ algebra}{matrix_z2_algebra} is the following unital associative algebra in square Schur-complement tensors:
\begin{itemize}
\item It is \emph{not} commutative.
\item The 0-data is the set $\{0\}$.
\item The product is
\begin{equation}
\begin{tikzpicture}
\atoms{algebra}{c/p={0,0}}
\draw[]  (c)edge[ind=c]++(-90:0.5) (c)edge[mark={ar,s},ind=b]++(30:0.5) (c)edge[mark={ar,s},ind=a]++(150:0.5);
\end{tikzpicture}
=
\begin{vmatrix}
&a&b&c\\
a&0&1&1\\
b&-1&0&1\\
c&-1&-1&0
\end{vmatrix}
\;.
\end{equation}
\item The unit is
\begin{equation}
\begin{tikzpicture}
\atoms{algebra}{x/p={0,0}}
\draw (x)edge[ind=a]++(0.5,0);
\end{tikzpicture}
=
\begin{vmatrix}
&a\\
a&0
\end{vmatrix}
\;.
\end{equation}
\end{itemize}
\end{mydef}

As we will see in the next section, level tensors for algebras which are special commutative Frobenius algebras can be emulated by ground-type tensors. In the complex or real case, there are many examples of commutative unital associative algebras which cannot be extended to special Frobenius algebras, in fact any non-semisimple algebra provides an example.

\begin{myobs}
Consider level tensors for the counting algebra for some number $n$. Every level tensor can be seen as a collection of $n$ different array tensors, labeled by a ``level'' $0\leq l<n$. Imagine a level tensor network, and turn it into an array tensor network by choosing a configuration of levels for each tensor,
\begin{equation}
\mathbf{l}=(l_0,l_1,\ldots)\;.
\end{equation}
Let us refer to the evaluation of this array tensor network by $E_{\mathbf{l}}$, and let us refer to the evaluation of the level tensor network at level $L$ by $X_L$. We find
\begin{equation}
X_L = \sum_{\mathbf{l}: l_0+l_1+\ldots = L} E_{\mathbf{l}}\;.
\end{equation}
E.g., $X_0$ is the evaluation of the array tensor network of all level-$0$ tensors, $X_0 = E_{(0,0,\ldots)}$. $X_1$ is the sum of array tensor network evaluations, where all tensors are level $0$, except for one tensor which is level $1$, $X_1=E_{(1,0,\ldots)}+E_{(0,1,\ldots)}+\ldots$. $X_2$ is a sum of $m(m+1)/2$ terms, where $m$ is the number of tensors in the network, and so on.
\end{myobs}

\subsection{Mappings}
\label{sec:level_mappings}
The first mapping in this section shows that for certain unital associative algebras, level tensors can be emulated via ground type tensors. More precisely, this is possible if the liquid model can be extended to a model of the following more complex liquid.

\begin{mydef}
\tdef{Symmetric Frobenius algebras}{symmetric_frobenius_algebra} are the following liquid extending the liquid of unital associative algebras:
\begin{itemize}
\item There are two more elements. The first is the \emph{co-unit},
\begin{equation}
\begin{tikzpicture}
\atoms{algebra}{0/p={0,0}}
\draw (0)edge[mark={ar,s}]++(0:0.5);
\end{tikzpicture}\;.
\end{equation}
\item The second is the \emph{dual Frobenius form},
\begin{equation}
\begin{tikzpicture}
\atoms{algebra}{0/p={0,0}}
\draw (0)edge[mark={ar,s}]++(180:0.5)(0)edge[mark={ar,s}]++(0:0.5);
\end{tikzpicture}\;.
\end{equation}
\item There are three additional moves, namely the \emph{form symmetry},
\begin{equation}
\begin{tikzpicture}
\atoms{algebra}{0/p={0,0}, 1/p={0,0.7}}
\draw (1)edge[mark={ar,s}](0) (0)edge[ind=a,mark={ar,s}]++(180:0.6)(0)edge[ind=b,mark={ar,s}]++(0:0.6);
\end{tikzpicture}
=
\begin{tikzpicture}
\atoms{algebra}{0/p={0,0}, 1/p={0,0.7}}
\draw (1)edge[mark={ar,s}](0) (0)edge[ind=b,mark={ar,s}]++(180:0.6)(0)edge[ind=a,mark={ar,s}]++(0:0.6);
\end{tikzpicture}\;,
\end{equation}
\item the \emph{inversion relation},
\begin{equation}
\begin{tikzpicture}
\atoms{algebra}{0/p={0,0}, 1/p={0.8,0}, 2/p={0,0.7}}
\draw (2)edge[mark={ar,s}](0) (0)edge[mark={ar,s}](1) (0)edge[mark={ar,s},ind=a]++(180:0.5) (1)edge[ind=b]++(0:0.5);
\end{tikzpicture}
=
\begin{tikzpicture}
\draw (0,0)edge[startind=a,ind=b](0.5,0);
\end{tikzpicture}
\end{equation}
\item and the \emph{Frobenius property},
\begin{equation}
\begin{tikzpicture}
\atoms{algebra}{0/p={0,0}, 1/p={0.7,0}, 2/p={0.7,0.7}}
\draw (2)edge[mark={ar,s}](1) (0)edge[mark={ar,e}](1) (0)edge[mark={ar,s},ind=a]++(180:0.5) (0)edge[mark={ar,s},ind=c]++(-90:0.5) (1)edge[mark={ar,s},ind=b]++(0:0.5);
\end{tikzpicture}
=
\begin{tikzpicture}
\atoms{algebra}{0/p={0,0}, 1/p={-0.7,0}, 2/p={-0.7,0.7}}
\draw (2)edge[mark={ar,e}](1) (0)edge[mark={ar,e}](1) (0)edge[mark={ar,s},ind=b]++(0:0.5) (0)edge[mark={ar,s},ind=c]++(-90:0.5) (1)edge[mark={ar,s},ind=a]++(180:0.5);
\end{tikzpicture}\;.
\end{equation}
\item We might optionally add the \emph{commutativity} move in Eq.~\eqref{eq:product_commutativity}, in which case we again use a smaller circle as shape.
\item Another optional property we can add is the \emph{special} move,
\begin{equation}
\label{eq:move_special}
\begin{tikzpicture}
\atoms{algebra}{0/p={0,0}, 1/p={0.6,0}}
\draw (0)edge[mark={ar,e},bend left=80,looseness=2](1) (0)edge[mark={ar,e},bend right=80,looseness=2](1) (1)edge[ind=a]++(0:0.5);
\end{tikzpicture}
=
\begin{tikzpicture}
\atoms{algebra}{c/p={0,0}}
\draw (c)edge[ind=a]++(0:0.4);
\end{tikzpicture}\;.
\end{equation}
\end{itemize}
\end{mydef}

\begin{myobs}
The moves of symmetric Frobenius algebras might seem a bit arbitrary at first glance. However, the set of moves they generate have a very clear diagrammatic interpretation: Every equation between two connected, planar, loop-free networks whose open indices (as well as whether they are input/output) match up, can be derived from the moves. E.g., the following move can be derived,
\begin{equation}
\begin{tikzpicture}
\atoms{algebra}{0/p={0,0}, 1/p={0.8,0}, 2/p={1.6,0}, 3/p={1.6,0.7}}
\draw (0)edge[mark={ar,e}](1) (0)edge[ind=a,mark={ar,s}]++(135:0.5) (0)edge[ind=b,mark={ar,s}]++(-135:0.5) (1)edge[ind=c,mark={ar,s}]++(90:0.5) (3)edge[mark={ar,s}](2) (1)edge[mark={ar,e}](2) (2)edge[mark={ar,s},ind=d]++(0:0.5);
\end{tikzpicture}=
\begin{tikzpicture}
\atoms{algebra}{0/p={0,0}, 1/p={0,0.6}, 2/p={0,1.2}, 3/p={0.7,0.6}}
\draw (0)edge[ind=d,mark={ar,s}]++(-45:0.5) (0)edge[ind=b,mark={ar,s}]++(-135:0.5) (2)edge[ind=c,mark={ar,s}]++(45:0.5) (2)edge[ind=a,mark={ar,s}]++(135:0.5) (3)edge[mark={ar,s}](1) (1)edge[mark={ar,s}](2) (1)edge[mark={ar,s}](0);
\end{tikzpicture}
\;.
\end{equation}
We can make use of this to define additional elements with arbitrary input/output indices, e.g.,
\begin{equation}
\begin{tikzpicture}
\atoms{algebra}{0/p={0,0}}
\draw (0)edge[mark={ar,s},ind=a]++(-150:0.5)(0)edge[ind=b]++(150:0.5)(0)edge[mark={ar,s},ind=c]++(-90:0.5)(0)edge[mark={ar,s},ind=d]++(90:0.5)(0)edge[ind=e]++(0:0.5);
\end{tikzpicture}
\coloneqq
\begin{tikzpicture}
\atoms{algebra}{0/p={0,0}, 1/p={0.8,0}, 2/p={1.6,0}, 3/p={0.8,0.7}}
\draw (0)edge[mark={ar,e}](1) (0)edge[ind=a,mark={ar,s}]++(135:0.5) (0)edge[ind=c,mark={ar,s}]++(-135:0.5) (2)edge[ind=d,mark={ar,s}]++(90:0.5) (3)edge[mark={ar,e}](1) (3)edge[ind=b]++(90:0.5) (1)edge[mark={ar,e}](2) (2)edge[ind=e]++(0:0.5);
\end{tikzpicture}
\;.
\end{equation}

If we additionally add the special move in Eq.~\eqref{eq:move_special}, the equivalence of planar diagrams extends to ones that are not loop-free. If we additionally add the commutativity move, this even extends to non-planar networks, that is, all connected networks with matching open indices are equivalent under moves.
\end{myobs}

\begin{myobs}
The group algebra can be extended to a Frobenius algebra, with
\begin{equation}
\begin{tikzpicture}
\atoms{algebra}{x/p={0,0}}
\draw[](x)--++(-0.5,0)node[left]{$a$};
\draw[](x)--++(0.5,0)node[right]{$b$};
\end{tikzpicture}=
\frac{1}{|G|} \delta_{a,b^{-1}}\;,
\end{equation}
and
\begin{equation}
\begin{tikzpicture}
\atoms{algebra}{x/p={0,0}}
\draw (x)edge[mark={ar,s},ind=a]++(0.5,0);
\end{tikzpicture}
=
|G| \delta_{a,1}\;.
\end{equation}
\end{myobs}

\begin{myobs}
The delta algebra can be extended to a Frobenius algebra, with
\begin{equation}
\begin{tikzpicture}
\atoms{comalg}{x/p={0,0}}
\draw (x)edge[ind=b]++(0.5,0) (x)edge[ind=a]++(-0.5,0);
\end{tikzpicture}
=
\delta_{a,b}\;,
\end{equation}
and
\begin{equation}
\begin{tikzpicture}
\atoms{comalg}{x/p={0,0}}
\draw (x)edge[mark={ar,s},ind=a]++(0.5,0);
\end{tikzpicture}
=
1\quad\forall a\;.
\end{equation}
\end{myobs}

\begin{myobs}
The complex number algebra can be extended to a Frobenius algebra, with
\begin{equation}
\begin{tikzpicture}
\atoms{comalg}{x/p={0,0}}
\draw (x)edge[ind=b]++(0.5,0) (x)edge[ind=a]++(-0.5,0);
\end{tikzpicture}
=
\frac{1}{2}
\begin{pmatrix}
1&0\\0&-1
\end{pmatrix}\;,
\end{equation}
and
\begin{equation}
\begin{tikzpicture}
\atoms{comalg}{x/p={0,0}}
\draw (x)edge[mark={ar,s},ind=a]++(0.5,0);
\end{tikzpicture}
=
2
\begin{pmatrix}
1&0
\end{pmatrix}
\;.
\end{equation}
\end{myobs}

\begin{mycom}
Consider a (special Frobenius unital associative) algebra, and multiply each of its tensors by an invertible scalar $\alpha$ for each input index and $\alpha^{-1}$ for each output index. As contractions are always between an input- and an output index, this leaves all the moves invariant, so the rescaled algebra is a (special Frobenius) algebra again.
\end{mycom}

\begin{mydef}
Consider level tensors whose ground type has duals and whose algebra can be extended to a special Frobenius algebra. In this case, level tensors can be mapped back to their ground type, by a mapping called the \tdef{frobenification mapping}{frobenification_mapping}:
\begin{itemize}
\item It has weakened homomorphicity. It \emph{not} a full mapping as it is not compatible with the tensor product. It is however compatible with all other 2-functions, and with the combination of the tensor product and a contraction, just like the direct sum mapping.
\item The mapping of a 0-data $a$ is given by
\begin{equation}
m(a)=a_x\otimes a\;.
\end{equation}
\item The mapping of a 1-data $A\in \dat_1(a)$ is that same 1-data, as a 1-data of the ground type,
\begin{equation}
\begin{tikzpicture}
\atoms{labbox=$M(A)$,bdastyle=rc}{t/}
\draw ([sy=0.15]t-l)edge[ind=a]++(-0.4,0) ([sy=-0.15]t-l)edge[level,ind=l]++(-0.4,0);
\end{tikzpicture}
=
\begin{tikzpicture}
\atoms{labbox=$A$}{t/}
\draw (t-l)edge[ind=a]++(-0.4,0);
\draw[level] (t-r)edge[ind=l]++(0.4,0);
\end{tikzpicture}\;.
\end{equation}
\item The product homomorphor of a 1-data $A\in \dat_1(a\otimes m(b\otimes c))$, expressed in the graphical calculus of the ground type, is given by
\begin{equation}
\begin{tikzpicture}
\atoms{labbox=$M_\otimes(A)$, bdastyle=rc}{0/}
\draw (0-t)edge[ind=a]++(90:0.4) ([sy=0.15]0-l)edge[ind=b]++(180:0.4) ([sy=-0.15]0-l)edge[level,ind=l_b]++(180:0.4) ([sy=0.15]0-r)edge[ind=c]++(0:0.4) ([sy=-0.15]0-r)edge[level,ind=l_c]++(0:0.4);
\end{tikzpicture}
=
\begin{tikzpicture}
\atoms{labbox=$A$, bdastyle=rc}{0/}
\atoms{prod}{p/p={0,-0.7}}
\draw (0-t)edge[ind=a]++(90:0.4) (0-l)edge[ind=b]++(180:0.4) (p)edge[level,ind=l_b]++(180:0.6) (0-r)edge[ind=c]++(0:0.4) (p)edge[level,ind=l_c]++(0:0.6) (0-b)edge[level,mark={ar,e}](p);
\end{tikzpicture}
\;.
\end{equation}
\item The unit homomorphor of a 1-data $A\in \dat_1(a\otimes m(1))$ is given by
\begin{equation}
\begin{tikzpicture}
\atoms{labbox=$M_1(A)$, bdastyle=rc}{0/}
\draw (0-l)edge[ind=a]++(180:0.4);
\end{tikzpicture}
=
\begin{tikzpicture}
\atoms{labbox=$A$, bdastyle=rc}{0/}
\atoms{prod}{p/p={0.7,0}}
\draw (0-l)edge[ind=a]++(180:0.4) (0-r)edge[level,mark={ar,e}](p);
\end{tikzpicture}
\;.
\end{equation}
\item The dual homomorphor of a 1-data $A\in \dat_1(a\otimes m(b^*))$ is given by
\begin{equation}
\label{eq:level_dual_homomorphor}
\begin{tikzpicture}
\atoms{labbox=$M_*(A)$, bdastyle=rc}{0/}
\draw (0-t)edge[ind=a]++(90:0.4) ([sy=0.15]0-r)edge[ind=b]++(0:0.4) ([sy=-0.15]0-r)edge[level,ind=l_b]++(0:0.4);
\end{tikzpicture}
=
\begin{tikzpicture}
\atoms{labbox=$A$, bdastyle=rc}{0/}
\atoms{prod}{p/p={0.7,-0.15}}

\draw (0-t)edge[ind=a]++(90:0.4) ([sy=0.15]0-r)edge[ind=b]++(0:0.4) ([sy=-0.15]0-r)edge[level,mark={ar,e}](p) (p)edge[level,ind=l_b,mark={ar,s}]++(0:0.5);
\end{tikzpicture}
\;.
\end{equation}
\end{itemize}
\end{mydef}

\begin{myrem}
The requirement that the ground type needs to have a dual is not a very essential one: Every tensor type can be equipped with a trivial dual. The introduction of such a dual corresponds to an in/out bookkeeping of the indices, allowing only contractions between an in- and out index. The result of the frobenification mapping depends on whether an index is in or out.
\end{myrem}

\begin{myrem}
The fact that the frobenification mapping is not compatible with the tensor product alone but only with a combination of tensor product and contraction implies that the mapping only commutes with the evaluation of tensor networks as long as the latter are connected.
\end{myrem}

\begin{myobs}
The compatibility of the mapping with associator, commutor, and unitor are easily seen to hold due to the associativity, commutativity and unitality of the algebra. The ``special'' property is needed for the compatibility with the contraction. For a 1-data $A\in \dat_1(a\otimes (b\otimes b^*))$ we find
\begin{equation}
\begin{gathered}
\begin{tikzpicture}
\atoms{labbox=$M([A])$}{t1/}
\draw (t1.west)edge[ind=a]++(-0.4,0) (t1.east)edge[level, ind=l]++(0.5,0);
\end{tikzpicture}
=
\begin{tikzpicture}
\atoms{labbox=$A$,bdastyle={wid=1}}{t/}
\draw (t-l)edge[ind=a]++(180:0.4);
\draw[rc,arr=+] ([sx=-0.3]t-t)--++(0,0.4)-|([sx=0.3]t-t);
\draw[level] (t-r)edge[ind=l]++(0.4,0);
\end{tikzpicture}
\\=
\begin{tikzpicture}
\atoms{labbox=$A$,bdastyle={wid=1}}{t/}
\atoms{prod}{0/p={0.9,0}, 1/p={1.3,-0.4}, 2/p={1.9,-0.7}}
\draw (t-l)edge[ind=a]++(180:0.4);
\draw[rc,arr=+] ([sx=-0.3]t-t)--++(0,0.4)-|([sx=0.3]t-t);
\draw[level,rc] (0)edge[ind=l]++(45:0.6) (t-r)edge[mark={ar,e}](0) (0)edge[mark={ar,e}](1);
\draw[level,rc,mark={ar,e}] (1)--++(0.3,-0.3)--(2);
\draw[level,rc,mark={ar,s}] (2)--++(0.4,0)--++(0,0.6)--++(-0.6,0)--(1);
\end{tikzpicture}
=
\begin{tikzpicture}
\atoms{labbox=$[M_X(A)]$}{t1/}
\draw (t1.west)edge[ind=a]++(-0.4,0) (t1.east)edge[level, ind=l]++(0.5,0);
\end{tikzpicture}\;,
\end{gathered}
\end{equation}
where the $M_X$ is short for $M$ followed by a sequence of homomorphors, $M_*M_\otimes M_{\otimes 0} M$. For the compatibility of the combination of tensor product and contraction applied to two level 1-data $A\in \dat_1(a\otimes b)$ and $B\in \dat_1(b^*\otimes c)$, we find
\begin{equation}
\begin{gathered}
\begin{tikzpicture}
\atoms{labbox=$M_X([A\otimes B])$}{t1/}
\draw ([sx=0.3]t1-t)edge[ind=b]++(90:0.4) ([sx=-0.3]t1-t)edge[ind=a]++(90:0.4) ([sy=-0.2]t1-r)edge[level, ind=l_a]++(0.5,0) ([sy=0.2]t1-r)edge[level, ind=l_b]++(0.5,0);
\end{tikzpicture}
=
\begin{tikzpicture}
\atoms{labbox=$[A\otimes B]$}{t1/}
\atoms{prod}{p/p={1.3,0}}
\draw ([sx=0.3]t1-t)edge[ind=b]++(90:0.4) ([sx=-0.3]t1-t)edge[ind=a]++(90:0.4) (t1-r)edge[level,mark={ar,e}](p) (p)edge[level, ind=l_b]++(45:0.5) (p)edge[level, ind=l_a]++(-45:0.5);
\end{tikzpicture}
\\=
\begin{tikzpicture}
\atoms{labbox=$A$}{a/}
\atoms{labbox=$B$}{b/p={0,1}}
\atoms{prod}{p/p={0.9,0.5}, p1/p={1.5,0.5}}
\draw (a-l)edge[ind=a]++(180:0.3) (b-l)edge[ind=b]++(180:0.3) (a-t)--(b-b);
\draw[level] (a-r)edge[arr=+,mark={ar,e}](p) (b-r)edge[arr=+,mark={ar,e}](p) (p)edge[mark={ar,e}](p1)  (p1)edge[level, ind=l_b]++(45:0.5) (p1)edge[level, ind=l_a]++(-45:0.5);
\end{tikzpicture}
=
\begin{tikzpicture}
\atoms{labbox=$A$}{a/}
\atoms{labbox=$B$}{b/p={0,1}}
\atoms{prod}{p/p={1,0}, p1/p={1,1}, pm/p={1,0.5}}
\draw (a-l)edge[ind=a]++(180:0.3) (b-l)edge[ind=b]++(180:0.3) (a-t)--(b-b);
\draw[level] (a-r)edge[arr=+,mark={ar,e}](p) (b-r)edge[arr=+,mark={ar,e}](p1) (p1)edge[mark={ar,e}](pm) (p)edge[mark={ar,e}](pm) (p1)edge[level, ind=l_b]++(0:0.5) (p)edge[level, ind=l_a]++(0:0.5);
\end{tikzpicture}
\\=
\begin{tikzpicture}
\atoms{labbox=$[M_X(A)\otimes M_Y(B)])$}{t1/}
\draw ([sx=0.3]t1-t)edge[ind=b]++(90:0.4) ([sx=-0.3]t1-t)edge[ind=a]++(90:0.4) ([sy=-0.2]t1-r)edge[level, ind=l_a]++(0.5,0) ([sy=0.2]t1-r)edge[level, ind=l_b]++(0.5,0);
\end{tikzpicture}
\;,
\end{gathered}
\end{equation}
where $M_X$ is short for $M_{\otimes 0}M$, $M_Y$ for $M_*M_{\otimes 0}M$, and we neglected invertible 2-functions of the source and target tensor type.
\end{myobs}

For the next mapping we need to introduce another extension of the algebra liquid.
\begin{mydef}
We can extend the (unital associative) algebra liquid by a further 1-index element called the \tdef{Hopf co-unit}{hopf_counit},
\begin{equation}
\begin{tikzpicture}
\atoms{hopfcounit}{0/p={1,0}}
\draw (0)edge[mark={ar,s},ind=a]++(0.5,0);
\end{tikzpicture}
\end{equation}
together with the following moves
\begin{equation}
\begin{tikzpicture}
\atoms{hopfcounit}{0/p={0,0}}
\atoms{algebra}{1/p={1,0}}
\draw (0)edge[mark={ar,s}](1);
\end{tikzpicture}=
\hspace{2cm}\;,
\end{equation}
and
\begin{equation}
\begin{tikzpicture}
\atoms{hopfcounit}{0/p={0,0}}
\atoms{algebra}{1/p={1,0}}
\draw (0)edge[mark={ar,s}](1) (1)edge[mark={ar,s},ind=a]++(-45:0.5) (1)edge[mark={ar,s},ind=b]++(45:0.5);
\end{tikzpicture}=
\begin{tikzpicture}
\atoms{hopfcounit}{0/p={1,0}, 1/p={1,0.7}}
\draw (0)edge[mark={ar,s},ind=a]++(0.5,0) (1)edge[mark={ar,s},ind=b]++(0.5,0);
\end{tikzpicture}
\;.
\end{equation}
\end{mydef}

\begin{mydef}
The \tdef{projection mapping}{projection_mapping} is the following tensor mapping from level tensors to their ground type:
\begin{itemize}
\item It can be defined for every choice of Hopf co-unit for its commutative associative unital algebra. 
\item It has strict homomorphicity 1-axioms.
\item The mapping of 0-data is the identity.
\item The mapping applied to a 1-data $A\in \dat_1(a)$ consists in contracting the level index with the Hopf co-unit,
\begin{equation}
\begin{tikzpicture}
\atoms{labbox=$M(A)$,bdastyle=rc}{t/}
\draw (t-l)edge[ind=a]++(-0.4,0);
\end{tikzpicture}
=
\begin{tikzpicture}
\atoms{labbox=$A$}{t/}
\atoms{hopfcounit}{p/p={0.8,0}}
\draw (t-l)edge[ind=a]++(-0.4,0);
\draw[level] (t-r)edge[mark={ar,e}](p);
\end{tikzpicture}\;.
\end{equation}
\end{itemize}
\end{mydef}

\begin{myexmp}
Consider array level tensors for the delta algebra for a basis $B$. Every element of $b\in B$ defines a $1$-index tensor (with entry $1$ for $b$ and $0$ elsewhere), which is a projection. The product of all the corresponding projection mappings is nothing but reinterpreting the corresponding level tensor as a set of $|B|$ array tensors that are contracted and tensor-producted independently.
\end{myexmp}

\begin{mydef}
The \tdef{levelification}{levelification} is the following mapping from complex array tensors to real array level tensors for the complex number algebra:
\begin{itemize}
\item It has strict homomorphicity 1-axioms.
\item The mapping for the 0-data is trivial,
\begin{equation}
m(a)=a\;.
\end{equation}
\item The mapping of a 1-data $A\in \dat_1(a)$ consists in decomposing the complex tensor into real and imaginary part,
\begin{equation}
\begin{gathered}
M(A)((\mathbf{1},b))=\mathop{Real}(A(b))\\
M(A)((\mathbf{i},b))=\mathop{Imag}(A(b))\;.
\end{gathered}
\end{equation}
\end{itemize}
This mapping is a simple bijection between 0-data and 1-data compatible with all 1-functions and 2-functions, so it is really just a reinterpretation of complex arrays in terms of two real arrays.
\end{mydef}

\begin{mydef}
The \tdef{zero-level mapping}{zero_level_mapping} is the mapping from ground type tensors to level tensors which sends everything to the $0$th level:
\begin{itemize}
\item It has strict homomorphicity 1-axioms.
\item The mapping for the 0-data is the identity.
\item The mapping for the 1-data consists in taking the tensor product with the algebra unit,
\begin{equation}
\begin{tikzpicture}
\atoms{labbox=$\calm(A)$}{l/}
\draw (l-r)edge[level,ind=l]++(0:0.4) (l-l)edge[ind=a]++(180:0.4);
\end{tikzpicture}
=
\begin{tikzpicture}
\atoms{labbox=$A$,bdastyle=rc}{l/}
\atoms{prod}{p/p={0.7,0}}
\draw (p)edge[level,ind=l]++(0:0.5) (l-l)edge[ind=a]++(180:0.4);
\end{tikzpicture}
\;.
\end{equation}
\end{itemize}
\end{mydef}

\subsection{Use in physics}
Imagine a classical thermal ensemble with a fixed ``particle number''. That is, for every degree of freedom, there is a function that associates a particle number $n(c)$ to each configuration $c$. The ensemble is a probability distribution over all global configurations whose individual particle numbers sum up to a fixed global particle number $N$. Apart from that, the probabilities are determined by a local Hamiltonian. This situation can be formalized using non-negative real array level tensors for the counting algebra with the number $N$. To this end, we associate to each degree of freedom the following level tensor,
\begin{equation}
\begin{gathered}
A((l,c_1,c_2,\ldots)) \\=
\begin{cases}
1& \text{if}\quad n(c_1)=l \quad \text{and}\quad c_1=c_2=\ldots\\
0& \text{otherwise}
\end{cases}\;,
\end{gathered}
\end{equation}
where $l$ is the level. Each Boltzmann weight is represented by a level tensor which is non-zero only at level $0$. Otherwise, the network has the same geometry as the network representing a classical thermal system without constant particle number.

As we have seen in Section~\ref{sec:level_special_dependencies}, the evaluation of the network at level $N$ consists only of level configurations whose levels sum to $N$. So, by restricting to the level-$N$ part of the evaluation, we get the desired measurement statistics restricted to the subspace with exactly $N$ particles.

The situation is slightly different for quantum ensembles with a fixed particle number. In a quantum spin system, the particle number is a $U(1)$ representation. This representation has to be a symmetry of the Hamiltonian for the ensemble to make sense. A level tensor network representing the ensemble can be constructed in the following way: Consider the original compactified-time tensor network representing the ensemble without fixed particle number. Turn each array tensor into a level tensor with only level $0$ non-zero. Consider the hyperplane $S\times 0$ (where $S$ is the space, and $0$ represents a point in compactified time), and all the bonds this hyperplane cuts. At every such bond intersection, insert the following $2$-index level tensor: Decompose the $U(1)$-representation into a sum of components, each corresponding to a fixed irrep. Take the irrep $l$ component of the representation as the level-$l$ component of the level tensor. Here, we assumed that all irreps $l$ are positive integers, i.e., ``there are no anti-particles''.

Tensor-network models can also formalize perturbation theory calculations in a very neat way: Consider an array tensor-network model $M_0$, and a (first-order) perturbation $M_1$. The perturbed model with perturbation strength $\epsilon$ is given by $M_0+\epsilon M_1$, recalling that $M_0$ and $M_1$ are just collections of tensors, and $+$ denotes element-wise addition of the array tensors. The evaluation of a network $N$ for $M_0+\epsilon M_1$ for small $\epsilon$ is then given by
\begin{equation}
\begin{gathered}
\operatorname{Eval}(N,M_0+\epsilon M_1)\\
= \operatorname{Eval}(N,M_0)+\epsilon \sum_{a\in N} \operatorname{Eval}(N_a, M_0\sqcup M_1)+ \mathcal{O}(\epsilon^2)\;.
\end{gathered}
\end{equation}
Here, $a$ runs over all atoms of the network $N$, and $N_a$ is the network where we take the $M_1$ tensor for $a$ and the $M_0$ tensor for all other atoms.

We can combine $M_0$ and $M_1$ into a model in array level tensors for the $n=2$ counting algebra, by taking $M_0$ at level $0$ and $M_1$ at level $1$. As we have seen in Section~\ref{sec:level_special_dependencies}, the evaluation of $N$ for this model yields a level tensor, which has the $0$th order approximation at level $0$ and the $1$st order approximation at level $1$. If we use the analogous construction for the counting algebra for a higher $n$, the evaluation contains all perturbations up to the $n-1$th order in its different levels.

As we have seen in Section~\ref{sec:level_mappings}, complex array tensors can be seen as level tensors for real array tensors and the complex number algebra. Combining this with the Frobenification mapping, we get a mapping from complex array tensors to real array tensors which we'll refer to as \emph{realification}. Recall that the mapping is not a full mapping but only commutes with the evaluation of connected networks. Apart from that, the mapping gives a way to emulate complex tensors with real tensors, which involves multiplying all bond dimensions by $2$. We see that complex numbers are not a fundamental ingredient of quantum mechanics but merely a matter of convention. The fact that real quantum mechanics is universal for quantum computation has been known for a long time \cite{Rudolph2002}.

To be more precise, realification is a mapping from complex array tensors to real array tensors with an artificial dual, i.e., the result of the mapping applied to a tensor depends on a choice of input/output for each index. Such an input/output structure equips every network with an \emph{orientation}. In Eq.~\eqref{eq:level_dual_homomorphor}, we saw that swapping input/output is done via the Frobenius form, which for the complex number algebra is complex conjugation. Likewise, swapping input/output for all indices of a realified tensor is the same as complex conjugation of the underlying array tensor. On the combinatorial side, swapping input/output means changing the orientation of the network. Thus, if the realified model is orientation independent, then the underlying complex model is invariant under simultaneous complex conjugation and orientation reversal. The latter is exactly the \emph{Hermiticity} condition for imaginary-time evolution tensor networks formalizing ground state physics.

\section{Projective tensors}
\subsection{Motivation}
Projective tensors are a particularly simple way to construct a new tensor type from an arbitrary ground tensor type by introducing an equivalence relation, namely equality up to products with scalars. In other words, they are tensors modulo a normalization.

Physically, such tensor types are natural in any theory with a statistical interpretation where predictions are normalized probability distributions over measurement outcomes, that is, real array tensors modulo normalization by a scalar. Instead of imposing normalization in the final probability distribution obtained from a model, we can already disregard the normalization at the level of the individual tensors.
\subsection{Definition}
\begin{myobs}
The scalars of a tensor type $\calg$ with unit 1-data $1$ are elements in $\dat_1(1)$. They form a commutative semigroup under
\begin{equation}
R\cdot S = U(R\otimes S)\;,
\end{equation}
with $R,S\in \dat_1(1)$. The unitor $U$ can be dropped if $\calg$ is strictly unital. If $\calg$ has a trivial tensor, then the latter acts as a unit of the semigroup making it into a commutative monoid. If we further restrict to the subset of scalars which are invertible we get a commutative group. The subset is closed as $a^{-1}\otimes b^{-1}$ is an inverse of $a\otimes b$. For fixed 0-data $a$, there is an action of the semigroup/monoid/group of scalars on the 1-data $A\in \dat_1(a)$ via
\begin{equation}
\label{eq:projective_scalar_action}
S\cdot A = U(S\otimes A)\;.
\end{equation}
\end{myobs}

\begin{mydef}
\tdef{Projective tensors}{projective_tensor} are the following tensor type:
\begin{itemize}
\item They can be defined for any choice of 1) a ground tensor type $\calg$ with a unit 0-data and trivial tensor, and 2) a subgroup $\mathcal{S}$ of the group of invertible scalars.
\item The flavor is that of $\calg$, that is, projective tensors are strictly associative if $\calg$ is, and so on.
\item The 0-data is that of $\calg$.
\item A 1-data $A\in \dat_1(a)$ is an equivalence class of ground type 1-data under the action of scalars in $\mathcal{S}$ from Eq.~\eqref{eq:projective_scalar_action},
\begin{equation}
\begin{gathered}
A\in \dat_1^\calg(a)/\sim\;,\\
A_1\sim A_2\in \dat_1^\calg(a) \quad \text{if}\\
A_1= S\cdot A_2 = U(S \otimes A_2), \quad \text{for some} \quad S\in \mathcal{S}\;.
\end{gathered}
\end{equation}
In terms of the graphical calculus of $\calg$, the latter equation becomes
\begin{equation}
\begin{tikzpicture}
\node[roundbox](0)at(0,0){$A_2$};
\draw (0.east)edge[ind=a]++(0.3,0);
\end{tikzpicture}
=
\begin{tikzpicture}
\node[roundbox](0)at(0,0){$A_1$};
\node[roundbox](h)at(0,0.8){$S$};
\draw (0.east)edge[ind=a]++(0.3,0);
\end{tikzpicture}\;.
\end{equation}
\item The 2-functions are the 2-functions of $\calg$ acting on the corresponding equivalence classes.
\end{itemize}
\end{mydef}

\begin{myobs}
Let us quickly check whether the previous definitions make sense. First, $\sim$ is an equivalence relation, as it is defined via an action of $\mathcal{S}$ on $\dat_1(a)$. Second, $\sim$ is compatible with all 2-functions due to the 2-axioms of $\calg$. When expressed in the graphical calculus of $\calg$, the compatibility becomes trivial. Written out explicitly for, e.g., the tensor product, we have
\begin{equation}
\begin{multlined}
A_1\sim A_2 \quad \text{and}\quad B_1\sim B_2\\
\Rightarrow A_2\otimes B_2=U(S_A\otimes A_1)\otimes U(S_B\otimes B_1) \\
=U(U(S_A\otimes S_B)\otimes (A_1 \otimes B_1))\\ \quad \Rightarrow A_1\otimes B_1\sim A_2\otimes B_2\;.
\end{multlined}
\end{equation}
\end{myobs}

\subsection{Specific dependencies}

\begin{myexmp}
Consider projective tensors where the ground type is array tensors. The monoid of scalars is given by the product and multiplicative unit of the depending commutative semiring. Consider the following examples of subgroups of invertible semiring elements:
\begin{itemize}
\item Every field defines a commutative semiring. The subgroup of all invertible elements is given by the set of all non-zero elements. Specific examples for this are the subgroup of non-zero real or complex numbers.
\item The positive reals numbers form a multiplicative subgroup of the group of all invertible real or complex numbers.
\item $\pm 1$ form a $\zz_2$ subgroup of the invertible real or complex numbers.
\item The complex numbers of modulus $1$ form a $U(1)$ subgroup of all invertible complex numbers.
\item For any positive integer $n$, the $n$th complex roots of unity form a $\zz_n$ subgroup of all invertible complex numbers.
\end{itemize}
\end{myexmp}

\begin{myexmp}
Consider level tensors as ground type. The scalars are one-index tensors of the ground type of the level tensors. If the ground type of the level tensors is real array tensors, then the (invertible) scalars are (invertible) elements of the underlying commutative unital associative algebra.

Consider the case where the latter is semi-simple, i.e., equivalent to a direct sum of trivial and complex number algebras. The group of all invertible elements is just the product of the group of real/complex invertible numbers for each component. We can construct subgroups that couple the different components, for example, there's a $\zz_2$ subgroup, whose non-trivial element is represented by $-1$ (as real/complex number) on each component.

For non-semisimple algebras, things are a little more interesting. As an example, consider the $n=1$ counting algebra. An vector $(a,b)$ of the algebra is invertible if $a\neq 0$, with the inverse given by $(1/a, -b/a^2)$. For every non-zero $x$, a non-trivial subgroup of all invertible elements is given by $\zz \ni n \rightarrow (1,nx)$.
\end{myexmp}

\subsection{Mappings}
The fact that projective tensors ``neglect scalar prefactors'' can be formalized as a tensor mapping.
\begin{mydef}
The \tdef{projective mapping}{projective_mapping} is the following tensor mapping from ground type tensors to projective tensors:
\begin{itemize}
\item The homomorphor 2-functions are strict.
\item The mapping of 0-data is trivial.
\item The mapping on 1-data is given by mapping a representative to the corresponding equivalence class.
\end{itemize}
\end{mydef}

\begin{myrem}
The direct sum mapping for array tensors is lost for projective array tensors. This is because the equivalence under scalar multiplication is not compatible with the direct sum. Consider two representatives of projective array 1-data $A\in \dat_1(a)$ and $B\in \dat_1(b)$ and a second representative $A'=S_A\otimes A$. If both $A$ and $B$ are non-zero, we find that for any scalar $S_{AB}$,
\begin{equation}
A \oplus B \neq S_{AB}\cdot (A'\oplus B)\;.
\end{equation}
\end{myrem}

\subsection{Use in physics}
\label{sec:projective_physical}

Consider a classical thermal Gibbs state for a local Hamiltonian, such as the classical Ising model. As we have explained in Section~\ref{sec:array_physics}, this can be formalized as a tensor-network model. This means that any probe of such a model by measurements can be translated into a tensor network, such that the evaluation of the tensor network yields the predicted probability distribution over the measurement outcomes. However, the positive real array tensor resulting from the evaluation will in general not be normalized. In order to make it an actual probability distribution, we will have to normalize it by hand. As we choose the normalization by hand anyways, we might as well work with real array tensors modulo normalization from the beginning on. That is, we can use projective positive real array tensors as tensor type.

The analogous holds for quantum systems. On the pure-state level, we can work with complex array tensors modulo arbitrary invertible complex numbers, including amplitude and phase. This is not much different from that in the conventional formalism, where state vectors are said to be ``physical'' only up to a complex phase, even when normalized in amplitude.

Note that for models with a flow if time, the ``problem'' of normalization can be solved more elegantly. In this case, we use a symmetric monoidal category with a \emph{terminal object}, the latter imposing the normalization. E.g., a classical stochastic time evolution such as in probabilistic cellular automata can be formalized using the terminal symmetric monoidal category of stochastic matrices.

For systems without flow of time such as thermal or ground state physics, however, we need to use tensor types which aren't compatible with the definition of a terminal object. In this case, ``modding out'' the normalization is the best we can do.

Surely, projective tensors do not have an immediate use in practice for concrete computations, as we would anyways work with representatives which are ground type tensors. However, they help to understand things in a very systematic way. For example, consider models of some liquid in projective tensors. Those are equivalent to models in the ground type for which the equations only need to hold up to scalar prefactors. This might be relevant for defining fixed-point models for topological phases with a chiral anomaly \cite{universal_state_sums}.


\section{Symmetric tensors}
\label{sec:symmetric}
\subsection{Motivation}
Symmetric tensors are a construction that generates a new tensor type by restricting some ground tensor type $\calg$ by a \emph{symmetry constraint}, stating that the 1-data is invariant under the representation of a group or Hopf algebra. We provide different axiomatizations of what is meant by symmetry, via groups, Hopf algebras, and gradings.

Physically, symmetric tensors describe systems with a global on-site symmetry, such as spin-, particle-number, or more abstract finite-group symmetries.

\subsection{Definition: Via group representation}
\label{sec:group_symmetric}
Let us start with the simplest definition of symmetric tensors, via group symmetries. We first need to define a liquid whose models are the group representations.

\begin{mydef}
For every group $G$, \tdef{group representations}{group_representation} are the following liquid:
\begin{itemize}
\item There is one single binding.
\item For every group element $g\in G$, there is one 2-index element,
\begin{equation}
\begin{tikzpicture}
\atoms{representation,lab={t=g,p=-90:0.4}}{g/p={0,0}}
\draw (g-l)--++(-0.4,0) (g-r)--++(0.4,0);
\end{tikzpicture}\;.
\end{equation}
If $\calg$ has duals, then the index on the white side in an input and the index on the black side is an output.
\item For every pair of group elements $g,h\in G$, there is a move
\begin{equation}
\begin{tikzpicture}
\atoms{lab={t=g,p=-90:0.4},representation}{g/p={0,0}}
\atoms{lab={t=h,p=-90:0.4},representation}{h/p={1,0}}
\draw (g-r)--(h-l) (g-l)edge[ind=a]++(-0.4,0) (h-r)edge[ind=b]++(0.4,0);
\end{tikzpicture}=
\begin{tikzpicture}
\atoms{lab={t=gh,p=-90:0.4},representation}{g/p={0,0}}
\draw (g-l)edge[ind=a]++(-0.4,0) (g-r)edge[ind=b]++(0.4,0);
\end{tikzpicture}\;.
\end{equation}
\item We also have the move
\begin{equation}
\begin{tikzpicture}
\atoms{lab={t=1,p=-90:0.4},representation}{g/p={0,0}}
\draw (g-l)edge[ind=a]++(-0.4,0) (g-r)edge[ind=b]++(0.4,0);
\end{tikzpicture}=
\begin{tikzpicture}
\draw[startind=a,ind=b] (0,0)--(1,0);
\end{tikzpicture}\;,
\end{equation}
where $1$ denotes the unit of the group.
\end{itemize}
\end{mydef}

\begin{mydef}
One can define a unit, dual and product operation for the representations of a fixed group:
\begin{itemize}
\item The \tdef{trivial representation}{grouprep_trivial_representation} associates the trivial tensor of $\mathcal{G}$ for every $g\in G$,
\begin{equation}
\begin{tikzpicture}
\atoms{lab={t=1:g,p=-90:0.4},representation}{g/p={0,0}}
\draw (g-l)edge[ind=a]++(-0.4,0) (g-r)edge[ind=b]++(0.4,0);
\end{tikzpicture}=
\hspace{2cm}\;.
\end{equation}
Note that the label before the colon is the name of the representation, whereas the label after the colon is the name of the group element.
\item The \tdef{dual representation}{grouprep_dual_representation} is given by exchanging the input with the output and inverting the group element,
\begin{equation}
\begin{tikzpicture}
\atoms{lab={t=R^*:g,p=-90:0.4},representation}{g/p={0,0}}
\draw (g-l)edge[ind=a]++(-0.4,0) (g-r)edge[ind=b]++(0.4,0);
\end{tikzpicture}=
\begin{tikzpicture}
\atoms{lab={t=R:g^{-1},p=-90:0.4},representation}{g/p={0,0}}
\draw (g-l)edge[ind=b]++(-0.4,0) (g-r)edge[ind=a]++(0.4,0);
\end{tikzpicture}\;.
\end{equation}
\item The \tdef{tensor product}{grouprep_tensor_product} of two representations $R$ and $S$ is given by
\begin{equation}
\begin{tikzpicture}
\atoms{lab={t=R\otimes S:g,p=-90:0.4},representation}{g/p={0,0}}
\draw (g-l)edge[ind=aa']++(-0.4,0) (g-r)edge[ind=bb']++(0.4,0);
\end{tikzpicture}=
\begin{tikzpicture}
\atoms{lab={t=R:g,p=-90:0.4},representation}{g/p={0,0}}
\atoms{lab={t=S:g,p=-90:0.4},representation}{h/p={0,-0.9}}
\draw (g-l)edge[ind=a]++(-0.4,0) (g-r)edge[ind=b]++(0.4,0);
\draw (h-l)edge[ind=a']++(-0.4,0) (h-r)edge[ind=b']++(0.4,0);
\end{tikzpicture}\;.
\end{equation}
\end{itemize}
\end{mydef}

\begin{mydef}
\tdef{Group-symmetric tensors}{group_symmetric_tensor} are the following tensor type:
\begin{itemize}
\item They can be defined for any ground tensor type $\mathcal{G}$ with symmetric contraction and a group $G$.
\item The flavor depends on the ground type. They always need a dual. (They can only be defined without dual if $\calg$ doesn't have a dual and $G=\mathbb{Z}_2^n$.)
\item A 0-data is given by a representation of $G$ in $\calg$, thus it consists of both 0-data and 1-data of $\calg$.
\item The 1-functions are given by the trivial, dual and tensor product representation defined above.
\item A 1-data $A\in \dat_1(R)$ is a 1-data $A^\calg\in \dat_1^\calg(R_0)$, where $R_0$ is the 0-data of the representation $R$. It has to fulfill the following \tdef{symmetry constraint}{group_symmetry_constraint}, which we can express within the graphical calculus of $\calg$,
\begin{equation}
\begin{tikzpicture}
\atoms{labbox=$A$}{x/}
\atoms{lab={t=R:g,p=-90:0.4},representation}{r/p={0.8,0}}
\draw (x.east)--(r-l) (r-r)edge[ind=i]++(0.3,0);
\end{tikzpicture}=
\begin{tikzpicture}
\atoms{labbox=$A$}{x/}
\draw (x.east)edge[ind=i]++(0.3,0);
\end{tikzpicture}\;,
\end{equation}
for all $g\in G$.
\item The 2-functions are those of the ground type $\mathcal{G}$.
\end{itemize}
\end{mydef}

\begin{myobs}
In order to check whether the above definition makes sense, we need to check that the 2-functions of $\mathcal{G}$ are compatible with the symmetry constraint. For most 2-functions (including all invertible ones), this compatibility is trivial when expressed in the diagrammatic calculus of the ground tensor type. This is also true for the compatibility with the tensor product,
\begin{equation}
\begin{gathered}
\begin{tikzpicture}
\atoms{labbox=$A\otimes B$}{x/}
\atoms{lab={t=R\otimes S:g,p=-90:0.4},representation}{r/p={1.3,0}}
\draw (x.east)--(r-l) (r-r)edge[ind=ab]++(0.3,0);
\end{tikzpicture}=
\begin{tikzpicture}
\atoms{labbox=$A$}{x/}
\atoms{lab={t=R:g,p=-90:0.4},representation}{r/p={0.8,0}}
\draw (x.east)--(r-l) (r-r)edge[ind=a]++(0.3,0);
\atoms{labbox=$B$}{y/p={0,-0.9}}
\atoms{lab={t=S:g,p=-90:0.4},representation}{s/p={0.8,-0.9}}
\draw (y.east)--(s-l) (s-r)edge[ind=b]++(0.3,0);
\end{tikzpicture}\\
=
\begin{tikzpicture}
\atoms{labbox=$A$}{x/}
\draw (x.east)edge[ind=a]++(0.3,0);
\atoms{labbox=$B$}{y/p={0,-0.8}}
\draw (y.east)edge[ind=b]++(0.3,0);
\end{tikzpicture}=
\begin{tikzpicture}
\atoms{labbox=$A\otimes B$}{x/}
\draw (x.east)edge[ind=ab]++(0.3,0);
\end{tikzpicture}\;.
\end{gathered}
\end{equation}
The compatibility of the contraction is the only place where the moves of the liquid are needed,
\begin{equation}
\label{eq:contraction_symmetry_constraint}
\begin{gathered}
\begin{tikzpicture}
\atoms{labbox=$A$,bdastyle={wid=1.6}}{x/}
\draw[rounded corners] ([xshift=-0.4cm]x.north)--++(0,0.3)-|([xshift=0.4cm]x.north);
\atoms{lab={t=S:g,p=-90:0.35},representation}{r/p={1.4,0}}
\draw (x.east)--(r-l) (r-r)edge[ind=a]++(0.3,0);
\end{tikzpicture}=
\begin{tikzpicture}
\atoms{labbox=$A$,bdastyle={wid=1.6}}{x/}
\atoms{lab={t=R:1,p=90:0.35},representation}{r1/p={0,0.8}}
\draw (r1-l)edge[rounded corners,-|]([xshift=-0.4cm]x.north) (r1-r)edge[rounded corners,-|]([xshift=0.4cm]x.north);
\atoms{lab={t=S:g,p=-90:0.35},representation}{r/p={1.4,0}}
\draw (x.east)--(r-l) (r-r)edge[ind=a]++(0.3,0);
\end{tikzpicture}\\=
\begin{tikzpicture}
\atoms{labbox=$A$,bdastyle={wid=1.6}}{x/}
\atoms{rotlab,lab={t=R:g,p=180:0.6},rot=90,representation}{r1/p={-0.4,0.8}}
\atoms{rotlab, lab={t=R:g^{-1},p=0:0.75},rot=-90,representation}{r2/p={0.4,0.8}}
\draw (r1-l)--([xshift=-0.4cm]x.north) (r2-r)--([xshift=0.4cm]x.north);
\draw[rounded corners] (r1-r)--++(0,0.4)-|(r2-l);
\atoms{lab={t=S:g,p=-90:0.35},representation}{r/p={1.4,0}}
\draw (x.east)--(r-l) (r-r)edge[ind=a]++(0.3,0);
\end{tikzpicture}=
\begin{tikzpicture}
\atoms{labbox=$A$,bdastyle={wid=1.6}}{x/}
\atoms{rotlab,lab={t=R:g,p=180:0.6},rot=90,representation}{r1/p={-0.4,0.8}}
\atoms{rotlab,lab={t=R^*:g,p=0:0.7},rot=90,representation}{r2/p={0.4,0.8}}
\draw (r1-l)--([xshift=-0.4cm]x.north) (r2-l)--([xshift=0.4cm]x.north);
\draw[rounded corners] (r1-r)--++(0,0.4)-|(r2-r);
\atoms{lab={t=S:g,p=-90:0.35},representation}{r/p={1.4,0}}
\draw (x.east)--(r-l) (r-r)edge[ind=a]++(0.3,0);
\end{tikzpicture}\\=
\begin{tikzpicture}
\atoms{labbox=$A$,bdastyle={wid=1.6}}{x/}
\draw[rounded corners] ([xshift=-0.4cm]x.north)--++(0,0.3)-|([xshift=0.4cm]x.north);
\draw (x.east)edge[ind=a]++(0.3,0);
\end{tikzpicture}\;.
\end{gathered}
\end{equation}
\end{myobs}

\subsection{Definition: Via \texorpdfstring{$\zz_2$}{Z2}-extended group representation}
\label{sec:z2_extended_symmetric}
In this section we generalize group symmetries to ground tensor types with asymmetric contraction. To this end we need to introduce some extra structure, namely there has to be an involutive element of the group that is represented by the backwards-directed identity.

\begin{mydef}
A $\zz_2$-extended group is a group $G$ together with $\phi\in G$ such that $\phi^2=1$. For any $\zz_2$-extended group, \tdef{$\zz_2$-extended group representations}{zg_group_representation} are the following liquid.
\begin{itemize}
\item The tensor type flavor does not need to have a symmetric contraction.
\item As for ordinary group representations, there is one 2-index element for each group element.
\item The version of the main move with bond directions is
\begin{equation}
\begin{tikzpicture}
\atoms{lab={t=g,p=-90:0.4},representation}{g/p={0,0}}
\atoms{lab={t=h,p=-90:0.4},representation}{h/p={1,0}}
\draw (g-r)edge[arr=+](h-l) (g-l)edge[ind=a]++(-0.4,0) (h-r)edge[ind=b]++(0.4,0);
\end{tikzpicture}=
\begin{tikzpicture}
\atoms{lab={t=gh,p=-90:0.4},representation}{g/p={0,0}}
\draw (g-l)edge[ind=a]++(-0.4,0) (g-r)edge[ind=b]++(0.4,0);
\end{tikzpicture}\;.
\end{equation}
\item The version of the identity move is
\begin{equation}
\begin{tikzpicture}
\atoms{lab={t=1,p=-90:0.4},representation}{g/p={0,0}}
\draw (g-l)edge[ind=a]++(-0.4,0) (g-r)edge[ind=b]++(0.4,0);
\end{tikzpicture}=
\begin{tikzpicture}
\draw[] (0,0)edge[startind=a,ind=b,arr=+](1,0);
\end{tikzpicture}\;.
\end{equation}
\item Additionally, the element for $\phi$ has to be the backwards identity tensor,
\begin{equation}
\begin{tikzpicture}
\atoms{lab={t=\phi,p=-90:0.4},representation}{g/p={0,0}}
\draw (g-l)edge[ind=a]++(-0.4,0) (g-r)edge[ind=b]++(0.4,0);
\end{tikzpicture}=
\begin{tikzpicture}
\draw (0,0)edge[startind=a,ind=b,arr=-](1,0);
\end{tikzpicture}\;.
\end{equation}
\end{itemize}
The moves are consistent with $\phi^2=1$ due to the 2-axiom in Eq.~\eqref{eq:identity_opposite_square}.
\end{mydef}

\begin{mydef}
$\zz_2$-extended group representations will form the 0-data of $\zz_2$-extended-group-symmetric tensors. Trivial representation and tensor product of representations are the same as for group representations, only the dual operation is different,
\begin{equation}
\begin{tikzpicture}
\atoms{lab={t=R^*:g,p=-90:0.4},representation}{g/p={0,0}}
\draw (g-l)edge[ind=a]++(-0.4,0) (g-r)edge[ind=b]++(0.4,0);
\end{tikzpicture}=
\begin{tikzpicture}
\atoms{lab={t=R:g^{-1}\phi,p=-90:0.4},representation}{g/p={0,0}}
\draw (g-l)edge[ind=b]++(-0.4,0) (g-r)edge[ind=a]++(0.4,0);
\end{tikzpicture}\;.
\end{equation}
\end{mydef}

\begin{mydef}
\tdef{Super-group-symmetric tensors}{sg_symmetric_tensor} are the following tensor type:
\begin{itemize}
\item They can be defined for any ground tensor type $\mathcal{G}$ and $\zz_2$-extended group $(G,\phi)$.
\item The 0-data are $\zz_2$-extended group representations (of $(G,\phi)$ in $\mathcal{G}$).
\item Similarly to group-symmetric tensors, the 1-data $A\in \dat_1(R)$ is a tensor of type $\mathcal{G}$ obeying the symmetry constraint. The only difference is that we have to specify bond directions,
\begin{equation}
\begin{tikzpicture}
\atoms{labbox=$A$}{x/}
\atoms{lab={t=R:g,p=-90:0.4},representation}{r/p={0.8,0}}
\draw (x-r)edge[cdir=+](r-l) (r-r)edge[ind=a]++(0.3,0);
\end{tikzpicture}=
\begin{tikzpicture}
\atoms{labbox=$A$}{x/}
\draw (x-r)edge[ind=a]++(0.3,0);
\end{tikzpicture}\;.
\end{equation}
\item The 2-functions are those of $\mathcal{G}$.
\end{itemize}
\end{mydef}

\begin{myobs}
The modified dual is needed for the symmetry constraint to be consistent with contraction, even in the presence of bond directions,
\begin{equation}
\label{eq:extended_contraction_symmetry_constraint}
\begin{gathered}
\begin{tikzpicture}
\atoms{labbox=$A$,bdastyle={wid=1.6}}{x/}
\draw[rounded corners,cdir=+] ([xshift=-0.4cm]x.north)--++(0,0.3)-|([xshift=0.4cm]x.north);
\atoms{lab={t=S:g,p=-90:0.35},representation}{r/p={1.4,0}}
\draw (x.east)edge[cdir=+](r-l) (r-r)edge[ind=a]++(0.3,0);
\end{tikzpicture}=
\begin{tikzpicture}
\atoms{labbox=$A$,bdastyle={wid=1.6}}{x/}
\atoms{lab={t=R:\phi,p=90:0.35},representation}{r1/p={0,0.8}}
\draw (r1-l)edge[rounded corners,-|,arr=-]([xshift=-0.4cm]x.north) (r1-r)edge[rounded corners,-|,arr=-]([xshift=0.4cm]x.north);
\atoms{lab={t=S:g,p=-90:0.35},representation}{r/p={1.4,0}}
\draw (x.east)edge[cdir=+](r-l) (r-r)edge[ind=a]++(0.3,0);
\end{tikzpicture}\\=
\begin{tikzpicture}
\atoms{labbox=$A$,bdastyle={wid=1.6}}{x/}
\atoms{lab={t=R:g,p=180:0.6},rot=90,representation}{r1/p={-0.4,0.8}}
\atoms{lab={t=R:g^{-1}\phi,p=0:0.9},rot=-90,representation}{r2/p={0.4,0.8}}
\draw (r1-l)edge[arr=-]([xshift=-0.4cm]x.north) (r2-r)edge[arr=-]([xshift=0.4cm]x.north);
\draw[rounded corners,cdir=+] (r1-r)--++(0,0.4)-|(r2-l);
\atoms{lab={t=S:g,p=-90:0.35},representation}{r/p={1.4,0}}
\draw (x.east)edge[cdir=+](r-l) (r-r)edge[ind=a]++(0.3,0);
\end{tikzpicture}=
\begin{tikzpicture}
\atoms{labbox=$A$,bdastyle={wid=1.6}}{x/}
\atoms{lab={t=\rotatebox{90}{$R:g$},p=180:0.35},rot=90,representation}{r1/p={-0.4,0.8}}
\atoms{lab={t=\rotatebox{90}{$R^*:g$},p=0:0.35},rot=90,representation}{r2/p={0.4,0.8}}
\draw (r1-l)edge[arr=-]([xshift=-0.4cm]x.north) (r2-l)edge[arr=-]([xshift=0.4cm]x.north);
\draw[rounded corners,cdir=+] (r1-r)--++(0,0.4)-|(r2-r);
\atoms{lab={t=S:g,p=-90:0.35},representation}{r/p={1.4,0}}
\draw (x.east)edge[cdir=+](r-l) (r-r)edge[ind=a]++(0.3,0);
\end{tikzpicture}\\=
\begin{tikzpicture}
\atoms{labbox=$A$,bdastyle={wid=1.6}}{x/}
\draw[rounded corners,cdir=+] ([xshift=-0.4cm]x.north)--++(0,0.3)-|([xshift=0.4cm]x.north);
\draw (x.east)edge[ind=a]++(0.3,0);
\end{tikzpicture}\;,
\end{gathered}
\end{equation}
where we used Eq.~\eqref{eq:identity_opposite_square} in the first step.
\end{myobs}

\begin{myobs}
The symmetry constraint for $g=\phi$ holds automatically due to Eq.~\eqref{eq:inverse_identity}.
\end{myobs}

\subsection{Definition: Via co-commutative Hopf algebras}
In the previous definitions, we formalized group representations as liquid models in $\calg$, whereas the group itself was a fixed structure independent of $\calg$. In this section we will formalize the group axioms themselves as a liquid, namely co-commutative Hopf algebras. For not too simple $\calg$ such as array tensors, every finite group can be represented by a model of that liquid, such that the family of tensor types in this section is more general than the one in Section~\ref{sec:group_symmetric} for finite groups.

\begin{mydef}
\tdef{Hopf algebras}{hopf_algebra} are the following liquid extending unital associative algebras.
\begin{itemize}
\item There is an additional co-product and a co-unit element of the same shape as the product and unit, just that input indices are output indices and vice versa. To distinguish it from the product and unit, will draw them using filled instead of empty circles,
\begin{equation}
\begin{tikzpicture}
\atoms{coalgebra}{a/p={0,0}}
\draw (a)edge[mark={ar,s}]++(0,0.5) (a)--++(-30:0.5) (a)--++(-150:0.5);
\end{tikzpicture}\;,
\qquad
\begin{tikzpicture}
\atoms{coalgebra}{a/p={0,0}}
\draw (a)edge[mark={ar,s}]++(0,0.5);
\end{tikzpicture}\;.
\end{equation}
\item There is an additional $2$-index element called the \tdef{antipode}{antipode},
\begin{equation}
\begin{tikzpicture}
\atoms{antipode}{a/p={0,0}}
\draw (a-l)edge[dualind]++(-0.5,0) (a-r)--++(0.5,0);
\end{tikzpicture}\;.
\end{equation}
\item The co-product and co-unit can be identified with the product and unit, and we add the unitality and associativity moves of algebras with this identification.
\item Additionally, there are the following four \emph{bi-algebra} moves,
\begin{equation}
\label{eq:move_bialgebra1}
\begin{tikzpicture}
\atoms{algebra}{0/p={0,0}, 1/p={0,0.7}}
\atoms{coalgebra}{2/p={1,0}, 3/p={1,0.7}}
\draw (0)edge[mark={ar,s}](2) (0)edge[mark={ar,s}](3) (1)edge[mark={ar,s}](2) (1)edge[mark={ar,s}](3) (0)edge[ind=a]++(-0.5,0) (1)edge[ind=b]++(-0.5,0) (2)edge[mark={ar,s},ind=c]++(0.5,0) (3)edge[mark={ar,s},ind=d]++(0.5,0);
\end{tikzpicture}=
\begin{tikzpicture}
\atoms{coalgebra}{0/p={0,0}}
\atoms{algebra}{1/p={1,0}}
\draw (0)edge[mark={ar,s}](1) (0)edge[ind=a]++(-135:0.5) (0)edge[ind=b]++(135:0.5) (1)edge[mark={ar,s},ind=c]++(-45:0.5) (1)edge[mark={ar,s},ind=d]++(45:0.5);
\end{tikzpicture}\;,
\end{equation}
\item
\begin{equation}
\label{eq:move_bialgebra2}
\hspace{2cm}=
\begin{tikzpicture}
\atoms{algebra}{0/p={0,0}}
\atoms{coalgebra}{1/p={1,0}}
\draw (0)edge[mark={ar,s}](1);
\end{tikzpicture}\;,
\end{equation}
\item
\begin{equation}
\label{eq:move_bialgebra3}
\begin{tikzpicture}
\atoms{coalgebra}{0/p={1,0}, 1/p={1,0.7}}
\draw (0)edge[mark={ar,s},ind=a]++(0.5,0) (1)edge[mark={ar,s},ind=b]++(0.5,0);
\end{tikzpicture}=
\begin{tikzpicture}
\atoms{coalgebra}{0/p={0,0}}
\atoms{algebra}{1/p={1,0}}
\draw (0)edge[mark={ar,s}](1) (1)edge[mark={ar,s},ind=a]++(-45:0.5) (1)edge[mark={ar,s},ind=b]++(45:0.5);
\end{tikzpicture}\;,
\end{equation}
\item and
\begin{equation}
\label{eq:move_bialgebra4}
\begin{tikzpicture}
\atoms{algebra}{0/p={1,0}, 1/p={1,0.7}}
\draw (0)edge[ind=a]++(-0.5,0) (1)edge[ind=b]++(-0.5,0);
\end{tikzpicture}=
\begin{tikzpicture}
\atoms{algebra}{0/p={0,0}}
\atoms{coalgebra}{1/p={-1,0}}
\draw (0)edge[mark={ar,e}](1) (1)edge[ind=a]++(-135:0.5) (1)edge[ind=b]++(135:0.5);
\end{tikzpicture}\;.
\end{equation}
\item Moreover, there are the two \emph{Hopf} moves,
\begin{equation}
\label{eq:move_hopf}
\begin{gathered}
\begin{tikzpicture}
\atoms{coalgebra}{0/p={0,0}}
\atoms{algebra}{1/p={1,0}}
\atoms{antipode}{s/p={0.5,-0.5}}
\draw (0)edge[mark={ar,s},ind=a]++(-0.6,0) (1)edge[ind=b]++(0.6,0);
\draw[rounded corners,mark={ar,e}] (0)--++(0.2,0.5)--++(0.6,0)--(1);
\draw[rounded corners] (0)--++(0.2,-0.5)--(s-l);
\draw[rounded corners,mark={ar,s}] (1)--++(-0.2,-0.5)--(s-r);
\end{tikzpicture}=
\begin{tikzpicture}
\atoms{coalgebra}{0/p={0,0}}
\atoms{algebra}{1/p={1,0}}
\atoms{antipode}{s/p={0.5,0.5}}
\draw (0)edge[mark={ar,s},ind=a]++(-0.6,0) (1)edge[ind=b]++(0.6,0);
\draw[rounded corners,mark={ar,e}] (0)--++(0.2,-0.5)--++(0.6,0)--(1);
\draw[rounded corners] (0)--++(0.2,0.5)--(s-l);
\draw[rounded corners,mark={ar,s}] (1)--++(-0.2,0.5)--(s-r);
\end{tikzpicture}
\\=
\begin{tikzpicture}
\atoms{coalgebra}{0/p={0,0}}
\atoms{algebra}{1/p={0.8,0}}
\draw (0)edge[mark={ar,s},ind=a]++(-0.6,0) (1)edge[ind=b]++(0.6,0);
\end{tikzpicture}\;.
\end{gathered}
\end{equation}
\end{itemize}
\end{mydef}

\begin{myobs}
The bi-algebra moves might appear rather arbitrary, but they generate a systematic family of moves: For every two integers $n,m\geq 0$ the following networks 1) and 2) define a move with $n+m$ open indices. 1) is a network with $n$ algebra atoms with $m+1$ indices each and $m$ co-algebra atoms with $n+1$ indices each. There is a bond between the $x$th index of the $y$th algebra atom and the $y$th index of the $x$th co-algebra axiom. 2) is a network with $1$ algebra atom with $m+1$ indices and $1$ co-algebra atom with $n+1$ indices. There is one bond between the atoms. E.g., for $n=3,m=2$, we get
\begin{equation}
\begin{tikzpicture}
\atoms{algebra}{0/p={0,0}, 1/p={0,0.7}, 2/p={0,1.4}}
\atoms{coalgebra}{3/p={1,0.35}, 4/p={1,1.05}}
\draw (0)edge[mark={ar,s}](3) (0)edge[mark={ar,s}](4) (1)edge[mark={ar,s}](3) (1)edge[mark={ar,s}](4) (2)edge[mark={ar,s}](3) (2)edge[mark={ar,s}](4) (0)edge[ind=a]++(-0.5,0) (1)edge[ind=b]++(-0.5,0) (2)edge[ind=c]++(-0.5,0) (3)edge[mark={ar,s},ind=d]++(0.5,0) (4)edge[mark={ar,s},ind=e]++(0.5,0);
\end{tikzpicture}=
\begin{tikzpicture}
\atoms{coalgebra}{0/p={0,0}}
\atoms{algebra}{1/p={1,0}}
\draw (0)edge[mark={ar,s}](1) (0)edge[ind=a]++(-135:0.5) (0)edge[ind=b]++(180:0.5) (0)edge[ind=c]++(135:0.5) (1)edge[mark={ar,s},ind=d]++(-45:0.5) (1)edge[mark={ar,s},ind=e]++(45:0.5);
\end{tikzpicture}\;.
\end{equation}
The $4$ bi-algebra moves correspond to $n=2$, $m=2$, and $n=0$, $m=0$, and $n=0$,$m=2$, and $n=2$,$m=1$.
\end{myobs}

\begin{myrem}
The following move can be derived from the moves of Hopf algebras,
\begin{equation}
\label{eq:move_hopf_antihom}
\begin{tikzpicture}
\atoms{algebra}{0/p={0,0}}
\atoms{antipode}{s1/p={0.8,0}}
\draw (0)edge[](s1-l) (s1-r)edge[ind=c]++(0.3,0) (0)edge[mark={ar,s},ind=a]++(135:0.6) (0)edge[mark={ar,s},ind=b]++(-135:0.6);
\end{tikzpicture}=
\begin{tikzpicture}
\atoms{algebra}{0/p={0,0}}
\atoms{antipode}{s0/p={-0.8,-0.4}, s1/p={-0.8,0.4}}
\draw (0)edge[ind=c]++(0.6,0) (s1-l)edge[ind=b]++(-0.3,0) (s0-l)edge[ind=a]++(-0.3,0);
\draw[rounded corners, mark={ar,s}] (0)--++(-0.4,-0.4)--(s0-r);
\draw[rounded corners, mark={ar,s}] (0)--++(-0.4,0.4)--(s1-r);
\end{tikzpicture}\;.
\end{equation}
This corresponds to the well-known fact that for every Hopf algebra, the antipode is a homomorphism between the algebra and the opposite algebra \cite{Sweedler1969}.
\end{myrem}

\begin{mydef}
For a \tdef{commutative}{hopf_commutative} Hopf algebra we add the commutativity move Eq.~\eqref{eq:product_commutativity} for the product. For a \tdef{co-commutative}{hopf_co_commutative} Hopf algebra, we add that move for the co-product. If we have commutativity (or co-commutativity), we will use a smaller circle to denote the product and unit (co-product and co-unit).
\end{mydef}

\begin{myrem}
The following move can be derived from the moves of a co-commutative Hopf algebra,
\begin{equation}
\label{eq:hopf_involutive}
\begin{tikzpicture}
\atoms{antipode}{0/p={0,0}, 1/p={1,0}}
\draw (0-r)edge[cdir=+](1-l) (0-l)edge[ind=a]++(-0.4,0) (1-r)edge[ind=b]++(0.4,0);
\end{tikzpicture}=
\begin{tikzpicture}
\draw (0,0)edge[startind=a,ind=b,cdir=+](1,0);
\end{tikzpicture}\;.
\end{equation}
This corresponds to the fact that for any co-commutative Hopf algebra, the antipode is \tdef{involutive}{hopf_involutive} \cite{Sweedler1969}.
\end{myrem}

So far we have defined co-commutative Hopf algebras as the liquid generalizing groups. We also need to define a liquid of representations of Hopf algebras.

\begin{mydef}
\tdef{Representations}{algebra_representation} are the following liquid extending the liquid of unital associative algebras:
\begin{itemize}
\item There is one additional binding which we'll call $r$.
\item There is one additional element,
\begin{equation}
\begin{tikzpicture}
\atoms{representation}{r/p={0,0}}
\draw (r-l)edge[ind=r,dualind]++(-0.4,0) (r-r)edge[ind=r]++(0.4,0) (r-t)edge[dualind]++(0,0.4);
\end{tikzpicture}\;.
\end{equation}
\item There are $2$ additional moves,
\begin{equation}
\label{eq:representation_ax1}
\begin{tikzpicture}
\atoms{representation}{r1/p={0,0}}
\atoms{representation}{r2/p={0.7,0}}
\draw (r1-r)--(r2-l) (r1-l)edge[ind=a]++(-0.4,0) (r2-r)edge[ind=b]++(0.4,0) (r1-t)edge[ind=x]++(0,0.4) (r2-t)edge[ind=y]++(0,0.4);
\end{tikzpicture}=
\begin{tikzpicture}
\atoms{representation}{r1/p={0,0}}
\atoms{algebra}{s/p={0,0.8}}
\draw (r1-l)edge[ind=a]++(-0.4,0) (r1-r)edge[ind=b]++(0.4,0) (s)edge[mark={ar,s},ind=x]++(135:0.5) (s)edge[mark={ar,s},ind=y]++(45:0.5) (r1-t)--(s);
\end{tikzpicture}\;,
\end{equation}
\item and
\begin{equation}
\begin{tikzpicture}
\atoms{representation}{r1/p={0,0}}
\atoms{algebra}{s/p={0,0.8}}
\draw (r1-l)edge[ind=a]++(-0.4,0) (r1-r)edge[ind=b]++(0.4,0) (r1-t)--(s);
\end{tikzpicture}=
\begin{tikzpicture}
\draw (0,0)edge[startind=a,ind=b](1,0);
\end{tikzpicture}\;.
\end{equation}
\end{itemize}
\end{mydef}

\begin{mydef}
The representations of a fixed Hopf algebra (or better, of the underlying unital associative algebra) will form the 0-data of Hopf-symmetric tensors. The co-unit and co-product are needed to define the unit, dual and product for this 0-data:
\begin{itemize}
\item The \tdef{trivial representation}{hopf_trivial_representation} associates the trivial 0-data to the binding $r$ and the co-unit to the representation element,
\begin{equation}
\begin{tikzpicture}
\atoms{representation}{r1/p={0,0}}
\node[below=0cm of r1-b]{$1$};
\draw (r1-l)edge[ind=a]++(-0.4,0) (r1-r)edge[ind=b]++(0.4,0) (r1-t)edge[ind=x]++(0,0.4);
\end{tikzpicture}=
\begin{tikzpicture}
\atoms{comcoalg}{s/p={0,0}}
\draw (s)edge[mark={ar,s},ind=x]++(0,0.5);
\end{tikzpicture}\;.
\end{equation}
The representation move with this definition can be derived from the bialgebra move in Eq.~\eqref{eq:move_bialgebra3}.
\item The \tdef{dual}{hopf_dual} $R^*$ of a representation $R$ is given by
\begin{equation}
\begin{tikzpicture}
\atoms{representation,lab={t=R^*,p=-90:0.35}}{r1/p={0,0}}
\draw (r1-l)edge[ind=a]++(-0.4,0) (r1-r)edge[ind=b]++(0.4,0) (r1-t)edge[ind=x]++(0,0.4);
\end{tikzpicture}=
\begin{tikzpicture}
\atoms{representation,lab={t=R,p=-90:0.35}}{r1/p={0,0}}
\atoms{rot=-90,antipode}{s/p={0,0.9}}
\draw (r1-l)edge[ind=b]++(-0.4,0) (r1-r)edge[ind=a]++(0.4,0) (s-l)edge[ind=x]++(90:0.3) (r1-t)--(s-r);
\end{tikzpicture}\;,
\end{equation}
not overlooking that the index labels $a$ and $b$ swap position. The representation move with this definition can be derived with the help of Eq.~\eqref{eq:move_hopf_antihom}.
\item The \tdef{tensor product}{hr_tensor_product} of two representations $R$ and $S$ is given by
\begin{equation}
\begin{tikzpicture}
\atoms{representation}{r1/p={0,0}}
\node[below=0cm of r1-b]{$R\otimes S$};
\draw (r1-l)edge[ind=aa']++(-0.6,0) (r1-r)edge[ind=bb']++(0.6,0) (r1-t)edge[ind=x]++(0,0.4);
\end{tikzpicture}=
\begin{tikzpicture}
\atoms{representation}{r1/p={0,0}, r2/p={0,1}}
\node[below=0cm of r1-b]{$R$};
\node[below=0cm of r2-b]{$S$};
\atoms{comcoalg}{s/p={1.4,0.9}}
\draw (r1-l)edge[ind=a]++(-0.4,0) (r1-r)edge[ind=b]++(0.4,0) (r2-l)edge[ind=a']++(-0.4,0) (r2-r)edge[ind=b']++(0.4,0);
\draw[rounded corners] (r1-t)--++(0,0.2)-|(s);
\draw[rounded corners] (r2-t)--++(0,0.2)-|(s);
\draw (s)edge[mark={ar,s},ind=x]++(0.4,0);
\end{tikzpicture}\;.
\end{equation}
The representation move with this definition can be derived from the move Eq.~\eqref{eq:move_bialgebra1}.
\end{itemize}
\end{mydef}

After introducing the necessary algebraic structures, we are now ready to define Hopf-symmetric tensors.
\begin{mydef}
\tdef{Hopf-symmetric tensors}{hopf_symmetric_tensor} are the following tensor type:
\begin{itemize}
\item They can be defined for every ground tensor type $\mathcal{G}$ and co-commutative Hopf algebra $H$ in $\mathcal{G}$. $\calg$ needs to have a symmetric contraction and identity.
\item The precise flavor depends on the flavor of the ground type. However, we always have a non-trivial dual.
\item The 0-data are representations of (the algebra part of) $H$.
\item A 1-data $A\in \dat_1(R)$ is a 1-data of $\calg$ with 0-data $R_{0r}$ (which is the 0-data associated to the binding $r$ by $R$). It has to fulfill the following \tdef{symmetry constraint}{hopf_symmetry_constraint}, which can be expressed in the graphical calculus of $\calg$,
\begin{equation}
\begin{tikzpicture}
\atoms{labbox=$A$}{x/}
\atoms{representation}{r/p={0.8,0}}
\draw (x.east)--(r-l) (r-r)edge[ind=a]++(0.3,0) (r-t)edge[ind=x]++(0,0.4);
\end{tikzpicture}=
\begin{tikzpicture}
\atoms{labbox=$A$}{x/}
\draw (x.east)edge[ind=a]++(0.3,0);
\atoms{comcoalg}{s/p={1.4,0}}
\draw (s)edge[mark={ar,s},ind=x]++(0,0.5);
\end{tikzpicture}
\;.
\end{equation}
\item The 2-functions are those of $\calg$.
\end{itemize}
\end{mydef}

\begin{myobs}
In order for the above definitions to make sense we need to check that all 2-functions of $\mathcal{G}$ are compatible with the symmetry constraint:
\begin{itemize}
\item The symmetry constraint is preserved by the commutor. This is due to the commutativity of the co-algebra, and is in the graphical calculus implied by the symmetry of the shape for the co-product.
\item The symmetry constraint is preserved by the unitor,
\begin{equation}
\label{eq:unitor_symmetry_constraint}
\begin{tikzpicture}
\atoms{labbox=$\mathbf{1}\otimes A$}{x1/}
\atoms{lab={t=1\otimes S,p=0:0.7},rot=90,representation}{r1/p={0,0.7}}
\draw (r1-l)--(x1.north);
\draw (r1-r)edge[ind=a]++(0,0.3);
\draw (r1-t)edge[ind=x]++(-0.3,0);
\end{tikzpicture}=
\begin{tikzpicture}
\atoms{labbox=$A$}{x2/p={1,0}}
\atoms{comcoalg}{s/p={0.3,1.7}, t/p={0,1.2}}
\atoms{lab={t=S,p=0:0.35},rot=90,representation}{r2/p={1,0.7}}
\draw (r2-l)--(x2.north);
\draw (r2-r)edge[ind=a]++(0,0.3);
\draw[rounded corners,mark={ar,s}] (t)--(s);
\draw[rounded corners] (r2-t)--++(-0.2,0)--++(0,0.7)--(s);
\draw (s)edge[mark={ar,s},ind=x]++(0,0.5);
\end{tikzpicture}=
\begin{tikzpicture}
\atoms{labbox=$A$}{x2/p={0.5,0}}
\draw (x2.north)edge[ind=a]++(0,0.3);
\atoms{comcoalg}{s/p={0.5,1.8}}
\atoms{comcoalg}{s1/p={0.1,1.3}}
\atoms{comcoalg}{s2/p={0.9,1.3}}
\draw (s)edge[mark={ar,e}](s1) (s)edge[mark={ar,e}](s2);
\draw (s)edge[mark={ar,s},ind=x]++(0,0.5);
\end{tikzpicture}=
\begin{tikzpicture}
\atoms{labbox=$A$}{x1/p={0.5,0}}
\draw (x1.north)edge[ind=a]++(0,0.3);
\atoms{comcoalg}{s/p={-0.2,0}}
\draw (s)edge[mark={ar,s},ind=x]++(0,0.5);
\end{tikzpicture}
\;.
\end{equation}
\item The symmetry constraint is preserved by the associator,
\begin{equation}
\label{eq:associator_symmetry_constraint}
\begin{tikzpicture}
\atoms{labbox=$A$,bdastyle={wid=2}}{x/}
\atoms{comcoalg}{s1/p={-0.3,1.8}, s2/p={0.3,2.2}}
\atoms{rot=90,representation}{r1/p={-0.8,0.8}, r2/p={0,0.8}, r3/p={0.8,0.8}}
\draw (r1-l)--([xshift=-0.8cm]x.north) (r2-l)--(x.north) (r3-l)--([xshift=0.8cm]x.north);
\draw (r1-r)edge[ind=a]++(0,0.3) (r2-r)edge[ind=b]++(0,0.3) (r3-r)edge[ind=c]++(0,0.3);
\draw[rounded corners] (r1-b)--++(0.2,0)--++(0,0.7)--(s1);
\draw[rounded corners] (r2-b)--++(0.2,0)--++(0,0.7)--(s1);
\draw[rounded corners] (r3-b)--++(0.2,0)--++(0,0.7)--(s2);
\draw (s1)edge[mark={ar,s}](s2)  (s2)edge[mark={ar,s},ind=x]++(0,0.5);
\end{tikzpicture}=
\begin{tikzpicture}
\atoms{labbox=$A$,bdastyle={wid=2}}{x/p={0,0.1}}
\atoms{comcoalg}{s1/p={-0.3,2.2}, s2/p={0.3,1.8}}
\atoms{rot=90,representation}{r1/p={-0.8,0.8}, r2/p={0,0.8}, r3/p={0.8,0.8}}
\draw (r1-l)--([xshift=-0.8cm]x.north) (r2-l)--(x.north) (r3-l)--([xshift=0.8cm]x.north);
\draw (r1-r)edge[ind=a]++(0,0.3) (r2-r)edge[ind=b]++(0,0.3) (r3-r)edge[ind=c]++(0,0.3);
\draw[rounded corners] (r1-b)--++(0.2,0)--++(0,0.7)--(s1);
\draw[rounded corners] (r2-b)--++(0.2,0)--++(0,0.7)--(s2);
\draw[rounded corners] (r3-b)--++(0.2,0)--++(0,0.7)--(s2);
\draw (s1)edge[mark={ar,e}](s2)  (s1)edge[mark={ar,s},ind=x]++(0,0.5);
\end{tikzpicture}
\;.
\end{equation}
This is due to the co-associativity move.
\item The symmetry constraint is preserved by the dual involutor,
\begin{equation}
\label{eq:involutor_symmetry_constraint}
\begin{tikzpicture}
\atoms{labbox=$A$,bdastyle={wid=1.5}}{x/p={0,0.1}}
\atoms{comcoalg}{s1/p={0.5,2.4}}
\atoms{rot=90,representation}{r1/p={-0.4,0.8}, r2/p={0.4,0.8}}
\atoms{rot=-90,antipode}{a0/p={0.8,1.2}, a1/p={0.8,1.8}}
\draw (r1-l)--([xshift=-0.4cm]x.north) (r2-l)--([xshift=0.4cm]x.north) (a0-l)--(a1-r) (a1-l)edge[](s1);
\draw (r1-r)edge[ind=a]++(0,0.3) (r2-r)edge[ind=b]++(0,0.3);
\draw[rounded corners] (r1-b)--++(0.2,0)--++(0,0.7)--(s1);
\draw[rounded corners] (r2-b)-|(a0-r);
\draw (s1)edge[mark={ar,s},ind=x]++(0,0.4);
\end{tikzpicture}=
\begin{tikzpicture}
\atoms{labbox=$A$,bdastyle={wid=1.5}}{x/p={0,0.1}}
\atoms{comcoalg}{s1/p={0.5,2}}
\atoms{rot=90,representation}{r1/p={-0.4,0.8}, r2/p={0.4,0.8}}
\draw (r1-l)--([xshift=-0.4cm]x.north) (r2-l)--([xshift=0.4cm]x.north);
\draw (r1-r)edge[ind=a]++(0,0.3) (r2-r)edge[ind=b]++(0,0.3);
\draw[rounded corners] (r1-b)--++(0.2,0)--++(0,0.7)--(s1);
\draw[rounded corners] (r2-b)--++(0.2,0)--++(0,0.7)--(s1);
\draw (s1)edge[mark={ar,s},ind=x]++(0,0.4);
\end{tikzpicture}
\;.
\end{equation}
This is due to the involutivity move.
\item The symmetry constraint is preserved by the dual automorphor,
\begin{equation}
\label{eq:homomorphor_symmetry_constraint}
\begin{gathered}
\begin{tikzpicture}
\atoms{labbox=$A$}{x1/}
\atoms{lab={t=A\otimes B^*\otimes C^*,p=0:1.2},rot=90,representation}{r1/p={0,0.7}}
\draw (r1-l)--(x1.north);
\draw (r1-r)edge[ind=abc]++(0,0.3);
\draw (r1-t)edge[ind=x]++(-0.3,0);
\end{tikzpicture}
=
\begin{tikzpicture}
\atoms{labbox=$A$,bdastyle={wid=2}}{x/p={0,0.1}}
\atoms{comcoalg}{s1/p={0,2}}
\atoms{representation}{{r0/p={-0.8,0.8},rot=90}, {r1/p={0,0.8},rot=-90}, {r2/p={0.8,0.8},rot=-90}}
\atoms{rot=-90,antipode}{a0/p={1.2,1.2}, a1/p={0.4,1.2}}
\draw (r0-l)--([xshift=-0.8cm]x.north) (r1-r)--([xshift=0cm]x.north) (r2-r)--([xshift=0.8cm]x.north) (a0-l)edge[](s1) (a1-l)edge[](s1);
\draw (r0-r)edge[ind=a]++(0,0.3) (r1-l)edge[ind=b]++(0,0.3) (r2-l)edge[ind=c]++(0,0.3);
\draw[rounded corners] (r2-t)-|(a0-r);
\draw[rounded corners] (r1-t)-|(a1-r);
\draw[rounded corners] (r0-b)--++(0.2,0)--++(0,0.7)--(s1);
\draw (s1)edge[mark={ar,s},ind=x]++(0,0.4);
\end{tikzpicture}\\
=
\begin{tikzpicture}
\atoms{labbox=$A$,bdastyle={wid=2}}{x/p={0,0.1}}
\atoms{comcoalg}{s1/p={0.2,2.7}, s2/p={0.7,1.7}}
\atoms{rot=-90,antipode}{a0/p={0.7,2.2}}
\atoms{representation}{{r1/p={-0.8,0.8},rot=90}, {r2/p={0,0.8},rot=-90}, {r3/p={0.8,0.8},rot=-90}}
\draw (r1-l)--([xshift=-0.8cm]x.north) (r2-r)--(x.north) (r3-r)--([xshift=0.8cm]x.north);
\draw (r1-r)edge[ind=a]++(0,0.3) (r2-l)edge[ind=b]++(0,0.3) (r3-l)edge[ind=c]++(0,0.3);
\draw[rounded corners] (r1-b)--++(0.2,0)--++(0,0.7)--(s1);
\draw[rounded corners] (r2-t)--++(0.2,0)--++(0,0.7)--(s2);
\draw[rounded corners] (r3-t)--++(0.2,0)--++(0,0.7)--(s2);
\draw (s2)edge[mark={ar,s}](a0-r) (a0-l)edge[](s1);
\draw (s1)edge[mark={ar,s},ind=x]++(0,0.5);
\end{tikzpicture}
=
\begin{tikzpicture}
\atoms{labbox=$A$}{x1/}
\atoms{lab={t=A\otimes (B\otimes C)^*,p=0:1.25},rot=90,representation}{r1/p={0,0.7}}
\draw (r1-l)--(x1.north);
\draw (r1-r)edge[ind=abc]++(0,0.3);
\draw (r1-t)edge[ind=x]++(-0.3,0);
\end{tikzpicture}
\;.
\end{gathered}
\end{equation}
This is due to the property of the antipode from Eq.~\eqref{eq:move_hopf_antihom}, co-associativity and co-commutativity.
\item The symmetry constraint is preserved by the trivial tensor,
\begin{equation}
\label{eq:trivial_symmetry_constraint}
\begin{tikzpicture}
\atoms{labbox=$\mathbf{1}$}{x1/}
\atoms{lab={t=1,p=0:0.35},rot=90,representation}{r1/p={0,0.7}}
\draw (r1-t)edge[ind=x]++(-0.3,0);
\end{tikzpicture}=
\begin{tikzpicture}
\atoms{labbox=$\mathbf{1}$}{x1/p={0,-0.7}}
\atoms{comcoalg}{s/p={0,0}}
\draw (s)edge[mark={ar,s},ind=x]++(-0.5,0);
\end{tikzpicture}
\;.
\end{equation}
\item The symmetry constraint is preserved by the tensor product,
\begin{equation}
\label{eq:product_symmetry_constraint}
\begin{gathered}
\begin{tikzpicture}
\atoms{labbox=$A\otimes B$}{x1/}
\atoms{lab={t=R\otimes S,p=0:0.7},rot=90,representation}{r1/p={0,0.7}}
\draw (r1-l)--(x1.north);
\draw (r1-r)edge[ind=ab]++(0,0.3);
\draw (r1-t)edge[ind=x]++(-0.3,0);
\end{tikzpicture}=
\begin{tikzpicture}
\atoms{labbox=$A$}{x1/}
\atoms{labbox=$B$}{x2/p={1,0}}
\atoms{comcoalg}{s/p={0.3,1.7}}
\atoms{lab={t=R,p=0:0.35},rot=90,representation}{r1/p={0,0.7}}
\atoms{lab={t=S,p=0:0.35},rot=90,representation}{r2/p={1,0.7}}
\draw (r1-l)--(x1.north) (r2-l)--(x2.north);
\draw (r1-r)edge[ind=a]++(0,0.3) (r2-r)edge[ind=b]++(0,0.3);
\draw[rounded corners] (r1-t)--++(-0.2,0)--++(0,0.7)--(s);
\draw[rounded corners] (r2-t)--++(-0.2,0)--++(0,0.7)--(s);
\draw (s)edge[mark={ar,s},ind=x]++(0,0.5);
\end{tikzpicture}=
\begin{tikzpicture}
\atoms{labbox=$A$}{x1/}
\atoms{labbox=$B$}{x2/p={1,0}}
\draw (x1.north)edge[ind=a]++(0,0.3) (x2.north)edge[ind=b]++(0,0.3);
\atoms{comcoalg}{s/p={0.5,1.8}}
\atoms{comcoalg}{s1/p={0.1,1.3}}
\atoms{comcoalg}{s2/p={0.9,1.3}}
\draw (s)edge[mark={ar,e}](s1) (s)edge[mark={ar,e}](s2);
\draw (s)edge[mark={ar,s},ind=x]++(0,0.5);
\end{tikzpicture}\\=
\begin{tikzpicture}
\atoms{labbox=$A$}{x1/}
\atoms{labbox=$B$}{x2/p={1,0}}
\draw (x1.north)edge[ind=a]++(0,0.3) (x2.north)edge[ind=b]++(0,0.3);
\atoms{comcoalg}{s/p={0.5,1.3}}
\draw (s)edge[mark={ar,s},ind=x]++(0,0.5);
\end{tikzpicture}=
\begin{tikzpicture}
\atoms{labbox=$A\otimes B$}{x1/}
\draw (x1.north)edge[ind=ab]++(0,0.3);
\atoms{comcoalg}{s/p={-0.5,1}}
\draw (s)edge[mark={ar,s},ind=x]++(0,0.5);
\end{tikzpicture}
\;.
\end{gathered}
\end{equation}
This is due co-unitality.
\item The symmetry constraint is preserved by the contraction,
\begin{equation}
\label{eq:hopf_contraction_symmetry_constraint}
\begin{gathered}
\begin{tikzpicture}
\atoms{labbox=$[A]$}{x/}
\atoms{lab={t=S,p=-90:0.4},rot=180,representation}{r0/p={-0.8,0}}
\draw (r0-r)edge[ind=a]++(-0.3,0) (r0-b)edge[ind=x]++(0,0.3) (x.west)--(r0-l);
\end{tikzpicture}=
\begin{tikzpicture}
\atoms{labbox=$A$,bdastyle={wid=1.2}}{x/}
\atoms{lab={t=S,p=-90:0.4},rot=180,representation}{r0/p={-1,0}}
\draw (r0-r)edge[ind=a]++(-0.3,0) (r0-b)edge[ind=x]++(0,0.3) (x.west)--(r0-l);
\draw[rounded corners] ([xshift=-0.4cm]x.north)--++(0,0.3)-|([xshift=0.4cm]x.north);
\end{tikzpicture}\\=
\begin{tikzpicture}
\atoms{labbox=$A$,bdastyle={wid=1.6}}{x/}
\atoms{comcoalg}{s/p={0,2}, s0/p={-0.9,2.2}}
\atoms{algebra}{s1/p={0,1.4}}
\atoms{lab={t=S,p=-90:0.4},rot=180,representation}{r0/p={-1.3,0}}
\atoms{lab={t=R,p=-90:0.35},rot=0,representation}{r1/p={0,0.8}}
\draw[rounded corners] (r0-b)--++(0,1.5)--(s0);
\draw (r0-r)edge[ind=a]++(-0.3,0);
\draw (r1-l)edge[rounded corners,-|]([xshift=-0.4cm]x.north) (r1-r)edge[rounded corners,-|]([xshift=0.4cm]x.north) (x.west)--(r0-l);
\draw (r1-t)edge[](s1);
\draw (s0)edge[mark={ar,s},ind=x]++(0,0.4) (s)edge[mark={ar,s}](s0);
\end{tikzpicture}=
\begin{tikzpicture}
\atoms{labbox=$A$,bdastyle={wid=1.6}}{x/}
\atoms{comcoalg}{s/p={0,2.5}, s0/p={-0.9,2.6}}
\atoms{algebra}{s1/p={0,1.4}}
\atoms{rot=-90,antipode}{a/p={0.4,2}}
\atoms{lab={t=R,p=-90:0.35},rot=0,representation}{r1/p={0,0.8}}
\atoms{lab={t=S,p=-90:0.4},rot=180,representation}{r0/p={-1.3,0}}
\draw (r1-l)edge[rounded corners,-|]([xshift=-0.4cm]x.north) (r1-r)edge[rounded corners,-|]([xshift=0.4cm]x.north) (x.west)--(r0-l);
\draw (r1-t)edge[](s1);
\draw (r0-r)edge[ind=a]++(-0.3,0);
\draw (s1)edge[mark={ar,s}](a-r) (a-l)edge[](s);
\draw[rounded corners,mark={ar,s}] (s1)--++(-0.4,0.4)--++(0,0.3)--(s);
\draw[rounded corners] (r0-b)--++(0,1.5)--(s0);
\draw (s0)edge[mark={ar,s},ind=x]++(0,0.4) (s)edge[mark={ar,s}](s0);
\end{tikzpicture}\\=
\begin{tikzpicture}
\atoms{labbox=$A$,bdastyle={wid=1.6}}{x/}
\atoms{comcoalg}{s/p={-0.3,2}, s0/p={-0.9,2.2}}
\atoms{rot=-90,antipode}{a/p={0.05,1.5}}
\atoms{lab={t=S,p=-90:0.4},rot=180,representation}{r0/p={-1.3,0}}
\atoms{lab={t=R,p=-45:0.4},rot=90,representation}{r1/p={-0.4,0.8}}
\atoms{lab={t=R,p=-45:0.4},rot=-90,representation}{r2/p={0.4,0.8}}
\draw (r1-l)edge[]([xshift=-0.4cm]x.north) (r2-r)edge[]([xshift=0.4cm]x.north) (x.west)--(r0-l);
\draw (r0-r)edge[ind=a]++(-0.3,0);
\draw[rounded corners] (r1-r)--++(0,0.2)-|(r2-l);
\draw[rounded corners] (r1-t)--++(-0.2,0)--++(0,0.7)--(s);
\draw[rounded corners] (r2-b)-|(a-r);
\draw[rounded corners] (a-l)|-(s);
\draw[rounded corners] (r0-b)--++(0,1.5)--(s0);
\draw (s0)edge[mark={ar,s},ind=x]++(0,0.4) (s)edge[mark={ar,s}](s0);
\end{tikzpicture}=
\begin{tikzpicture}
\atoms{labbox=$A$,bdastyle={wid=1.6}}{x/}
\atoms{comcoalg}{s/p={-0.5,1.8}, s0/p={-0.9,2.2}}
\atoms{lab={t=S,p=-90:0.4},rot=180,representation}{r0/p={-1.3,0}}
\atoms{lab={t=R,p=-45:0.4},rot=90,representation}{r1/p={-0.4,0.8}}
\atoms{lab={t=R^*,p=-45:0.4},rot=90,representation}{r2/p={0.4,0.8}}
\draw (r1-l)edge[]([xshift=-0.4cm]x.north) (r2-l)edge[]([xshift=0.4cm]x.north) (x.west)--(r0-l);
\draw (r0-r)edge[ind=a]++(-0.3,0);
\draw[rounded corners] (r1-r)--++(0,0.4)-|(r2-r);
\draw[rounded corners] (r1-t)--++(-0.2,0)--++(0,0.7)--(s);
\draw[rounded corners] (r2-t)--++(-0.2,0)--++(0,0.7)--(s);
\draw[rounded corners] (r0-b)--++(0,1.5)--(s0);
\draw (s0)edge[mark={ar,s},ind=x]++(0,0.4) (s)edge[mark={ar,s}](s0);
\end{tikzpicture}\\=
\begin{tikzpicture}
\atoms{labbox=$A$,bdastyle={wid=1.2}}{x/}
\draw (x.west)edge[ind=a]++(-0.3,0);
\draw[rounded corners] ([xshift=-0.4cm]x.north)--++(0,0.3)-|([xshift=0.4cm]x.north);
\atoms{comcoalg}{s0/p={-0.9,0.5}}
\draw (s0)edge[mark={ar,s},ind=x]++(0,0.4);
\end{tikzpicture}=
\begin{tikzpicture}
\atoms{labbox=$[A]$}{x/}
\draw (x.west)edge[ind=a]++(-0.3,0);
\atoms{comcoalg}{s0/p={-0.9,0.5}}
\draw (s0)edge[mark={ar,s},ind=x]++(0,0.4);
\end{tikzpicture}
\;.
\end{gathered}
\end{equation}
This is due to the representation move, the definition of the dual representation, the Hopf move, and co-unitality and co-associativity.
\item The symmetry constraint is preserved by the identity tensor,
\begin{equation}
\begin{gathered}
\begin{tikzpicture}
\atoms{labbox=$\idop$}{x1/}
\atoms{lab={t=S^*\otimes S,p=0:0.8},rot=90,representation}{r1/p={0,0.7}}
\draw (r1-l)--(x1.north);
\draw (r1-r)edge[ind=ab]++(0,0.3);
\draw (r1-t)edge[ind=x]++(-0.3,0);
\end{tikzpicture}=
\begin{tikzpicture}
\atoms{lab={t=S^*,p=-90:0.4},rot=180,representation}{r1/p={-0.4,0}}
\atoms{lab={t=S,p=-90:0.4},representation}{r2/p={0.4,0}}
\atoms{comcoalg}{c/p={0,0.7}}
\draw (r1-l)--(r2-l) (r1-r)edge[ind=a]++(-0.3,0) (r2-r)edge[ind=b]++(0.3,0) (c)edge[mark={ar,s},ind=x]++(0,0.4);
\draw[rounded corners] (r1-b)--++(0,0.3)--(c);
\draw[rounded corners] (r2-t)--++(0,0.3)--(c);
\end{tikzpicture}\\=
\begin{tikzpicture}
\atoms{lab={t=S,p=-90:0.4},representation}{r1/p={0,0}}
\atoms{comcoalg}{c/p={0,1.5}}
\atoms{algebra}{m/p={0,0.5}}
\atoms{rot=-90,antipode}{a/p={-0.4,1}}
\draw (r1-t)edge[](m) (m)edge[bend right=45,mark={ar,s}](c) (m)edge[mark={ar,s}](a-r) (a-l)edge[](c) (r1-l)edge[ind=a]++(-0.3,0) (r1-r)edge[ind=b]++(0.3,0) (c)edge[mark={ar,s},ind=x]++(0,0.4);
\end{tikzpicture}=
\begin{tikzpicture}
\atoms{lab={t=S,p=-90:0.4},representation}{r1/p={0,0}}
\atoms{comcoalg}{c/p={0,1}}
\atoms{algebra}{m/p={0,0.5}}
\draw (r1-t)edge[](m) (r1-l)edge[ind=a]++(-0.3,0) (r1-r)edge[ind=b]++(0.3,0) (c)edge[mark={ar,s},ind=x]++(0,0.4);
\end{tikzpicture}=
\begin{tikzpicture}
\atoms{labbox=$\idop$}{x1/}
\atoms{comcoalg}{c/p={0.8,0}}
\draw (x1.north)edge[ind=ab]++(0,0.3) (c)edge[mark={ar,s},ind=x]++(0,0.4);
\end{tikzpicture}
\;.
\end{gathered}
\end{equation}
\end{itemize}
\end{myobs}

\subsection{Specific dependencies}
\label{sec:symmetric_specific_dependencies}
Let us look at the classic example for co-commutative Hopf algebras in complex array tensors.
\begin{mydef}
For every finite group $G$ and every commutative semiring $K$ with zero, the \tdef{group Hopf algebra}{group_hopf_algebra} the following co-commutative Hopf algebra in array tensors, extending the group algebra for $G$:
\begin{itemize}
\item The co-product and co-unit are given by
\begin{equation}
\begin{tikzpicture}
\atoms{comcoalg}{c/p={0,0}}
\draw[]  (c)edge[mark={ar,s},ind=c]++(90:0.5) (c)edge[ind=b]++(-30:0.5) (c)edge[ind=a]++(-150:0.5);
\end{tikzpicture}=
\delta_{a,b,c}\;.
\end{equation}
\item The unit is
\begin{equation}
\begin{tikzpicture}
\atoms{comcoalg}{c/p={0,0}}
\draw[] (c)edge[mark={ar,s},ind=a]++(90:0.5);
\end{tikzpicture}=
1\forall a\;.
\end{equation}
\item The antipode is
\begin{equation}
\begin{tikzpicture}
\atoms{antipode}{a/p={0,0}}
\draw (a-l)edge[ind=a]++(-0.5,0) (a-r)edge[ind=b]++(0.5,0);
\end{tikzpicture}
=
\delta_{a^{-1},b}\;.
\end{equation}
\end{itemize}
\end{mydef}

\begin{myrem}
For any finite group $G$, group-symmetric array tensors for $G$ are equivalent to Hopf-symmetric array tensors for the group Hopf algebra of $G$. Conversely, it is known that any complex finite-dimensional co-commutative Hopf algebra is isomorphic to a group algebra \cite{Milnor1965}. So, for complex array tensors and finite groups, group-symmetric and Hopf-symmetric tensors are equivalent.

The same is not true for real array tensors: Consider the group algebra for an abelian group, exchange product with co-product and unit with co-unit, and swap the indices of the involution. This yields the dual group algebra which for an abelian group is co-commutative as well. The group algebra and the delta algebra are isomorphic as complex algebras, but not as real algebras. Symmetry under a dual abelian group algebra is the same as a grading by that group, which is equivalent to the group symmetry in the complex case, but not in the real case.

Conversely, the definition of group-symmetry also works for infinite groups. The latter cannot be directly formalized as array tensor Hopf algebra models. This includes important physical symmetries such as Lie group symmetries.
\end{myrem}

Let us also consider an example for $\zz_2$-extended-group-symmetric tensors.
\begin{myobs}
The simplest example of a $\zz_2$-extended group is the product with an independent group,
\begin{equation}
G=\zz_2\times G/\phi\;.
\end{equation}
$\zz_2$ can also be non-trivially embedded, for example as $\pm 1$ in $U(1)$, corresponding to a $\zz$-grading with a $\zz_2$ sub-grading. This case is of particular physical relevance when the base type is $\zz_2$-twisted-symmetric tensors as introduced below in Section~\ref{sec:twisted_z2_grading}.
\end{myobs}

\subsection{Mappings}
Every ground type 1-data automatically fulfills the symmetry constraint if we choose the trivial representation as 0-data. This can be formalized as a tensor mapping.
\begin{mydef}
The \tdef{trivial representation mapping}{trivial_representation_mapping} is the following mapping from the ground type $\calg$ to Hopf-symmetric tensors:
\begin{itemize}
\item It has strict homomorphicity.
\item A 0-data $a$ of $\calg$ is mapped to the representation consisting of the tensor product of the identity tensor with the co-unit,
\begin{equation}
\begin{tikzpicture}
\atoms{representation,lab={t=m(a),p=-90:0.35}}{0/}
\draw (0-l)edge[ind=a]++(180:0.4) (0-r)edge[ind=b]++(0:0.4) (0-t)edge[ind=x]++(90:0.4);
\end{tikzpicture}
=
\begin{tikzpicture}
\draw[ind=b,startind=a] (0,0)--(1,0);
\atoms{comcoalg}{0/p={0.5,0.5}}
\draw (0)edge[mark={ar,s},ind=x]++(0,0.4);
\end{tikzpicture}
\;.
\end{equation}
\item A 1-data $A\in \dat_1(a)$ is mapped to that same 1-data. It trivially obeys the symmetry constraint.
\end{itemize}
There is an analogous version of the mapping for group-symmetric tensors where each representation matrix is an identity tensor.
\end{mydef}

Obviously, every 1-data of $\calg$ fulfilling the symmetry constraint is also a 1-data, which is another mapping.
\begin{mydef}
The \tdef{symmetry-forgetting mapping}{symmetry_forgetting_mapping} is the following mapping from symmetric tensors to $\calg$:
\begin{itemize}
\item It has strict homomorphicity.
\item A 0-data $R$ is mapped to the $\calg$ 0-data associated to the $r$-binding by $R$,
\begin{equation}
m(R)=R_{0r}\;.
\end{equation}
\item The mapping on the 1-data is the identity.
\end{itemize}
\end{mydef}

\subsection{Use in physics}
\label{sec:symmetric_physics}
Imagine a tensor $A$ with many indices, i.e., a 1-data whose input 0-data is a tensor product of many 0-data, equipped with different representations. Then the symmetry constraint reads
\begin{equation}
\begin{tikzpicture}
\atoms{labbox=$A$,bdastyle={wid=3}}{x/}
\atoms{rotlab,lab/p={180:0.35},rot=90,representation}{{r0/p={-1,0.8},lab={t=g}}, {r1/p={-0.2,0.8},lab={t=g}}, {r2/p={0.6,0.8},lab={t=g}}}
\draw (r0-l)edge[cdir=-]([xshift=-1cm]x.north) (r1-l)edge[cdir=-]([xshift=-0.2cm]x.north) (r2-l)edge[cdir=-]([xshift=0.6cm]x.north) (r0-r)edge[ind=a]++(0,0.3) (r1-r)edge[ind=b]++(0,0.3) (r2-r)edge[ind=c]++(0,0.3);
\node at ($(x.north)+(1.1,0.2)$){$\ldots$};
\end{tikzpicture}=
\begin{tikzpicture}
\atoms{labbox=$A$,bdastyle={wid=3}}{x/}
\draw ([xshift=-1cm]x.north)edge[ind=a]++(0,0.4) ([xshift=-0.2cm]x.north)edge[ind=b]++(0,0.4) ([xshift=0.6cm]x.north)edge[ind=c]++(0,0.4);
\node at ($(x.north)+(1.1,0.2)$){$\ldots$};
\end{tikzpicture}
\end{equation}
for all $g\in G$. For array tensors, this means that the array is invariant under a set of linear operators forming a group, each of which is a tensor product of operators acting on one degree of freedom only.

Such a symmetry acting on, e.g., the Hamiltonian, or ground states of a quantum mechanics model is known as an \emph{on-site} symmetry. If we have a tensor-network model describing the imaginary- or real-time evolution or thermal ensemble of a local quantum spin Hamiltonian, then the tensors of this tensor-network model will inherit the symmetry constraint from the Hamiltonian. It will thus be a tensor-network model of symmetric array tensors.

$\zz_2$-extended-group-symmetric tensors are uninteresting for ground types with a symmetric contraction, such as array tensors. They have however a very relevant physical interpretation if the ground type is the one describing many-body models with fermionic degrees of freedom, namely $\zz_2$-twisted-symmetric tensors as introduced below in Section~\ref{sec:twisted_z2_grading}. The backwards-directed identity representing the group element $\phi$ is then given by $(-1)^{P_f}$ with $P_f$ being the fermion parity. So $\zz_2$-extended-group-symmetric tensors describe fermionic models with on-site symmetries, where the fermion parity can be a non-trivial $\zz_2$ subgroup of the overall symmetry group. For example, the case $G=U(1)$ with $\phi=-1$ mentioned in Section~\ref{sec:symmetric_specific_dependencies} describes fermionic systems with \emph{particle-number conservation}.

\section{Twisted-symmetric tensors}
\label{sec:twisted_symmetric}
\subsection{Motivation}
Twisted-symmetric tensors are a generalization of symmetric tensors. The 2-functions of symmetric tensors are just the 2-functions of the ground type. The 2-axioms such as commutativity or associativity directly follow from the related moves of co-algebra and co-unit. We can weaken those liquid moves in the sense of replacing them by elements for which further moves hold. We can define an according tensor type whose 2-functions are now non-trivial in the graphical calculus of the ground type.

Physically most relevant is the case of twisted $\zz_2$-symmetry, which is the tensor type we need for describing quantum models with fermionic degrees of freedom. We will therefore single out this case of particular importance in a first definition. The much more general definition also allows us to describe fermionic systems with additional symmetries which interact non-trivially with the fermion parity, such as particle-number conservation.

\subsection{Definition: Twisted \texorpdfstring{$\zz_2$}{Z2}-grading}
\label{sec:twisted_z2_grading}
\begin{mydef}
\tdef{$\mathbb{Z}_2$-twisted-symmetric tensors}{zt_symmetric_tensor} are the following tensor type:
\begin{itemize}
\item They can be defined for any field $K$.
\item They have an identity tensor, a unit, and a trivial tensor. They need a dual (if we want to have block-compatibility). They have strict dual involutivity and unit automorphicity 1-axioms, but a weak product automorphor. They have \emph{neither} a symmetric contraction nor symmetric identity.
\item A 0-data is a finite set $a$ together with a function
\begin{equation}
|\cdot|_a: a\rightarrow \zz_2
\end{equation}
called the \tdef{grading}{z2_grading}. We will usually drop the subscript $a$, denote $\zz_2$ additively, and also refer to $0$ and $1$ as even and odd \mdef{parity}.
\item The product of 0-data is the cartesian product of sets, and the parities are combined via the $\mathbb{Z}_2$ group product,
\begin{equation}
\begin{gathered}
a\otimes b=a\times b\;,\\
|(i,j)|=|i|+|j|\;.
\end{gathered}
\end{equation}
\item The unit 0-data consists of a single even-parity element,
\begin{equation}
1=\{0\},\quad|0|=0\;.
\end{equation}
\item The dual 1-function is the identity function. (However, the dual automorphors will be non-trivial).
\item A 1-data $A\in \dat_1(a)$ is a functions that associates a number in $K$ to every even configuration,
\begin{equation}
A: \{i\in a: |i|=0\}\rightarrow K\;.
\end{equation}
\item The tensor product of two 1-data $A\in \dat_1(a)$ and $B\in\dat_1(b)$ is just the tensor product of arrays,
\begin{equation}
(A\otimes B)((i,j))= A(i) B(j)\;.
\end{equation}
\item Also the contraction of a 1-data $A\in\dat_1(a\otimes (b\otimes b^*))$ is the usual contraction of arrays,
\begin{equation}
[A](i)=\sum_{j}A((i,(j,j)))\;.
\end{equation}
\item The commutor of a 1-data $A\in \dat_1(a\otimes (b\otimes c))$ is non-canonical and given by
\begin{equation}
\sigma(A)((i,(j,k))) = (-1)^{|j||k|} \cdot A((i,(k,j)))\;,
\end{equation}
where $|j||k|$ denotes the product of two $\zz_2$-elements interpreted as numbers $0$ or $1$, i.e., the product of $\zz_2$ as a ring or finite field.
\item The product dual-automorphor of a 1-data $A\in \dat_1(a\otimes (b\otimes c)^*)$ is given by
\begin{equation}
D_\otimes(A)((i,(j,k)) =  (-1)^{|j||k|} \cdot A((i,(j,k)))\;.
\end{equation}
\item All other invertible 2-functions are canonical, that is, they are the 2-functions of array tensors after ignoring the grading.
\end{itemize}
\end{mydef}

\begin{myrem}
Let us check that the above definitions are consistent and fulfill the 2-axioms. Many 2-axioms are fulfilled trivially because the corresponding 2-functions are canonical. We would like to highlight a few non-trivial cases:
\begin{itemize}
\item The commutativity of the tensor product is fulfilled because each individual tensor only associates numbers to even configurations,
\begin{equation}
\begin{multlined}
\sigma_0(A\otimes B)((i,j))= (-1)^{|i||j|}A(i) B(j)\\ = A(i)B(j)=(A\otimes B)((i,j))\;.
\end{multlined}
\end{equation}
This holds because $A(i)$ is only non-zero if $|i|=0$, and the same for $B(j)$.
\item The hexagon equation holds, as
\begin{equation}
\begin{multlined}
\alpha\sigma\alpha\sigma\alpha\sigma(A)((i,(j,k)))\\
=(-1)^{|i|(|j|+|k|)}(-1)^{|j|(|k|+|i|)}(-1)^{|k|(|i|+|j|)}A((i,(j,k)))\\
=A((i,(j,k)))\;.
\end{multlined}
\end{equation}
\item The commutor is involutive, as
\begin{equation}
\begin{multlined}
\sigma_0\sigma_0(A)((i,j))\\= (-1)^{|i||j|}(-1)^{|j||i|} A((i,j))=A((i,j))\;.
\end{multlined}
\end{equation}
\item The contraction is block-compatible, as
\begin{equation}
\begin{multlined}
[XD_\otimes^{-1} Y\sigma Z(A)](a)\\=(-1)^{|j||k|} (-1)^{|j||k|} \sum_{(j,k)} A(((i,(j,j)),(k,k)))\\
= \sum_j\sum_k A(((i,(j,j)),(k,k)))=[[A]](a)\;.
\end{multlined}
\end{equation}
Here, $X$, $Y$ and $Z$ are sequences of canonical 2-functions (associator, unitor, etc.) such that $XD_\otimes^{-1} Y\sigma Z$ effectively exchanges the second $j$ with the first $k$ and then swaps the dual with the product of the second $(j,k)$-pair.
\end{itemize}
\end{myrem}

\begin{myobs}
Due to the non-canonical commutor, $\mathbb{Z}_2$-twisted-symmetric tensors do \emph{not} have a symmetric contraction. In general, we can have
\begin{equation}
\begin{multlined}
[\widetilde{C}\alpha_0^{-1}\widetilde{D_I}\alpha_0\sigma(A)](i) \\=\sum_j (-1)^{|j||j|} A((i,(j,j)))
=\sum_j (-1)^{|j|} A((i,(j,j))) \\\neq \sum_j A((i,(j,j)))=[A](i)\;.
\end{multlined}
\end{equation}
\end{myobs}

\begin{mydef}
\label{def:ztwisted_symmetric_tensors}
\mdef{$\zz$-twisted-symmetric tensors} are the following variant of $\zz_2$-twisted-symmetric tensors:
\begin{itemize}
\item Instead of a $\zz_2$-grading, the 0-data $a$ have a $\zz$-grading,
\begin{equation}
|\cdot|: a\rightarrow \zz
\end{equation}
\item The dual is non-trivial, given by inverting the grading,
\begin{equation}
|i|_{a^*} = -|i|_a\;.
\end{equation}
\item Apart from that, all the formulas are the same, e.g., the 1-data associates numbers to $i\in a$ with $|i|=0$, and the commutor involves a sign $(-1)^{|j||k|}$, where $|\cdot|$ denotes a $\zz$-grading instead of a $\zz_2$-grading.
\end{itemize}
\end{mydef}

\subsection{Definition: Via triangular Hopf algebras}
In this section we give a tensor type which is both a generalization of $\zz_2$-twisted symmetric and Hopf-symmetric tensors. This tensor type will be constructed from an arbitrary ground tensor type $\calg$ with the help of a liquid.

\begin{mydef}
\tdef{Triangular Hopf algebras}{triangular_hopf_algebra} are the following liquid extending Hopf algebras:
\begin{itemize}
\item There is one additional $2$-index element that we'll refer to as the \tdef{braiding}{hopf_braiding}. We'll represent it by a crossed-out square,
\begin{equation}
\begin{tikzpicture}
\atoms{hopfbraiding}{r/p={0,0}}
\draw (r-l)--++(-0.3,0) (r-r)--++(0.3,0);
\end{tikzpicture}\;,
\end{equation}
with two output indices.
\item There are $4$ additional moves. Conjugation with the braiding exchanges the ordering of the co-algebra,
\begin{equation}
\label{eq:triangular_ax1}
\begin{tikzpicture}
\atoms{coalgebra}{d/p={0,0}}
\atoms{algebra}{s0/p={-1,1}, s1/p={1,1}}
\atoms{hopfbraiding}{{r1/p={-0.5,1.4},hflip}, r2/p={0.5,1.4}}
\draw[mark={ar,s},rounded corners] (s0)--++(0,-0.4)--(d);
\draw[mark={ar,s},rounded corners] (s1)--++(0,-0.4)--(d);
\draw[mark={ar,s},rounded corners] (s0)--++(0.8,0)--++(0.4,0.4)--(r2-l);
\draw[mark={ar,s},rounded corners] (s1)--++(-0.8,0)--++(-0.4,0.4)--(r1-l);
\draw[mark={ar,s},rounded corners] (s0)--++(-0.4,0)|-(r1-r);
\draw[mark={ar,s},rounded corners] (s1)--++(0.4,0)|-(r2-r);
\draw (s0)edge[ind=a]++(0,0.8) (s1)edge[ind=b]++(0,0.8) (d)edge[mark={ar,s},ind=c]++(0,-0.5);
\end{tikzpicture}=
\begin{tikzpicture}
\atoms{coalgebra}{d/p={0,0}}
\draw (d)edge[ind=b]++(135:0.5) (d)edge[ind=a]++(45:0.5) (d)edge[mark={ar,s},ind=c]++(0,-0.5);
\end{tikzpicture}\;.
\end{equation}
\item The braiding acts as a homomorphism between the algebra and the co-algebra,
\begin{equation}
\label{eq:triangular_ax2}
\begin{tikzpicture}
\atoms{coalgebra}{d/p={0,0}}
\atoms{hopfbraiding,rot=90}{r/p={0,-0.7}}
\draw (d)edge[mark={ar,s}](r-r) (r-l)edge[ind=c]++(0,-0.3) (d)edge[ind=a]++(135:0.5) (d)edge[ind=b]++(45:0.5);
\end{tikzpicture}=
\begin{tikzpicture}
\atoms{algebra}{d/p={0,0}}
\atoms{hopfbraiding,rot=90}{r0/p={-0.5,0.8}, r1/p={0.5,0.8}}
\draw[rounded corners,mark={ar,e}] (r0-l)--++(0,-0.3)--(d);
\draw[rounded corners,mark={ar,e}] (r1-l)--++(0,-0.3)--(d);
\draw (d)edge[ind=c]++(0,-0.5) (r0-r)edge[ind=a]++(0,0.3) (r1-r)edge[ind=b]++(0,0.3);
\end{tikzpicture}\;,
\end{equation}
\item and
\begin{equation}
\label{eq:triangular_ax4}
\begin{tikzpicture}
\atoms{coalgebra}{d/p={0,0}}
\atoms{hopfbraiding}{r/p={-0.7,0}}
\draw (d)edge[mark={ar,s}](r-r) (r-l)edge[ind=a]++(-0.3,0);
\end{tikzpicture}=
\begin{tikzpicture}
\atoms{algebra}{d/p={0,0}}
\draw (d)edge[ind=a]++(-0.5,0);
\end{tikzpicture}\;.
\end{equation}
\item The \emph{triangularity} move ensures that the braiding and its transpose are inverses, when interpreted as elements of two copies of the Hopf algebra,
\begin{equation}
\label{eq:triangular_ax3}
\begin{tikzpicture}
\atoms{hopfbraiding}{r0/p={0,0}, {r1/p={0,0.8},hflip}}
\atoms{algebra}{s0/p={-0.6,0.4}, s1/p={0.6,0.4}}
\draw[rounded corners,mark={ar,e}] (r0-l)-|(s0);
\draw[rounded corners,mark={ar,e}] (r0-r)-|(s1);
\draw[rounded corners,mark={ar,e}] (r1-r)-|(s0);
\draw[rounded corners,mark={ar,e}] (r1-l)-|(s1);
\draw (s0)edge[ind=a]++(-0.5,0) (s1)edge[ind=b]++(0.5,0);
\end{tikzpicture}=
\begin{tikzpicture}
\atoms{algebra}{s0/p={-0.3,0.4}, s1/p={0.3,0.4}}
\draw (s0)edge[ind=a]++(-0.5,0) (s1)edge[ind=b]++(0.5,0);
\end{tikzpicture}\;.
\end{equation}
\end{itemize}
\end{mydef}

\begin{myrem}
For every triangular Hopf algebra we have
\begin{equation}
\begin{tikzpicture}
\atoms{rot=-90,antipode}{0/p={0,0}, 1/p={0,0.6}}
\draw (0-l)--(1-r) (0-r)edge[ind=a]++(0,-0.2) (1-l)edge[ind=b]++(0,0.2);
\end{tikzpicture}
=
\begin{tikzpicture}
\atoms{algebra}{0/p={0,0}}
\atoms{drinfeldelement,lab/ang=180}{{u/p={-0.6,0}}}
\atoms{idrinfeldelement,lab/ang=0}{ux/p={0.6,0}}
\draw (0)edge[mark={ar,s}](u-r) (0)edge[mark={ar,s}](ux-l) (0)edge[mark={ar,s},ind=b]++(0,0.6) (0)edge[ind=a]++(0,-0.6);
\end{tikzpicture}
\end{equation}
and
\begin{equation}
\begin{tikzpicture}
\atoms{algebra}{0/p={0,0}}
\atoms{drinfeldelement,lab/ang=180}{u/p={-0.6,0}}
\atoms{idrinfeldelement,lab/ang=0}{ux/p={0.6,0}}
\draw (0)edge[mark={ar,s}](u-r) (0)edge[mark={ar,s}](ux-l) (0)edge[ind=a]++(0,-0.6);
\end{tikzpicture}
=
\begin{tikzpicture}
\atoms{algebra}{0/p={0,0}}
\draw (0)edge[ind=a]++(0,-0.6);
\end{tikzpicture}
\end{equation}
and
\begin{equation}
\begin{tikzpicture}
\atoms{algebra}{0/p={0,0}}
\atoms{idrinfeldelement,lab/ang=180}{u/p={-0.6,0}}
\atoms{drinfeldelement,lab/ang=0}{ux/p={0.6,0}}
\draw (0)edge[mark={ar,s}](u-r) (0)edge[mark={ar,s}](ux-l) (0)edge[ind=a]++(0,-0.6);
\end{tikzpicture}
=
\begin{tikzpicture}
\atoms{algebra}{0/p={0,0}}
\draw (0)edge[ind=a]++(0,-0.6);
\end{tikzpicture}
\end{equation}
where
\begin{equation}
\begin{tikzpicture}
\atoms{drinfeldelement,lab/ang=90}{u/p={0,0}}
\draw (u-b)edge[ind=a]++(0,-0.3);
\end{tikzpicture}
=
\begin{tikzpicture}
\atoms{hopfbraiding,hflip}{r/p={0,0}}
\atoms{rot=-90,antipode}{s/p={-0.5,-0.5}}
\atoms{algebra}{m/p={0,-1}}
\draw[rounded corners] (r-r)-|(s-l);
\draw[rounded corners,mark={ar,s}] (m)--++(0.5,0.5)|-(r-l);
\draw (m)edge[mark={ar,s}](s-r) (m)edge[ind=a]++(0,-0.5);
\end{tikzpicture}
\end{equation}
and
\begin{equation}
\begin{tikzpicture}
\atoms{idrinfeldelement,lab/ang=90}{u/p={0,0}}
\draw (u-b)edge[ind=a]++(0,-0.3);
\end{tikzpicture}
=
\begin{tikzpicture}
\atoms{hopfbraiding,hflip}{r/p={0,0}}
\atoms{rot=-90,antipode}{s1/p={0.5,-0.4}}
\atoms{rot=-90,antipode}{s2/p={0.5,-0.9}}
\atoms{algebra}{m/p={0,-1.5}}
\draw[rounded corners] (r-l)-|(s1-l);
\draw[rounded corners,mark={ar,s}] (m)--++(-0.5,0.5)|-(r-r);
\draw (s1-r)--(s2-l) (m)edge[mark={ar,s}](s2-r) (m)edge[ind=a]++(0,-0.5);
\end{tikzpicture}
\end{equation}
are known as the \emph{Drinfeld element} and its inverse \cite{Drinfeld1989}.
\end{myrem}

\begin{mydef}
\tdef{Twisted-symmetric tensors}{twisted_symmetric_tensor} are the following tensor type:
\begin{itemize}
\item They are defined with respect to a ground tensor type $\calg$, and a triangular (non-co-commutative) Hopf algebra $H$ in this ground tensor type.
\item They need duals, have an identity and a unit. The contraction and identity are \emph{not} symmetric.
\item As for symmetric tensors, the 0-data are given by representations of the algebra of $H$. The 1-functions are the trivial representation, dual representation and tensor product of representations.
\item As for symmetric tensors, a 1-data $A\in \dat_1(R)$ is a ground type 1-data whose 0-data is the one associated to the $r$-binding by the representation $R$. Additionally, $A$ has to fulfill the symmetry constraint,
\begin{equation}
\label{eq:twisted_symmetry_constraint}
\begin{tikzpicture}
\atoms{labbox=$A$}{x/}
\atoms{representation}{r/p={0.8,0}}
\draw (x.east)--(r-l) (r-r)edge[ind=a]++(0.3,0) (r-t)edge[ind=x]++(0,0.4);
\end{tikzpicture}=
\begin{tikzpicture}
\atoms{labbox=$A$}{x/}
\draw (x.east)edge[ind=a]++(0.3,0);
\atoms{coalgebra}{s/p={1.4,0}}
\draw (s)edge[mark={ar,s},ind=x]++(0,0.5);
\end{tikzpicture}\;.
\end{equation}
\item The commutor is defined via the braiding of the triangular Hopf algebra,
\begin{equation}
\label{eq:twisted_symm_commutor}
\begin{tikzpicture}
\atoms{labbox=$\sigma(A)$,bdastyle={wid=1.4}}{x/p={0,-0.7}}
\draw ([xshift=-0.4cm]x.north)edge[ind=b]++(0,0.3) ([xshift=0.4cm]x.north)edge[ind=c]++(0,0.3) (x.west)edge[ind=a]++(-0.3,0);
\end{tikzpicture}=
\begin{tikzpicture}
\atoms{labbox=$A$,bdastyle={wid=1.8}}{x/p={0,-0.9}}
\atoms{rot=90,representation}{r1/p={-0.6,-0.2}, r2/p={0.6,-0.2}}
\atoms{hopfbraiding}{b/p={0,-0.2}}
\draw (b-l)--(r1-b) (b-r)--(r2-t) (r1-l)--([xshift=-0.6cm]x.north) (r2-l)--([xshift=0.6cm]x.north) (r1-r)edge[ind=b]++(0,0.3) (r2-r)edge[ind=c]++(0,0.3) (x.west)edge[ind=a]++(-0.3,0);
\end{tikzpicture}\;.
\end{equation}
\item The dual-involutor is defined via the Drinfeld element of the triangular Hopf algebra,
\begin{equation}
\label{eq:twisted_involutor}
\begin{tikzpicture}
\atoms{labbox=$D_I(A)$,bdastyle={wid=1.4}}{x/p={0,-0.7}}
\draw ([xshift=-0.4cm]x.north)edge[ind=c]++(0,0.3) ([xshift=0.4cm]x.north)edge[ind=b]++(0,0.3);
\end{tikzpicture}=
\begin{tikzpicture}
\atoms{labbox=$A$,bdastyle={wid=1.4}}{x/p={0,-0.9}}
\atoms{rot=90,representation}{r2/p={0.4,-0.2}}
\atoms{lab/ang=0,drinfeldelement}{u/p={1,-0.2}}
\draw (r2-b)--(u-l) (r2-l)--([xshift=0.4cm]x.north) ([xshift=-0.4cm]x.north)edge[ind=a]++(0,0.8) (r2-r)edge[ind=b]++(0,0.3);
\end{tikzpicture}\;.
\end{equation}
The inverse is given accordingly by replacing $u$ by $u^{-1}$.
\item The dual product-automorphor looks the same as the commutor in the graphical calculus of $\calg$,
\begin{equation}
\label{eq:twisted_automorphor}
\begin{tikzpicture}
\atoms{labbox=$D_\otimes(A)$,bdastyle={wid=1.4}}{x/p={0,-0.7}}
\draw ([xshift=-0.4cm]x.north)edge[ind=b]++(0,0.3) ([xshift=0.4cm]x.north)edge[ind=c]++(0,0.3) (x.west)edge[ind=a]++(-0.3,0);
\end{tikzpicture}=
\begin{tikzpicture}
\atoms{labbox=$A$,bdastyle={wid=1.8}}{x/p={0,-0.9}}
\atoms{rot=90,representation}{r1/p={-0.6,-0.2}, r2/p={0.6,-0.2}}
\atoms{hopfbraiding}{b/p={0,-0.2}}
\draw (b-l)--(r1-b) (b-r)--(r2-t) (r1-l)--([xshift=-0.6cm]x.north) (r2-l)--([xshift=0.6cm]x.north) (r1-r)edge[ind=b]++(0,0.3) (r2-r)edge[ind=c]++(0,0.3) (x.west)edge[ind=a]++(-0.3,0);
\end{tikzpicture}\;.
\end{equation}
\item All other 2-functions are those of $\calg$.
\end{itemize}
\end{mydef}

\begin{myobs}
Let us check whether all the above definitions are consistent. Many of the consistency equations are identical with those of (non-twisted) Hopf-symmetric tensors.
\begin{itemize}
\item The commutor is consistent with the symmetry constraint, due to
\begin{equation}
\begin{gathered}
\begin{tikzpicture}
\atoms{labbox=$\sigma(A)$,bdastyle={wid=1.2}}{x/p={0,-0.7}}
\atoms{rot=90,representation}{r1/p={-0.4,0}, r2/p={0.4,0}}
\atoms{coalgebra}{d/p={1.4,0.2}}
\draw (r1-l)--([xshift=-0.4cm]x.north) (r2-l)--([xshift=0.4cm]x.north) (d)edge[mark={ar,s},ind=x]++(0.5,0) (r1-r)edge[ind=b]++(0,0.5) (r2-r)edge[ind=c]++(0,0.5);
\draw[rounded corners] (r2-b)--++(0.2,0)--++(0.2,0.4)--(d);
\draw[rounded corners] (r1-b)--++(0.2,0)--++(0.2,0.4)--++(0.6,0)--++(0.2,-0.4)--(d);
\atoms{representation}{rxx/p={$(x.east)+(0.5,0)$}}
\draw (x.east)--(rxx-l) (rxx-r)edge[ind=a]++(0.3,0);
\draw[rounded corners] (d)--++(0,0.4)--++(0.4,0)--++(0,-0.8)--($(rxx-t)+(0,0.3)$)--(rxx-t);
\end{tikzpicture}=
\begin{tikzpicture}
\atoms{labbox=$A$,bdastyle={wid=1.8}}{x/p={0,-0.8}}
\atoms{rot=90,representation}{r1b/p={-0.6,0}, r2b/p={0.6,0}, r1/p={-0.6,0.5}, r2/p={0.6,0.5}}
\atoms{hopfbraiding}{b/, {b0/p={1.2,0},rot=-90}, {b1/p={1.2,1.6},rot=90}}
\atoms{coalgebra}{d/p={2,0.8}}
\atoms{algebra}{a0/p={1.2,0.5}, a1/p={1.2,1.1}}
\draw (r1b-r)--(r1-l) (r2b-r)--(r2-l) (b-l)--(r1b-b) (b-r)--(r2b-t) (r1b-l)--([xshift=-0.6cm]x.north) (r2b-l)--([xshift=0.6cm]x.north) (d)edge[mark={ar,s},ind=x]++(0.5,0) (r2-b)edge[](a0) (r1-r)edge[ind=b]++(0,0.5) (r2-r)edge[ind=c]++(0,0.5) (a0)edge[mark={ar,s}](b0-l) (a1)edge[mark={ar,s}](b1-l);
\draw[rounded corners,mark={ar,s}] (a0)--++(0.4,0)--(d);
\draw[rounded corners,mark={ar,s}] (a1)--++(0.4,0)--(d);
\draw[rounded corners] (r1-b)--++(0.2,0)--++(0.2,0.6)--(a1);
\draw[rounded corners,mark={ar,e}] (b0-r)|-++(0.4,-0.2)--++(0,0.8)--(a1);
\draw[rounded corners,mark={ar,e}] (b1-r)|-++(0.4,0.2)--++(0,-0.8)--(a0);
\atoms{representation}{rxx/p={$(x.east)+(1,0)$}}
\draw (x.east)--(rxx-l) (rxx-r)edge[ind=a]++(0.3,0);
\draw[rounded corners] (d)--++(0,0.4)--++(0.4,0)--++(0,-0.8)--($(rxx-t)+(0,0.3)$)--(rxx-t);
\end{tikzpicture}\\
=
\begin{tikzpicture}
\atoms{labbox=$A$,bdastyle={wid=1.2}}{x/p={0,-1}}
\atoms{rot=90,representation}{r1/p={-0.4,0}, r2/p={0.4,0}}
\atoms{hopfbraiding}{{b0/p={2,-0.5},rot=90}, {b1/p={2,1.1},rot=-90}, {b0x/p={1.2,-0.5},rot=-90}}
\atoms{coalgebra}{d/p={2.8,0.3}}
\atoms{algebra}{a0/p={2,0}, a1/p={2,0.6}, a0x/p={1.2,0}, a1x/p={1.2,0.6}}
\draw (r1-l)--([xshift=-0.4cm]x.north) (r2-l)--([xshift=0.4cm]x.north) (d)edge[mark={ar,s},ind=x]++(0.5,0) (r2-b)edge[](a0x) (r1-r)edge[ind=b]++(0,0.5) (r2-r)edge[ind=c]++(0,0.5) (a0)edge[mark={ar,s}](b0-r) (a1)edge[mark={ar,s}](b1-r) (a0x)edge[mark={ar,s}](a0) (a1x)edge[mark={ar,s}](a1) (a0x)edge[mark={ar,s}](b0x-l);
\draw[rounded corners,mark={ar,s}] (a0)--++(0.4,0)--(d);
\draw[rounded corners,mark={ar,s}] (a1)--++(0.4,0)--(d);
\draw[rounded corners] (r1-b)--++(0.2,0)--++(0.2,0.6)--(a1x);
\draw[rounded corners,mark={ar,e}] (b0-l)|-++(0.4,-0.2)--++(0,0.8)--(a1);
\draw[rounded corners,mark={ar,e}] (b1-l)|-++(0.4,0.2)--++(0,-0.8)--(a0);
\draw[rounded corners,mark={ar,e}] (b0x-r)|-++(0.4,-0.2)--++(0,0.8)--(a1x);
\atoms{representation}{rxx/p={$(x.east)+(2,0)$}}
\draw (x.east)--(rxx-l) (rxx-r)edge[ind=a]++(0.3,0);
\draw[rounded corners] (d)--++(0,0.4)--++(0.4,0)--++(0,-0.8)--($(rxx-t)+(0,0.3)$)--(rxx-t);
\end{tikzpicture}=
\begin{tikzpicture}
\atoms{labbox=$A$,bdastyle={wid=1.2}}{x/p={0,-0.8}}
\atoms{rot=90,representation}{r1/p={-0.4,0}, r2/p={0.4,0}}
\atoms{rot=-90,hopfbraiding}{b1/p={1,1.1}}
\atoms{coalgebra}{d/p={2,0.3}}
\atoms{algebra}{a0/p={1,0}, a1/p={1,0.6}}
\draw (r1-l)--([xshift=-0.4cm]x.north) (r2-l)--([xshift=0.4cm]x.north) (d)edge[mark={ar,s},ind=x]++(0.5,0) (r2-b)edge[](a0) (r1-r)edge[ind=b]++(0,0.5) (r2-r)edge[ind=c]++(0,0.5) (a1)edge[mark={ar,s}](b1-r);
\draw[rounded corners,mark={ar,s}] (a0)--++(0.4,0)--(d);
\draw[rounded corners,mark={ar,s}] (a1)--++(0.4,0)--(d);
\draw[rounded corners] (r1-b)--++(0.2,0)--++(0.2,0.6)--(a1);
\draw[rounded corners,mark={ar,e}] (b1-l)|-++(0.4,0.2)--++(0,-0.8)--(a0);
\atoms{representation}{rxx/p={$(x.east)+(0.5,0)$}}
\draw (x.east)--(rxx-l) (rxx-r)edge[ind=a]++(0.3,0);
\draw[rounded corners] (d)--++(0,0.4)--++(0.4,0)--++(0,-0.8)--($(rxx-t)+(0,0.3)$)--(rxx-t);
\end{tikzpicture}\\
=
\begin{tikzpicture}
\atoms{labbox=$A$,bdastyle={wid=1.8}}{x/p={0,-0.8}}
\atoms{rot=90,representation}{r1/p={-0.6,-0.2}, r2/p={0.6,-0.2}, r1b/p={-0.6,0.5}, r2b/p={0.6,0.5}}
\atoms{hopfbraiding}{b/p={0,0.5}}
\atoms{coalgebra}{d/p={1.2,0}}
\draw (r1-r)--(r1b-l) (r2-r)--(r2b-l) (b-l)--(r1b-b) (b-r)--(r2b-t) (r1-l)--([xshift=-0.6cm]x.north) (r2-l)--([xshift=0.6cm]x.north) (d)edge[mark={ar,s},ind=x]++(0.5,0) (r1b-r)edge[ind=b]++(0,0.3) (r2b-r)edge[ind=c]++(0,0.3);
\draw[rounded corners] (r1-b)--++(0.2,0)--++(0.2,0.4)--++(0.8,0)--(d);
\draw[rounded corners] (r2-b)--++(0.1,0)--(d);
\atoms{representation}{rxx/p={$(x.east)+(0.3,0)$}}
\draw (x.east)--(rxx-l) (rxx-r)edge[ind=a]++(0.3,0);
\draw[rounded corners] (d)--++(0,0.4)--++(0.4,0)--++(0,-0.6)--($(rxx-t)+(0,0.3)$)--(rxx-t);
\end{tikzpicture}
=
\begin{tikzpicture}
\atoms{labbox=$\sigma(A)$,bdastyle={wid=1.2}}{x/p={0,-0.7}}
\atoms{coalgebra}{d/p={1,0}}
\draw ([xshift=-0.4cm]x.north)edge[ind=b]++(0,0.3) ([xshift=0.4cm]x.north)edge[ind=c]++(0,0.3) (d)edge[mark={ar,s},ind=x]++(0.4,0) (x.east)edge[ind=a]++(0.3,0);
\end{tikzpicture}
\;.
\end{gathered}
\end{equation}
In the first step, we used the definition of the commutor in Eq.~\eqref{eq:twisted_symm_commutor}, and the triangular Hopf move Eq.~\eqref{eq:triangular_ax1}. In the second step, we used twice the representation move in Eq.~\eqref{eq:representation_ax1}. In the third step, we used the triangular Hopf move in Eq.~\eqref{eq:triangular_ax3} together with the associativity move of the Hopf algebra. In step 4 we used again twice the representation move in Eq.~\eqref{eq:representation_ax1}. In the last step, we used the symmetry constraint Eq.~\eqref{eq:twisted_symmetry_constraint} on $A$, and the definition of the braiding.
\item We need to check the 2-axioms involving the commutor. The involutivity of the commutor holds due to
\begin{equation}
\begin{gathered}
\begin{tikzpicture}
\atoms{labbox=$\sigma\sigma(A)$,bdastyle={wid=1.2}}{x/p={0,-0.7}}
\draw ([xshift=-0.4cm]x.north)edge[ind=b]++(0,0.3) ([xshift=0.4cm]x.north)edge[ind=c]++(0,0.3);
\draw (x.west)edge[ind=a]++(-0.3,0);
\end{tikzpicture}=
\begin{tikzpicture}
\atoms{labbox=$A$,bdastyle={wid=1.8}}{x/p={0,-0.8}}
\atoms{rot=90,representation}{r1/p={-0.6,-0.2}, r2/p={0.6,-0.2}, r1b/p={-0.6,0.3}, r2b/p={0.6,0.3}}
\atoms{hopfbraiding}{bx/p={0,-0.2}, {b/p={0,0.3},rot=180}}
\draw (r1-r)--(r1b-l) (r2-r)--(r2b-l) (bx-l)--(r1-b) (bx-r)--(r2-t) (b-r)--(r1b-b) (b-l)--(r2b-t) (r1-l)--([xshift=-0.6cm]x.north) (r2-l)--([xshift=0.6cm]x.north) (r1b-r)edge[ind=b]++(0,0.3) (r2b-r)edge[ind=c]++(0,0.3);
\draw (x.west)edge[ind=a]++(-0.3,0);
\end{tikzpicture}\\
=
\begin{tikzpicture}
\atoms{labbox=$A$,bdastyle={wid=2.5}}{x/p={0,-0.8}}
\atoms{rot=90,representation}{r1/p={-1,0}, r2/p={1,0}}
\atoms{algebra}{0/p={0.5,0}, 1/p={-0.5,0}}
\atoms{hopfbraiding}{bx/p={0,-0.3}, {b/p={0,0.3},rot=180}}
\draw (1)edge[](r1-b) (0)edge[](r2-t) (r1-l)--([xshift=-1cm]x.north) (r2-l)--([xshift=1cm]x.north) (r1-r)edge[ind=b]++(0,0.3) (r2-r)edge[ind=c]++(0,0.3);
\draw[rounded corners,mark={ar,e}] (bx-r)--++(0.1,0)--(0);
\draw[rounded corners,mark={ar,e}] (b-l)--++(0.1,0)--(0);
\draw[rounded corners,mark={ar,e}] (bx-l)--++(-0.1,0)--(1);
\draw[rounded corners,mark={ar,e}] (b-r)--++(-0.1,0)--(1);
\draw (x.west)edge[ind=a]++(-0.3,0);
\end{tikzpicture}
=
\begin{tikzpicture}
\atoms{labbox=$A$,bdastyle={wid=2.1}}{x/p={0,-0.8}}
\atoms{rot=90,representation}{r1/p={-0.8,0}, r2/p={0.8,0}}
\atoms{algebra}{0/p={0.3,0}, 1/p={-0.3,0}}
\draw (1)edge[](r1-b) (0)edge[](r2-t) (r1-l)--([xshift=-0.8cm]x.north) (r2-l)--([xshift=0.8cm]x.north) (r1-r)edge[ind=b]++(0,0.3) (r2-r)edge[ind=c]++(0,0.3);
\draw (x.west)edge[ind=a]++(-0.3,0);
\end{tikzpicture}\\
=
\begin{tikzpicture}
\atoms{labbox=$A$,bdastyle={wid=1.2}}{x/p={0,-0.7}}
\draw ([xshift=-0.4cm]x.north)edge[ind=b]++(0,0.3) ([xshift=0.4cm]x.north)edge[ind=c]++(0,0.3);
\draw (x.west)edge[ind=a]++(-0.3,0);
\end{tikzpicture}
\;.
\end{gathered}
\end{equation}
Here we used the representation move in Eq.~\eqref{eq:representation_ax1}, the triangular Hopf move Eq.~\eqref{eq:triangular_ax3}, as well as the representation move in Eq.~\eqref{eq:representation_ax1}.
\item The 2-axiom given in Eq.~\eqref{eq:2axiom_hexagon} for strict ground types holds due to
\begin{equation}
\begin{gathered}
\begin{tikzpicture}
\atoms{labbox=$A$,bdastyle={wid=2.5}}{g/p={1,-0.7}}
\atoms{rot=90,representation}{r0/p={0,0.5}, r2/p={1,0}, r4/p={2,0}, r5/p={2,0.5}}
\atoms{hopfbraiding}{b0/p={1.5,0}, b1/p={1.5,0.5}}
\draw (r4-r)--(r5-l) ([xshift=-1cm]g.north)--(r0-l) ([xshift=-0.0cm]g.north)--(r2-l) ([xshift=1cm]g.north)--(r4-l) (r0-r)edge[ind=b]++(0,0.3) (r2-r)edge[ind=c]++(0,0.8) (r5-r)edge[ind=d]++(0,0.3) (g.west)edge[ind=a]++(-0.3,0);
\draw (b0-l)--(r2-b) (b0-r)--(r4-t) (b1-l)--(r0-b) (b1-r)--(r5-t);
\end{tikzpicture}
=
\begin{tikzpicture}
\atoms{labbox=$A$,bdastyle={wid=2.5}}{g/p={1,-0.7}}
\atoms{rot=90,representation}{r0/p={0,0.5}, r2/p={0.5,0}, r4/p={2,0.25}}
\atoms{hopfbraiding}{b0/p={1,0}, b1/p={1,0.5}}
\atoms{algebra}{m/p={1.5,0.25}}
\draw ([xshift=-1cm]g.north)--(r0-l) ([xshift=-0.5cm]g.north)--(r2-l) ([xshift=1cm]g.north)--(r4-l) (r0-r)edge[ind=b]++(0,0.3) (r2-r)edge[ind=c]++(0,0.8) (r4-r)edge[ind=d]++(0,0.55) (g.west)edge[ind=a]++(-0.3,0);
\draw (b0-l)--(r2-b) (b1-l)--(r0-b) (m)edge[](r4-t);
\draw[rounded corners,mark={ar,e}] (b0-r)--++(0.1,0)--(m);
\draw[rounded corners,mark={ar,e}] (b1-r)--++(0.1,0)--(m);
\end{tikzpicture}\\
=
\begin{tikzpicture}
\atoms{labbox=$A$,bdastyle={wid=2.5}}{g/p={1,-0.7}}
\atoms{rot=90,representation}{r0/p={0,0.5}, r2/p={0.5,0}, r4/p={2,0.25}}
\atoms{hopfbraiding}{b0/p={1.5,0.25}}
\atoms{coalgebra}{m/p={1,0.25}}
\draw ([xshift=-1cm]g.north)--(r0-l) ([xshift=-0.5cm]g.north)--(r2-l) ([xshift=1cm]g.north)--(r4-l) (r0-r)edge[ind=b]++(0,0.3) (r2-r)edge[ind=c]++(0,0.8) (r4-r)edge[ind=d]++(0,0.55) (g.west)edge[ind=a]++(-0.3,0);
\draw (m)edge[mark={ar,s}](b0-l) (b0-r)--(r4-t);
\draw[rounded corners] (r0-b)--++(0.6,0)--(m);
\draw[rounded corners] (r2-b)--++(0.1,0)--(m);
\end{tikzpicture}
\;.
\end{gathered}
\end{equation} 
In the first equation we used the representation move Eq.~\eqref{eq:representation_ax1}. In the second equation, we used the move in Eq.~\eqref{eq:triangular_ax2}.
\item The tensor product is commutative,
\begin{equation}
\begin{gathered}
\begin{tikzpicture}
\atoms{labbox=$A\otimes B$,bdastyle={wid=1.4}}{x/p={0,-0.7}}
\draw ([xshift=-0.4cm]x.north)edge[ind=b]++(0,0.3) ([xshift=0.4cm]x.north)edge[ind=c]++(0,0.3);
\end{tikzpicture}=
\begin{tikzpicture}
\atoms{labbox=$A$}{g1/p={-0.6,-0.9}}
\atoms{labbox=$B$}{g2/p={0.6,-0.9}}
\atoms{rot=90,representation}{r1/p={-0.6,-0.2}}
\atoms{rot=90,representation}{r2/p={0.6,-0.2}}
\atoms{hopfbraiding}{x/p={0,-0.2}}
\draw (x-l)--(r1-b) (x-r)--(r2-t) (r1-l)--(g1.north) (r2-l)--(g2.north) (r1-r)edge[ind=b]++(0,0.3) (r2-r)edge[ind=c]++(0,0.3);
\end{tikzpicture}
=
\begin{tikzpicture}
\atoms{labbox=$A$}{g1/p={-0.6,-0.9}}
\atoms{labbox=$B$}{g2/p={0.6,-0.9}}
\atoms{coalgebra}{m0/p={-0.6,0.4}, m1/p={0.6,0.4}}
\atoms{hopfbraiding}{x/p={0,0.4}}
\draw (x-l)edge[mark={ar,e}](m0) (x-r)edge[mark={ar,e}](m1) (g1.north)edge[ind=b]++(0,0.3) (g2.north)edge[ind=c]++(0,0.3);
\end{tikzpicture}\\
=
\begin{tikzpicture}
\atoms{labbox=$A$}{g1/p={-0.6,-0.9}}
\atoms{labbox=$B$}{g2/p={0.6,-0.9}}
\atoms{algebra}{m0/p={-0.4,0.4}}
\atoms{coalgebra}{m1/p={0.4,0.4}}
\draw (m0)edge[mark={ar,e}](m1) (g1.north)edge[ind=b]++(0,0.3) (g2.north)edge[ind=c]++(0,0.3);
\end{tikzpicture}
=
\begin{tikzpicture}
\atoms{labbox=$A$}{g1/p={-0.6,-0.9}}
\atoms{labbox=$B$}{g2/p={0.6,-0.9}}
\draw (g1.north)edge[ind=b]++(0,0.3) (g2.north)edge[ind=c]++(0,0.3);
\end{tikzpicture}=
\begin{tikzpicture}
\atoms{labbox=$B\otimes A$,bdastyle={wid=1.4}}{x/p={0,-0.7}}
\draw ([xshift=-0.4cm]x.north)edge[ind=b]++(0,0.3) ([xshift=0.4cm]x.north)edge[ind=c]++(0,0.3);
\end{tikzpicture}
\;.
\end{gathered}
\end{equation}
Here, we used the symmetry constraint Eq.~\eqref{eq:twisted_symmetry_constraint}, the triangular Hopf move Eq.~\eqref{eq:triangular_ax4}, and the Hopf move in Eq.~\eqref{eq:move_bialgebra2}.
\item $D_I$ and $D_I^{-1}$ are inverses,
\begin{equation}
\begin{gathered}
\begin{tikzpicture}
\atoms{labbox=$A$,bdastyle={wid=1.4}}{x/p={0,-0.9}}
\atoms{rot=90,representation}{r1/p={0.4,-0.2}}
\atoms{rot=90,representation}{r2/p={0.4,0.4}}
\atoms{lab/ang=0,drinfeldelement}{u1/p={1,-0.2}}
\atoms{lab/ang=0,idrinfeldelement}{u2/p={1,0.4}}
\draw (r1-b)--(u1-l) (r2-b)--(u2-l) (r1-l)--([xshift=0.4cm]x.north) (r1-r)--(r2-l) ([xshift=-0.4cm]x.north)edge[ind=a]++(0,0.3) (r2-r)edge[ind=b]++(0,0.3);
\end{tikzpicture}
=
\begin{tikzpicture}
\atoms{labbox=$A$,bdastyle={wid=1.4}}{x/p={0,-0.9}}
\atoms{rot=90,representation}{r/p={0.4,-0.2}}
\atoms{algebra}{m/p={1,-0.2}}
\atoms{lab/ang=0,drinfeldelement}{u1/p={1.5,-0.5}}
\atoms{lab/ang=0,idrinfeldelement}{u2/p={1.5,0.1}}
\draw (r-b)edge[](m) (m)edge[mark={ar,s}](u1-l) (m)edge[mark={ar,s}](u2-l) (r-l)--([xshift=0.4cm]x.north) ([xshift=-0.4cm]x.north)edge[ind=a]++(0,0.3) (r-r)edge[ind=b]++(0,0.3);
\end{tikzpicture}
\\=
\begin{tikzpicture}
\atoms{labbox=$A$,bdastyle={wid=1.4}}{x/p={0,-0.9}}
\atoms{rot=90,representation}{r/p={0.4,-0.2}}
\atoms{algebra}{m/p={1,-0.2}}
\draw (r-b)edge[](m) (r-l)--([xshift=0.4cm]x.north) ([xshift=-0.4cm]x.north)edge[ind=a]++(0,0.3) (r-r)edge[ind=b]++(0,0.3);
\end{tikzpicture}
=
\begin{tikzpicture}
\atoms{labbox=$A$,bdastyle={wid=1.4}}{x/p={0,-0.9}}
\draw ([xshift=-0.4cm]x.north)edge[ind=a]++(0,0.3) ([xshift=0.4cm]x.north)edge[ind=b]++(0,0.3);
\end{tikzpicture}
\;,
\end{gathered}
\end{equation}
using the properties of the Drinfeld element as well as the representation moves.
\end{itemize}
\end{myobs}

\subsection{Specific dependencies}
\label{sec:twisted_symmetric_specific}
\begin{myobs}
Consider the group Hopf algebra for $\mathbb{Z}_2$, together with the following braiding,
\begin{equation}
\begin{tikzpicture}
\atoms{hopfbraiding}{r/p={0,0}}
\draw (r-l)edge[ind=a]++(-0.3,0) (r-r)edge[ind=b]++(0.3,0);
\end{tikzpicture}
=
(-1)^{ab}\;,
\end{equation}
where the product $ab$ in the exponent is not the group product but the product of $a$ and $b$ as integers. It is easy to check that this defines a triangular Hopf algebra, in particular the braiding is nothing but the $\zz_2$ group Fourier transform. After a basis change such that the algebra is represented by a delta tensor, the $\zz_2$-symmetry becomes a $\zz_2$-grading. In this case we have
\begin{equation}
\begin{tikzpicture}
\atoms{labbox=$A$,bdastyle={wid=1.8}}{x/p={0,-0.8}}
\atoms{rot=90,representation}{r1/p={-0.6,-0.2}}
\atoms{rot=90,representation}{r2/p={0.6,-0.2}}
\atoms{hopfbraiding}{b/p={0,-0.2}}
\draw (b-l)--(r1-b) (b-r)--(r2-t) (r1-l)--([xshift=-0.6cm]x.north) (r2-l)--([xshift=0.6cm]x.north) (r1-r)edge[ind=b]++(0,0.3) (r2-r)edge[ind=c]++(0,0.3) (x-l)edge[ind=a]++(-0.4,0);
\end{tikzpicture}=
(-1)^{|b||c|}A(a,(b,c))\;.
\end{equation}
Thus, twisted-symmetric tensors for this choice of Hopf algebra are equivalent to $\zz_2$-twisted-symmetric tensors as defined in the section before.
\end{myobs}

\begin{myobs}
For every finite $\zz_2$-extended group $(G,\phi)$ as defined in Section~\ref{sec:z2_extended_symmetric}, consider the group Hopf algebra of $G$ together with the following braiding,
\begin{equation}
\begin{tikzpicture}
\atoms{hopfbraiding}{r/p={0,0}}
\draw (r-l)edge[ind=a]++(-0.3,0) (r-r)edge[ind=b]++(0.3,0);
\end{tikzpicture}
=
\begin{cases}
0  &\text{if} \quad a\notin \{1,\phi\} \quad\text{or}\quad b\notin \{1,\phi\}\\
-1 &\text{if} \quad a=\phi \quad \text{and} \quad b=\phi\\
1 &\text{otherwise}
\end{cases}.
\end{equation}
Twisted-symmetric tensors for such a Hopf algebra are equivalent to $\zz_2$-extended symmetric tensors as defined in Section~\ref{sec:z2_extended_symmetric} starting with the $\zz_2$-twisted-symmetric tensors from Section~\ref{sec:twisted_z2_grading} as ground tensor type.
\end{myobs}

\subsection{Mappings}
\begin{mydef}
Twisted-symmetric tensors have a \tdef{trivial-representation mapping} from the ground type to this type, analogously to (untwisted) symmetric tensors.
\end{mydef}

\begin{myrem}
In contrast to symmetric tensors, there is no symmetry-forgetting mapping. The non-trivial commutor depends on the representation, and is therefore not compatible with discarding the symmetry.
\end{myrem}

\begin{mydef}
Every co-commutative Hopf algebra can be equipped with a \tdef{trivial braiding}{trivial_braiding} which makes it into a triangular Hopf algebra,
\begin{equation}
\begin{tikzpicture}
\atoms{hopfbraiding}{r/p={0,0}}
\draw (r-l)edge[ind=a]++(-0.3,0) (r-r)edge[ind=b]++(0.3,0);
\end{tikzpicture}=
\begin{tikzpicture}
\atoms{algebra}{s0/p={-0.3,0.4}, s1/p={0.3,0.4}}
\draw (s0)edge[ind=a]++(-0.5,0) (s1)edge[ind=b]++(0.5,0);
\end{tikzpicture}\;.
\end{equation}
Thus, symmetric tensors for co-commutative Hopf algebras are a special case of twisted-symmetric tensors for triangular Hopf algebras.
\end{mydef}

\subsection{Use in physics}
Clearly, the main physical significance of twisted-symmetric array tensors is that they are able to describe physical models with fermionic degrees of freedom. Time evolution, ground state or thermal properties of local many-body quantum models with fermions can be formalized by tensor-network models of $\zz_2$-twisted-symmetric tensors. In usual condensed-matter many-body models, we have degrees of freedom of \emph{modes} that which can be either occupied by a fermion or not. An operator acting on $n$ modes labeled $0,\ldots,n-1$, can be expanded in terms of creation and annihilation operators,
\begin{equation}
\label{eq:fermionic_operator}
\begin{multlined}
\sum_{\substack{s_0,\ldots s_{n-1}\\s_0'\ldots,s_{n-1}'}} A^{s_0,\ldots,s_{n-1}}_{s_0',\ldots,s_{n-1}'}\\
(c_0^\dagger)^{s_0} \cdots (c_{n-1}^\dagger)^{s_{n-1}} \ket{0} \bra{0} (c_y)^{s_{n-1}'}\cdots (c_0)^{s_0'}\;,
\end{multlined}
\end{equation}
where $s_i$ and $s_i'$ take values in $\{0,1\}$ depending on whether the fermionic degree of freedom is occupied or not. We observe the following.

Each physical operator must preserve fermion parity. That is, $A$ can have non-zero entries only when
\begin{equation}
\label{eq:mode_fermion_parity}
\sum_i s_i+\sum_i s_i'=0\;,
\end{equation}
where the summation is understood mod $2$. Furthermore, the coefficients in the array $A$ depend on an ordering of the modes, since creation and annihilation operators at different sites anti-commute instead of commute. If we change the ordering of, e.g., the $0$th and $1$st mode in the output, the same operator is represented by a different array $A'$, given by
\begin{equation}
\label{eq:mode_fermion_commutor}
(A')^{s_1,s_0,\ldots,s_{n-1}}_{s_0',\ldots,s_{n-1}'}=A^{s_0,\ldots,s_{n-1}}_{s_0',\ldots,s_{n-1}'}\quad (-1)^{s_0 s_1}\;.
\end{equation}

We see that $A$ is nothing but a $\zz_2$-twisted-symmetric tensor, with the 0-data of each index given by $\{0,1\}$ and $|s|=s$. With this identification, Eq.~\eqref{eq:mode_fermion_parity} is the symmetry constraint, whereas Eq.~\eqref{eq:mode_fermion_commutor} is the formula for the commutor. It is well-known that tensor networks can be equipped with a fermionic interpretation by introducing an ordering bookkeeping, which is often denoted using \emph{Grassmann variables}, see for example Ref.~\cite{Kraus2009} or Ref.~\cite{Gaiotto2015}. The formulation as a tensor types gives a conceptually very clear, rigorous and compact formalization of such a bookkeeping.

In Section~\ref{sec:symmetric_physics}, we mentioned how $\zz_2$-extended-group-symmetric with $\zz_2$-twisted-symmetric tensors as ground type describe physics with a symmetry group which contains fermion parity as a subgroup, such as particle-number conservation. In Section~\ref{sec:twisted_symmetric_specific} we found that for any $\zz_2$-extended group we can define a triangular Hopf algebra, yielding an equivalent tensor type from array tensors in only one step.

\section{Frobenius-positive tensors}
\subsection{Motivation}
Just like symmetric tensors, Frobenius-positive tensors are a method to construct a new tensor type from any ground tensor type by imposing a constraint. The constraint is defined via models of a certain liquid in the ground tensor type. We present two variants of this tensor type based on slightly different liquids.

Frobenius-positive tensors for array tensors as ground type can be seen as a generalization of array tensors with non-negative entries. They are the tensor type that one has to use for dealing with full quantum mechanics on a ket-bra level, including measurements, controlled operations, or dissipative channels. They are closely related to the symmetric monoidal category of \emph{CPTP maps} and \emph{categorical quantum mechanics}.

\subsection{Definition: Via \texorpdfstring{$T$}{T}-Frobenius algebras}

Let us start by defining a liquid whose models will be the 0-data of Frobenius-positive tensors (similarly to how representations were the 0-data of symmetric tensors).

\begin{mydef}
\tdef{$T$-Frobenius algebras}{t_frobenius_algebra} are the following liquid extending symmetric Frobenius algebras by the following move,
\begin{equation}
\label{eq:t_frobenius_move}
\begin{tikzpicture}
\atoms{algebra}{0/p={0,0}, 1/p={0.8,0}}
\draw (0)edge[mark={ar,s}](1) (0)edge[ind=a]++(120:0.5) (0)edge[mark={ar,s},ind=c]++(-120:0.5) (1)edge[ind=b]++(0:0.5);
\end{tikzpicture}
=
\begin{tikzpicture}
\atoms{algebra,hflip}{0/p={0,0}}
\draw (0)edge[mark={ar,s},ind=a]++(120:0.5) (0)edge[ind=c]++(-120:0.5) (0)edge[mark={ar,s},ind=b]++(0:0.5);
\end{tikzpicture}
\;.
\end{equation}
Be aware that the shape on the right-hand side is reflected, so the indices $a$ and $b$ are interchanged as opposed to the non-reflected case. Note that this move can only be interpreted in tensor types that do not have a dual, in contrast to the pure Frobenius algebras.

The move above says that we can reflect the tensor shape and at the same time interchange ``input'' with ``output'' indices. This can be elegantly incorporated into the notation by replacing ingoing/outgoing arrows by clockwise- or counter-clockwise pointing ``flags''. So we write
\begin{equation}
\begin{tikzpicture}
\atoms{circ}{c/p={0,0}}
\draw  (c)edge[mark={flag,s,r}]++(-90:0.6) (c)edge[mark={flag,s}]++(30:0.6) (c)edge[mark={flag,s}]++(150:0.6);
\end{tikzpicture}
,\qquad
\begin{tikzpicture}
\atoms{circ}{c/p={0,0}}
\draw  (c)edge[mark={flag,s}]++(0:0.6) (c)edge[mark={flag,s}]++(180:0.6);
\end{tikzpicture}
,\qquad
\begin{tikzpicture}
\atoms{circ}{c/p={0,0}}
\draw  (c)edge[mark={flag,s}]++(0:0.6);
\end{tikzpicture}
\end{equation}
For the product, Frobenius form, and unit. If we mirror a shape, the flags change from clockwise to counter-clockwise, according to the move above.

We can also add the commutative and special move.
\end{mydef}

\begin{myrem}
As for ordinary symmetric Frobenius algebras, we can define elements with arbitrarily many indices and flags for $T$-Frobenius algebras. Again, any move between planar (cycle-free) networks can be derived from the move. For $T$-Frobenius algebras, we can also reflect the networks before we equate them.
\end{myrem}

\begin{mydef}
\tdef{$T$-Frobenius-positive tensors}{tf_positive_tensor} are the following tensor type:
\begin{itemize}
\item They can be defined for any ground tensor type $\calg$ without dual. They have an identity if we restrict to special $T$-Frobenius algebras.
\item The flavor is the flavor of the ground type $\calg$.
\item The 0-data are given by (special) $T$-Frobenius algebras. The unit 0-data is the trivial algebra, the product of 0-data is the tensor product of algebras.
\item A 1-data $A\in \dat_1(F)$ is a 1-data of the ground type,
\begin{equation}
A\in \dat_1^{\mathcal{G}}(F_0)
\end{equation}
fulfilling the \tdef{positivity constraint}{positivity_constraint}
\begin{equation}
\label{eq:positivity_constraint}
\begin{tikzpicture}
\atoms{labbox=$A$}{t/}
\draw (t-r)edge[ind=a]++(0.4,0);
\end{tikzpicture}=
\begin{tikzpicture}
\atoms{lab={t=F,p=-60:0.35},tfrobenius}{s/p={0.7,0}}
\atoms{labbox=$R$,bdastyle=rc}{r1/p={0,0.4}, r2/p={0,-0.4}}
\draw[rc,internal] (r1.west)--++(-0.3,0)|-(r2.west);
\draw[rounded corners] (r1.east)edge[mark={flag,e}](s) (r2.east)edge[mark={flag,e,r}](s) (s)edge[mark={flag,s,r},ind=a]++(0.6,0);
\end{tikzpicture}\;,
\end{equation}
for some
\begin{equation}
R\in \dat_1^\calg(R_0\otimes F_0)\;.
\end{equation}
$R$ will be called a \emph{root tensor} for $A$ and the dotted index contracted between the two copies of $R$ in Eq.~\eqref{eq:positivity_constraint} will be called the \emph{internal index}. $F_0$ is the 0-data of the $T$-Frobenius algebra $F$.
\item The 2-functions are those of the ground type $\calg$.
\end{itemize}
\end{mydef}

\begin{myrem}
Note that there can be multiple root tensors $R$ for the same 1-data $A$. Also note that what we call ``positive'' in this document would be usually rather referred to as ``non-negative'', or ``positive semidefinite''.
\end{myrem}

\begin{myobs}
In order to show that the above definition actually defines a tensor type, we have to show that the constraint in Eq.~\eqref{eq:positivity_constraint} is compatible with all the 2-functions of the ground type $\calg$:
\begin{itemize}
\item The compatibility of the invertible 2-functions is trivial when formulated within the graphical calculus of $\calg$.
\item Consider two 1-data $A\in \dat_1(a)$ and $B\in \dat_1(b)$ with root tensors $R$ and $S$. Then the following tensor $T$ defines a root tensor for $A\otimes B$,
\begin{equation}
\begin{tikzpicture}
\atoms{labbox=$T$,bdastyle=rc}{r1/p={0,0.5}}
\draw (r1.west)edge[internal]node[pos=1,left]{$xy$}++(-0.3,0) (r1.east)--node[pos=1,right]{$ab$}++(0.3,0);
\end{tikzpicture}=
\begin{tikzpicture}
\atoms{labbox=$R$,bdastyle=rc}{r1/p={0,0.5}}
\atoms{labbox=$S$,bdastyle=rc}{r2/p={0,1.2}}
\draw (r1.west)edge[internal]node[pos=1,left]{$x$}++(-0.3,0) (r1.east)--node[pos=1,right]{$a$}++(0.3,0);
\draw (r2.west)edge[internal]node[pos=1,left]{$y$}++(-0.3,0) (r2.east)--node[pos=1,right]{$b$}++(0.3,0);
\end{tikzpicture}\;.
\end{equation}
\item For a 1-data $A\in \dat_1(F\otimes(G^*\otimes G))$ with root tensor $R$, the contraction $[A]$ has the following root tensor $S$,
\begin{equation}
\begin{tikzpicture}
\atoms{labbox=$S$,bdastyle=rc}{r1/p={0,0.5}}
\draw (r1.west)edge[internal]node[pos=1,left]{$xy$}++(-0.3,0) (r1.east)--node[pos=1,right]{$a$}++(0.3,0);
\end{tikzpicture}=
\begin{tikzpicture}
\atoms{labbox=$R$,bdastyle={rc,hei=1}}{r1/}
\atoms{lab={t=G,p=60:0.35},tfrobenius}{s1/p={0.9,0.15}}
\draw[internal] (r1.west)edge[ind=x]++(-0.3,0);
\draw ([yshift=0.3cm]r1.east)edge[bend left, mark={flag,e,r}](s1) ([yshift=0cm]r1.east)edge[bend right,mark={flag,e}](s1) ([yshift=-0.3cm]r1.east)edge[ind=a]++(0.3,0);
\draw (s1)edge[mark={flag,s,r},ind=y]++(0.5,0);
\end{tikzpicture}\;.
\end{equation}
With this choice we indeed find
\begin{equation}
\begin{gathered}
\begin{tikzpicture}
\atoms{labbox=$[A]$}{t/}
\draw (t-r)edge[ind=a]++(0.3,0);
\end{tikzpicture}=
\begin{tikzpicture}
\atoms{labbox=$A$,bdastyle={hei=1}}{t/}
\draw[rounded corners] ([yshift=0.3cm]t.east)--++(0.5,0)|-([yshift=0cm]t.east);
\draw ([yshift=-0.3cm]t.east)edge[ind=a]++(0.3,0);
\end{tikzpicture}=
\begin{tikzpicture}
\atoms{tfrobenius}{{s1/p={0.9,0.6},lab={t=G,p=45:0.4}}, {s2/p={0.9,0},lab={t=G,p=45:0.4}}, {s3/p={0.9,-0.6},lab={t=F,p=45:0.4}}}
\atoms{labbox=$R$,bdastyle={rc,hei=1}}{r1/p={0,0.7}, r2/p={0,-0.7}}
\draw[internal,rounded corners] (r1.west)--++(-0.3,0)|-(r2.west);
\draw ([yshift=0.3cm]r1.east)edge[mark={flag,e}](s1) ([yshift=0.2cm]r2.east)edge[mark={flag,e,r}](s1) ([yshift=0cm]r1.east)edge[mark={flag,e}](s2) ([yshift=0cm]r2.east)edge[mark={flag,e,r}](s2) ([yshift=-0.3cm]r1.east)edge[mark={flag,e}](s3) ([yshift=-0.3cm]r2.east)edge[mark={flag,e,r}](s3);
\draw[rc,mark={flag,s,r},mark={flag,e,r}] (s1)--++(0.6,0)|-(s2);
\draw (s3)edge[mark={flag,s,r},ind=a]++(0.5,0);
\end{tikzpicture}\\=
\begin{tikzpicture}
\atoms{tfrobenius}{{s1/p={0.9,0.85},lab={t=G,p=60:0.35}}, {s2/p={0.9,-0.55},lab={t=G,p=60:0.35}}, {s3/p={1.8,-0.3},lab={t=F,p=60:0.35}}}
\atoms{labbox=$R$,bdastyle={rc,hei=1}}{r1/p={0,0.7}, r2/p={0,-0.7}}
\draw[internal,rounded corners] (r1.west)--++(-0.3,0)|-(r2.west);
\draw ([yshift=0.3cm]r1.east)edge[bend left,mark={flag,e,r}](s1) ([yshift=0cm]r1.east)edge[bend right,mark={flag,e}](s1) ([yshift=0.3cm]r2.east)edge[bend left,mark={flag,e,r}](s2) ([yshift=0cm]r2.east)edge[bend right,mark={flag,e}](s2);
\draw[mark={flag,e},rc] ([yshift=-0.3cm]r1.east)--++(0.5,0)--(s3);
\draw[mark={flag,e,r},rc] ([yshift=-0.3cm]r2.east)--++(0.5,0)--(s3);
\draw[rc,mark={flag,s,r},mark={flag,e,r}] (s1)--++(0.4,0)|-(s2);
\draw (s3)edge[mark={flag,s,r},ind=a]++(0.5,0);
\end{tikzpicture}
=
\begin{tikzpicture}
\atoms{tfrobenius,lab={t=F,p=60:0.35}}{s/p={0.7,0}}
\atoms{labbox=$S$,bdastyle={rc}}{r1/p={0,0.4}, r2/p={0,-0.4}}
\draw[rounded corners,internal] (r1.west)--++(-0.3,0)|-(r2.west);
\draw[rounded corners] (r1.east)edge[mark={flag,e}](s) (r2.east)edge[mark={flag,e}](s) (s)edge[mark={flag,s},ind=a]++(0.5,0);
\end{tikzpicture}\;.
\end{gathered}
\end{equation}
In the third step we used the $T$-Frobenius move in Eq.~\eqref{eq:t_frobenius_move} and the associativity move of Frobenius algebras.
\item The trivial tensor has a root tensor, namely itself, which is trivial in the graphical calculus of $\calg$.
\item The root tensor $R$ for the identity tensor $\idop \in \dat_1(G^*\otimes G)$ is given by the identity tensor itself,
\begin{equation}
\begin{tikzpicture}
\atoms{labbox=$R$,bdastyle={rc,hei=0.7}}{r1/}
\draw ([yshift=-0.2cm]r1.east)edge[ind=a]++(0.3,0) ([yshift=0.2cm]r1.east)edge[ind=b]++(0.3,0);
\end{tikzpicture}
=
\begin{tikzpicture}
\draw[rounded corners,startind=a, ind=b] (0,0)--++(-0.5,0)|-(0,0.5);
\end{tikzpicture}\;.
\end{equation}
We indeed find
\begin{equation}
\begin{tikzpicture}
\atoms{tfrobenius}{{0/p={0,0},lab={t=G,p=-60:0.35}}, {1/p={0,0.6},lab={t=G,p=60:0.35}}}
\draw[rc,mark={flag,s,r},mark={flag,e,r}] (0)--++(-0.6,-0.6)--++(0,0.6)--(1);
\draw[rc,mark={flag,s},mark={flag,e}] (0)--++(-0.6,0.6)--++(0,0.6)--(1);
\draw (0)edge[mark={flag,s,r},ind=a]++(0.5,0) (1)edge[mark={flag,s,r},ind=b]++(0.5,0);
\end{tikzpicture}
=
\begin{tikzpicture}
\draw[rounded corners,startind=a, ind=b] (0,0)--++(-0.5,0)|-(0,0.5);
\end{tikzpicture}
\end{equation}
which holds due to the ``special'' move.
\end{itemize}
\end{myobs}

\subsection{Definition: Via \texorpdfstring{$\dagger$}{t}-Frobenius algebras}
In this section we discuss a variant of Frobenius-positive tensors that is closer to conventional formulations of quantum mechanics. What we describe in this section is equivalent to categorical quantum mechanics, just that we do not have a flow of time.

\begin{mydef}
\tdef{$\dagger$-Frobenius algebras}{dagger_frobenius_algebra} are not a liquid in a tensor 2-scheme, but in a 2-scheme consisting of the tensor 2-scheme together with an involutive tensor mapping $\calm$ from this tensor type to itself. More precisely, the mapping is special in that the mapping 1-function is equal to the dual 1-function (if there exists one). It can be obtained by extending symmetric Frobenius algebras by the following move,
\begin{equation}
\begin{tikzpicture}
\atoms{algebra}{0/p={0,0}, 1/p={0.8,0}}
\draw (0)edge[mark={ar,s}](1) (0)edge[ind=a]++(120:0.5) (0)edge[mark={ar,s},ind=c]++(-120:0.5) (1)edge[ind=b]++(0:0.5);
\end{tikzpicture}
=
\begin{tikzpicture}
\atoms{algebra,hflip, shadecirc}{0/p={0,0}}
\draw (0)edge[mark={ar,s},ind=a]++(120:0.5) (0)edge[ind=c]++(-120:0.5) (0)edge[mark={ar,s},ind=b]++(0:0.5);
\end{tikzpicture}
\;.
\end{equation}
Note that the gray shade around the tensor denotes the mapping $\calm$ applied to the 1-data, as introduced in Section~\ref{sec:mapping_effective_scheme}.
\end{mydef}

\begin{mydef}
\tdef{$\dagger$-Frobenius-positive tensors}{df_positive_tensor} are the following modification of $T$-Frobenius tensors:
\begin{itemize}
\item The 0-data are given by (special) $\dagger$-Frobenius algebras instead of $T$-Frobenius algebras.
\item There is a non-trivial dual 1-function. The dual of a $\dagger$-Frobenius algebra $F$ is given by reversing the arrows,
\begin{equation}
\begin{tikzpicture}
\atoms{algebra,hflip,lab={t=F,p=90:0.35}}{0/p={0,0}}
\draw (0)edge[mark={ar,s},ind=a]++(150:0.5) (0)edge[mark={ar,s},ind=b]++(30:0.5) (0)edge[ind=c]++(-90:0.5);
\end{tikzpicture}
\quad\rightarrow\quad
\begin{tikzpicture}
\atoms{algebra,hflip,lab={t=F^*,p=90:0.4}}{0/p={0,0}}
\draw (0)edge[ind=b]++(150:0.5) (0)edge[ind=a]++(30:0.5) (0)edge[mark={ar,s},ind=c]++(-90:0.5);
\end{tikzpicture}
\;.
\end{equation}
\item The positivity constraint for a 1-data $A\in \dat_1(F)$ becomes
\begin{equation}
\label{eq:dagger_positivity_constraint}
\begin{tikzpicture}
\atoms{labbox=$A$}{t/}
\draw (t-r)edge[ind=a]++(0.4,0);
\end{tikzpicture}=
\begin{tikzpicture}
\atoms{lab={t=F,p=-60:0.35},algebra}{s/p={0.7,0}}
\atoms{labbox=$R$,bdastyle=rc}{r1/p={0,0.4}, {r2/p={0,-0.4},shaderect={0.4,0.4}}}
\draw[rc,internal] (r1-l)--++(-0.3,0)|-(r2-l);
\draw[rounded corners] (r1-r)edge[mark={ar,e}](s) (r2-r)--(s) (s)edge[ind=a]++(0.5,0);
\end{tikzpicture}\;.
\end{equation}
\end{itemize}
\end{mydef}

\begin{myobs}
The compatibility of the positivity with the 2-functions is similar to the case of $T$-Frobenius algebras. E.g., for the contraction of a 1-data $A\in \dat_1(F\otimes(G^*\otimes G))$ with root tensor $R$, we can define a root tensor
\begin{equation}
\begin{tikzpicture}
\atoms{labbox=$S$,bdastyle=rc}{r1/p={0,0.5}}
\draw (r1.west)edge[internal]node[pos=1,left]{$xy$}++(-0.3,0) (r1.east)--node[pos=1,right]{$a$}++(0.3,0);
\end{tikzpicture}=
\begin{tikzpicture}
\atoms{labbox=$R$,bdastyle={rc,hei=1}}{r1/}
\atoms{lab={t=G,p=60:0.35},algebra,hflip}{s1/p={0.9,0.15}}
\draw[internal] (r1.west)edge[ind=x]++(-0.3,0);
\draw ([yshift=0.3cm]r1.east)edge[bend left,mark={ar,e}](s1) ([yshift=0cm]r1.east)edge[bend right](s1) ([yshift=-0.3cm]r1.east)edge[ind=a]++(0.3,0);
\draw (s1)edge[ind=y]++(0.5,0);
\end{tikzpicture}\;.
\end{equation}
Be aware that the shape for the $G$-tensor is reflected. With this choice we indeed find
\begin{equation}
\begin{gathered}
\begin{tikzpicture}
\atoms{labbox=$[A]$}{t/}
\draw (t-r)edge[ind=a]++(0.3,0);
\end{tikzpicture}=
\begin{tikzpicture}
\atoms{labbox=$A$,bdastyle={hei=1}}{t/}
\draw[rounded corners] ([yshift=0.3cm]t.east)--++(0.5,0)|-([yshift=0cm]t.east);
\draw ([yshift=-0.3cm]t.east)edge[ind=a]++(0.3,0);
\end{tikzpicture}=
\begin{tikzpicture}
\atoms{algebra}{{s1/p={0.9,0.6},lab={t=G,p=45:0.4}}, {s2/p={0.9,0},lab={t=G^*,p=45:0.4}}, {s3/p={0.9,-0.6},lab={t=F,p=45:0.4}}}
\atoms{labbox=$R$,bdastyle={rc,hei=1}}{r1/p={0,0.7}, {r2/p={0,-0.7},shaderect={0.4,0.7}}}
\draw[internal,rounded corners] (r1.west)--++(-0.3,0)|-(r2.west);
\draw ([yshift=0.3cm]r1.east)edge[mark={ar,e}](s1) ([yshift=0.2cm]r2.east)edge[](s1) ([yshift=0cm]r1.east)edge[mark={ar,e}](s2) ([yshift=0cm]r2.east)edge[](s2) ([yshift=-0.3cm]r1.east)edge[mark={ar,e}](s3) ([yshift=-0.3cm]r2.east)edge[](s3);
\draw[rc] (s1)--++(0.6,0)|-(s2);
\draw (s3)edge[ind=a]++(0.5,0);
\end{tikzpicture}\\=
\begin{tikzpicture}
\atoms{algebra}{{s1/p={0.9,0.85},hflip,lab={t=G,p=60:0.35}}, {s2/p={0.9,-0.55},hflip,shadecirc,lab={t=G,p=60:0.35}}, {s3/p={1.8,-0.3},lab={t=F,p=60:0.35}}}
\atoms{labbox=$R$,bdastyle={rc,hei=1}}{r1/p={0,0.7}, {r2/p={0,-0.7},shaderect={0.4,0.7}}}
\draw[internal,rounded corners] (r1.west)--++(-0.3,0)|-(r2.west);
\draw ([yshift=0.3cm]r1.east)edge[bend left,mark={ar,e}](s1) ([yshift=0cm]r1.east)edge[bend right](s1) ([yshift=0.3cm]r2.east)edge[bend left,mark={ar,e}](s2) ([yshift=0cm]r2.east)edge[bend right](s2);
\draw[mark={ar,e},rc] ([yshift=-0.3cm]r1.east)--++(0.5,0)--(s3);
\draw[rc] ([yshift=-0.3cm]r2.east)--++(0.9,0)--(s3);
\draw[rc] (s1)--++(0.4,0)|-(s2);
\draw (s3)edge[ind=a]++(0.5,0);
\end{tikzpicture}
=
\begin{tikzpicture}
\atoms{algebra,lab={t=F,p=60:0.35}}{s/p={0.7,0}}
\atoms{labbox=$S$,bdastyle={rc}}{r1/p={0,0.4}, {r2/p={0,-0.4},shaderect={0.4,0.4}}}
\draw[rounded corners,internal] (r1.west)--++(-0.3,0)|-(r2.west);
\draw[rounded corners] (r1.east)edge[mark={ar,e}](s) (r2.east)edge[](s) (s)edge[ind=a]++(0.5,0);
\end{tikzpicture}\;.
\end{gathered}
\end{equation}
\end{myobs}

\subsection{Specific dependencies}
In this section we will look in more detail at the physically most relevant cases. For $T$-Frobenius-positive tensors, this is when the ground type $\calg$ is equal to real array tensors. For $\dagger$-Frobenius-positive tensors, this is when $\calg$ is complex array tensors and the mapping $\calm$ is complex conjugation. Let us start by giving a few examples for real $T$-Frobenius algebras, which at the same time are also examples for complex $\dagger$-Frobenius algebras.

\begin{myobs}
\begin{itemize}
\item The group algebra is a special $T$-Frobenius algebra, after we change the prefactors.
\item The delta algebra is a special $T$-Frobenius algebra.
\item The complex number algebra is a special $T$-Frobenius algebra.
\end{itemize}
\end{myobs}

In contrast to level tensors, where we first introduced Frobenius algebras, the $T$-Frobenius algebras in this section need not be commutative. Thus, let us introduce two more non-commutative examples.
\begin{mydef}
The \tdef{quaternion algebra}{quaternion_star_algebra} is the following special $T$-Frobenius algebra:
\begin{itemize}
\item The 0-data is the set $\{\mathbf{1},\mathbf{i},\mathbf{j},\mathbf{k}\}$.
\item The product is
\begin{equation}
\begin{tikzpicture}
\atoms{tfrobenius}{c/p={0,0}}
\draw[]  (c)edge[mark={flag,s,r},ind=c]++(-90:0.5) (c)edge[mark={flag,s},ind=b]++(30:0.5) (c)edge[mark={flag,s},ind=a]++(150:0.5);
\end{tikzpicture}=
\frac12
\begin{pmatrix}
\mathbf{1}:1 & \mathbf{i}:1 & \mathbf{j}:1 & \mathbf{k}:1\\
\mathbf{i}:1 & \mathbf{1}:-1 & \mathbf{k}:1 & \mathbf{j}:-1\\
\mathbf{j}:1 & \mathbf{k}:-1 & \mathbf{1}:-1 & \mathbf{i}:1\\
\mathbf{k}:1 & \mathbf{j}:1 & \mathbf{i}:-1 & \mathbf{1}:-1
\end{pmatrix}\;.
\end{equation}
Here row and column correspond to $a$ and $b$, respectively, in the order $(\mathbf{1},\mathbf{i},\mathbf{j},\mathbf{k})$. The index $c$ is denoted in sparse notation as a list with items ``index configuration : tensor entry'' containing only items for non-zero tensor entries (in the above case there's only one such non-zero entry always).
\item The unit is
\begin{equation}
\begin{tikzpicture}
\atoms{tfrobenius}{x/p={0,0}}
\draw (x)edge[mark={flag,s},ind=a]++(0.5,0);
\end{tikzpicture}
=
2
\begin{pmatrix}
1&0&0&0
\end{pmatrix}
\;.
\end{equation}
\item The Frobenius form is
\begin{equation}
\begin{tikzpicture}
\atoms{tfrobenius}{x/p={0,0}}
\draw (x)edge[mark={flag,s},ind=b]++(0.5,0) (x)edge[mark={flag,s},ind=a]++(-0.5,0);
\end{tikzpicture}
=
\begin{pmatrix}
1&0&0&0\\0&-1&0&0\\0&0&-1&0\\0&0&0&-1
\end{pmatrix}\;.
\end{equation}
\end{itemize}
\end{mydef}

\begin{mydef}
For every $n$, the \tdef{matrix algebra}{matrix_algebra} is the following special $T$-Frobenius algebra:
\begin{itemize}
\item The 0-data is the set $\{0,\ldots,n-1\}\times \{0,\ldots,n-1\}$.
\item The product is given by
\begin{equation}
\begin{tikzpicture}
\atoms{tfrobenius}{c/p={0,0}}
\draw[]  (c)edge[mark={flag,s,r},ind=c_0c_1]++(-90:0.5) (c)edge[mark={flag,s},ind=b_0b_1]++(30:0.5) (c)edge[mark={flag,s},ind=a_0a_1]++(150:0.5);
\end{tikzpicture}=
\frac{1}{\sqrt{n}}
\begin{tikzpicture}
\atoms{void}{m1/p={-30:0.2}, m2/p={90:0.2}, m3/p={-150:0.2}};
\draw[rounded corners,startind=a_1,ind=a_2] ($(m1)+(30:0.8)$)--(m1)--++(-90:0.8);
\draw[rounded corners,startind=b_1,ind=a_0] ($(m2)+(30:0.8)$)--(m2)--++(150:0.8);
\draw[rounded corners,startind=b_0,ind=b_2] ($(m3)+(150:0.8)$)--(m3)--++(-90:0.8);
\end{tikzpicture}\;.
\end{equation}
The right hand side is a network consisting of three free bonds representing three $\{0,\ldots,n-1\}\times \{0,\ldots,n-1\}$ identity matrices whose indices are blocked in a non-trivial way.
\item The unit is given by
\begin{equation}
\begin{tikzpicture}
\atoms{tfrobenius}{x/p={0,0}}
\draw (x)edge[mark={flag,s},ind=a_0a_1]++(0.5,0);
\end{tikzpicture}
=
\sqrt{n}
\quad
\begin{tikzpicture}
\draw[rc,startind=a_0,ind=a_1] (0,0)--++(-0.7,0)--++(0,0.4)--++(0.7,0);
\end{tikzpicture}
\;.
\end{equation}
\item The Frobenius form is given by
\begin{equation}
\begin{tikzpicture}
\atoms{tfrobenius}{x/p={0,0}}
\draw (x)edge[mark={flag,s},ind=b_0b_1]++(0.5,0) (x)edge[mark={flag,s},ind=a_0a_1]++(-0.5,0);
\end{tikzpicture}
=
\begin{tikzpicture}
\draw[startind=a_1,ind=b_2] (0,0)--++(1,0);
\draw[startind=b_1,ind=a_2] (0,0.4)--++(1,0);
\end{tikzpicture}
\;.
\end{equation}
\end{itemize}
\end{mydef}

\begin{myrem}
\label{rem:matrix_general}
Note that, apart from the prefactors, the matrix algebra is defined purely in terms of the network graphical calculus, and thus works for any 0-data $a$ of any ground type with an identity tensor. For the normalization to work out, we need two scalars $|a|^{1/2}$ and $|a|^{-1/2}$ such that
\begin{equation}
\begin{tikzpicture}
\atoms{circ,lab={t=$|a|^{1/2}$,p=90:0.35}}{0/, 1/p={0.8,0}}
\end{tikzpicture}
=
\begin{tikzpicture}
\draw (0,0)circle(0.5);
\end{tikzpicture}\;,\qquad
\begin{tikzpicture}
\atoms{circ,lab={t=$|a|^{1/2}$,p=90:0.35}}{0/}
\atoms{circ,lab={t=$|a|^{-1/2}$,p=90:0.35}}{1/p={1,0}}
\end{tikzpicture}
=
\hspace{1cm}\;,
\end{equation}
where the right hand side of the first equation is a loop of the binding in consideration, and the right hand side of the second equation is the trivial tensor.
\end{myrem}

\begin{myrem}
$T$-Frobenius algebras in real array tensors are equivalent to what is known as \emph{finite-dimensional real C*-algebras}. It is known \cite{Li2003} that every such algebra is isomorphic to a direct sum of irreducible blocks. Each of these blocks is $M_n(X)$, which is the algebra of $n\times n$ matrices whose entries are real numbers, complex numbers, or quaternions. In other words, each $T$-Frobenius algebra $A$ is isomorphic to one of the following form,
\begin{equation}
A=\bigoplus_{i} M_{n_i}\otimes X_i\;.
\end{equation}
Here $M_{n_i}$ is the matrix algebra for $n=n_i$, and $X_i$ is either 1) the trivial algebra, 2) the complex number algebra or 3) the quaternionic algebra, depending on $i$. To be precise, the matrix, trivial, complex or quaternion algebras above were presented for a particular choice of $\alpha$ normalization. This normalization could be different for each component of the direct sum.

Similarly, $\dagger$-Frobenius algebras in complex array tensors with complex conjugation are equivalent to what is known as \emph{finite-dimensional C*-algebras}, and it is known that those can be decomposed as
\begin{equation}
A=\bigoplus_{i} M_{n_i}\;.
\end{equation}

It might seem surprising that the classification for the complex case is easier than for the real case. However, every finite-dimensional C*-algebra can be made real via isomorphism. The complex number algebra is complex-isomorphic to a direct sum of two trivial algebras, but not real-isomorphic. The quaternion algebra is complex-isomorphic to the $n=2$ matrix algebra, but not real-isomorphic.
\end{myrem}

\begin{myobs}
\label{obs:delta_positivity}
Consider a $T$-Frobenius-positive 1-data $A\in \dat_1(F)$ with $F$ being a delta algebra. In this case, Eq.~\eqref{eq:positivity_constraint} becomes
\begin{equation}
\begin{multlined}
A(i)=\sum_j R((j,i)) R((j,i))\\=
\sum_j (R(j,i))^2\geq 0 \quad \forall a\;.
\end{multlined}
\end{equation}
Conversely, for every $A$ with
\begin{equation}
A(a)\geq 0 \quad \forall a
\end{equation}
we can take
\begin{equation}
R(a)=\sqrt{A(a)}
\end{equation}
as a root tensor with trivial internal index. So we found that $T$-Frobenius-positive tensors with delta algebra as 0-data are the same as array tensors with non-negative reals as entries.
\end{myobs}

\begin{myobs}
\label{obs:matrix_positivity}
Consider a $T$-Frobenius-positive 1-data $A\in \dat_1(F)$ with $F$ being a matrix algebra. In this case, Eq.~\eqref{eq:positivity_constraint} becomes
\begin{equation}
A((a,a'))=
\sum_{i,a''} R((i,(a'',a))) R((i,(a'',a')))\;.
\end{equation}
When we interpret $A$ and $R$ as matrices, blocking the first two indices of $R$, we can write this as
\begin{equation}
A=RR^T\;,
\end{equation}
so $A$ as a matrix is positive semidefinite.

Conversely, for any positive semidefinite matrix $A$ we can take
\begin{equation}
R=\sqrt{A}
\end{equation}
as a root tensor, where the square root denotes the square root of $A$ as a matrix. The internal index is trivial in this case.
\end{myobs}

\begin{myobs}
\label{obs:complex_positivity}
Consider a $T$-Frobenius-positive 1-data $A\in \dat_1(F)$ with $F$ being the complex number algebra. In this case, Eq.~\eqref{eq:positivity_constraint} becomes
\begin{equation}
\begin{gathered}
A(\mathbf{1})=
\sum_i \left(R((i,\mathbf{1})) R((i,\mathbf{1}))+ R((i,\mathbf{i})) R((i,\mathbf{i}))\right) \geq 0\;,\\
A(\mathbf{i})=
\sum_i \left(R((i,\mathbf{1})) R((i,\mathbf{i}))- R((i,\mathbf{i})) R((i,\mathbf{1}))\right) = 0\;.
\end{gathered}
\end{equation}
In other words, if we interpret $A$ as a complex number with real part $A(\mathbf{1})$ and imaginary part $A(\mathbf{i})$, it is purely real and positive.

Conversely, for every real and positive number $A$, we can take
\begin{equation}
\begin{gathered}
R(\mathbf{1})=\sqrt{A(\mathbf{1})}\;,\\
R(\mathbf{i})=0
\end{gathered}
\end{equation}
as a root tensor with trivial internal index.

Analogously, if $F$ is the quaternion algebra, $A$ interpreted as a quaternion is purely-real and positive.
\end{myobs}

\begin{myrem}
Consider a $\dagger$-Frobenius-positive 1-data $A\in \dat_1(M)$ where $M$ is a matrix algebra. Eq.~\eqref{eq:dagger_positivity_constraint} becomes
\begin{equation}
A((a,a'))=
\sum_{i,a''} R(i,(a'',a)) \overline{R(i,(a'',a'))}\;,
\end{equation}
or,
\begin{equation}
A=RR^\dagger\;,
\end{equation}
so $A$ is hermitian and positive semi-definite as a matrix. Conversely, any positive semidefinite matrix has a (positive semidefinite) square root which we can pick as the root tensor.
\end{myrem}

\begin{myrem}
We have seen that for real or complex array tensors as ground type we can always choose a root tensor whose internal index is trivial. Note that this might not be true for other ground tensor types.
\end{myrem}

\subsection{Mappings}
\begin{mydef}
There's the obvious \tdef{positivity-forgetting mapping}{positivity_forgetting_mapping} from Frobenius-positive tensors to ground type tensors:
\begin{itemize}
\item It works for both $T$-Frobenius algebras and $\dagger$-Frobenius algebras.
\item It has strict homomorphicity.
\item The mapping of 0-data $F$ simply forgets the 1-data of the algebra $F$,
\begin{equation}
m(F)=F_0\;.
\end{equation}
\item The mapping of 1-data is the identity.
\end{itemize}
\end{mydef}

\begin{mydef}
Consider a ground tensor type where each 0-data $a$ is equipped with a loop square root $|a|^{1/2}$, an inverse thereof $|a|^{-1/2}$, and we also demand that a fourth square root $|a|^{1/4}$ exists. The \tdef{doubling mapping}{doubling_mapping} is the following mapping from the ground type to Frobenius-positive tensors:
\begin{itemize}
\item A 0-data $a$ is mapped to the generalized matrix algebra over that 0-data as defined in Remark~\ref{rem:matrix_general},
\begin{equation}
m(a)=\text{Matrix}[a]\;.
\end{equation}
\item The mapping of a 1-data $A\in \dat_1(a)$ is the tensor product of twice that 1-data,
\begin{equation}
\begin{tikzpicture}
\atoms{labbox=$M(A)$}{m/}
\draw (m-r)edge[ind=ij]++(0.4,0);
\end{tikzpicture}
=
\begin{tikzpicture}
\atoms{labbox=$A$, bdastyle={rc}}{d1/, d2/p={0,0.6}}
\draw[rc,ind=j] (d2.east)--++(0.4,0);
\draw[rc,ind=i] (d1.east)--++(0.4,0);
\end{tikzpicture}\;.
\end{equation}
\item The mapping is defined within the graphical calculus of $\calg$, and the homomorphicity 2-functions are obtained from the invertible 2-functions of $\calg$ accordingly.
\item There is an analogous $\dagger$-Frobenius version of the mapping,
\begin{equation}
\begin{tikzpicture}
\atoms{labbox=$M(A)$}{m/}
\draw (m-r)edge[ind=ij]++(0.4,0);
\end{tikzpicture}
=
\begin{tikzpicture}
\atoms{labbox=$A$, bdastyle={rc}}{{d1/shaderect={0.35,0.35}}, d2/p={0,0.7}}
\draw[rc,ind=j] (d2.east)--++(0.4,0);
\draw[rc,ind=i] (d1.east)--++(0.4,0);
\end{tikzpicture}\;.
\end{equation}
\end{itemize}
\end{mydef}

\begin{myobs}
In order to show that the doubling mapping is well defined, we need to find a root tensor for $A\otimes A$. It is given by
\begin{equation}
\begin{tikzpicture}
\atoms{labbox=$R$, bdastyle={rc,hei=0.7}}{r1/}
\draw[internal] (r1.west)edge[ind=i]++(-0.3,0);
\draw ([yshift=-0.2cm]r1.east)edge[ind=j]++(0.3,0) ([yshift=0.2cm]r1.east)edge[ind=k]++(0.3,0);
\end{tikzpicture}
=
\begin{tikzpicture}
\atoms{labbox=$A$, bdastyle={rc}}{d/}
\draw[ind=j] (d.east)--++(0.3,0) coordinate(x);
\draw[ind=i,startind=k] ([sy=0.4]x)--++(-1,0);
\end{tikzpicture}
\quad
\left(|a|^{1/4}\right)
\end{equation}
We find
\begin{equation}
\begin{gathered}
\begin{tikzpicture}
\atoms{labbox=$R$, bdastyle={rc}}{r1/, r2/p={0,0.8}}
\atoms{lab={t=\text{Matrix}[a],p=-30:0.7},tfrobenius}{0/p={0.8,0.4}}
\draw[rc] (r1.west)--++(-0.4,0)|-(r2.west) (r1.east)--(0) (r2.east)--(0) (0)edge[ind=ij]++(0.5,0);
\end{tikzpicture}\\
=
\begin{tikzpicture}
\atoms{labbox=$A$, bdastyle={rc}}{d1/, d2/p={0,1}}
\draw[rc,ind=j] (d2.east)--++(0.4,0)--++(0,-0.6)--++(-1.2,0)--++(0,1)--++(1.6,0)--++(0,-0.5)--++(0.3,0);
\draw[rc,ind=i] (d1.east)--++(0.8,0)--++(0,0.5)--++(0.3,0);
\end{tikzpicture}
\left(|a|^{1/4}\right) \left(|a|^{1/4}\right) \left(|a|^{-1/2}\right)
\\
=
\begin{tikzpicture}
\atoms{labbox=$A$, bdastyle={rc}}{d1/, d2/p={0,0.6}}
\draw[rc,ind=j] (d2.east)--++(0.4,0);
\draw[rc,ind=i] (d1.east)--++(0.4,0);
\end{tikzpicture}\;.
\end{gathered}
\end{equation}
\end{myobs}
\subsection{Use in physics}
\label{sec:uf_positive_physical}
Of particular physical importance are $\dagger$-Frobenius-positive tensors with complex array tensors as ground type. Tensor-network models for this tensor type describe full quantum many-body physics, including controlled state preparation, dissipative channels and measurements. As explained in \cite{cstar_qmech}, $\dagger$-Frobenius-positive tensors (there called ``positive *-tensors'') describe different objects in a quantum model on a mixed-state level. Examples for this are the super-operators representing quantum channels, density matrices representing (mixed) states, POVMs or instruments representing measurements. In general, every quantum degree of freedom of a state or at the input or output of a channel or measurement correspond to an index whose 0-data is a matrix $\dagger$-Frobenius algebra. The two components of that 0-data correspond to the ket- and bra level. Every classical degree of freedom, e.g., at the output of a measurement or the input to a controlled quantum operation is an index of the corresponding tensor whose 0-data is the delta $\dagger$-Frobenius algebra.

Surely, for any $\dagger$-Frobenius-positive tensor-network model, we can forget about the positivity constraint and obtain a conventional array tensor-network model, as described in Section~\ref{sec:array_physics}. So $\dagger$-Frobenius-positive tensors are not necessary for writing down tensor-network models for quantum mechanics. Their significance is to state in a very concise fashion which models we are allowed to write down: Array tensor-network models not fulling the positivity constraint do not have a quantum mechanical interpretation, and their evaluations will not in general yield probability distributions.

The categorical version of this tensor type is the symmetric monoidal category of CPTP maps with the trace as terminal object. More precisely, we can consider the generalization thereof based on $\dagger$-Frobenius algebras which also features classical indices as defined in Ref.~\cite{Selinger2007}. This category is the natural way to describe quantum mechanics with a flow of time. However, thermal or ground state properties of quantum systems are more naturally formalized as tensor-network models in a tensor type. E.g., as demonstrated in \cite{cstar_qmech}, measurements on a thermal state for a local quantum spin Hamiltonian constitute $\dagger$-Frobenius-positive tensor-network model, apart from an approximation error due to Trotterization.

Most formulations of quantum mechanics are on the level of pure states and unitary time evolution. Unitary time evolution can be discretized into a circuit which is a tensor network of ordinary complex array tensors. This tensor network is not a full tensor-network model as it does not include measurements or classical controls. The doubling mapping is exactly the mapping that implements the transition from the pure-state tensor network to the full $\dagger$-Frobenius-positive tensor-network model. Physically, this means going from pure states to the corresponding mixed states, and from unitaries to the corresponding quantum channels.



\section{Geometric and combinatorial tensor types}
In this section we give a few examples for tensor types based on geometric intuition. The 1-data are geometric structures of some kind ``between some boundaries'' which are given by the 0-data. The (tensor) product of 0-data (1-data) is the disjoint union of the geometric structures and contraction consists in some kind of ``gluing'' of boundary parts. A simple case of this is when the 1-data are manifolds with boundary.

\begin{mydef}
\tdef{$n$-dimensional bordism tensors}{bordism_tensor} are the following tensor type:
\begin{itemize}
\item They can be defined in any dimension $n$.
\item They don't need duals, have a unit 0-data, trivial tensor, and identity tensor.
\item A 0-data $a$ is a topological $(n-1)$-manifold. The product is the disjoint union of manifolds, and the unit 0-data is the empty manifold.
\item A 1-data $A\in \dat_1(a)$ consists of 1) a topological $n$-manifold $A_M$ with boundary and 2) a homeomorphism of $A_H: a \rightarrow \partial A_M$.
\item The trivial tensor is given by the empty manifold, with the homeomorphism given by the unique function from the empty set to the empty set.
\item The tensor product of two 1-data $A\in \dat_1(a)$ and $B\in \dat_1(b)$ is the disjoint union of the corresponding $n$-manifolds $M$ and the concatenation of the homeomorphisms $H$,
\begin{equation}
\begin{gathered}
(A\otimes B)_M=A_M\sqcup B_M\\
(A\otimes B)_H=A_H\hat\sqcup B_H\;.
\end{gathered}
\end{equation}
Note that the first $\sqcup$ denotes the disjoint union of manifolds, whereas the second $\hat\sqcup$ denotes the concatenation of functions, as defined in Eq.~\eqref{eq:function_concatenation}.
\item The contraction applied to a 1-data $A\in \dat_1(a\sqcup (b\sqcup b))$ is given by gluing the two $b$-parts of the boundary according to $H$,
\begin{equation}
\begin{gathered}
[A]_M=A_M/\{A_H((1,(0,j)))\sim A_H((1,(1,j))) \forall j\in b\}\;.\\
[A]_H(i)=A_H((0,i))\;.
\end{gathered}
\end{equation}
\item The identity tensor $\idop \in \dat_1(a\sqcup a)$ is given by $a$ times an interval,
\begin{equation}
\begin{gathered}
\idop_M=a\times [0,1]\;,\\
\idop_H((0,i))=(i,0)\;,\\
\idop_H((1,i))=(i,1)\;.
\end{gathered}
\end{equation}
For this to fulfill the identity 2-axiom, we need to define two 1-data as equivalent if the manifolds are homeomorphic. Another possibility would be to allow for ``infinitely thin'' manifolds as 1-data and let the identity be $a\times [0]$.
\item The invertible 2-functions are all canonical, in the sense that they are defined via the canonical bijections $\Phi_\sqcup$ introduced in Section~\ref{sec:schur_preliminaries}.
\end{itemize}
\end{mydef}

\begin{mycom}
The bordism tensors are the tensor-type formulation of the compact closed category of cobordisms. As in general, our language never makes a distinction between source and target boundary components, whereas the categorical language starts with this distinction but effectively reverts it by adding a (co-) evaluation. We find that this makes the tensor language a bit more natural in the context of axiomatic TQFT, where there is no notion of a flow of time from the very beginning on.
\end{mycom}

\begin{myrem}
The bordism tensor type can be enriched with different kinds of structure. For example we can consider a version for oriented manifolds with oriented boundary, spin manifolds with spin boundary, manifolds with corners with manifolds with boundary as boundary, and so on. E.g., for oriented manifolds, the definition is completely analogous, but we need a non-trivial dual 1-function consisting of orientation reversal.
\end{myrem}

Even simpler than continuum manifolds are discrete combinatorial geometric structures, such as the following one.
\begin{mydef}
\tdef{Graph tensors}{graph_tensor} are the following tensor type:
\begin{itemize}
\item They do not need a dual. They have a unit 0-data, trivial tensor, and identity tensor.
\item The 0-data are finite sets with disjoint union.
\item A 1-data $A \in \dat_1(a)$ consists of a graph $A_G$ (with multiple edges and loops allowed) with vertex set $A_V$ and a map $A_H: a \rightarrow A_V$ (which neither needs to be an injection nor a surjection).
\item The tensor product of two 1-data $A\in \dat_1(a)$ and $B\in \dat_1(b)$ is the disjoint union,
\begin{equation}
\begin{gathered}
(A\otimes B)_G= A_G\sqcup B_G\;,\\
(A\otimes B)_H= A_H\hat\sqcup B_H\;.\\
\end{gathered}
\end{equation}
\item The contraction of a 1-data $A\in \dat_1(a\sqcup (b\sqcup b))$ is given by identifying the $b$-vertices according to $H$,
\begin{equation}
\begin{gathered}
[A]_G=A_G/\{A_H((0,(0,j)))\sim A_H((0,(1,j)))\forall j\in b\}\;,\\
[A]_H(i)=A_H((0,i))\;.\\
\end{gathered}
\end{equation}
\item The trivial tensor is the empty graph with no vertices or edges.
\item The identity tensor $\idop\in \dat_1(a\sqcup a)$ is given a graph $G_{\text{NE}}(a)$ with vertex set $a$ and no edges,
\begin{equation}
\begin{gathered}
\idop_V = a\;,\\
\idop_G = G_{\text{NE}}(a)\;,\\
\idop_H((0,j))=j\;,\\
\idop_H((1,j))=j\;.
\end{gathered}
\end{equation}
\item The invertible 2-functions are canonical.
\end{itemize}
\end{mydef}

Let us give one more example for a tensor type describing a combinatorial geometric structure, this time a discretized version of (piece-wise linear) bordism tensors.
\begin{mydef}
\tdef{Simplicial tensors}{simplicial_tensor} are the following tensor type:
\begin{itemize}
\item They can be defined in any dimension $n$. They do not require a dual, have a unit 0-data, a trivial tensor and identity tensor.
\item The 0-data are finite sets with disjoint union.
\item A 1-data $A \in \dat_1(a)$ consists of 1) an $n$-dimensional simplicial complex $A_C$ with boundary, where $A_S$ is the set of boundary $(n-1)$-simplices, 2) a bijection $A_H:a\rightarrow A_S$. By simplicial complex, we basically mean a piece-wise linear triangulation with a \emph{branching structure}, that is, an orientation of the edges that is non-cyclic around every triangle. However, we allow the manifold to be ``infinitely thin'' at certain places, such that there can be two copies of the same $n-1$-simplex in $A_S$.
\item The tensor product is the disjoint union analogous to bordism tensors or graph tensors.
\item The contraction applied to a 1-data $A\in \dat_1(a\sqcup (b\sqcup b))$ is given by gluing the boundary $n-1$-simplices of the $b$-components, according to $A_H$,
\begin{equation}
\begin{gathered}
[A]_C=A_C/\{A_H((0,(0,j)))\sim A_H((0,(1,j)))\forall j\in b\}\;,\\
[A]_H(i)=A_H((0,i))\;.\\
\end{gathered}
\end{equation}
Thereby, the branching structure determines which of the vertices of $A_H((0,(0,j)))$ are identified with which of the vertices of $A_H((0,(1,j)))$.
\item The trivial tensor is the empty triangulation of the empty manifold.
\item Consider the simplicial complex $C_2^a$ consisting of one isolated $n-1$-simplex for each $i\in a$. This is one of the ``infinitely thin'' manifolds mentioned above, and for each $i\in a$, the corresponding $n-1$-simplex gives rise to two boundary simplices $C_2^a(i,0)\in A_S$ and $C_2^a(i,1)\in A_S$. The identity tensor $\idop\in \dat_1(a\sqcup a)$ is given by
\begin{equation}
\begin{gathered}
\idop_C=C_2^a\;,\\
\idop_H((\chi,i))=C_2^a(i,\chi)\;.
\end{gathered}
\end{equation}
\item All invertible 2-functions are canonical.
\end{itemize}
\end{mydef}

\begin{mycom}
As a variant of simplicial tensors, we could introduce an equivalence of 1-data under so-called \emph{Pachner moves} which are local deformations which map between any two triangulations of the same topology. Tensor mappings from this tensor type to a ``physical'' tensor type like array tensors then correspond to concrete microscopic fixed-point models for TQFTs, analogously to how tensor mappings from bordism tensors to array tensors are abstract axiomatic TQFTs.
\end{mycom}

We have mentioned that algebraic structures such as TQFTs can often be formalized as tensor mappings from a geometric tensor type to another tensor type such as array tensors. In fact, any algebraic structure, that is, models of any liquid in a tensor type $\mathcal{X}$, can be interpreted as tensor mappings from a certain combinatorial tensor type to $\mathcal{X}$. This combinatorial tensor type is given as follows.
\begin{mydef}
The \tdef{free tensor type}{free_tensor_type} is the following tensor type:
\begin{itemize}
\item It can be defined for any liquid.
\item The flavor is the same as the flavor of tensor type for which the liquid is defined. E.g., if the liquid is for tensor types with duals, the free tensor type has duals. Here, we describe the simplest case without duals or identity tensor.
\item A 0-data $a$ consists of a finite set $a(i)$ for each binding $i$ (recall that bindings are placeholders for the different 0-data needed to specify liquid models). The product is component-wise disjoint union,
\begin{equation}
(a\otimes b)(i)=a(i)\sqcup b(i)\;.
\end{equation}
The unit 0-data (if needed) is
\begin{equation}
1(i)=\{\}\forall i\;.
\end{equation}
\item A 1-data $A\in\dat_1(a)$ consists of 1) a network diagram $A_N$ involving the elements of the liquid, with an open index set $A_O(i)$ for each binding $i$, and 2) a set of bijections
\begin{equation}
A_H^i: a(i)\rightarrow A_O(i)\;.
\end{equation}
Two 1-data are considered equivalent if they differ by applying the moves of the liquid to the network. The actual 1-data are such equivalence classes.
\item The tensor product is the disjoint union of the network diagrams.
\item The contraction of a 1-data $A\in \dat_1(a\sqcup (b\sqcup b))$ is given by
\begin{equation}
\begin{gathered}
[A]_N=A_N/\{A_H^i((0,(0,j)))\sim A_H^i((0,(1,j)))\forall i, \forall j\in b(i)\}\;,\\
[A]_H(i)=A_H((0,i))\;,\\
\end{gathered}
\end{equation}
where the equivalence $\sim$ denotes that we connect the two open indices $A_H^i((0,(0,j)))$ and $A_H^i((0,(1,j)))$ by a bond.
\item The invertible 2-functions are canonical.
\end{itemize}
\end{mydef}

\begin{myobs}
A tensor mapping from the free tensor type over a liquid is specified by what it associates to networks consisting of individual elements. Such a mapping is consistent if it associates the same to the two sides of any move of the liquid. So we see that tensor mappings from the free tensor type to another tensor type $\mathcal{X}$ are equivalent to liquid models in the tensor type $\mathcal{X}$.
\end{myobs}

\begin{mycom}
A liquid is similar to what is known as a \emph{PROP} in category theory. A \emph{model} of a PROP is a \emph{functor} from the PROP to another (symmetric monoidal) category, which is analogous to how a model of a liquid is equivalent to a tensor mapping from the free tensor type of the liquid to another tensor type.
\end{mycom}

\section{Labeled-graph tensors}
\subsection{Motivation}
In this section we describe a tensor type whose 1-data consists of a mix of a geometric combinatorial object and an array. A tensor of such a type carries the information about a combinatorial structure possibly describing some $n$-dimensional space, as well as an array whose open indices are associated to different places in the combinatorial structure. The bond dimensions of the array tensor are allowed to depend on the surrounding combinatorial structure.


\subsection{Definition}
\begin{mydef}
\tdef{Labelled-graph tensors}{labelled_graph_tensor} are the following tensor type:
\begin{itemize}
\item They can be defined for any (finite) \tdef{label set}{label_set} $L$, and any commutative semiring $K$.
\item They don't need a dual 1-function. They have an identity and a trivial tensor.
\item A 0-data $a$ consists of 1) a finite set $a_X$ of \tdef{dependencies}{dependency}, and 2) for every \tdef{labelling}{dependency_labelling} of dependencies $l: a_X\rightarrow L$, a finite set $a_M(l)$, called \tdef{multiplicity}{graph_labelled_multiplicity}.
\item The product of two 0-data $a$ and $b$ is given by the disjoint union of the dependencies, together with the cartesian products of the multiplicities,
\begin{equation}
\begin{gathered}
(a \otimes b)_X = a_X \sqcup b_X\;,\\
(a \otimes b)_M(l_1 \sqcup l_2) = a_M(l_1) \times b_M(l_2)\;.
\end{gathered}
\end{equation}
\item The unit 0-data has no dependencies, $\tpart{1}{X} = \{\}$, with $1_M(l)=\{0\}$ for $l$ the unique labeling of the empty set.
\item A 1-data $A \in \dat_1(a)$ consists of 1) a set $A_V$ of \tdef{label vertices}{label_vertex}, 2) a surjective (but not necessarily injective) map that associates to every dependency a label vertex,
\begin{equation}
A_\psi: \tpart{a}{X} \rightarrow A_V\;,
\end{equation}
and, 3) for each \mdef{graph labeling} $l: A_V\rightarrow L$, an array 1-data whose 0-data is the corresponding multiplicity,
\begin{equation}
A_A(l) \in \dat_1^{\text{Array}}(a_M(l\circ \psi))\;.
\end{equation}
\item The tensor product of two 1-data $A\in \dat_1(a)$ and $B\in \dat_1(b)$ is given by the disjoint union of label vertices, and the tensor product of the two array tensors,
\begin{equation}
\begin{gathered}
(A\otimes B)_V= A_V \sqcup B_V\;,\\
(A\otimes B)_\psi= A_\psi \hat\sqcup B_\psi\;,\\
(A\otimes B)_A(l_1\sqcup l_2)=A_A(l_1) \otimes B_A(l_2)\;.
\end{gathered}
\end{equation}
\item The contraction applied to a 1-data $A \in \dat_1(a\otimes (b\otimes b))$ is given by the following. We have
\begin{equation}
[A]_V = A_V/\sim/V_{\text{disconnect}}\;,
\end{equation}
where the first backslash denotes modulo an equivalence relation, whereas the second backslash denotes the complement with respect to a subset. The equivalence relation and subset are given by
\begin{equation}
A_\psi((1,(0,x))) \sim A_\psi((1,(1,x))) \text{ for any } x\in b_X\;,
\end{equation}
and
\begin{equation}
V_{\text{disconnect}} = \{v \in A_V/\sim \big| \quad \nexists x\in a_X: A_\psi((0,x))\in v\}\;.
\end{equation}
In other words, we fuse together label vertices that correspond to the same dependency of the two contracted components, and then remove label vertices that are not associated to any dependency. Moreover, we have
\begin{equation}
[A]_\psi = \sim \circ A_\psi\rvert_0\;,
\end{equation}
where $\sim$ denotes the map from elements of $A_V$ to the corresponding equivalence class. For underlying array 1-data, we have
\begin{equation}
A_A(l)=\sum_{l_d\in L^{V_{\text{disconnect}}}} [A_A]((l\sqcup l_d)\circ \sim)\;.
\end{equation}
In other words, we push back the labeling of the equivalence classes to a labeling of the original label vertices, and contract the corresponding array tensor. Thereby we sum over all labelings of $V_{\text{disconnect}}$.

\item The trivial tensor is given by $\mathbf{1}_V = \{\}$ and $\mathbf{1}_A(l) = \mathbf{1}^{\text{Array}}$.

\item The identity tensor $\idop\in \dat_1(a\otimes a)$ is given by
\begin{equation}
\begin{gathered}
\idop_V = a_X\;,\\
\idop_\psi((0,x)) = \idop_\psi((1,x)) = x \quad \forall x\in a_X\;,\\
\idop_A(l) = \idop^{\text{Array}}(a_M(l)\times a_M(l))\;.
\end{gathered}
\end{equation}

\item All invertible 2-functions are canonical.
\end{itemize}
\end{mydef}

\subsection{Mappings}
In this section we define a tensor mapping from labeled-graph tensors to array tensors: Labeled-graph tensors can be seen as array tensors whose 0-data is a disjoint union of different sets, indexed by labelings of the dependencies. However, multiple dependencies can refer to the same label vertex, and therefore labelings can be forced to have the same value on different dependencies. Alternatively, we could use ordinary array tensors whose entries we set to $0$ for configurations with incompatible labelings of dependencies. This is what the following label-copy mapping formalizes.

\begin{mydef}
The \tdef{label-copy mapping}{label_copy_mapping} is the following mapping from labeled-graph tensors to array tensors:
\begin{itemize}
\item It has strict homomorphicity.
\item The mapping of a 0-data $a$ is given by taking the disjoint union of all original 0-data for arbitrary labelings,
\begin{equation}
m(a)=\bigsqcup_{l\in L^{a_X}} a_M(l)\;.
\end{equation}
\item The mapping of a 1-data $A\in \dat_1(b)$ is given by
\begin{equation}
\begin{gathered}
M(A)((l,m))
=
\begin{cases}
A(l')(m) & \text{if} \quad \exists l': l = l'\circ a_\psi\\
0 & \text{otherwise}
\end{cases}
\\
\forall l\in L^{a_X}, m\in a_M(l)\;.
\end{gathered}
\end{equation}
Here, $(l,m)$ denotes the element $m$ in the $l$-component of the disjoint union.
\end{itemize}
\end{mydef}

\subsection{Use in physics}
Labeled-graph tensors are based on array tensors, which formalize classical statistical or quantum many-body models. Usually, we assume that the degrees of freedom of such a quantum or classical statistical system can take configurations completely independently. However, in some situations, only certain local configurations may be allowed.

One example for this are \emph{dimer models}, which are defined on some grid. The total set of configurations of such a model consists of valid \emph{dimer coverings}, i.e. colorings of edges (where the colored edges correspond to the ``dimers''), such that each vertex is adjacent to (exactly, or at most) one colored edge. E.g., consider a square lattice, and let the configuration space consist of all edge colorings such that every vertex is adjacent to exactly one colored edge. A quantum state or classical probability distribution of such a model can be written as a labeled-graph tensor: Each edge corresponds to a label vertex, where the label is the corresponding coloring. Each vertex corresponds to an index whose 0-data has $4$ dependencies. The 1-data associates those dependencies to the $4$ label vertices at the adjacent edges. The multiplicity of the 0-data is a one-element set if exactly one adjacent edge is colored, and the empty set otherwise. Also operators acting locally on dimer configurations are graph-labeled tensors.

Another example are fixed-point models of topological order, such as the Levin-Wen string-net model \cite{Levin2004}. This is a Hamiltonian model in $2+1$ dimensions with degrees of freedom on the edges of a trivalent lattice. The Hamiltonian contains a ``vertex term'', which is mathematically more naturally formulated as a hard constraint on the Hilbert space: We only allow edge configurations with $N^{ij}_k=1$ (where $N$ is part of the fusion category defining the model) around every vertex, where $i,j,k$ are the configurations of the edges adjacent to a fixed tri-valent vertex. As for dimer models, a state of this model can thus be written as a labeled-graph tensor with label vertices on the edges and indices at the vertices. The multiplicity of the 0-data at a vertex whose surrounding edges have labels $i,j,k$ is an $N^{ij}_k$-element set. This formulation naturally allows for multiplicities greater than $1$. The space-time picture of the Levin-Wen model is known as Turaev-Viro state-sum \cite{Turaev1992, Barrett1993}. This state-sum is nothing but a tensor network of labeled-graph tensors, with one $4$-index tensor at every tetrahedron of the triangulation of space-time.

Another way to get constraints on a many-body state space are commuting, but not on-site local symmetries that we impose on the system. By block-diagonalizing the symmetry, we obtain a Hilbert space consisting of irrep configurations (playing the role of the labelings), internal indices for multi-dimensional irreps, and multiplicity indices. Internal and multiplicity indices depend on the irrep configuration, such that the state space is elegantly described by a labeled-graph tensor. In fact, the Levin-Wen model is a special case of this, starting from a Kitaev quantum double \cite{Kitaev2003}, and imposing the vertex terms as local symmetries.

We can always forget about constraints on the state space, and use tensor-network models for ordinary array tensors. This is formalized by the label-copy mapping in the previous section. However, using labeled-graph of unconstrained array tensors has the benefit that the hard constraints greatly reduce the bond dimensions of the arrays we have to contract. Although labeled-graph tensors do not have a drastic advantage over array tensors in terms of computational complexity such as Schur-complement tensors, they can make life much easier for computations by hand or brute-force numerics.

\chapter{Outlook}
\label{sec:outlook}
In this chapter we present a few future directions and sketch new tensor types that we didn't have time to thoroughly investigate in the previous chapter.
\section{Stabilizer tensors}
We omitted one family of tensor types which is very important in physics because it can be seen as a subset of array tensors that is efficiently contractible, similar to Schur-complement tensors. \emph{Stabilizer tensors} are a formalization of the well-known \emph{stabilizer formalism} which is a major tool in quantum error correction. In essence, the 1-data of a stabilizer tensor is a stabilizer code with a unique code state, and this state is the corresponding array tensor.

In this section, we will give a brief sketch of how to define them in a tensor-type language. As for Schur-complement tensors, the 0-data consists of a set (or number) of ``elementary degrees of freedom'', or ``modes'' instead of individual configurations, and the product of 0-data is based on the disjoint union (or sum) rather than the cartesian product (or product). Thus, the 1-data and 2-functions require only polynomial resources in storage size and computation time. A key difference to Schur-complement tensors is that the 1-data stabilizer tensors form a discrete set, not a continuous one.

The formal definition goes along the following lines.
\begin{mydef}
\tdef{Stabilizer tensors}{stabilizer_tensor} are the following tensor type:
\begin{itemize}
\item The 0-data are finite sets with disjoint union.
\item They have a trivial tensor and identity tensor. They don't need a dual.
\item A 1-data $A\in \dat_1(a)$ consists of an $a\times a$ matrix
\begin{equation}
A_S: a\times a\rightarrow \{1,X,Y,Z\}
\end{equation}
called \tdef{stabilizer matrix}{stabilizer_matrix} together with a \tdef{sign vector}{stabilizer_sign_vector}
\begin{equation}
A_\sigma: a\rightarrow \{1, -1\}\;.
\end{equation}

$1$, $X$, $Y$, and $Z$ can be interpreted as the corresponding Pauli operators acting on a 2-dimensional Hilbert space of a qubit. The elements of $a$ can be interpreted as qubits, and $A_S(\cdot,i)$ for $i\in A_g$ can be interpreted as a tensor product of Pauli operators acting on those $|a|$ qubits. The stabilizers of a stabilizer code must commute, thus for all $i,j\in a$, we have
\begin{equation}
A_S(\cdot,i) A_S(\cdot,j) = A_S(\cdot,j) A_S(\cdot,i)\;.
\end{equation}
We also wanted there to be a unique code state. For this, we view the Pauli operators as a projective representation of $\zz_2\times \zz_2$, and interpret $A_S(\cdot,i)$ as a vector vectors in $\zz^{2|a|}$. Those vectors for all $i\in a$ to be linearly independent.

The sign $A_\sigma(i)$ can be viewed as an overall prefactor for the tensor-product Pauli operator $A_S(\cdot,i)$. For stabilizer codes, the code operators can be linearly recombined, so two 1-data $A\in\dat_1(a)$ and $B\in\dat_1(a)$ are equivalent if they are related by an invertible $\zz_2$-valued matrix
\begin{equation}
O: a\times a \rightarrow \{1,-1\}
\end{equation}
such that
\begin{equation}
B_\sigma(i) \cdot B_S(\cdot, i) = \prod_j O(i, j) \cdot A_\sigma(j)\cdot A_S(\cdot, j)\;.
\end{equation}

So, more precisely, 1-data are equivalence classes of the 1-data described above. We will not always explicitly distinguish between the equivalence classes and their representatives.

\item The tensor product of two 1-data $A\in\dat_1(a)$ and $B\in\dat_1(b)$ is given by concatenating $S$ and $\sigma$,
\begin{equation}
\begin{gathered}
(A\otimes B)_S = A_S\oplus B_S\;,\\
(A\otimes B)_\sigma((\beta,g))=
\begin{cases}
A_\sigma(g) & \text{if}\ \beta=0\\
B_\sigma(g) & \text{if}\ \beta=1\\
\end{cases}\;.
\end{gathered}
\end{equation}
Recall the notation for the disjoint union and direct sum from Eq.~\eqref{eq:disjoint_union_definition} and Eq.~\eqref{eq:direct_sum_definition}.
\item To compute the contraction of a 1-data $A\in \dat_1(a\sqcup (b\sqcup b))$ we first use the equivalence transformations via an invertible matrix $O$ above to bring it into a form where
\begin{equation}
A_S((1,(0,i)),(0,j))=A_S((1,(1,i)),(0,j))\quad \forall j\in a, i\in b\;.
\end{equation}
Then, we simply restrict to the left-over modes,
\begin{equation}
\begin{gathered}
[A]_S(i,j)=A_S((0,i),(0,j))\;,\\
[A]_\sigma(i)=A_\sigma((0,i))\;.\\
\end{gathered}
\end{equation}
\item The trivial tensor is the unique empty stabilizer matrix and sign vector.
\item The identity tensor $\idop\in \dat_1(a\sqcup a)$ for a single-element set $a$ is given by the matrix
\begin{equation}
\begin{gathered}
\idop_S=
\begin{pmatrix}
X&X\\Z&Z
\end{pmatrix}\;,\\
\idop_\sigma=
\begin{pmatrix}
1&1
\end{pmatrix}\;.
\end{gathered}
\end{equation}
where the columns of $\idop_S$ correspond to its first index, whereas the rows are labeled by the second index.
\item The invertible 2-functions are all canonical, i.e., determined by the bijections $\Phi_\sqcup$ introduced in Section~\ref{sec:schur_preliminaries}.
\end{itemize}
\end{mydef}

The interpretation of stabilizer matrix and sign vector $(A_S, A_\sigma) \in \dat_1(a)$ in terms of its unique code state on $|a|$ qubits is a tensor mapping from stabilizer to array tensors. Formally, each element of $a$ gets mapped to a qubit,
\begin{equation}
m(a)=\{0,1\}^a\;.
\end{equation}
The array 1-data $M(A)$ is the $+1$ eigenstate of all the operators
\begin{equation}
A_\sigma(i)\cdot A_S(\cdot, j), \quad j\in a\;.
\end{equation}
Due to the commutativity and $\zz_2$-linear independence of the operators, this eigenstate is unique, but only up to a global prefactor. There is no way to consistently fix this prefactor, which can be seen by e.g. considering the tensor given by the stabilizer operators $-X\otimes X$ and $-Z\otimes Z$, which would be mapped to the Pauli-$Y$ matrix interpreted as an array with two $2$-dimensional indices. The stabilizer operators are symmetric under exchanging the two modes, whereas the Pauli-$Y$ array is anti-symmetric. So, all one can hope for is a mapping to \emph{projective} array tensors. But even in this case, contracting the two $2$-dimensional indices of the Pauli-$Y$ array (i.e., taking the trace of that matrix) results in a scalar $0$ which is distinct from non-zero scalars for projective array tensors, whereas stabilizer tensors have a unique scalar, namely the trivial tensor. This can be fixed by additionally bookkeeping a scalar prefactor together with the stabilizer data.

\section{Quasi-Hopf algebras}
In the Section~\ref{sec:symmetric} we defined symmetric tensors with respect to co-commutative Hopf algebras. The consistency of the symmetry constraint with commutor, associator, and unitor followed immediately from the co-commutativity, co-associativity, and co-unitality. In Section~\ref{sec:twisted_symmetric}, we saw that we can weaken the co-commutativity requirement and instead introduce a braiding such that the co-commutativity only holds up to conjugation with that braiding. The commutor of twisted-symmetric tensor types then also involves not only the ground type commutor but also applying the braiding. We can also weaken the co-associativity and co-unitality requirements. That is, we extend the Hopf-algebra liquid by ``associator'' and ``unitor'' elements (not to be confused with the associator and unitor 2-functions of tensor types or monoidal categories), and demand that co-associativity and co-unitality only need to hold up to conjugation with those elements. Those liquids are well-known in algebra under the name \emph{(triangular) quasi-Hopf algebras}. The associator 2-function of the resulting tensor type then involves contraction of the corresponding indices with the associator element, via the corresponding representations, and analogously for the unitor.

We have already seen that twisted-symmetric tensors elegantly combine fermions and ordinary symmetries. It will be interesting to look for simple examples of triangular quasi-Hopf algebras and to look for what kind of physics they describe.

\section{Anti-unitary symmetries}
Sometimes, symmetry representations in quantum physics are allowed to be anti-linear operators on a complex Hilbert space, instead of linear ones. This is often referred to as the presence of a \emph{time-reversal symmetry}. For the case of spin systems described by ordinary array tensors, on-site anti-unitaries that square to the identity operator become entry-wise complex conjugation after a basis change. Thus, such models can be described by real array tensors. However, if the anti-unitary does not square to the identity, such a symmetry will correspond to a new tensor type. This is for example the case for spin-full fermions, where a time-reversal symmetry is defined to square to $(-1)^{P_f}$ where $P_f$ is the fermion parity. It will be interesting to see how symmetries involving time-reversal can be formulated as tensor types.

\section{Graded symmetries}
For tensor types to be practically useful and directly implementable on a computer, we want them to be as ``skeletal'' as possible. That is, we want the 0-data to be as simple as possible, and the 1-data to contain all the actual data. Ideally, the 0-data form a discrete set, and any two different 0-data are inequivalent. In this section we sketch how to arrive at a skeletal version of (twisted) Hopf-symmetric tensors whose ground type are real/complex array tensors. In that case, the 0-data (namely, representations) form a continuous family and not a discrete set, and many of them are equivalent, i.e., isomorphic. The skeletal formulation does not only come with discrete 0-data, but also corresponds to choosing a more efficient parametrization of the 1-data, instead of taking the ground type 1-data and imposing an ad hoc symmetry constraint on it. So whereas the Hopf-symmetric formulation is sufficient to formalize symmetries in physics, the skeletal formulation has concrete practical usage, e.g., to speed up concrete numerical calculations.

For array tensors, direct sums of representations are representations, and by construction, any representation is isomorphic to a direct sum of indecomposible ones. Thus, to construct the skeletal formulation, we need to classify all isomorphism classes of indecomposible representations. That means, we want an efficient procedure that constructs representatives of each isomorphism class from some combinatorial information, which is clearly the difficult part of constructing the skeleton. Then, the 0-data $a$ is given by assigning a multiplicity $a_i$ to each indecomposible representation $i$, such that the sum of all multiplicities if finite. The situation is much simpler if the (Hopf) algebra in question is finite-dimensional semisimple. That is, it is isomorphic to a direct sum of irreducible algebras, and the indecomposible representations correspond to those irreducible components.

Note that in general it can happen that there are different direct sums of indecomposible representations yielding isomorphic representations. In this case, the procedure described makes the tensor type only ``more skeletal'' and not completely skeletal. This doesn't matter from a practical point of view though, and it also doesn't occur in the semi-simple case.

The next step consists in computing/choosing the ``fusion ring'' for the indecomposible representations. That is, for all isomorphism classes $i$, $j$ and $k$, there is an integer $N_{ij}^k$ and an isomorphism
\begin{equation}
I_{i,j}: i\otimes j \rightarrow \bigoplus_k N_{ij}^k k\;,
\end{equation}
where $N_{ij}^k k$ means that the direct sum contains the representation $k$ $N_{ij}^k$ times. The isomorphisms $I$ are known as $3j$-symbols. The numbers $N$ determine the product of 0-data,
\begin{equation}
(a\otimes b)_k = \sum_{i,j} N_{ij}^k a_i b_j\;.
\end{equation}
$I$ can be extended to decomposible representations,
\begin{equation}
I_{a,b}: a\otimes b \rightarrow \bigoplus_{i,j} N_{ij}^k a_i b_j k\;.
\end{equation}
The co-unit of the Hopf algebra is a representation $1$ which defines the unit 0-data,
\begin{equation}
1_i=
\begin{cases}
1 &\text{if}\ i=1\\
0 &\text{otherwise}
\end{cases}\;.
\end{equation}
Be aware that the first $1$ is the unit 0-data, the second $1$ is the integer $1$, and the third $1$ the co-unit representation. The 1-data for some 0-data $a$ is simply a vector whose dimension is the multiplicity of the co-unit in $a$.

In the next step we note that we have canonical isomorphisms between representations such as
\begin{equation}
C_{\alpha 0}: a\otimes (b\otimes c) \rightarrow (a\otimes b)\otimes c\;,
\end{equation}
or
\begin{equation}
C_\sigma :a\otimes (b\otimes c) \rightarrow a\otimes (c\otimes b)\;.
\end{equation}
If the Hopf algebra is not co-commutative (but triangular), $C_\sigma$ is not so canonical anymore but involves the braiding. The analogous is true for $C_{\alpha 0}$ if we do not have co-associative but only a quasi-Hopf algebra. Now, the invertible 2-functions are given by those canonical isomorphisms conjugated with the isomorphisms $I$ above. E.g., the associator is given by
\begin{equation}
\alpha_0^{a,b,c} = (I_{a,b}\otimes \idop) I_{a\otimes b, c} C_{\alpha 0} (I_{a,b\otimes c})^{-1} (\idop \otimes (I_{b,c})^{-1})\;.
\end{equation}
Note that the so-defined $\alpha_0$ acts on the vectorspace of the complete representation
\begin{equation}
\bigoplus_{x,y} N_{ij}^x N_{xk}^y a_i b_j c_k y\;,
\end{equation}
whereas the 1-data is only defined within that vector space restricted to $y=1$. As $\alpha_0$ preserves that subspace, we can simply restrict it.

The tensor product of 1-data $A\in \dat_1(a)$ and $B\in \dat_1(b)$ is simply the tensor product of array tensors combined with $I_{a,b}$. The contraction is given by twice applying $I$ (like in the definition of $\alpha_0$ above with $b=c$) and then performing the contraction of array tensors.

We would like to remark that instead of starting with Hopf-symmetric tensors, we can also directly look at what axioms the fusion coefficients $N$ and $3j$-symbols $I$ have to fulfill such that they imply the 2-axioms of tensor types. This way it is possible to obtain new tensor types that do not come from skeletonization of Hopf algebras. It seems to be the case that any example with a finite number of isomorphism classes of indecomposible representations comes from some suitable generalization of Hopf algebras (such as quasi- or triangular Hopf algebras). However, the case of infinite symmetry groups is not technically covered by Hopf-symmetric tensors, as the corresponding group algebra would be infinite-dimensional and thus cannot be formalized as an array tensor liquid model. This contains a type of symmetry groups that is very important in physics, namely Lie groups such as $SU(2)$ or $U(1)$.

\subsection*{Acknowledgments}
We would like to thank Alexander Jahn, Marek Gluza, Juani Bermejo-Vega, Ryan Sweke, and Markus Kesselring for helpful discussions, and Frederik Hahn for proofreading parts of the paper. AB thanks EI 519/15-1 for support.

\bibliography{tensor_types_refs}{}
\bibliographystyle{habbrv}

\end{document}